\documentclass[showpacs,preprintnumbers,amsmath,amssymb]{revtex4}

\usepackage{epsfig}
\usepackage{graphicx}
\usepackage{dcolumn}
\usepackage{bm}
\usepackage{epstopdf}
\usepackage{subfigure}
\DeclareMathOperator{\arccot}{arccot}

\newcommand{\EQ}{\begin{equation}}
\newcommand{\EN}{\end{equation}}
\newcommand{\ea}{\end{eqnarray}}
\newcommand{\ba}{\begin{eqnarray}}

\newcommand{\bear}{\begin{eqnarray}}
\newcommand{\ear}{\end{eqnarray}}

\begin{document}


\title{1D Hubbard model elementary objects scattering}
\author{J. M. P. Carmelo$^{\rm a,b,c}$ and P. D. Sacramento$^{\rm b,d}$}
\affiliation{$^{\rm a}$Center of Physics, University of Minho, Campus Gualtar, P-4710-057 Braga, Portugal}
\affiliation{$^{\rm b}$Beijing Computational Science Research Center, Beijing 100084, China}
\affiliation{$^{\rm c}$Institut f\"ur Theoretische Physik III, Universit\"at Stuttgart, D-70550 Stuttgart, Germany}
\affiliation{$^{\rm d}$CFIF, Instituto Superior T\'ecnico, TU Lisbon, P-1049-001 Lisboa, Portugal}

\date{30 August 2013}


\begin{abstract}
In terms of electron processes, the 1D Hubbard model is a nonperturbative problem. 
That renders the description in terms of electron scattering of the microscopic processes that control the model 
properties a very difficult task. In this paper we study the corresponding scattering processes of the
elementary objects whose occupancy configurations generate the energy eigenstates from the electron
vacuum. Due to the related occurrence of an infinite set of conservation laws associated with the model
integrability, such objects are found to undergo only zero-momentum forward-scattering collisions.
The description of the model dynamical properties in terms of such elementary objects scattering
events then drastically simplifies. The corresponding 1D Hubbard model scattering theory refers to 
arbitrary values of the densities and finite repulsive interaction $U>0$. 
Each ground-state - excited-state transition is associated with 
a well defined set of elementary zero-momentum forward-scattering events. 
The elementary-object scatterers dressed $S$ matrix is expressed as a {\it commutative} product
of $S$ matrices, each corresponding to a two-object scattering event.
This commutative factorization is stronger than the factorization associated with
Yang-Baxter equation for the original spin-$1/2$ electron bare $S$ matrix. 
The power-law singularities exponents in the finite-energy correlation-functions 
of the metallic phases of a wide class of 1D integrable and non-integrable systems 
are momentum dependent. In the present exactly solvable model such an exponent
momentum dependence is controlled by the phase shifts and
corresponding dressed $S$ matrix considered in this paper. 
\end{abstract}


\maketitle
\section{Introduction}
\label{Introduction}

On a one-dimensional (1D) lattice, the Hubbard model \cite{Gutzwiller,Hubbard} 
for correlated electrons with effective on-site repulsion $U$ and nearest-neighbor transfer integral $t$ 
is a non-perturbative many-electron problem. Fortunately, it is solvable by Bethe-ansatz (BA) \cite{Lieb,Takahashi,Woy,Martins}.
In the low-energy limit, the model behavior is described by a two-component Luttinger-Tomonaga-liquid theory
\cite{Schulz,Voit,Lederer}. Within that limit, the BA solution may be combined with bosonization \cite{Schulz,Voit,Lederer} or 
the model conformal invariance \cite{Woy-89,Frahm} to evaluate the asymptotics 
of correlation functions. 

A pseudofermion dynamical theory (PDT) \cite{V,VI,LE,TTF} for the 1D Hubbard model,
which is a generalization to arbitrary $U/t>0$ values of the $U/t\gg 1$
method of Refs. \cite{Penc-96,Penc-97}, has been used to calculate finite-energy spectral and correlation functions.
More recently, alternative methods valid for both integrable and non-integrable 1D correlated problems have
reached similar results \cite{Glazman,Glazman-09,Glazman-10,Glazman-12,Affleck,Affleck-09,Essler-10,DSF-n1}.
(While here {\it finite energy} means an energy larger than the typical linear Luttinger-Tomonaga-liquid theory
excitation energy, in this paper finite energy often refers to both low energy and high energy in the
sense that there are no restrictions to its values.)

In the case of the 1D Hubbard model, all the above methods rely in part on the combination of the
BA solution with suitable invariances, symmetries, and/or numerical computations. 
Which microscopic elementary processes are behind both the low-energy and
finite-energy spectral and correlation function behaviors obtained by such techniques is
an issue that is not well understood and thus deserves further investigations. It is an extremely complex
issue in terms of the non-perturbative many-electron microscopic processes. On the other hand, it is shown in this
paper that it drastically simplifies in terms of microscopic processes of the elementary objects
whose occupancy configurations generate exact energy eigenstates from the electron vacuum.
The operator algebra associated with the pseudoparticle representation of Refs. \cite{I,II} 
has in a recent paper, Ref. \cite{paper-I}, been uniquely related to the electron creation and annihilation operators. The ``bridge'' between the electrons
and the elementary objects of such a representation has been found to refer to rotated electrons. Those are generated
from the electrons by a unitary transformation performed by the BA solution. The studies of this paper focus on 
the properties of the related scatterers and scattering centers, which are some of the elementary objects 
of the rotated-electron related representation of Refs. \cite{I,II,paper-I}.  We find that such a 
representation leads to a uniquely defined choice of scattering states basis. For $U/t>0$ 
the states of that basis are found to be in one-to-one correspondence to the excited energy eigenstates of ground 
states with arbitrary values of the electronic density $n$ and spin density $m$. 

The present studies are an extension of the preliminary results on the PDT related pseudofermion 
scattering mechanisms presented in short form in Ref. \cite{PST-05}. That preliminary study has 
identified correctly the pseudofermion dressed phase shifts and $S$ matrices that control the PDT 
one- and two-electron spectral-weight distributions \cite{V,VI,LE,TTF}. However, it lacked many aspects of the pseudofermion 
scattering theory introduced in this paper needed both to justify 
its validity and to clarify the physics behind it. It did not include figures illustrating the
dependence of the important two-pseudofermion phase shifts that control the PDT spectral weight distributions 
on the scatterer and scattering center momenta and on-site repulsion ratio $U/t$.
Furthermore, the relation of the theory scatterers and scattering centers to the rotated electrons 
that are generated from the electrons by a unitary transformation performed by the BA solution 
was not discussed. Hence the effects of the pseudofermion transformation laws under the 
electron - rotated-electron unitary transformation were not investigated.
The operator that counts the number of rotated electrons singly
occupied sites is for $U/t>0$ the generator of the $c$ hidden $U(1)$ symmetry in the
model global $[SU(2)\times SU(2)\times U(1)]/Z_2^2$ symmetry more recently
found in Ref. \cite{bipartite}. The preliminary studies of Ref. \cite{PST-05} have not
accounted for the interplay of that symmetry with the model scattering properties.
Moreover, important issues such as a clear definition of the one-pseudofermion 
``in" and ``out" asymptote pseudofermion scattering states was lacking. The relation of the latter states 
to the corresponding many-pseudofermion ``in" and ``out" states was neither given. 
In this paper all such issues are addressed and clarified.

From the relation of the PDT to the scattering theory studied in the following, one
confirms that the corresponding exotic scattering centers of the 1D Hubbard model 
are observed by angle-resolved photoelectron spectroscopy in quasi-1D metals \cite{spectral0,spectral,spectral-06}. 
The model or its extensions describe as well the effects of correlations in semiconductor - metal
transitions of doped quasi-1D materials \cite{polyace} and are of interest for systems of ultra-cold 
atoms on 1D optical lattices \cite{Zoller}.

In Ref. \cite{paper-I} it is shown that 1D Hubbard model elementary-object representations other than those
used in the studies of this paper refer to alternative sets of degenerate energy
eigenstates that span well-defined model reduced subspaces. This applies
for instance to the holon and spinon representations of Refs. \cite{Natan,S-0,S}.
The corresponding  holon and spinon scattering theories of Ref. \cite{Natan} 
and Refs. \cite{S-0,S}, respectively, are actually different. However, both such theories use
the ground state whose excited states span the subspaces wherein such
theories are defined as the holon and spinon vacuum. Alike in the elementary-object scattering theory
studied in this paper, for such holon and spinon scattering theories 
the elementary objects created under the transitions to the excited states play the
role of scattering centers. Consistent with the ground state playing the role of
vacuum of the latter theories elementary objects, the scatterers are as well holons and 
spinons created under such transitions. Hence the holon and spinon scattering theories
only account for the phase shifts of such scatterers. As a result, the theories of Ref. \cite{Natan} 
and Refs. \cite{S-0,S} do not account
for most phase shifts found in this paper to control the one- and two-electron spectral-weight
distributions. Indeed most of the latter phase shifts are of scatterers
that pre-existed in the ground state of the excited state under consideration.

The relation of the elementary-object representations used in the studies of this paper to
the holons and spinons of Refs. \cite{Natan,S-0,S} is an issue that has been clarified \cite{paper-I}. 
That relation is consistent with the major advantage of the scattering theory considered in this paper 
relative to the holon and spinon scattering theories referring indeed to the explicit description
of the microscopic processes that control the model dynamical and spectral properties. Such an advantage
follows from the occupancy configurations of the elementary objects of the former theory generating
the excited energy eigenstate from the electron vacuum rather then from a ground state. Therefore,
the pseudofermion scattering theory accounts for both the phase shifts of the scatterers that pre-exist in the ground
state and those that are created under the transitions from it to the excited states. Furthermore, the pseudofermion scattering theory refers to excited
states of ground states with arbitrary electronic density $n$ and spin density $m$. On the other hand, the scattering
theory of Ref. \cite{Natan} and that of Refs. \cite{S-0,S} applies to densities $n\in [0,1];m=0$ and $n=1;m=0$,
respectively. 

Due to the nonperturbative character of the many-electron problem, the microscopic elementary processes 
studied in this paper are in terms of electron scattering very involved. 
The simplicity of the elementary-objects scattering events studied in this paper stems both from
the occupancy configurations of such elementary objects generating exact energy eigenstates and from the related occurrence of an infinite 
number of conservation laws \cite{CM-86,CM,Prosen}. Those are associated with the 1D Hubbard model integrability 
\cite{Lieb,Takahashi,Woy,Martins}. The elementary-object representation used in our investigations \cite{I,II,paper-I} naturally emerges 
from the interplay of the model global $[SU(2)\times SU(2)\times U(1)]/Z_2^2$ symmetry \cite{bipartite} with its exact BA solution \cite{paper-I}. 
Consistent, such conservation laws are explicit in their scattering theory, in that the scatterers and scattering centers 
are only allowed to undergo zero-momentum forward scattering events. 

The pseudofermion scattering theory goes beyond the BA solution in that it accounts for both excited
states inside and outside that solution subspace, which is spanned by the Bethe states. For Bethe states
it is meant the energy eigenstates inside the BA solution subspace. Those 
can be chosen to be either lowest-weight states (LWSs) or highest-weight states (HWSs) of the $\eta$-spin 
and spin $SU(2)$ algebras algebras \cite{Completeness} in the
model global $[SU(2)\times SU(2)\times U(1)]/Z_2^2$ symmetry \cite{bipartite}. 
The $\eta$-spin (and spin) and $\eta$-spin projection (and spin projection) of the energy
eigenstates are denoted by $S_{\eta}$ and $S_{\eta}^{x_3}$ (and $S_s$ and
$S_s^{x_3}$), respectively. The $S_{\alpha}$ and $S_{\alpha}^{x_3}$ values of the
LWSs and HWSs are such that $S_{\alpha}= -S_{\alpha}^{x_3}$ and $S_{\alpha}=S_{\alpha}^{x_3}$, respectively.
Here $\alpha =\eta$ for $\eta$-spin and $\alpha =s$ for spin. In this
paper we use the BA solution LWS representation.

The paper is organized as follows: The model and the elementary objects arising from the
rotated electrons as defined in Refs. \cite{I,paper-I} are the topics addressed in Section \ref{model-ele-obj}.
In Section \ref{Pseudofermion} it is found that a pseudofermion scattering theory 
naturally emerges from such elementary objects by means of a unitary transformation. It slightly shifts the
pseudoparticle discrete momentum values, which renders the corresponding pseudofermion spectrum 
without energy interaction terms. The corresponding pseudofermion dressed
phase shifts are studied in Section \ref{Phaseshifts}. The effects of the pseudofermion transformation laws 
on their scattering properties and the relation of the theory dressed phase shifts
to the spectral weights of the PDT are issues also addressed in that section.
Finally, the concluding remarks are presented in Section \ref{Concluding}.
Complementary useful information is given in four appendices:
Appendix \ref{pseudo-repre} provides further information on the related pseudoparticle representation. The integral
equations that define the two-pseudofermion phase shifts are given in Appendix \ref{Ele2PsPhaShi}.
In that Appendix some other phase-shift related issues are also addressed.
Appendix \ref{Consequences} discusses the extension of the pseudofemion
scattering theory to excited states of ground states with densities $n=1$ and/or $m=0$. 
Finally, in Appendix \ref{energies} information of the interaction and densities dependence of 
energy scales useful for the studies of this paper is reported.

\section{The model and the elementary objects emerging from the electron - rotated-electron unitary transformation}
\label{model-ele-obj}

The Hubbard model Hamiltonian, under periodic boundary conditions, on a 1D lattice with a site number 
$N_a\gg 1$ very large and even and in a chemical potential $\mu $ and magnetic field $H$ is given by,
\begin{equation}
\hat{H}={\hat{H}}_{symm} - \sum_{\beta =\eta,s}\mu_{\alpha}\, {\hat{S}}_{\alpha}^{x_3} \, ,
\label{H}
\end{equation}
where
\begin{eqnarray}
{\hat{H}}_{symm} & = & t\,[\hat{T}+4u\,\hat{V}_D]   \, ;
\hspace{0.35cm} u = U/4t \, ,
\hspace{0.25cm} {\hat{S }}_{\eta}^{x_3}=-{1\over
2}[N_a-\hat{N}]  \, ; \hspace{0.35cm} 
{\hat{S }}_s^{x_3}= -{1\over 2}[{\hat{N}}_{\uparrow}-{\hat{N}}_{\downarrow}] \, ,
\nonumber \\
\hat{T} & = & -\sum_{\sigma=\uparrow
,\downarrow }\sum_{j=1}^{N_a}\Bigl[c_{j,\sigma}^{\dag}\,c_{j+1,\sigma} + c_{j+1,\sigma}^{\dag}\,c_{j,\sigma}\Bigr]  \, ;
\hspace{0.35cm}
\hat{V}_D = \sum_{j=1}^{N_a}\left(\hat{n}_{j,\uparrow} -1/2\right)
\left(\hat{n}_{j,\downarrow}-1/2\right) \, .
\label{HH}
\end{eqnarray}
Here $\hat{T}$ is the kinetic-energy operator in units of $t$, $\hat{V}_D$ 
the electron on-site repulsion operator in units of $U$, $u=U/4t$ is the electron on-site interaction
in units of $4t$, which is often used in this paper, 
$\mu_c=2\mu$, $\mu_s=2\mu_B H$, $\mu_B$ is the Bohr magneton, and 
${\hat{S }}_{\eta}^{x_3}$ and ${\hat{S }}_s^{x_3}$ are the diagonal generators of the 
$\eta$-spin and spin $SU(2)$ symmetry algebras \cite{HL,Lieb89,Yang}, respectively.
The operator $c_{j,\sigma}^{\dagger}$ (and $c_{j,\sigma}$) that appears in the above equations
creates (and annihilates) a spin-projection $\sigma $ electron at lattice site
$j=1,...,N_a$. The operator $\hat{n}_{j,\sigma} = c_{j,\sigma }^{\dagger }\,c_{j,\sigma }$ counts the number of
spin-projection $\sigma$ electrons at such a lattice site. The electronic number operators read
${\hat{N}}=\sum_{\sigma=\uparrow ,\downarrow }\,\hat{N}_{\sigma}$ and
${\hat{N}}_{\sigma}=\sum_{j=1}^{N_a}\hat{n}_{j,\sigma}$.
The momentum operator is given by $\hat{P} = \sum_{\sigma=\uparrow
,\downarrow }\sum_{k}\, \hat{n}_{\sigma} (k)\, k$, where the spin-projection $\sigma$
momentum distribution operator reads $\hat{n}_{\sigma} (k) = c_{k,\sigma }^{\dagger
}\,c_{k,\sigma }$ and the operator $c_{k,\sigma}^{\dagger}$ (and $c_{k,\sigma}$)
creates (and annihilates) a spin-projection $\sigma $ electron of momentum $k$.

Throughout this paper we use in general units of both Planck constant $\hbar$ and lattice constant
$a$ one. We denote the lattice length by $L=N_a\,a=N_a$.
The LWSs have electronic densities $n=N/L$ and spin densities $m=[N_{\uparrow}-N_{\downarrow}]/L$ 
whose ranges obey the inequalities $0\leq n \leq 1$ and $0\leq m \leq n$, respectively. The description 
of the states corresponding to densities such that $0\leq n \leq 1$\, ;
$1\leq n \leq 2$ and $-n\leq m \leq n$\, ; $-(2-n)\leq m \leq (2-n)$, respectively, is
achieved by application onto the LWSs of off-diagonal generators of the
$\eta$-spin and spin $SU(2)$ symmetry algebras. 

The two global $SU(2)$ symmetries of the Hubbard model on a bipartite lattice, including the present 1D lattice, 
have been known for a long time \cite{HL,Lieb89}. The studies of Ref. \cite{Yang} revealed that 
the model global symmetry was at least $SO(4) =[SU(2)\otimes SU(2)]/Z_2$. The recent investigations of Ref. 
\cite{bipartite} found that for finite on-site interaction values it is larger and given by $[SO(4)\otimes U(1)]/Z_2=SO(3)\otimes SO(3)\otimes U(1)$.
That global symmetry may be rewritten as $[SU(2)\times SU(2)\times U(1)]/Z_2^2$. It stems from the
local gauge $SU(2)\times SU(2) \times U(1)$ symmetry of the Hubbard model on a bipartite lattice with vanishing 
transfer integral, $t=0$ \cite{U(1)-NL}. For finite $U$ and $t$ values the latter local symmetry becomes a group of permissible unitary transformations. 
The corresponding local $U(1)$ canonical transformation is not the ordinary gauge $U(1)$ subgroup of electromagnetism. 
It is rather a ``nonlinear" transformation \cite{U(1)-NL}. 

The BA solution accounts for the quantum number occupancy configurations that generate the representations
of the $c$ hidden $U(1)$ symmetry algebra in $[SU(2)\otimes SU(2)\otimes U(1)]/Z_2^2=[SO(4)\otimes U(1)]/Z_2$
beyond $SO(4)$ \cite{paper-I}. The energy eigenstates outside the BA solution subspace 
have exactly the same $c$ hidden $U(1)$ symmetry algebra representations as the Bethe states from which
they are generated by the off-diagonal $\eta$-spin and spin operator algebras.
This is why there is no contradiction whatsoever between the global
$[SU(2)\otimes SU(2)\otimes U(1)]/Z_2^2=[SO(4)\otimes U(1)]/Z_2$ symmetry 
found in Ref. \cite{bipartite} for the Hubbard model on any bipartite lattice and
the results of Ref. \cite{Completeness}, concerning the counting of the $4^{N_a}$ 
energy eigenstates of the Hubbard model on the bipartite 1D lattice. 
The studies of that reference have not explicitly considered
the model $c$ hidden $U(1)$ symmetry algebra beyond $SO(4)$, yet have used
the BA solution, which includes the quantum-number occupancy configurations that 
generate the representations of the $c$ hidden $U(1)$ symmetry algebra \cite{paper-I}. 

In the following we shortly report how the elementary objects used in our studies
emerge from the electrons. Here and in in Appendix \ref{pseudo-repre}, $\vert l_{\rm r},l_{\eta s},u\rangle$ denotes 
the model energy eigenstates whose LWSs of both $SU(2)$ symmetry algebras are the Bethe states,
$\vert l_{\rm r},l_{\eta s}^0,u\rangle$. General non-LWS energy eigenstates are
generated from the Bethe states as given in Eq. (\ref{Gstate-BAstate}) of that Appendix.
The state labels $l_{\eta s}$ and $l_{\rm r}$ stand for the set of numbers $[S_{\eta},S_{s},M_{\eta,-1/2}^{un},M_{s,-1/2}^{un}]$ and
all remaining quantum numbers, respectively, needed to uniquely define an energy eigenstate
whereas the Bethe-state label $l_{\eta s}^0$ refers to the specific $l_{\eta s}$ values $[S_{\eta},S_{s},0,0]$.

Within the $u\rightarrow\infty$ limit, the energy eigenstates $\vert l_{\rm r},l_{\eta s},u\rangle$
adiabatically correspond to states $\vert l_{\rm r},l_{\eta s},\infty \rangle$,
which refer to one of the many choices of $u\rightarrow\infty$ energy eigenstates. In this paper we consider 
such $u\rightarrow\infty$ energy eigenstates. They can be generated from the electron vacuum $\vert 0_{\rm elec}\rangle$
by application onto it of a uniquely defined operator $\hat{G}^{\dag}$, {\it i.e.}
$\vert l_{\rm r},l_{\eta s},\infty \rangle=\hat{G}^{\dag}\,\vert 0_{\rm elec}\rangle$. The complete set of $4^{N_a}$ energy 
eigenstates, $\{\vert l_{\rm r},l_{\eta s},\infty \rangle\}$, uniquely defines 
an electron - rotated-electron unitary operator $\hat{V}=\hat{V}(u)$ such that the states 
$\vert l_{\rm r},l_{\eta s},u\rangle ={\hat{V}}^{\dag}\vert  l_{\rm r},l_{\eta s},\infty \rangle$ are energy eigenstates 
for $u>0$. That unitary operator is constructed inherently to the rotated-electron occupancy configurations that generate the corresponding
$4^{N_a}$ energy eigenstates from the electron vacuum, $\vert  l_{\rm r},l_{\eta s},u \rangle=\tilde{G}^{\dag}\,\vert 0_{\rm elec}\rangle$,
being exactly the same as the corresponding electron occupancy configurations that 
generate the state $\vert l_{\rm r},l_{\eta s},\infty \rangle$ from that same vacuum. Hence
the generator $\tilde{G}^{\dag}={\hat{V}}^{\dag}\,\hat{G}^{\dag}\,{\hat{V}}$ has the same expression in terms
of the rotated-electron operators,
\begin{equation}
{\tilde{c}}_{j,\sigma}^{\dag} =
{\hat{V}}^{\dag}\,c_{j,\sigma}^{\dag}\,{\hat{V}}
\, ; \hspace{0.50cm}
{\tilde{c}}_{j,\sigma} =
{\hat{V}}^{\dag}\,c_{j,\sigma}\,{\hat{V}} 
\, ; \hspace{0.50cm}
{\tilde{n}}_{j,\sigma} = {\tilde{c}}_{j,\sigma}^{\dag}\,{\tilde{c}}_{j,\sigma} \, ,
\label{rotated-operators}
\end{equation} 
as $\hat{G}^{\dag}$ in terms of electron creation and annihilation operators.
Such a unitary operator ${\hat{V}}$ is that associated with the electron - rotated-electron
unitary transformation performed by the BA solution. (It is uniquely defined in Ref. \cite{paper-I} in terms
of its $4^{N_a}\times 4^{N_a}$ matrix elements between energy eigenstates.)
The rotated electrons have been contracted inherently to rotated-electron single and double occupancies
being good quantum numbers for $u>0$. 

The elementary objects considered in Ref. \cite{I,II} naturally emerge from the 
degrees of freedom separation of the rotated-electron occupancy configurations
that generate the energy eigenstates from the electron vacuum \cite{paper-I}. Such
degrees of freedom correspond to the state representations of the $c$ hidden
$U(1)$ symmetry, $\eta$-spin $SU(2)$ symmetry, and spin $SU(2)$ symmetry algebras
in the model global $[SU(2)\otimes SU(2)\otimes U(1)]/Z_2^2$ symmetry algebra. 
The first step of such elementary objects emergence is the electron - rotated-electron unitary transformation
performed by the BA, which is directly related to that symmetry. Indeed, the
generator of the global $c$ hidden $U(1)$ symmetry beyond $SO(4)$ is the number 
of rotated-electron singly occupied sites operator \cite{paper-I,bipartite},
\begin{equation}
2{\tilde{S}}_c \equiv {\hat{V}}^{\dag}\,2{\hat{S}}_c\,{\hat{V}} 
= \sum_{j=1}^{N_a}{\tilde{s}}_{j,c} \, ; \hspace{0.5cm}
{\tilde{s}}_{j,c} = \sum_{\sigma =\uparrow
,\downarrow}\,{\tilde{n}}_{j,\sigma}\,(1- {\tilde{n}}_{j,-\sigma}) \, .
\label{Or-ope-Q}
\end{equation}
Here $2{\hat{S}}_c$ and the related operator ${\hat{D}}$ count the number of electron singly-occupied sites and
doubly-occupied sites and read,
\begin{equation}
2{\hat{S}}_c = \sum_{j=1}^{N_a}{\hat{s}}_{j,c} \ ;
\hspace{0.50cm} {\hat{D}}= ({\hat{N}}-2{\hat{S}}_c)/2 \ ;
\hspace{0.50cm} 
{\hat{s}}_{j,c}= \sum_{\sigma =\uparrow,\downarrow}\,{\hat{n}}_{j,\sigma}\,(1- {\hat{n}}_{j,-\sigma}) \, ,
\label{DQ}
\end{equation}
respectively. The eigenvalues $2S_c=0,1,...$ of the operator $2{\tilde{S}}_c$, Eq. (\ref{Or-ope-Q}), 
are thus the number of rotated-electron singly occupied sites \cite{bipartite}.

It is convenient to express the local generator ${\hat{s}}_{j,c}$ of the Hamiltonian 
electron-interaction term gauge $U(1)$ symmetry in $SU(2)\otimes SU(2) \otimes U(1)$,
which is given in Eq. (\ref{Or-ope-Q}), and the alternative local generator 
${\hat{s}}_{j,c}^h = (1 - {\hat{s}}_{j,c})$, as follows,
\begin{equation}
{\hat{s}}_{j,c} ={\hat{n}}_{j,c} = {\hat{f}}_{j,c}^{\dag}\,{\hat{f}}_{j,c} 
\, ; \hspace{0.5cm}
{\hat{s}}_{j,c}^h = (1-{\hat{n}}_{j,c}) = {\hat{f}}_{j,c}\,{\hat{f}}_{j,c}^{\dag} \, , \hspace{0.25cm} j = 1,...,N_a \, .
\label{n-r-c}
\end{equation}
Here ${\hat{f}}_{j,c}^{\dag}$ and ${\hat{f}}_{j,c}$ stand for the following creation and annihilation
operators, respectively, which obey a fermionic algebra, 
\begin{eqnarray}
{\hat{f}}_{j,c}^{\dag} & = & c_{j,\uparrow}^{\dag}\,(1-{\hat{n}}_{j,\downarrow})
+ (-1)^j\,c_{j,\uparrow}\,{\hat{n}}_{j,\downarrow} \, ,
\nonumber \\
{\hat{f}}_{j,c} & = & c_{j,\uparrow}\,(1-{\hat{n}}_{j,\downarrow})
+ (-1)^j\,c_{j,\uparrow}^{\dag}\,{\hat{n}}_{j,\downarrow} \, , \hspace{0.25cm} j = 1,...,N_a \, .
\label{fc-unrot-0}
\end{eqnarray}

In terms of the operators obtained from the electron rotation of those given in Eq. (\ref{fc-unrot-0}), 
\begin{eqnarray}
f_{j,c}^{\dag} & = & {\hat{V}}^{\dag}\,{\hat{f}}_{j,c}^{\dag}\,{\hat{V}} = {\tilde{c}}_{j,\uparrow}^{\dag}\,
(1-{\tilde{n}}_{j,\downarrow}) + (-1)^j\,{\tilde{c}}_{j,\uparrow}\,{\tilde{n}}_{j,\downarrow} \, ,
\nonumber \\
f_{j,c} & = & {\hat{V}}^{\dag}\,{\hat{f}}_{j,c}\,{\hat{V}} = {\tilde{c}}_{j,\uparrow}\,
(1-{\tilde{n}}_{j,\downarrow}) + (-1)^j\,{\tilde{c}}_{j,\uparrow}^{\dag}\,{\tilde{n}}_{j,\downarrow} 
\, , \hspace{0.25cm} j = 1,...,N_a \, ,
\label{fc+}
\end{eqnarray}
the two alternative generators of the global $c$ hidden $U(1)$ symmetry 
have the following very simple expressions,
\begin{equation}
2{\tilde{S}}_c = \sum_{j=1}^{N_a}n_{j,c} = \sum_{j=1}^{N_a} f_{j,c}^{\dag}\,f_{j,c} 
\, ; \hspace{0.5cm}
2{\tilde{S}}_c^h = \sum_{j=1}^{N_a}(1-n_{j,c}) = \sum_{j=1}^{N_a} f_{j,c}\,f_{j,c}^{\dag} \, .
\label{Or-ope-cf}
\end{equation}
Here,
\begin{equation}
n_{j,c} = f_{j,c}^{\dag}\,f_{j,c} \, .
\label{n-r-c-rot}
\end{equation}
Hence the operators, Eq. (\ref{fc+}), have been obtained from electron rotation of those given in Eq. (\ref{fc-unrot-0})
inherently to be associated with the $c$ hidden $U(1)$ symmetry degrees of freedom of the rotated electrons. 
In this paper we use in general a notation within which ${\tilde{O}}$ stands for the operator that
in terms of rotated-electron creation and annihilation operators, Eq. (\ref{rotated-operators}), has exactly
the same expression as a given operator ${\hat{O}}$ in terms of electron creation and annihilation 
operators, respectively. However, for simplicity, no upper index is used onto 
the electron-rotated $c$ pseudoparticle operators $f_{j,c}^{\dag}$ and $f_{j,c}$, 
Eq. (\ref{fc+}), and $n_{j,c}$, Eq. (\ref{n-r-c-rot}).

One may introduce corresponding momentum-dependent operators,
\begin{equation}
f_{q_j,c}^{\dag} = (f_{q_j,c})^{\dag} = {1\over{\sqrt{N_a}}}\sum_{j'=1}^{N_a}\,e^{+iq_j j'}\,
f_{j',c}^{\dag} \, , \hspace{0.25cm} j = 1,...,N_a \, .
\label{fc-q-x-1D}
\end{equation}
where the discrete momentum values $q_j = (2\pi/N_a)\,I^c_j$, such that $q_{j+1}-q_j=2\pi/N_a$, are 
defined in Eqs. (\ref{q-j}) and (\ref{Ic-an}) of Appendix \ref{pseudo-repre}.

The unitary operator ${\hat{V}}$ in Eq. (\ref{fc+}) is chosen to be that associated with the electron - rotated-electron 
unitary transformation performed by the BA solution. For that specific choice of unitary operator, 
the operators $f_{q_j,c}^{\dag}$ and $f_{q_j,c}$, Eq. 
(\ref{fc-q-x-1D}), create and annihilate the $c$ pseudoparticles previously considered in Refs. 
\cite{I,II}. In such references the $c$ pseudoparticles emerged from an
empirical association with the BA quantum numbers $q_j = (2\pi/N_a)\,I^c_j$,
Eqs. (\ref{q-j}) and (\ref{Ic-an}) of Appendix \ref{pseudo-repre}. On the other hand, here we use
the operator representation of Refs. \cite{I,paper-I} within which their operators emerge naturally from
the rotated-electron $c$ hidden $U(1)$ symmetry degrees of freedom \cite{paper-I}.

The six generators of the global $\eta$-spin and spin $SU(2)$ symmetry algebras
commute with the electron - rotated-electron unitary operator \cite{paper-I,bipartite}. Hence they have
the same expressions when expressed in terms of electron and rotated-electron, respectively, creation and
annihilation operators, so that the following equality holds,
\begin{equation}
{\hat{S}}_{\alpha}^{l} = {\tilde{S}}_{\alpha}^{l} = \sum_{j=1}^{N_a}{\tilde{s}}_{j,\alpha}^{l}  
\, , \hspace{0.25cm} \alpha =\eta,s \, , \hspace{0.25cm} l =\pm,x_3 \, .
\label{Scs-rot}
\end{equation}
The electron-rotated $\alpha =\eta,s$ local operators ${\tilde{s}}_{j,\alpha}^{l}$ read,
\begin{equation}
{\tilde{s}}^l_{j,\eta} = (1-n_{j,c})\,{\tilde{q}}^l_{j}
\, ; \hspace{0.50cm}
{\tilde{s}}^l_{j,s} = n_{j,c}\,{\tilde{q}}^l_{j} \, , 
\hspace{0.25cm} l =\pm,x_3 \, ,
\label{sir-pir}
\end{equation}
where $n_{j,c}$ is the $c$ pseudoparticle local density operator, Eq. (\ref{n-r-c-rot}). 
Moreover, ${\tilde{q}}^{\pm}_{j}= {\tilde{q}}^{x_1}_{j}\pm i\,{\tilde{q}}^{x_2}_{j}$ and ${\tilde{q}}^{x_3}_{j}$, 
where $x_1,x_2,x_3$ denotes the Cartesian coordinates, are the
following $\eta s$ quasi-spin operators,
\begin{equation}
{\tilde{q}}^-_{j} = 
({\tilde{c}}_{j,\uparrow}^{\dag}
+ (-1)^j\,{\tilde{c}}_{j,\uparrow})\,
{\tilde{c}}_{j,\downarrow} 
\, ; \hspace{0.50cm}
{\tilde{q}}^+_{j} = ({\tilde{q}}^-_{j})^{\dag} \, ;
\hspace{0.50cm}
{\tilde{q}}^{x_3}_{j} = ({\tilde{n}}_{j,\downarrow} - 1/2) \, .
\label{rotated-quasi-spin}
\end{equation}
The six local operators in Eq. (\ref{sir-pir}) are generated by electron rotating the
remaining six generators besides ${\hat{s}}_{j,c}$ of the Hamiltonian electron-interaction term
local gauge $SU(2)\otimes SU(2)\otimes U(1)$ symmetry.

As mentioned above, the local operators $f_{j,c}^{\dag}$ and $f_{j,c}$, Eq. (\ref{fc+}), refer to the
$c$ hidden $U(1)$ symmetry degrees of freedom of the rotated electrons. Similarly, the
three local operators ${\tilde{s}}^l_{j,s}$ and three local operators ${\tilde{s}}^l_{j,\eta}$
where $l =\pm,x_3$, Eqs. (\ref{sir-pir}) and (\ref{rotated-quasi-spin}), correspond to 
the rotated-electron spin $SU(2)$ symmetry and $\eta$-spin $SU(2)$ symmetry degrees of freedom, 
respectively. The latter operators are associated with the spin-$1/2$ spinons
and $\eta$-spin-$1/2$ $\eta$-spinons, respectively, as defined within the elementary-object representation
of Refs. \cite{I,II,paper-I}. Equations (\ref{rotated-operators}), (\ref{fc+}), (\ref{sir-pir}), 
and (\ref{rotated-quasi-spin}) together with the unitary operator ${\hat{V}}$
$4^{N_a}\times 4^{N_a}$ matrix elements between energy eigenstates given in Ref. \cite{paper-I}
define the $c$ pseudoparticle, $\eta$-spinon, and spinons operators in terms of the electron 
creation and annihilation operators.

The $\eta s$ quasi-spin operators ${\tilde{q}}^{l}_{j}$, Eq. (\ref{rotated-quasi-spin}),
may be rewritten as,
\begin{equation}
{\tilde{q}}^l_{j} = {\tilde{s}}^l_{j,s} + {\tilde{s}}^l_{j,\eta} \, ,
\hspace{0.25cm} l=\pm,x_3 \, .
\label{q-oper}
\end{equation}
This reveals that the three local spinon operators ${\tilde{s}}^l_{j,s}$ and the three local $\eta$-spinon operators 
${\tilde{s}}^l_{j,\eta}$ are particular cases of the general $SU(2)$ local $\eta s$ quasi-spin 
operators ${\tilde{q}}^l_{j}$. The latter operators refer to all lattice sites and are associated with a 
general quasi-spin $SU(2)$ symmetry. On the other hand, the three local spinon operators 
${\tilde{s}}^l_{j,s}$ and three local $\eta$-spinon operators ${\tilde{s}}^l_{j,\eta}$
are associated with the spin and $\eta$-spin $SU(2)$ symmetry algebra representations,
respectively. Such two sets of three local operators are defined in two independent sets of sites:
(i) The $2S_c$ spin-up and spin-down rotated-electron singly occupied sites and (ii) the $2S_c^h$ rotated-electron doubly-occupied
and unoccupied sites, respectively. The $c$ pseudoparticle and $c$ pseudoparticle hole local density operators $(1-n_{j,c})$ and
$n_{j,c}$  in the expressions of the operators ${\tilde{s}}^l_{j,\eta}$ and ${\tilde{s}}^l_{j,s}$, Eq. (\ref{sir-pir}), play the role
of projectors onto such two sets of lattice-site rotated-electron occupancies, respectively.

The electron - rotated-electron transformation associated with the operator $\hat{V}$
is unitary. Hence the rotated-electron operators ${\tilde{c}}_{j,\sigma}^{\dag}$ 
and ${\tilde{c}}_{j,\sigma}$, Eq. (\ref{rotated-operators}),
have the same anticommutation relations as the corresponding electron 
operators $c_{j,\sigma}^{\dag}$ and $c_{j,\sigma}$, respectively. It follows that
straightforward manipulations based on Eqs. (\ref{fc+}) and (\ref{rotated-quasi-spin}) lead
to the following algebra for the $c$ pseudoparticle operators,
\begin{equation}
\{f^{\dag}_{j,c}\, ,f_{j',c}\} = \delta_{j,j'} ;
\hspace{0.25cm}
\{f_{j,c}^{\dag}\, ,f_{j',c}^{\dag}\} =
\{f_{j,c}\, ,f_{j',c}\} = 0 \, ,
\label{albegra-cf-j}
\end{equation}
and the $c$ pseudoparticle operators and the local $\eta s$ quasi-spin operators,
\begin{equation}
[f_{j,c}^{\dag}\, ,{\tilde{q}}^l_{j'}] =
[f_{j,c}\, ,{\tilde{q}}^l_{j'}] = [f_{j,c}^{\dag}\, ,{\tilde{s}}^l_{j',\alpha}] =
[f_{j,c}\, ,{\tilde{s}}^l_{j',\alpha}] = 0 \, , \hspace{0.25cm} l=\pm,x_3
\, , \hspace{0.25cm} \alpha =\eta , s \, .
\label{albegra-cf-s-h}
\end{equation}
Alike the local $c$ pseudoparticle operators in Eq. (\ref{albegra-cf-j}), the
corresponding momentum dependent operators, Eq. (\ref{fc-q-x-1D}), obey 
an anticommuting algebra,
\begin{equation}
\{f^{\dag}_{q_j,c}\, ,f_{q_{j'},c}\} = \delta_{j,j'} ;
\hspace{0.25cm}
\{f_{q_j,c}^{\dag}\, ,f_{q_{j'},c}^{\dag}\} =
\{f_{q_j,c}\, ,f_{q_{j'},c}\} = 0 \, .
\label{albegra-cf}
\end{equation}
The first of such relations is valid provided that the two operators in the anticommutators 
act onto subspaces whose BA quantum numbers $I^{c}_j$ in
the $c$ band discrete momentum values $\{q_j\} = \{[2\pi/N_a]\,I^{c}_j\}$, 
Eqs. (\ref{q-j}) and (\ref{Ic-an}) of Appendix \ref{pseudo-repre}, are both integers or half-odd integers. 

From the use of Eqs. (\ref{sir-pir}) and (\ref{rotated-quasi-spin}) one confirms that the $SU(2)$ algebra obeyed by 
electron-rotated local quasi-spin operators ${\tilde{q}}^{l}_{j}$, where $l = x_3, \pm$, and corresponding 
$\eta$-spin ($\alpha =\eta$) and spin ($\alpha =s$) operators ${\tilde{s}}^{l}_{j,\alpha}$ is the usual one,
\begin{equation}
[{\tilde{q}}^{+}_{j},{\tilde{q}}^{-}_{j'}] = \delta_{j,j'}\,2\,{\tilde{q}}^{x_3}_{j}
\, ; \hspace{0.50cm}
[{\tilde{q}}^{\pm}_{j},{\tilde{q}}^{x_3}_{j'}] = \mp \delta_{j,j'}\,{\tilde{q}}^{\pm}_{j} \, ,
\label{albegra-q-com}
\end{equation}
and
\begin{equation}
[{\tilde{s}}^{+}_{j,\alpha},{\tilde{s}}^{-}_{j',\alpha'}] = \delta_{j,j'}\delta_{\alpha,\alpha'}\,2\,{\tilde{s}}^{x_3}_{j,\alpha}
\, ; \hspace{0.50cm}
[{\tilde{s}}^{\pm}_{j,\alpha},{\tilde{s}}^{x_3}_{j',\alpha'}] = \mp \delta_{j,j'}\delta_{\alpha,\alpha'}\,{\tilde{s}}^{\pm}_{j,\alpha} 
\, , \hspace{0.25cm} \alpha ,\alpha'=\eta ,s \, ,
\label{albegra-s-e-s-com}
\end{equation}
respectively. Moreover, one has that $[{\tilde{q}}^{l}_{j},{\tilde{q}}^{l}_{j'}]=0$
and $[{\tilde{s}}^{l}_{j,\alpha},{\tilde{s}}^{l}_{j',\alpha'}] = 0$ where $l= 0, \pm$ and
$\alpha ,\alpha'=\eta , s$. 

The $c$ pseudoparticle and $\eta s$ quasi-spin operator algebras refer to the whole Hilbert space.
On the other hand, those of the $\eta$-spinon and spinon operators correspond to well-defined subspaces.
Those are spanned by states whose number $2S_c$ of rotated-electron singly occupied sites is fixed. This assures
that the value of the corresponding $\eta$-spinon number, $M_{\eta}=[N_a -2S_c]$, and spinon number, $M_{s}=2S_c$, 
is fixed as well. 

Moreover, the studies of the Ref. \cite{paper-I} have confirmed that the $\eta\nu$ pseudoparticles (and $s\nu$ pseudoparticles) 
considered in Refs. \cite{I,II} are $\eta$-spin-neutral (and spin-neutral) composite objects containing $\nu=1,...,\infty$ 
pairs of $\eta$-spin-$1/2$ anti-bound $\eta$-spinons with opposite $\eta$-spin projection (and 
of spin-$1/2$ bound spinons with opposite spin projection). 
The $c$ and $\alpha\nu$ pseudoparticles where $\alpha =\eta,s$ of Refs. \cite{I,II,paper-I} 
are an extension to the whole Hilbert space of those of Refs. \cite{92-00,currents} for the 
$c$ and $s1$ pseudo particle two-component subspace.
An additional well-defined number of $\eta$-spinons (and spinons) remain invariant under the electron - rotated-electron unitary 
transformation performed by the BA solution considered below. Since they remain unbound, they are called 
unbound $\pm 1/2$ $\eta$-spinons (and unbound  $\pm 1/2$ spinons). The values of the numbers 
$M^{un}_{\eta,\pm 1/2}$ of unbound $\pm 1/2$ $\eta$-spinons 
and $M^{un}_{s,\pm 1/2}$ of unbound $\pm 1/2$ spinons 
are fully controlled by the $\eta$-spin $S_{\eta}$ and $\eta$-spin projection $S_{\eta}^{x_3}$ 
and spin $S_{s}$ and spin projection $S_{s}^{x_3}$, respectively,
of the subspace or state under consideration as follows,
\begin{equation}
M^{un}_{\alpha} = [M^{un}_{\alpha,-1/2}+M^{un}_{\alpha,+1/2}]=2S_{\alpha} \, ;
\hspace{0.25cm}
M^{un}_{\alpha,\pm 1/2} = [S_{\alpha}\mp S_{\alpha}^{x_3}]
\, , \hspace{0.25cm} \alpha = \eta \, , s \, .
\label{L-L}
\end{equation}
Thus the $\eta$-spin $S_{\eta}$, $\eta$-spin projection $S_{\eta}^{x_3}$, 
spin $S_{s}$, and spin projection $S_{s}^{x_3}$ of an energy eigenstate
are fully determined by the $\eta$-unbound spinons and unbound spinons occupancies. 
For Bethe states with finite spin $S_s$ and/or $\eta$-spin $S_{\eta}$ all unbound spinons 
and/or unbound $\eta$-spinons have spin up and $\eta$-spin up, respectively. 

One can introduce $\alpha\nu$ pseudoparticle operators $f^{\dag}_{q_{j},\alpha\nu}$ 
labeled by the discrete momentum values $q_j$, Eqs. (\ref{q-j}) and (\ref{Ic-an}) of Appendix \ref{pseudo-repre},
such that $j=1,...,N_{a_{\alpha\nu}}$. Here $\alpha =\eta,s$ and $\nu=1,...,\infty$ \cite{I,II,paper-I}. Those are the conjugate 
variables of the $\alpha\nu$ effective lattice real-space
coordinates of site index $j=1,...,N_{a_{\alpha\nu}}$ defined in Ref. \cite{paper-I}.
For subspaces for which the ratio $N_{a_{\alpha\nu}}/N_a$ involving the number
$N_{a_{\alpha\nu}}$ of sites of the $\alpha\nu$ effective lattice, Eqs. (\ref{N*}) and (\ref{N-h-an}) 
of Appendix \ref{pseudo-repre}, is finite, such operators are given by,
\begin{equation}
f_{q_j,\alpha\nu}^{\dag} = (f_{q_j,\alpha\nu})^{\dag} =
{1\over{\sqrt{N_{a_{\alpha\nu}}}}}\sum_{j'=1}^{N_{a_{\alpha\nu}}}\,e^{+iq_j a_{\alpha\nu}\,j'}\,
f_{j',\alpha\nu}^{\dag} \, , \hspace{0.25cm} j = 1,...,N_{a_{\alpha\nu}} \, .
\label{fan-q-x-1D}
\end{equation}
Such local operators have anticommuting relations,
\begin{equation}
\{f^{\dag}_{q_j,\alpha\nu}\, ,f_{q_{j'},\alpha\nu}\} =
\delta_{j,j'} 
\, ; \hspace{0.50cm}
\{f^{\dag}_{q_j,\alpha\nu}\, ,f^{\dag}_{q_{j'},\alpha\nu}\} =
\{f_{q_j,\alpha\nu}\, ,f_{q_{j'},\alpha\nu}\}  = 0 \, .
\label{1D-anti-com-1D}
\end{equation}
Again, the first of such relations is valid provided that the two operators in the anticommutators 
act onto subspaces whose BA quantum numbers $I^{\alpha\nu}_j$ in
the $\alpha\nu$ band discrete momentum values $\{q_j\} = \{[2\pi/N_a]\,I^{\alpha\nu}_j\}$, 
Eqs. (\ref{q-j}) and (\ref{Ic-an}) of Appendix \ref{pseudo-repre}, are both integers or half-odd integers. 

The transformation laws of the elementary objects that emerge from the
rotated electrons under the electron - rotated-electron unitary transformation play 
an important role in the identification of the scatterers and scattering centers of the theory studied in
this paper. The latter are found in the following to be the PDT $c$ pseudofermions and $\alpha\nu$ 
pseudofermions \cite{V,VI,LE,TTF}. Except for a slight shift of their discrete momentum values, which renders
such objects without energy interaction terms, they have exactly the same
properties as the corresponding $c$ pseudoparticles and $\alpha\nu$ pseudoparticles, respectively,
from which they are generated. 

\section{The pseudofermion scattering theory}
\label{Pseudofermion}

Scattering theories of BA solvable models involve dressed $S$ matrices \cite{Natan-79,S-Natan,S-Hein}. 
In this section we introduce the pseudoparticle - pseudofermion unitary transformation and
corresponding pseudofermion and pseudofermion-hole dressed $S$ matrices. The
$\eta$-spin $1/2$ unbound $\eta$-spinons and spin $1/2$ unbound spinons are found
to be scattering-less elementary objects as far as their internal $SU(2)$ degrees of freedom
is concerned. On the other hand, the theory scatterers and scattering centers have no internal degrees of freedom such as 
spin or $\eta$-spin. Hence their scattering is associated with fully diagonal
dressed $S$ matrices. Consistent, our analysis of the problem follows the standard quantum
non-relativistic scattering theory of spin-less particles \cite{Taylor}. 
For simplicity and without loss in generality, in this section we consider in general 
densities in the ranges $n\in [0,1[$ and $m \in ]0,n]$. In Appendix \ref{Consequences}, the pseudofermion 
scattering theory is extended to PSs of ground states with densities $n=1$ and/or $m=0$.

\subsection{The 1D Hubbard model in the pseudofermion subspace: The pseudofermion canonical momentum,
associated functionals, and exotic pseudofermion algebra}
\label{canonical-func}

In the remaining of this paper we use a label $\beta$ that refers both to the $\beta =c$ 
band and $\beta =\alpha\nu$ band excitation branches. Here $\alpha =\eta$ and $\alpha =s$ correspond to 
$\eta$-spin and spin and $\nu =1,...,\infty$ to the number of $\eta$-spinon and spinon pairs, respectively. 
The pseudofermion scattering theory studied in this paper refers to 
the 1D Hubbard model in the pseudofermion subspace (PS). Such a subspace is
spanned by a given ground state with arbitrary values of the electronic
density $n$ and spin density $m$ and all excited energy eigenstates whose generation
from it involve changes in the occupancy configurations of a finite number of $\beta$ pseudoparticles, unbound spinons, and
unbound $\eta$-spinons. 

For the excited states belonging to the PS, the
following ratios then vanish as $N_a\rightarrow\infty$: $\delta N_{\beta}/N_a\rightarrow 0$, $\delta M_{s}^{un}/N_a\rightarrow 0$,
and $\delta M_{\eta}^{un}/N_a\rightarrow 0$. Here $\delta N_{\beta}$, $\delta M_{s}^{un}$,
and $\delta M_{\eta}^{un}$ denote the deviations in the numbers $N_{\beta}$, $M_{s}^{un}=2S_s$, and $M_{\eta}^{un}=2S_{\eta}$
of $\beta$ pseudoparticles, unbound spinons, and unbound $\eta$-spinons, respectively, under the ground-state - excited-state
transitions. (The numbers of unbound $\eta$-spinons and unbound spinons are defined in Eq. (\ref{L-L}).) 
We emphasize that within the definition of the PS there are no restrictions on the value
of the excitation energy and excitation momentum. 

For the energy eigenstates that span the PS, the rapidity functionals defined by the 
thermodynamic BA equations, Eqs. (\ref{Tapco1})-(\ref{kcn}) of Appendix \ref{pseudo-repre},
have the following exact property,
\begin{eqnarray}
\Lambda_{c}(q_j) & = & \sin k_c (q_j) = \sin k^0_c\Bigl({\bar{q}} (q_j)\Bigr) 
\, ; \hspace{0.50cm} k_c (q_j) = k^0_c\Bigl({\bar{q}} (q_j)\Bigr)
\, , \hspace{0.25cm} j = 1,...,N_a \, ,
\nonumber \\
\Lambda_{\beta}(q_j) & = & \Lambda_{\beta}^0\Bigl({\bar{q}} (q_j)\Bigr) \, , \hspace{0.25cm} j = 1,...,N_{a_{\alpha\nu}} \, .
\label{FL}
\end{eqnarray}

The set of discrete numbers ${\bar{q}}_j = {\bar{q}} (q_j)$ where $j=1,...,N_{a_{\beta}}$ in the arguments of the functions 
$k^0_c\Bigl({\bar{q}} (q_j)\Bigr)$ for $\beta =c$ and $\Lambda_{\beta}^0\Bigl({\bar{q}} (q_j)\Bigr)$ for
$\beta =\alpha\nu$ appearing in Eq. (\ref{FL}) play a central role in the pseudofermion scattering theory.
They are the $\beta$ pseudofermion discrete canonical-momentum values. Their spacing is to first order in $1/N_a$, 
\begin{equation}
{\bar{q}}_{j+1}-{\bar{q}}_{j}= {2\pi\over N_a} + {\rm h.o.} \, ,
\label{spacing-bar}
\end{equation}
where h.o. stands for terms of second order in $1/N_a$.
Such pseudofermions have been previously used in the PDT studies \cite{V,VI,LE,TTF}.
A $\beta$ pseudofermion carries discrete canonical momentum ${\bar{q}}_j = {\bar{q}} (q_j)$ whereas
the corresponding $\beta$ pseudoparticle carries the discrete momentum $q_j$. 
Often in the remaining of this paper we call $q_j$ bare momentum, to distinguish it
from the corresponding canonical momentum, ${\bar{q}}_j = {\bar{q}} (q_j)$.

For the $\beta =\eta\nu$ (and $\beta =s\nu\neq s1$) branches, the relation
$\Lambda_{\beta}(q_j) = \Lambda_{\beta}^0\Bigl({\bar{q}} (q_j)\Bigr)$, Eq. (\ref{FL}), is valid provided
that the hole concentration $(1-n)$ (and the spin density $m$) is finite. 
For a $S_{\eta}=0;n=1$ (and $S_{\eta}=0;m=0$) ground state, the number of $\eta\nu$ (and $s\nu\neq s1$)
band discrete momentum values $N_{a_{\eta\nu}}$ (and $N_{a_{s\nu}}$), Eqs.
(\ref{N*}) and (\ref{N-h-an}) of Appendix \ref{pseudo-repre}, vanishes, $N_{a_{\eta\nu}}=0$ (and $N_{a_{s\nu}}=0$.) 
Therefore, the corresponding pseudofermion branch does not exist. Hence the ground-state
rapidity $\Lambda_{\beta}^0$ appearing in Eq. (\ref{FL}) is undefined. For the excited states of such 
a $S_{\eta}=0;n=1$ (and $S_{\eta}=0;m=0$) ground state the value of $\Lambda_{\beta}(q_j)$ for $\beta =\eta\nu$ 
(and $\beta =\alpha\nu\neq s1$) is an issue addressed in Appendix \ref{Consequences}.

By use of the expressions provided in Eq. (\ref{FL}) in the thermodynamic BA equations, 
Eqs. (\ref{Tapco1})-(\ref{kcn}) of Appendix \ref{pseudo-repre}, one uniquely finds that,
to leading order in $1/N_a$, the discrete canonical-momentum values have the following form,
\begin{equation}
{\bar{q}}_j = {\bar{q}} (q_j) = q_j + {Q^{\Phi}_{\beta} (q_j)\over N_a} = {2\pi\over
N_a}I^{\beta}_j + {Q^{\Phi}_{\beta} (q_j)\over N_a} \, , \hspace{0.25cm} \beta = c, \alpha\nu
\, , \hspace{0.25cm} j=1,...,N_{a_{\beta}} \, .
\label{barqan}
\end{equation}
Here $N_{a_{\beta}}$ is given in Eqs. (\ref{N*}) and (\ref{N-h-an}) 
of that Appendix for both $\beta = c$ and $\beta =\alpha\nu$.
The $\beta $ band discrete momentum value $q_j$ in Eq. (\ref{barqan})
can be written as $q_j = [2\pi/N_a] I^{\beta}_j$, where $I^{\beta}_j$ with $j=1,...,N_{a_{\beta}}$ 
are BA quantum numbers. As given in Eqs. (\ref{q-j}) and (\ref{Ic-an}) of Appendix \ref{pseudo-repre},
those can have either integer or half-odd integer values. 

The relation, Eq. (\ref{barqan}), uniquely defines a one-to-one
correspondence between the two sets $\{q_j\}$ and $\{{\bar{q}}_j \}$ such that
$j =1,...,N_{a_{\beta}}$. That correspondence defines the $\beta$ pseudoparticle - $\beta$ pseudofermion unitary
transformation. The corresponding $\beta$ pseudoparticle - $\beta$ pseudofermion unitary operator,
\begin{equation}
{\hat{S}}^{\Phi}_{\beta} = 
e^{\sum_{j=1}^{N_{a_{\beta}}}f^{\dag}_{q_{j} + Q^{\Phi}_{\beta}(q_j)/N_a,\beta}f_{q_{j},\beta}} 
\, ; \hspace{0.50cm}
\left({\hat{S}}^{\Phi}_{\beta} \right)^{\dag} = 
e^{\sum_{j=1}^{N_{a_{\beta}}}f^{\dag}_{q_{j} - Q^{\Phi}_{\beta}(q_j)/N_a,\beta}f_{q_{j},\beta}}
\label{P-P-UT}
\end{equation}
is such that,
\begin{equation}
f^{\dag}_{{\bar{q}}_j,\beta} = f^{\dag}_{q_j + Q^{\Phi}_{\beta}(q_j)/N_a,\beta} =
\left({\hat{S}}^{\Phi}_{\beta} \right)^{\dag}f^{\dag}_{q_j,\beta}\,{\hat{S}}^{\Phi}_{\beta} \, .
\label{f-f-Q}
\end{equation}
Here $f^{\dag}_{q_j,\beta}$ and $f_{q_j,\beta}$ are the $\beta$ pseudoparticle operators, Eqs. (\ref{fc-q-x-1D}) and (\ref{fan-q-x-1D}).
The corresponding $\beta$ pseudofermion operators are denoted by $f^{\dag}_{{\bar{q}}_j,\beta}$ and
$f_{{\bar{q}}_j,\beta}$, respectively. Except for the slightly shifted discrete canonical momentum values, Eq. (\ref{barqan}), which below
is shown to render the $\beta$ pseudofermion spectrum without energy interaction terms,
the $\beta$ pseudofermions have the same properties as the corresponding $\beta$ pseudoparticles. 

The quantity $Q^{\Phi}_{\beta} (q_j)$ in Eqs. (\ref{barqan}), (\ref{P-P-UT}), and (\ref{f-f-Q}) plays a key role in the
pseudofermion scattering theory. It reads \cite{V},
\begin{equation}
Q^{\Phi}_{\beta} (q_j) = \sum_{\beta'}\,
\sum_{j'=1}^{N_{a_{\beta'}}}\,2\pi\,\Phi_{\beta,\beta'}(q_j,q_{j'})\, \delta N_{\beta'}(q_{j'}) \, . 
\label{qcan1j}
\end{equation}
The $\beta'$ band momentum distribution function deviation $\delta N_{\beta'}(q_{j'})$ appearing
here is defined in Eq. (\ref{DNq}) of Appendix \ref{pseudo-repre}. The elementary-object
representation used in the studies of this paper refers to the limit of a very large system, $N_a\gg 1$. 
Thus we often approximate the discrete bare momentum values $q_j$, such that $[q_{j+1}-q_j]=2\pi/N_a$, by a continuum variable
$q$. The corresponding general $\beta$ band bare momentum distribution function deviation then reads,
\begin{eqnarray}
\delta N_{\beta} (q) & = & {2\pi\over N_a}\sum_{p=1}^{N_{\beta}^p}\delta (q-q_{p}) -
{2\pi\over N_a}\sum_{h=1}^{N_{\beta}^h}\delta (q-q_{h}) \, , \hspace{0.25cm}{\rm if}
\hspace{0.15cm}\delta G_{\beta}\hspace{0.1cm}{\rm even}\hspace{0.1cm}{\rm for}\hspace{0.1cm}\beta = c,s1  \, ,
\nonumber \\
 & = & {2\pi\over N_a}\sum_{p=1}^{N_{\beta}^p}\delta (q-q_{p}) -
{2\pi\over N_a}\sum_{h=1}^{N_{\beta}^h}\delta (q-q_{h}) 
\pm {\pi\over N_a}\sum_{j=\pm 1}j\,\delta (q - j\,q_{F\beta}) \, , \hspace{0.25cm}{\rm if}
\hspace{0.15cm}\delta G_{\beta}\hspace{0.1cm}{\rm odd}\hspace{0.1cm}{\rm for}\hspace{0.1cm}\beta = c,s1 \, ,
\nonumber \\
\delta N_{\beta} (q) & = & {2\pi\over N_a}\sum_{p=1}^{N_{\beta}^p}\delta (q-q_{p}) \, , \hspace{0.25cm}
{\rm for}\hspace{0.1cm}\beta =\alpha\nu \neq s1  \, ,
\label{deltaN-pq-cont}
\end{eqnarray} 
Here and throughout this paper, $\delta (x)$ denotes the usual Dirac delta-function distribution. 
Since there is a one-to-one correspondence between the canonical momentum ${\bar{q}} (q)$ and the 
bare-momentum $q$, we may refer to the {\it bare momentum $q$ of a pseudofermion}. By that it is meant 
the bare momentum $q$ corresponding to the pseudofermion canonical momentum ${\bar{q}} (q)$. 
For instance, $q_1,...,q_{N_{\beta}^p}$ are in Eq. (\ref{deltaN-pq-cont}) the bare momentum values 
of the $N_{\beta}^p$ $\beta$ pseudofermions added under the transition to the excited state and $q_1,...,q_{N_{\beta}^h}$ 
those of the $N_{\beta}^h$ $\beta$ pseudofermion holes added under that transition.
$\delta G_{\beta}$ stands in that equation for the deviation in the value of the number,
\begin{equation}
G_{\beta} =\delta_{\beta,c}\sum_{\beta =\eta,s} B_{\alpha}
+ \delta_{\beta,s1} [N_c + N_{s1}]  \, ; \hspace{0.5cm}
B_{\alpha}=\sum_{\nu=1}^{\infty}N_{\alpha\nu} \, , \hspace{0.25cm} \alpha = \eta, s \, .
\label{F-beta}
\end{equation}
The additional term in the Eq. (\ref{deltaN-pq-cont}) expression corresponding to the deviation $\delta G_{\beta}$ 
being an odd integer number rather than an even integer number results from the corresponding
$\beta =c,s1$ band momentum shift, $\pm \pi/N_a$.

Moreover, the quantity $\pi\,\Phi_{\beta,\beta'}(q_j,q_{j'})$ in Eq. (\ref{qcan1j}) is a
function of both the bare-momentum values $q_j$ and $q_{j'}$, given by,
\begin{equation}
\pi\,\Phi_{\beta,\beta'}(q_j,q_{j'}) = \pi\,\bar{\Phi }_{\beta,\beta'}
\left({\Lambda^{0}_{\beta}(q_j)\over u}, {\Lambda^{0}_{\beta'}(q_{j'})\over u}\right) \, . 
\label{Phi-barPhi}
\end{equation}
It is found below that $\pi\,\Phi_{\beta,\beta'}(q,q')$ [or
$-\pi\,\Phi_{\beta,\beta'}(q,q')$] is an elementary {\it two-pseudofermion
phase shift}. It is such that $q$ is the bare-momentum value of a $\beta$ pseudofermion or
$\beta$ pseudofermion hole scattered by a $\beta'$ pseudofermion [or
$\beta'$ pseudofermion hole] of bare-momentum $q'$ created under a ground-state -
excited-state transition. 

The function $\pi\,\bar{\Phi }_{\beta,\beta'} (r ,r')$, Eq. (\ref{Phi-barPhi}), is the unique
solution of the integral equations, Eqs. (\ref{Phis1c-m})-(\ref{Phisnsn-m}) of Appendix \ref{Ele2PsPhaShi}.
The ground-state rapidity functions $\Lambda_{\beta}^0 (q_j)$, where
$\Lambda^0_{c}(q_j)=\sin k_c^0 (q_j)$ for $\beta=c$, are defined  
in terms of their inverse functions in Eq. (\ref{GS-R-functions}) of that Appendix.
We recall that for $S_{\eta}=0;n=1$ (and $S_{\eta}=0;m=0$) ground states there are no $\beta=\eta\nu$ (and $\beta=s\nu\neq s1$) 
bare-momentum bands. Indeed, Eqs. (\ref{N*}) and (\ref{N-h-an}) of Appendix \ref{pseudo-repre} lead to
$N_{a_{\eta\nu}}=0$ (and $N_{a_{s\nu}}=0$) for such states. As reported in Appendix \ref{Consequences}, then
the ground-state rapidity functions $\Lambda^{0}_{\eta\nu}(q)$
(and $\Lambda^{0}_{s\nu}(q)$) must be replaced by those of the excited
state,  $\Lambda_{\eta\nu}(q)$ (and $\Lambda_{s\nu}(q)$), respectively.
It is found below that the functions, Eq. (\ref{Phi-barPhi}), are phase
shifts originated by well-defined ground-state - excited-state transitions.

In the $u\rightarrow\infty$ limit, the ground-state rapidity momentum function $k_{c}^{0} (q_j)$, 
Eq. (\ref{GS-R-functions}) of Appendix \ref{pseudo-repre}, is given by $k_{c}^{0} (q_j)  = q_j$. 
Hence, according to Eq. (\ref{FL}), for all PSs it reads, $k_c (q_j) = k^0_c\Bigl({\bar{q}} (q_j)\Bigr) = {\bar{q}}_j$. 
The $u\rightarrow\infty$ spinless fermions of
Refs. \cite{Penc-96,Penc-97} have been constructed inherently to carry the rapidity momentum
$k_j = k_c (q_j)$. Since $k_c (q_j) = {\bar{q}}_j$ as $u\rightarrow\infty$, such spinless fermions are
nothing but the $c$ pseudofermions as defined here for $u\rightarrow\infty$. The spinless fermions 
have creation and annihilation operators $b^{\dag}_{k_j}$ and $b_{k_j}$, respectively, which have the
same expression in terms of electron creation and annihilation operators as the corresponding $u>0$
$c$ pseudofermion operators $f^{\dag }_{{\bar{q}}_j,c}$ and $f_{{\bar{q}}_j,c}$ in terms of rotated-electron 
creation and annihilation operators, so that,
\begin{equation}
f^{\dag }_{{\bar{q}}_j,c} =
{\hat{V}}^{\dag}\,b^{\dag}_{k_j}\,{\hat{V}}
\, ; \hspace{0.50cm}
f_{{\bar{q}}_j,c} =
{\hat{V}}^{\dag}\,b_{k_j}\,{\hat{V}} \, .
\label{c-pseudo-ope}
\end{equation} 
Such relations hold except for unimportant phase factors provided that ${\hat{V}}$ is the 
electron - rotated-electron unitary operator associated with the specific
unitary transformation performed by the BA solution.

An important functional related to that defined in Eq. (\ref{qcan1j}) reads,
\begin{equation}
Q_{\beta}(q_j) = Q_{\beta}^0 + Q^{\Phi}_{\beta} (q_j) \, .
\label{Qcan1j}
\end{equation}
The related quantity, $Q_{\beta}^0/N_a$, is the shift in the $\beta =c,\alpha\nu$ band 
discrete bare-momentum value $q_j$ under the ground-state - excited-state transition. Its
values are,
\begin{eqnarray}
Q_{c}^0 & = & 0 \, ; \hspace{0.5cm} \sum_{\beta =\eta,s}\,\sum_{\nu=1}^{\infty} \delta
N_{\alpha\nu} \hspace{0.25cm} {\rm even} \, ;  \hspace{1.0cm} Q_{c}^0=\pm\pi \, ;
\hspace{0.5cm} \sum_{\beta =\eta,s}\,\sum_{\nu=1}^{\infty} \delta
N_{\alpha\nu} \hspace{0.25cm} {\rm odd} \, ; \nonumber \\
Q_{\alpha\nu}^0 & = & 0 \, ; \hspace{0.5cm} \delta N_{c}+\delta N_{\alpha\nu}
\hspace{0.25cm} {\rm even} \, ; \hspace{1.0cm} Q_{\alpha\nu}^0=\pm\pi \, ; \hspace{0.5cm}
\delta N_{c}+\delta N_{\alpha\nu} \hspace{0.25cm} {\rm odd} \, ; \hspace{0.5cm} \alpha = \eta,s \, . 
\label{pican}
\end{eqnarray}
When under such a transition the BA quantum numbers $I^{\beta}_j$ in  
$q_j = [2\pi/N_a] I^{\beta}_j$, Eqs. (\ref{q-j}) and (\ref{Ic-an}) of Appendix \ref{pseudo-repre}, change from 
integers (or half-odd integers) to half-odd integers (or integers), a finite shift, $Q_{\beta}^0/N_a=\pm\pi/N_a$ occurs. 

Analysis of the discrete canonical-momentum value expression, Eqs. (\ref{barqan}) and (\ref{qcan1j}), 
reveals that for the ground state the canonical momentum ${\bar{q}}_j$ and corresponding bare 
momentum $q_j$ have the same value, ${\bar{q}}_j = q_j$. Hence the ground-state limiting $\beta$ 
canonical momenta and $c$ and $s1$ Fermi canonical momenta equal the corresponding bare momenta. 
However, for the excited states the $c$ and $s1$ Fermi canonical momentum values are
shifted. The corresponding deviations play an important role in the spectral properties, as
discussed below in Section \ref{PDT}.

The bare-momentum distribution-function deviations second-order energy spectrum, Eq. (\ref{DE-fermions}) 
of Appendix \ref{pseudo-repre}, can be exactly expressed in terms of 
the corresponding $\beta$ pseudofermion canonical-momentum distribution-function deviations as follows,
\begin{equation}
\delta E_{l_{\rm r},l_{\eta s}} = \sum_{\beta =c,s1}\sum_{j=1}^{N_{a_{\beta}}}\varepsilon_{\beta} ({\bar{q}}_j)\delta {\cal{N}}_{\beta}({\bar{q}}_j)
+  \sum_{\beta\nu\neq s1}\sum_{j=1}^{N_{a_{\alpha\nu}}}\varepsilon^0_{\alpha\nu} ({\bar{q}}_j)\delta {\cal{N}}_{\alpha\nu}({\bar{q}}_j)
+ 2\vert\mu\vert D_r + 2\mu_B\vert H\vert S_{r} \, . 
\label{DE}
\end{equation}
The deviations second-order terms of the equivalent pseudoparticle energy spectrum, Eq. (\ref{DE-fermions}) of Appendix 
\ref{pseudo-repre}, refer to zero-momentum forward-scattering interactions. The unitary transformation relations, Eqs. (\ref{barqan}) 
and (\ref{qcan1j}), have absorbed such second-order energy terms into the pseudofermion canonical momentum. 
Consistent, the pseudoparticle $f$ functions, Eq. (\ref{ff}) of Appendix \ref{pseudo-repre}, and momentum-shift functional, 
Eq. (\ref{qcan1j}), contain the same two-pseudofermion phase shifts.

Importantly, the $\beta$-pseudoparticle - $\beta$-pseudofermion unitary transformation maps a quantum problem in 
terms of pseudoparticles with residual zero-momentum forward-scattering energy interaction terms,
the second-order terms in Eq. (\ref{DE-fermions}), into a non-interacting
pseudofermion problem. Indeed, the equivalent pseudofermion energy spectrum,
Eq. (\ref{DE}), has no interaction energy terms, so that the pseudofermions are not energy entangled. 
Such a lack of $\beta$ pseudofermion energy entanglement plays a key role in the PDT. As shortly discussed below
in Section \ref{PDT}, it allows the one- and two-electron spectral functions to be expressed as convolutions of $c$ and
$s1$ pseudofermion spectral functions \cite{V,VI,LE}. 

The energy dispersions $\varepsilon_{\beta} ({\bar{q}}_j)$ in the energy spectrum, Eq. (\ref{DE}), have 
exactly the same dependence on ${\bar{q}}_j$ as those defined by Eqs. (\ref{e-0-bands}) and (\ref{epsilon-q}) 
of Appendix \ref{pseudo-repre} on $q_j$. The limiting analytical behaviors of such energy dispersions are 
provided in Appendix \ref{energies}. Moreover, the finite-energy part of the energy spectrum,
Eq. (\ref{DE-fermions}) of Appendix \ref{pseudo-repre}, is expressed in Eq. (\ref{DE}) in terms of the numbers $D_r$
and $S_{r}$, Eq. (\ref{DrS-r}) of that Appendix. $D_r$ is the number of rotated-electron doubly occupied sites.
$S_{r}$ is that of those spin-down rotated-electron singly occupied sites whose spins are not
contained into two-spinon $s1$ pseudofermions. The numbers $D_r$ and $S_{r}$ vanish for all
ground states. Hence their deviations are given by $\delta D_r=D_r$ and $\delta S_r=S_r$. 

That for $N_a\gg 1$ and up to contributions of $1/N_a$ order the $c$ and $s1$ pseudofermions have no residual energy 
interactions is a result of major physical importance \cite{V,VI,LE}.
Hence in the following we confirm that such elementary objects have indeed no residual energy interactions.
The contributions to the PDT one- and two-electron spectral-weight distributions that are more involved to
be accounted for are those of the $c$ and $s1$ pseudofermions processes. Indeed, LWS ground states are not 
populated by $\alpha\nu\neq s1$ pseudofermions, unbound $-1/2$ $\eta$-spinons, and unbound $-1/2$ spinons. That
much simplifies accounting for the contribution of such elementary-objects creation under the transitions
to the excited states to the corresponding one- and two-electron spectral weights \cite{V,VI,LE}. 
Hence for simplicity and without loss in generality, in the following we consider PSs for which $M^{un}_{\eta,-1/2}=M^{un}_{s,-1/2}=0$ in 
Eq. (\ref{L-L}) and $\delta N_{\beta} (q_j)=0$ for the $\beta\neq c,s1$ branches in Eq. (\ref{DE-fermions}) of Appendix \ref{pseudo-repre}. 

Our analysis refers to the limit of a very large system, $N_a\gg 1$. Hence we approximate both the $\beta =c,s1$ band
discrete bare momentum values, $q_j$, and the corresponding discrete canonical momentum values, ${\bar{q}}_j$, 
by continuum variables, $q$ and ${\bar{q}} = q+Q^{\Phi}_{\beta} (q)/N_a$, respectively. The excited-state $\beta =c,s1$ 
deviations $\delta N_{\beta} (q)$ have the general form, Eq. (\ref{deltaN-pq-cont}). For the PSs under consideration, 
the energy spectrum, Eq. (\ref{DE-fermions}) of Appendix \ref{pseudo-repre}, becomes in terms of continuum momentum 
variables \cite{Carmelo91,Carmelo92},
\begin{eqnarray}
\delta E_{l_{\rm r},l_{\eta s}} & = &  {N_a\over 2\pi}\sum_{\beta =c,s1}\int_{-q_{\beta}}^{q_{\beta}}dq\,\varepsilon_{\beta} (q)\delta N_{\beta} (q)  
\nonumber \\
& + & {1\over N_a}\left({N_a\over 2\pi}\right)^2\sum_{\beta =c,s1}\sum_{\beta' =c,s}\int_{-q_{\beta}}^{q_{\beta}}dq\int_{-q_{\beta'}}^{q_{\beta'}}dq'
{1\over 2}\,f_{\beta,\beta'} (q,q')\,\delta N_{\beta} (q)\delta N_{\beta'} (q') \, .
\label{DE-fermions-ff}
\end{eqnarray}
The energy dispersions $\varepsilon_{\beta} (q)$ and $f$ functions appearing here are those defined in 
Eqs. (\ref{e-0-bands})-(\ref{eplev0}) and Eq. (\ref{ff}) of Appendix \ref{pseudo-repre}, respectively. 
Interestingly, such $f$ functions involve only the $\beta$ group velocities $v_{\beta} (q)$ and 
$v_{\beta}\equiv v_{\beta} (q_{F\beta})$, Eq. (\ref{velo}) of Appendix \ref{pseudo-repre}, and the
two-pseudofermion phase shifts, Eq. (\ref{Phi-barPhi}), of the scattering theory considered below.
For low-energy excitations, both the first-order and second-order energy terms, Eq. (\ref{DE-fermions-ff}),
contribute to the finite-size spectrum of conformal-field theory \cite{LE}. The corresponding finite-size corrections 
are of $1/N_a$ order \cite{Woy-89,Frahm,LE}.

The energy spectrum, Eq. (\ref{DE-fermions-ff}), can be rewritten in terms of the functional
$Q^{\Phi}_{\beta} (q)$, Eq. (\ref{qcan1j}), as follows \cite{Carmelo92}, 
\begin{eqnarray}
\delta E_{l_{\rm r},l_{\eta s}} & = & {N_a\over 2\pi}\sum_{\beta =c,s1}\int_{-q_{\beta}}^{q_{\beta}}dq\,\varepsilon_{\beta} (q)\,\delta N_{\beta} (q) 
\nonumber \\
& + & \sum_{\beta =c,s}\left[{1\over 2\pi}\int_{-q_{\beta}}^{q_{\beta}}dq\,v_{\beta} (q) \,
Q^{\Phi}_{\beta} (q) \delta N_{\beta} (q) + {v_{\beta}\over 2\pi N_a}\sum_{\iota = \pm 1} \left(Q^{\Phi}_{\beta} (\iota q_{F\beta})\right)^2\right]  \, .
\label{DE-fermions-HH-cont}
\end{eqnarray}
We now confirm that the expression of this energy functional in terms of the $\beta =c,s1$ pseudofermion canonical momenta
${\bar{q}}$, Eq. (\ref{barqan}), indeed simplies to,   
\begin{equation}
\delta E_{l_{\rm r},l_{\eta s}} = 
{N_a\over 2\pi}\sum_{\beta =c,s1}\int_{-q_{\beta}}^{q_{\beta}}d {\bar{q}}\,\varepsilon_{\beta} ({\bar{q}})\,\delta {\bar{N}}_{\beta} ({\bar{q}}) \, .
\label{DE-c-s}
\end{equation}

To first order in $1/N_a$, the deviation $\delta {\bar{N}}_{\beta} ({\bar{q}})$ in Eq. (\ref{DE-c-s}) accounts for two
types of contributions. The first contribution type results from the shift, $q\rightarrow q+Q^{\Phi}_{\beta} (q)/N_a$, in 
the arguments of the deviation $\delta N_{\beta} (q)$ $\delta$-functions,
Eq. (\ref{deltaN-pq-cont}). It is accounted for by considering that
$\varepsilon_{\beta} ({\bar{q}})=\varepsilon_{\beta} (q)$ and $\delta {\bar{N}}_{\beta} ({\bar{q}}) = \delta N_{\beta} (q+Q^{\Phi}_{\beta} (q)/N_a)$
in Eq. (\ref{DE-c-s}). This leads to the energy terms 
${N_a\over 2\pi}\sum_{\beta =c,s1}\int_{-q_{\beta}}^{q_{\beta}}dq\,\varepsilon_{\beta} (q)\,\delta N_{\beta} (q)$
and ${1\over 2\pi}\sum_{\beta =c,s}\int_{-q_{\beta}}^{q_{\beta}}dq\,v_{\beta} (q)\,Q^{\Phi}_{\beta} (q) \delta N_{\beta} (q)$ in Eq. (\ref{DE-fermions-HH-cont}). 
To first order in $1/N_a$, exactly the same energy terms are obtained if one uses instead
$\varepsilon_{\beta} ({\bar{q}})=\varepsilon_{\beta} (q+Q^{\Phi}_{\beta} (q)/N_a)$ and
$\delta {\bar{N}}_{\beta} ({\bar{q}}) = \delta N_{\beta} (q)$ in Eq. (\ref{DE-c-s}). This latter choice is the most convenient 
for adding to $\delta {\bar{N}}_{\beta} ({\bar{q}})$ the second contribution.

The first contribution type refers to small changes in the momenta of the original $\beta $ pseudoparticle creation and 
annihilation processes described by the deviations, Eq. (\ref{deltaN-pq-cont}). On the other
hand, the second contribution to the $\beta$ pseudofermion canonical-momentum distribution function deviation 
$\delta {\bar{N}}_{\beta} ({\bar{q}})$ involves the quantity 
$\iota \,Q^{\Phi}_{\beta} (\iota\,q_{F\beta})/N_a= \iota [\delta {\bar{q}}_{F\beta}^{\iota} 
- \delta q_{F\beta}^{\iota}]$. Here $\delta {\bar{q}}_{F\beta}^{\iota}$ and
$\delta q_{F\beta}^{\iota}$ are the excited-state deviations of the $\beta$ pseudofermion
and $\beta$ pseudoparticle right ($\iota =1$) and left ($\iota =-1$) Fermi points, which read,
\begin{equation}
\delta {\bar{q}}_{F\beta}^{\iota} = \delta q_{F\beta}^{\iota} +
{Q^{\Phi}_{\beta} (\iota q_{F\beta})\over N_a} \, ; \hspace{0.5cm}
\delta q_{F\beta}^{\iota} = \iota\, {2\pi\over N_a}\delta N^F_{\beta,\iota} 
\, , \hspace{0.25cm} \beta =c, s1 \, , \hspace{0.25cm} \iota = \pm 1 \, ,
\label{dbarqFb}
\end{equation}
respectively. In this equation, $\delta N^{F}_{\beta,\iota} =[\delta N^{0,F}_{\beta,\iota}
+\iota \,Q^0_{\beta}/2\pi]$ denotes the deviation in the number of $\beta =c,s1$
pseudofermions at the right ($\iota =+1$) and left ($\iota =+1$) $\beta =c,s1$ Fermi points, under
the ground-state - excited-state transition. Such a deviation includes the effects 
from the possible shifts, $\pm Q^0_{\beta}/N_a$ and $\pm Q^0_{\beta}/2\pi$,
of the bare discrete momentum values $q_j$, Eq. (\ref{q-j}) of Appendix \ref{pseudo-repre},
and corresponding BA quantum numbers $I^{\beta}_j$,
Eq. (\ref{Ic-an}) of that Appendix, respectively. (The quantity $Q^0_{\beta}$ is that
given in Eq. (\ref{pican}) for $\beta =c, s1$.)
On the other hand, the deviation $\delta N^{0,F}_{\beta,\iota}$ does not account for 
the effects from such shifts. Its value follows only from the net change in the
$\beta =c, s1$ pseudofermion occupancies at the corresponding $\iota =\pm 1$ Fermi points.

The second term, $Q^{\Phi}_{\beta} (\iota q_{F\beta})/N_a$, in the deviation
$\delta {\bar{q}}_{F\beta}^{\iota}$, Eq. (\ref{dbarqFb}), which
is that associated with the second contribution type to $\delta {\bar{N}}_{\beta} ({\bar{q}})$, occurs
in that expression even when $\delta q_{F\beta}^{\iota}=0$. In that case there is no change 
in the $\iota =\pm 1$ $\beta =c, s1$ Fermi-point occupancies under the ground-state - excited-state transition.  
To first order in $1/N_a$, the two types of contribution are accounted for provided that one uses the 
following expressions for $\varepsilon_{\beta} ({\bar{q}})$ and $\delta {\bar{N}}_{\beta} ({\bar{q}})$ in the energy 
functional, Eq. (\ref{DE-c-s}),
\begin{eqnarray}
\varepsilon_{\beta} ({\bar{q}}) & = & \varepsilon_{\beta} (q+Q^{\Phi}_{\beta} (q)/N_a) \approx
\varepsilon_{\beta} (q) + v_{\beta} (q)\,{Q^{\Phi}_{\beta} (q)\over N_a}
\nonumber \\
\delta {\bar{N}}_{\beta} ({\bar{q}}) & = & 
\delta N_{\beta} (q) + \iota {Q^{\Phi}_{\beta} (\iota\,q_{F\beta})\over N_a}\,\delta (q-\iota\,q_{F\beta}) \, .
\label{bar-N}
\end{eqnarray}
Consistent with our above discussion, the extra term $ \iota [Q^{\Phi}_{\beta} (\iota\,q_{F\beta})/N_a]\,\delta (q-\iota\,q_{F\beta})$ in
the $\delta {\bar{N}}_{\beta} ({\bar{q}})$ expression accounts for the second contribution type.
The use on the right-hand side of Eq. (\ref{DE-c-s}) of the expressions, Eq. (\ref{bar-N}), 
readily leads to the full energy expression, Eq. (\ref{DE-fermions-HH-cont}). It is
of second order in the $\beta$ pseudoparticle momentum distribution function deviations $\delta N_{\beta}(q)$. 
(We recall that $\varepsilon_{\beta} (\iota\,q_{F\beta}) = 0$, so that the second contribution type leads 
indeed only to the energy term $\sum_{\beta =c,s}{v_{\beta}\over 2\pi N_a}\sum_{\iota = \pm 1}(Q^{\Phi}_{\beta} (\iota q_{F\beta}))^2$
in Eq. (\ref{DE-fermions-HH-cont}).)

On the one hand, within the $\beta$ pseudoparticle representation
the quantity $Q^{\Phi}_{c} (q)/N_a$ (and $Q^{\Phi}_{\beta} (q)/N_a$)
in the argument of the rapidity function $k^0_c (q + Q^{\Phi}_{c} (q)/N_a)$
(and $\Lambda_{\beta}^0 (q + Q^{\Phi}_{\beta} (q)/N_a)$) 
is behind the $\beta$ pseudoparticle residual interactions. This is confirmed by
the form of the second-order energy terms in the first expression 
of Eq. (\ref{DE-fermions-HH-cont}). Those would vanish if
$Q^{\Phi}_{\beta} (q)=0$. On the other hand, 
within the $\beta$ pseudofermion representation that quantity is
rather incorporated in the canonical momentum. Consistent,
we have just confirmed that the $\beta =c,s1$ pseudofermions are not energy entangled:
The momentum shift $Q^{\Phi}_{\beta} (q)/N_a$ in the
$\beta$ pseudofermion canonical momentum
${\bar{q}}=q + Q^{\Phi}_{\beta} (q)/N_a$, Eq. (\ref{barqan}),
exactly cancels the second-order energy terms in Eq. (\ref{DE-fermions-HH-cont}).

That within the $\beta$ pseudofermion representation the
quantity $Q^{\Phi}_{\beta} (q)$ is incorporated in the canonical momentum
has consequences though in the exotic algebra obeyed by the $\beta$ pseudofermion creation and annihilation
operators. Consider a $\beta$ pseudofermion of canonical momentum ${\bar{q}}$ and a
$\beta'$ pseudofermion of canonical momentum ${\bar{q}'}$. Here ${\bar{q}}$ and ${\bar{q}'}=q'$ correspond to an excited energy
eigenstate $\beta$ band and the ground state $\beta'$ band, respectively. The $\beta$ band canonical momentum 
having the form ${\bar{q}}=q + Q^{\Phi}_{\beta} (q)/N_a$ implies that the effective anticommutators involving 
the creation and/or annihilation operators of these two pseudofermions have
the general form \cite{V},
\begin{equation}
\{f^{\dag }_{{\bar{q}},\beta},f_{{\bar{q}}',\beta'}\} =
\delta_{\beta,\beta'}{1\over N_{a_{\beta}}}\,e^{-i({\bar{q}}-{\bar{q}}')/
2}\,e^{iQ_{\beta}(q)/2}\,{\sin\Bigl(Q_{\beta} (q)/
2\Bigr)\over\sin ([{\bar{q}}-{\bar{q}}']/2)} \, , 
\label{pfacrGS}
\end{equation}
and $\{f^{\dag}_{{\bar{q}},\beta},f^{\dag}_{{\bar{q}}',\beta'}\} = \{f_{{\bar{q}},\beta},f_{{\bar{q}}',\beta'}\}=0$.
Here $Q_{\beta}(q_j) = Q_{\beta}^0 + Q^{\Phi}_{\beta} (q_j)$ is the functional,
Eq. (\ref{Qcan1j}), whose value is specific to the excited energy eigenstate under consideration.
The related quantity, $Q_{\beta}(q_j)/N_a= [Q_{\beta}^0 + Q^{\Phi}_{\beta} (q_j)]/N_a$, is the overall shift in the discrete canonical-momentum
value that results from the ground-state - excited-state transition. 

Note that for $\beta =\beta'$ the unitarity of the pseudoparticle - pseudofermion transition preserves the
pseudoparticle operator algebra provided that the canonical momentum values
${\bar{q}}$ and ${\bar{q}'}$ correspond to the same excited state for $\beta$ band. The exotic form
of the anticommutator in Eq. (\ref{pfacrGS}) follows from for $\beta =\beta'$
the canonical momenta ${\bar{q}}$ and ${\bar{q}'}$ corresponding
rather to the excited state $\beta$ band and the ground state $\beta$ band, respectively. 
Such an exotic $\beta$ pseudofermion algebra plays an important role in the 
one- and two-electron finite-energy spectral weight distributions \cite{V,VI,LE}.
Consistent with the form of the anticommutator in Eq. (\ref{pfacrGS}) for $\beta =\beta'$ and as discussed below in Section \ref{PDT}, 
the functional $Q_{\beta}(q)$ controls the quantum overlaps associated with such spectral weight distributions. 

The effective character for $\beta =\beta'$ of the anticommutator in Eq. (\ref{pfacrGS}) results from it
involving two operators acting onto subspaces with different $\beta$ band
discrete canonical momentum values. Such subspaces are those of the
ground state and excited state, respectively. The corresponding
shake-up effects makes it to be an effective anticommutator. 
Indeed, the standard operator commutators involve operators acting onto
the same Hilbert space. However, the effective anticommutators of such a form are physically
meaningful. They control the orthogonal-catrastrophe quantum overlaps
associated with the PDT one- and two-electron spectral-weight distributions \cite{V,VI,LE}.

\subsection{The ground-state - virtual-state transition} 

The momentum value of a unbound $\pm 1/2$ $\eta$-spinon (and a unbound $\pm 1/2$ spinon) internal degrees of freedom is
$q_{\eta,+1/2}=0$ or $q_{\eta,-1/2}=\pi$ (and $q_{s,\pm 1/2}=0$) for all energy
eigenstates with finite occupancy of such objects. That such momentum values
remain unchanged under the ground-state - excited-state transitions reveals 
that as far as their internal degrees of freedom is concerned such objects 
are not scatterers. Neither do the $\beta$ pseudofermions scatterer off on them, so that they are 
not scattering centers. Hence the scattering processes described in the following
involve only the $\beta$ pseudofermions.

Each transition from the ground state to a PS excited energy 
eigenstate can be divided into three steps. The first process is a scatter-less 
finite-energy and finite-momentum excitation that transforms the ground state onto 
a well defined virtual state. For the $\alpha\nu$ branches, that excitation can involve a change 
in the number of discrete bare-momentum values. Following Eqs. (\ref{N*}) and (\ref{N-h-an}) of Appendix \ref{pseudo-repre},
those are given by,
\begin{equation}
\delta N_{a_{s1}} = \delta N_{c} - \delta N_{s1} - 2\sum_{\nu =2}^{\infty} \delta N_{s\nu}
\, ; \hspace{0.5cm} \delta N_{a_{\alpha\nu}} = \delta M^{un}_{\alpha} + 2\sum_{\nu'=\nu
+1}^{\infty} (\nu' -\nu) \delta N_{\alpha\nu'} \, , \hspace{0.25cm} \alpha\nu\neq s1 \, . 
\label{DN*s1an}
\end{equation}
For the ground state, the numbers of discrete bare-momentum values read,
\begin{equation}
N^{0}_{a_{s1}}=N_{\uparrow} \, ; \hspace{0.5cm}
N^{0}_{a_{s\nu}}=(N_{\uparrow} -N_{\downarrow}) \, , \hspace{0.25cm} s\nu\neq s1
\, ; \hspace{0.5cm} N^{0}_{a_{\eta\nu}}=(N_a -N) \, ,
\label{N*csnu}
\end{equation}
and $N^{0}_{a_c}=N_{a_c}$ is given by $N_{a_c}=N_a$ for the whole Hilbert space.  
The $\beta\neq c, s1$ branches have no finite pseudofermion occupancy 
in the ground state. In spite of that, for the present densities ranges one can define the 
discrete momentum number values $N_{a_{\beta}}=N^h_{\beta}$ of the corresponding unoccupied bands. For the
$\beta\neq c, s1$ branches, those are the numbers $N^{0}_{a_{\eta\nu}}$ and
$N^{0}_{a_{s\nu}}$ given in Eq. (\ref{N*csnu}). Thus, for $\beta\neq c, s1$ branches
with finite pseudofermion occupancy in the virtual state the discrete bare-momentum shifts,
Eq. (\ref{pican}), and deviations, Eq. (\ref{DN*s1an}), are relative to the values of 
such unoccupied bands. 

In addition and following the change in the number of discrete bare-momentum values,
this excitation also involves the pseudofermion creation and annihilation processes and 
pseudofermion particle-hole processes that generate the PS excited states. The 
excitation momentum of the corresponding ground-state - virtual-state transition reads,
\begin{eqnarray}
\delta P_{l_{\rm r},l_{\eta s}}^0 &= & \sum_{j=1}^{N_a} q_j\, \delta N_c (q_j)
+ \sum_{\nu =1}^{\infty}\sum_{j'=1}^{N_{a_{s\nu}}}
q_{j}\, \delta N_{s\nu} (q_{j}) + \sum_{\nu =1}^{\infty}\sum_{j=1}^{N_{a_{\eta\nu}}}
[\pi -q_{j}]\, N_{\eta\nu} (q_{j}) 
+\pi\,M_{\eta,-1/2} \, ,
\nonumber \\ 
M_{\eta,-1/2} & = & M^{un}_{\eta,-1/2} +\sum_{\nu =1}^{\infty}\nu N_{\eta\nu}
\, ; \hspace{0.50cm} \delta N_{s\nu} (q_{j}) = N_{s\nu} (q_{j}) 
\hspace{0.15cm}{\rm for }\hspace{0.15cm}s\nu\neq s1 \, .
\label{DP}
\end{eqnarray}
Here $M^{un}_{\eta,-1/2}$ is the $\eta$-spin-projection $-1/2$ unbound $\eta$-spinon
number, Eq. (\ref{L-L}). 

The momentum spectrum is of first order in the $\beta$ momentum distribution
function deviations. Thus it is convenient to express it in terms of the corresponding
occupancies of the $\beta$ band bare momentum values $q_j$, as given here, rather than
of those of the $\beta$ band canonical momentum values ${\bar{q}}_j = {\bar{q}} (q_j)$. 
On the other hand, the excitation energy is provided in Eq. (\ref{DE}).

In this first scatter-less step, the pseudofermions acquire the excitation energy needed for the second and third steps.
The second step may give rise to a momentum contribution to be added to that given in Eq. (\ref{DP}).

\subsection{Pseudofermion scattering processes, dressed $S$ matrices, and phase shifts}
\label{PSPS}

In order to study the second and third processes of the ground-state
- excited-state transition, it is useful to express the
many-pseudofermion states and operators in terms of one-pseudofermion
states and operators, respectively. The PS energy and momentum eigenstates can be written as direct products of  
states. Those are generated by the occupancy configurations of each of the 
$\alpha\nu$ branches with finite pseudofermion occupancy. Moreover, the many-pseudofermion states generated
by occupancy configurations of each $\alpha\nu$ branch can be expressed 
as a direct product of $N_{a_{\beta}}$ one-pseudofermion states.
Each of the latter states refers to one discrete bare-momentum value $q_j$, where
$j=1,...,N_{a_{\beta}}$.

The Hamiltonian of the quantum problem described by the 1D Hubbard model in the PS,
whose energy spectrum is for the $\beta$ pseudofermion representation
given in Eq. (\ref{DE}), has within that representation a uniquely defined expression of the 
general form,
\begin{equation}
:\hat{H}: = \sum_{\beta}\sum_{j=1}^{N_{a_{\beta}}}\hat{H}_{\beta,{\bar{q}}_j} +
\sum_{\beta}\hat{H}_{\alpha} \, .
\label{Hexp}
\end{equation}
Here we have denoted the ground-state normal ordered Hamiltonian by $:\hat{H}:$,
$\hat{H}_{\beta,{\bar{q}}_j}$ is the one-pseudofermion Hamiltonian 
associated with excited-state $\beta$ pseudofermion or $\beta$ pseudofermion hole of canonical momentum ${\bar{q}}_j$,
and $\hat{H}_{\alpha}$ refers to the unbound $\eta$-spinons ($\alpha =\eta$) and unbound spinons 
($\alpha =s$) whose $SU(2)$ internal degrees of freedom are scatter-less. 

For each many-pseudofermion PS virtual state reached under the first step of the
transition from the ground state to the excited energy eigenstate,
the number of Hamiltonian terms, $\hat{H}_{\beta,{\bar{q}}_j}$, equals that of 
one-pseudofermion states of the virtual state. This number reads,
\begin{equation}
N_{a_c} + N_{a_{s1}} + \sum_{\beta\neq c ,s1} \theta (\vert\delta
N_{\beta}\vert)\, N_{a_{\beta}} \, . 
\label{DimSm}
\end{equation}
Here $\theta (x)=1$ for $x>0$ and $\theta (x)=0$ for $x= 0$. The numbers 
$N_{a_c} =N_{c}+N^h_{c}=N_a$, $N_{a_{s1}}=N_{s1}+N^h_{s1}$, and
$N_{a_{\beta}}=N_{\beta}+N^h_{\beta}$ refer to
the virtual state and corresponding excited energy eigenstate 
under consideration. The pseudofermion-hole number, $N^h_{\beta}$,
is provided in Eq. (\ref{N-h-an}) of Appendix \ref{pseudo-repre}.

The second scatter-less process generates the ``in" state. The one-pseudofermion states belonging 
to the many-pseudofermion ``in" state are the ``in" asymptote states of the 
pseudofermion scattering theory. The unitary operator ${\hat{S}}^{0}$ whose conjugate operator
$\left({\hat{S}}^{0}\right)^{\dag}$
generates the virtual-state - ``in"-state transition is of the form,
\begin{eqnarray}
{\hat{S}}^{0} & = & \prod_{\beta}{\hat{S}}^{0}_{\beta} 
\, ; \hspace{0.5cm} f^{\dag}_{q_j + Q^{0}_{\beta}/N_a,\beta} =
\left({\hat{S}}^{0}_{\beta}\right)^{\dag}f^{\dag}_{q_j,\beta}\,{\hat{S}}^{0}_{\beta} \, ,
\nonumber \\
{\hat{S}}^{0}_{\beta} & = & e^{i [Q^{0}_{\beta}/N_a]\hat{G}_{\beta}} 
= e^{\sum_{j=1}^{N_{a_{\beta}}}f^{\dag}_{q_{j} + Q^{0}_{\beta}/N_a,\beta}f_{q_{j},\beta}}  \, .
\label{S-0}
\end{eqnarray}
Here $f^{\dag}_{q_j,\beta}$ and $f_{q_j,\beta}$ are the $\beta$ pseudoparticle operators,
Eqs. (\ref{fc-q-x-1D}) and (\ref{fan-q-x-1D}), and the $Q^{0}_{\beta}/N_a$ shift Hermitian operator $\hat{G}_{\beta}$ 
is within a continuum bare-momentum representation given by,
\begin{equation}
\hat{G}_{\beta} = -i\sum_{q'}\left[{\partial\over\partial q'}f^{\dag}_{q',\beta}\right]f_{q',\beta} \, .
\label{G-beta}
\end{equation}

For simplicity and without loss in generality, consider that the many-pseudofermion ``in" state of the virtual 
state is a Bethe state. The virtual state can then be written as a $\beta$ pseudoparticle Slater determinant, 
\begin{equation}
\vert {\rm virt},l_{\rm r},l_{\eta s}^0,u\rangle =
\prod_{\beta}\prod_{j=1}^{N_{a_{\beta}}}f^{\dag}_{q_j,\beta}\vert 0_{\rm elec}\rangle \, ,
\label{virtual-state}
\end{equation}
where alike in Section \ref{model-ele-obj}, $\vert 0_{\rm elec}\rangle$ stands for the electron vacuum.
That vacuum remains invariant under the application of the unitary operator $\left({\hat{S}}^{0}_{\beta}\right)^{\dag}$.
Combination of that property with the operator $f^{\dag}_{q_j,\beta}$ transformation law under that unitary operator,
Eq. (\ref{S-0}), one finds that,
\begin{equation}
\vert {\rm in},l_{\rm r},l_{\eta s}^0,u\rangle=\prod_{\beta}\left({\hat{S}}^{0}_{\beta}\right)^{\dag}\vert {\rm virt},l_{\rm r},l_{\eta s}^0,u\rangle =
\prod_{\beta}\prod_{j=1}^{N_{a_{\beta}}}f^{\dag}_{q_j+ Q^{0}_{\beta}/N_a,\beta}\vert 0_{\rm elec}\rangle \, .
\label{in-state}
\end{equation}
Hence application of the unitary operator $\left({\hat{S}}^{0}\right)^{\dag}=\prod_{\beta}\left({\hat{S}}^{0}_{\beta}\right)^{\dag}$
onto the many-pseudofermion virtual state gives rise to the ``in"-state. Under
that process, the virtual state one-pseudofermion states discrete bare-momentum values $q_j$ 
are shifted to the excited-state discrete bare-momentum value $q_j+Q_{\beta}^0/N_a$, where $Q_{\beta}^0$ is given in 
Eq. (\ref{pican}). (Note that $Q_{\beta}^0$ may vanish, as given in that equation.) 

This second step may add a finite momentum to that given in Eq. (\ref{DP}) such
that the total excitation momentum reads,
\begin{equation}
\delta P_{l_{\rm r},l_{\eta s}} = \delta P_{l_{\rm r},l_{\eta s}}^0 + \iota\,2k_F\,{Q_{c}^0\over\pi}
+ \iota'\,k_{F\downarrow}\,{Q_{s1}^0\over\pi} 
\, , \hspace{0.25cm} \iota, \iota' = \pm 1 \, .
\label{P-Tot}
\end{equation}

Finally, the third step consists of a set of two-pseudofermion
scattering events. It corresponds to the ``in"-state - ``out"-state transition,
where the latter state is the PS excited energy eigenstate under
consideration. The generator of that transition is the conjugate of the following unitary operator,
\begin{equation}
{\hat{S}}^{\Phi} = \prod_{\beta}{\hat{S}}^{\Phi}_{\beta} \, ,
\label{Sphi}
\end{equation}
where ${\hat{S}}^{\Phi}_{\beta}$ is the $\beta$ pseudoparticle - $\beta$ pseudofermion
unitary operator, Eq. (\ref{P-P-UT}). 

The one-pseudofermion states belonging to the 
many-pseudofermion ``out" state are the ``out" asymptote pseudofermion
scattering states. Application of the unitary operator $\left({\hat{S}}^{\Phi}\right)^{\dag}=\prod_{\beta}\left({\hat{S}}^{\Phi}_{\beta}\right)^{\dag}$,
Eq. (\ref{P-P-UT}), onto the many-pseudofermion ``in"
state, shifts its one-pseudofermion states discrete bare-momentum values $q_j+Q_{\beta}^0/N_a$ to the
``out"-state discrete canonical-momentum values 
$q_j+Q_{\beta} (q_j)/N_a$. It follows that the generator of the overall
virtual-state - ``out"-state transition is the unitary operator $\left({\hat{S}}_T\right)^{\dag}$
whose conjugate reads,
\begin{eqnarray}
{\hat{S}}_T & \equiv & {\hat{S}}^{\Phi}{\hat{S}}^{0} = \prod_{\beta}
{\hat{S}}_{\beta}   \, ; \hspace{0.5cm}
f^{\dag}_{q_j + Q_{\beta}(q_j)/N_a,\beta} =
\left({\hat{S}}_{\beta}\right)^{\dag}f^{\dag}_{q_j,\beta}\,{\hat{S}}_{\beta} \, ,
\nonumber \\
{\hat{S}}_{\beta} & = & e^{\sum_{j=1}^{N_{a_{\beta}}}f^{\dag}_{q_{j} + Q_{\beta}(q_j)/N_a,\beta}f_{q_{j},\beta}} 
\, ; \hspace{0.50cm}
\left({\hat{S}}_{\beta} \right)^{\dag} = 
e^{\sum_{j=1}^{N_{a_{\beta}}}f^{\dag}_{q_{j} - Q_{\beta}(q_j)/N_a,\beta}f_{q_{j},\beta}} \, .
\label{Sope}
\end{eqnarray}
Thus application of the unitary operator $\left({\hat{S}}_T\right)^{\dag}=\prod_{\beta}\left({\hat{S}}_{\beta} \right)^{\dag}$ onto the 
corresponding many-pseudofermion virtual 
state, shifts its one-pseudofermion states discrete bare-momentum values $q_j$ directly into the ``out"-state 
discrete canonical-momentum values $q_j+Q_{\beta} (q_j)/N_a$,
\begin{equation}
\vert {\rm out},l_{\rm r},l_{\eta s}^0,u\rangle=\prod_{\beta}\left({\hat{S}}_{\beta}\right)^{\dag}\vert {\rm virt},l_{\rm r},l_{\eta s}^0,u\rangle =
\prod_{\beta}\prod_{j=1}^{N_{a_{\beta}}}f^{\dag}_{q_j+ Q_{\beta} (q_j),\beta}\vert 0_{\rm elec}\rangle 
= \prod_{\beta}\prod_{j=1}^{N_{a_{\beta}}}f^{\dag}_{{\bar{q}}_j,\beta}\vert 0_{\rm elec}\rangle \, .
\label{out-state}
\end{equation}

The ``in" state and ``out" state are different representations of the
same PS excited energy eigenstate. Specifically, they refer to
the alternative $\beta$ pseudoparticle and $\beta$ pseudofermion representations of that state.
Consistent, the canonical-momentum shift $Q^{\Phi}_{\beta}(q_j)/N_a$ does not contribute
to the physical momentum, Eq. (\ref{P-Tot}), and
the ``in" state and ``out" state can be shown to differ by a mere overall phase factor,
\begin{eqnarray}
\vert {\rm out},l_{\rm r},l_{\eta s}^0,u\rangle & = & S^{\Phi}_T\,\vert {\rm in},l_{\rm r},l_{\eta s}^0,u\rangle \, ,
\nonumber \\
S^{\Phi}_T & = & e^{i2\delta^{\Phi}_T} \, ; \hspace{0.5cm}
\delta^{\Phi}_T = \sum_{\beta}\sum_{j=1}^{N_{a_{\beta}}} 
Q^{\Phi}_{\beta}(q_j)/2 \, .
\label{S-Phi}
\end{eqnarray}

That the one-pseudofermion states of the 
many-pseudofermion ``in" state and ``out" state are the ``in" and ``out" asymptote 
pseudofermion scattering states, respectively, implies that the one-pseudofermion 
Hamiltonian $\hat{H}_{\beta ,{\bar{q}}_j}$ plays the role of the unperturbed 
Hamiltonian $\hat{H}_0$ of the spin-less one-particle nonrelativistic scattering theory \cite{Taylor}. 
It follows that the matrix elements between one-pseudofermion states
of ${\hat{S}}^{\Phi}_{\beta}$ are
diagonal. Therefore, these operators are fully defined by the set of their eigenvalues
of such states. The same applies to the generator ${\hat{S}}^{\Phi}$, Eq. (\ref{Sphi}). 
The matrix elements of that generator between many-pseudofermion ``in" states are 
also diagonal and thus it is fully defined by the set of 
its eigenvalues of such states. 

The unitarity of the operators ${\hat{S}}^{\Phi}_{\beta}$ and ${\hat{S}}_{\beta}$ implies that
each of their eigenvalues has modulus one. It can thus be written as the exponent of a 
purely imaginary number. In the case of a $\beta$ one-pseudofermion state of bare
momentum $q_j$, such eigenvalues are given by,
\begin{eqnarray}
S^{\Phi}_{\beta} (q_j) & = & e^{iQ^{\Phi}_{\beta}(q_j)} =
\prod_{\beta'}\,\prod_{j'=1}^{N_{a_{\beta'}}}\,S_{\beta ,\beta'} (q_j, q_{j'}) \, ;
\hspace{0.25cm} j=1,..., N_{a_{\beta}} \, ,
\nonumber \\
S_{\beta} (q_j) & = & e^{iQ_{\beta}(q_j)} =
e^{i\,Q_{\beta}^0} S^{\Phi}_{\beta} (q_j) = e^{i\,Q_{\beta}^0}\prod_{\beta'}\,\prod_{j'=1}^{N_{a_{\beta'}}}\,S_{\beta ,\beta'}  (q_j, q_{j'}) \, ;
\hspace{0.25cm} j=1,..., N_{a_{\beta}} \, , 
\label{San}
\end{eqnarray}
respectively.
Here $Q^{\Phi}_{\beta}(q_j)$ and $Q_{\beta}(q_j)$ are the functionals, Eq. (\ref{qcan1j}) and Eq. (\ref{Qcan1j}), respectively. 
By use of the functional $Q^{\Phi}_{\beta}(q_j)$ expression, Eq. (\ref{qcan1j}), we find that
the quantity $S_{\beta ,\beta'} (q_j, q_{j'})$ in Eq. (\ref{San}) reads,
\begin{equation}
S_{\beta ,\beta'} (q_j, q_{j'}) =
e^{\pm i2\pi\,\Phi_{\beta,\beta'}(q_j,q_{j'})\, \delta N_{\beta'}(q_{j'})} \, , 
\label{Sanan}
\end{equation}
where the functions $\pi\,\Phi_{\beta,\beta'}(q_j,q_{j'})$ are given in Eq. (\ref{Phi-barPhi}).

The effects produced by a ground-state - excited-state transition
beyond the ground-state - virtual-state transition occupancy configuration changes
are those of interest for the scattering theory. Except for the latter changes,
the only effect of, under such a transition, moving the 
$\beta$ pseudofermion or $\beta$ pseudofermion hole of virtual-state canonical-momentum 
${\bar{q}}_j=q_j$ once around the length $L$ lattice ring is that its wave function acquires 
the overall phase factor $S_{\beta} (q_j)$, Eq. (\ref{San}).
This property is consistent with the lack of interaction energy terms in the spectrum of the 
$\beta$ pseudofermions, Eq. (\ref{DE}). The procedure of moving the 
$\beta$ pseudofermion or $\beta$ pseudofermion hole once around the length $L$ lattice ring 
refers to a method to derive the corresponding dressed $S$ matrix. It is
precisely the overall phase factor $S_{\beta} (q_j)$, Eq. (\ref{San}),
acquired by its wave function.

The phase factor $S_{\beta ,\beta'} (q_j, q_{j'})$, Eq. (\ref{Sanan}), 
in the wave function of the $\beta$ pseudofermion or $\beta$ pseudofermion hole of bare-momentum
$q_j$ results from an elementary two-pseudofermion 
zero-momentum forward-scattering event. Its scattering
center is a $\beta'$ pseudofermion ($\delta N_{\beta'}(q_{j'})=1$)
or $\beta'$ pseudofermion hole ($\delta N_{\beta'}(q_{j'})=-1$) created
under the ground-state - excited-state transition. The third step 
of that transition involves a well-defined set of elementary two-pseudofermion 
scattering events. Under those, all $\beta$ pseudofermions and $\beta$ 
pseudofermion holes of bare-momentum $q_j+Q_{\beta}^0/N_a$ of the ``in"
state play the role of scatterers. This leads to the overall scattering phase factor 
$S^{\Phi}_{\beta} (q_j)$ in their wave function, Eq. (\ref{San}). 
On the other hand, the $\beta'$ pseudofermions or $\beta'$
pseudofermion holes of bare momentum $q_{j'}+Q_{\beta}^0/N_a$ created 
under the ground-state - ``in"-state transition play the role of scattering centers. This is confirmed 
by noting that $S_{\beta ,\beta'} (q_j,q_{j'})=1$ for $\delta N_{\beta'}(q_{j'}) =0$. 
(The latter elementary objects play as well the role of scatterers.) Thus, out of the scatterers whose 
number equals that of the one-pseudofermion states, Eq. (\ref{DimSm}), the scattering centers are only those whose 
bare-momentum distribution-function deviation is finite. 

That the many-pseudofermion ``in" and ``out" states, which are a
direct product of one-pseudofermion ``in" and ``out" asymptote pseudofermion 
scattering states, respectively, are PS excited energy eigenstates is behind the 
validity of the pseudofermion scattering theory. 
Indeed, the validity of any scattering theory requires that the excited states
associated with the asymptotic one-particle scattering states have a well-defined energy. 
For the present quantum problem, this requirement is fulfilled provided that
the excited states associated with the one-particle scattering states 
are model energy eigenstates. Such a requirement is obeyed by {\it all} the excited
states associated with the one-particle scattering states of the pseudofermion scattering theory.

The following properties play an important role in the pseudofermion scattering
theory:
\begin{enumerate}

\item
The elementary two-pseudofermion scattering processes associated with the phase
factors, Eq. (\ref{Sanan}), conserve the total energy and total momentum. This stems from
the occurrence of an infinite number of conservation laws \cite{CM-86,CM,Prosen}, which are associated
with the model integrability \cite{Lieb,Takahashi,Woy,Martins} 
being explicit in the present pseudofermion scattering theory. As a result, its scatterers and scattering 
centers only undergo zero-momentum forward scattering.

\item
That the elementary two-pseudofermion scattering processes are of zero-momentum 
forward-scattering type also implies that they conserve the individual ``in" asymptote $\beta$ 
pseudofermion bare momentum value $q_j+Q_{\beta}^0/N_a$ and energy. 
(The additional canonical-momentum scattering phase-shift term, $Q^{\Phi}_{\beta} (q_j)/N_a$,
does not contribute to the physical momentum, Eq. (\ref{P-Tot}).)

\item
These processes also conserve the $\beta$ branch, usually called {\it channel} in
the scattering language \cite{Taylor}.

\item
The scattering amplitude does not connect quantum objects with different $\eta$ spin or
spin. (All such objects have no internal degrees of freedom such as $\eta$-spin or spin.)

\item
For each $\beta$ pseudofermion or $\beta$ pseudofermion hole of virtual-state bare-momentum
$q_j$, the dressed $S$ matrix associated with the ground-state - excited-state transition 
is simply the phase factor $S_{\beta} (q_j)$, Eq. (\ref{San}).
\end{enumerate}

The one-$\beta$-pseudofermion or one-$\beta$-pseudofermion-hole phase factor $S^{\Phi}_{\beta} (q_j)$ 
of the present 1D quantum problem corresponds to the usual one-particle phase factor $s_l (E)$ of
similar three-dimensional quantum problems. The latter depends on the energy $E$ and 
angular-momentum quantum numbers $l$ and $m$. (See, for example, Eq. (6.9) of Ref. \cite{Taylor}.)
In the present 1D case, the energy $E$ and the quantum numbers $l$ and $m$ are replaced 
by the bare-momentum $q_j$ in the phase factor $S^{\Phi}_{\beta} (q_j)$. The $\beta$ pseudofermion 
or $\beta$-pseudofermion hole energy is uniquely defined by the absolute bare-momentum value $\vert q_j\vert$.
In 1D the sign of $q_j$ corresponds to the three-dimensional 
angular-momentum quantum numbers. Another difference is that $s_l (E)$ 
is associated with a single scattering event. Here, $S^{\Phi}_{\beta} (q_j)$ 
results in general from several scattering events. Each of such events 
corresponds to a well defined factor $S_{\beta ,\beta'} (q_j, q_{j'})$,
Eq. (\ref{Sanan}), in the $S^{\Phi}_{\beta} (q_j)$ expression, Eq. (\ref{San}). There are as many of such factors as 
$\beta'$ pseudofermion and $\beta'$ pseudofermion hole scattering centers created 
under the transition to the virtual state and corresponding
excited energy eigenstate under consideration. 
The factor $2$ in the phase factor of Eq. (6.9) 
of Ref. \cite{Taylor} corresponds to the phase-shift definition of the standard 
nonrelativistic scattering theory for spin-less particles. We use in general here such 
a definition. As discussed below in Section \ref{PhSh-def},
it introduces the overall scattering phase shift 
$\delta^{\Phi}_{\beta} (q_j)= Q^{\Phi}_{\beta} (q_j)/2$ and
overall phase shift $\delta_{\beta} (q_j)= Q_{\beta} (q_j)/2$. 
However, if instead we insert a factor $1$, we would have an 
overall scattering phase shift $Q^{\Phi}_{\beta} (q_j)$ and
overall phase shift $Q_{\beta} (q_j)$. (That is the
phase-shift definition used in Refs. \cite{Natan,S-0,S}.)

The factorization of the BA bare $S$ matrix for the original spin-$1/2$ electrons is
associated with the so called Yang-Baxter equation (YBE) \cite{Natan}. On
the other hand, the factorization of the $\beta$ pseudofermion or $\beta$ pseudofermion-hole dressed $S$ matrix
$S_{\beta} (q_j)$, Eq. (\ref{San}), in terms of the elementary 
two-pseudofermion $S$ matrices $S_{\beta ,\beta'} (q_j, q_{j'})$, 
Eq. (\ref{Sanan}), is {\it commutative}. Such commutativity is stronger than
the symmetry associated with the YBE. This results from the $c$, $\eta\nu$, and $s\nu$ pseudofermions
and $c$ and $s1$ pseudofermion holes,
which are the scatterers and scattering centers, having no internal degrees of freedom such as $\eta$-spin or spin.

The BA bare $S$ matrix refers to spin-$1/2$ electrons. Therefore its form
reflects the spin-$1/2$ $SU(2)$ symmetry of its scatterers and
scattering centers. That of the $\beta$ pseudofermions
refers to neutral particles. This is in spite of the $\eta$-spinons and spinons
carrying $\eta$-spin $1/2$ and spin $1/2$, respectively. On the other hand, the $M^{bo}_{\eta}$ anti-bound
$\eta$-spinons (and $M^{bo}_{s}$ bound spinons) have $\eta$-spin $1/2$ (and spin $1/2$) but are
anti-bound (and bound) within composite neutral objects, which play the role of scatterers and scattering centers. 
On the other hand, the $\eta$-spin (and spin) internal $SU(2)$ degrees of freedom of the
$\eta$-spin-$1/2$ unbound $\eta$-spinons and spin-$1/2$ unbound spinons have a scattering-less character.

Indeed, as mentioned above, the momentum values of the $\eta$-spin-$1/2$ unbound $\eta$-spinons and spin-$1/2$ 
unbound spinons remain unchanged under the transition from an ``in" state to the corresponding
``out" state (excited energy eigenstate). This confirms that as far as their $SU(2)$ internal degrees of freedom is concerned such objects 
are not scatterers. Neither do the $\beta$ pseudofermions scatterer off on them, so that they are not scattering centers. 
That is consistent with the unbound $\eta$-spinons (and unbound spinons)
playing the passive role of unoccupied sites of both the $\eta$-spin (and spin) effective lattice and the corresponding
$\eta\nu$ (and $s\nu$) effective lattices. 

The $\beta =c,\alpha\nu$ effective lattice elementary-object motion can alternatively be described in terms of $\beta$ pseudofermions 
or $\beta$ pseudoermion holes. Both the $\beta$ pseudofermions and $\beta$ pseudofermion 
holes are scatterers of the pseudofermion scattering theory. When created under transitions to
excited states, they play as well the role of scattering centers. Under transitions to excited states,
only $c$ pseudofermion holes and $s1$ pseudofermion holes are created. They 
play a major role in the transport of charge and spin, respectively \cite{paper-I}.
According to Eq. (\ref{N-h-an}) of Appendix \ref{pseudo-repre}, the numbers of 
$c$ pseudofermion holes and $s1$ pseudofermion holes can be written as,
\begin{equation}
N^h_{c} = M_{\eta} = M_{\eta}^{un}+\sum_{\nu=1}^{\infty}2\nu N_{\eta\nu}
\, ; \hspace{0.50cm}
N^h_{s1} = M_{s}^{un}+\sum_{\nu=1}^{\infty}2(\nu-1)N_{s\nu} \, ,
\label{N-h-c-s1}
\end{equation}
respectively. The studies of Ref. \cite{paper-I} reveal that, out of the $N^h_{c}$ $c$ pseudofermion holes 
(and $N^h_{s1}$ $s1$ pseudofermion holes), Eq. (\ref{N-h-c-s1}),
the $M_{\eta}$ unbound $+1/2$ $\eta$-spinons (and $M_{s}$ unbound $+1/2$ spinons)
use $M_{\eta}$ $c$ pseudofermion holes (and $M_{s}$ $s1$ pseudofermion holes) as hosts to couple to $\eta$-spin (and spin) proves.
As justified in that reference, they are {\it host shadows} of the corresponding $M_{\eta}$ 
unbound $+1/2$ $\eta$-spinons (and $M_{s}$ unbound $+1/2$ spinons.)
Indeed, they carry their $x_3$-component $\eta$-spin (and spin) $U(1)$ currents.
Such unbound $\eta$-spinon and unbound spinon degrees of freedom
are associated with the $\eta$-spin and spin $U(1)$ symmetry algebra state representations within
the $\eta$-spin and spin $SU(2)$ symmetry algebras, respectively. 

On the other hand, the remaining $\sum_{\nu=1}^{\infty}2\nu N_{\eta\nu}$ $c$ pseudofermion holes (and $\sum_{\nu=1}^{\infty}2(\nu-1)N_{s\nu}$ 
$s1$ pseudofermion holes), out of the $N^h_{c}$ $c$ pseudofermion holes 
(and $N^h_{s1}$ $s1$ pseudofermion holes), Eq. (\ref{N-h-c-s1}), are found in
Ref.  \cite{paper-I} to be neutral shadows of the excited-states $N_{\eta\nu}$ $\eta\nu$ pseudofermions
(and $N_{s\nu}$ $s\nu$ pseudofermions with $\nu>1$ spinon pairs.) By neutral shadows 
it is meant that the virtual elementary $\eta$-spin (and spin) currents carried by
the remaining $\sum_{\nu=1}^{\infty}2\nu N_{\eta\nu}$ $c$ pseudofermion holes (and $\sum_{\nu=1}^{\infty}2(\nu-1)N_{s\nu}$ 
$s1$ pseudofermion holes) are exactly cancelled by those of the corresponding $N_{\eta\nu}$ $\eta\nu$ pseudofermions
(and $N_{s\nu}$ $s\nu$ pseudofermions with $\nu>1$ spin on pairs.) That is consistent with a $2\nu$-$\eta$-spinon $\eta\nu$ 
pseudofermion (and $2\nu$-spinon $s\nu$ pseudofermion) being a $\eta$-spin-neutral (and spin-neutral) composite object. 
\begin{figure}
\subfigure{\includegraphics[width=7cm,height=7cm]{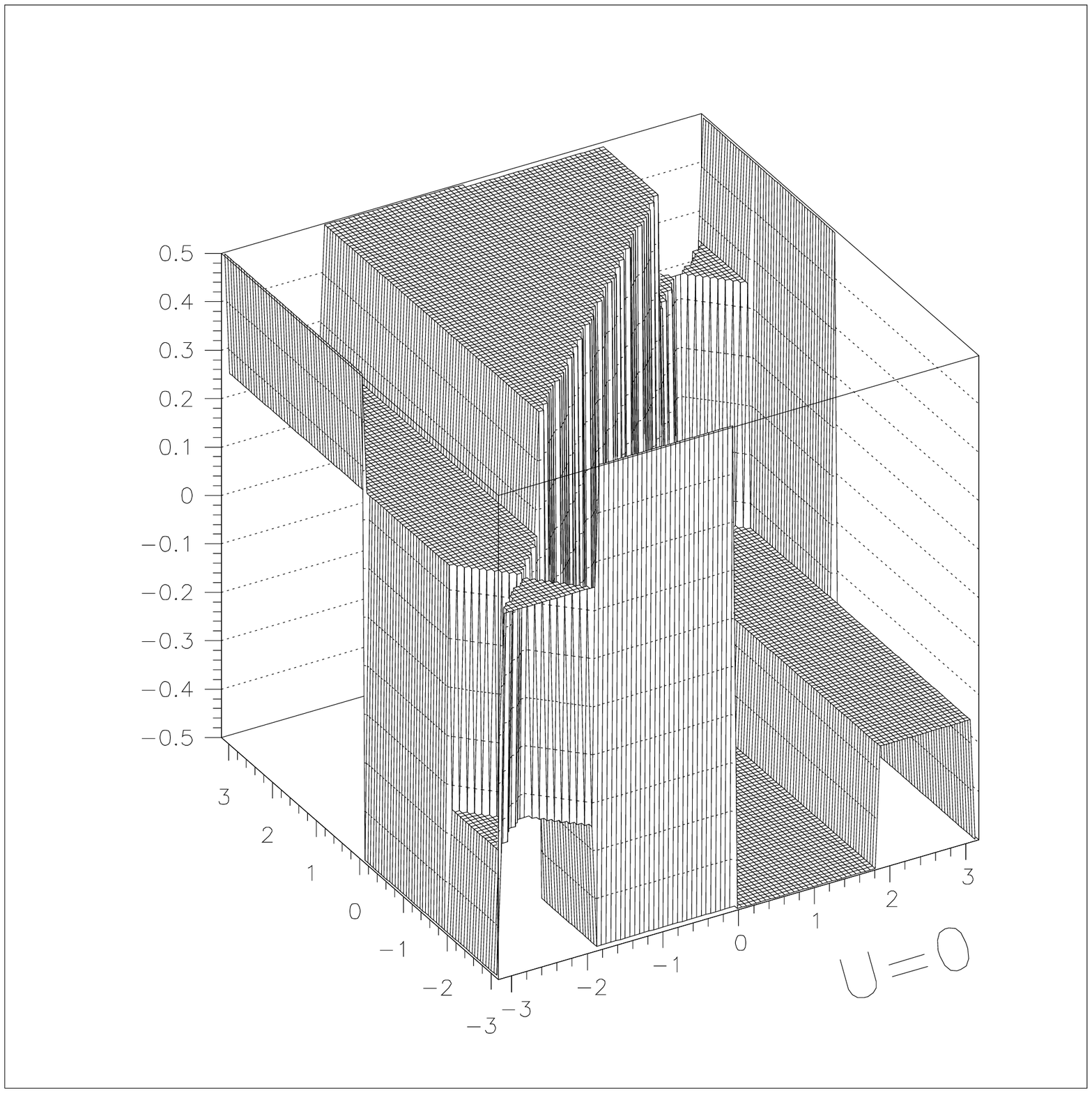}}
\subfigure{\includegraphics[width=7cm,height=7cm]{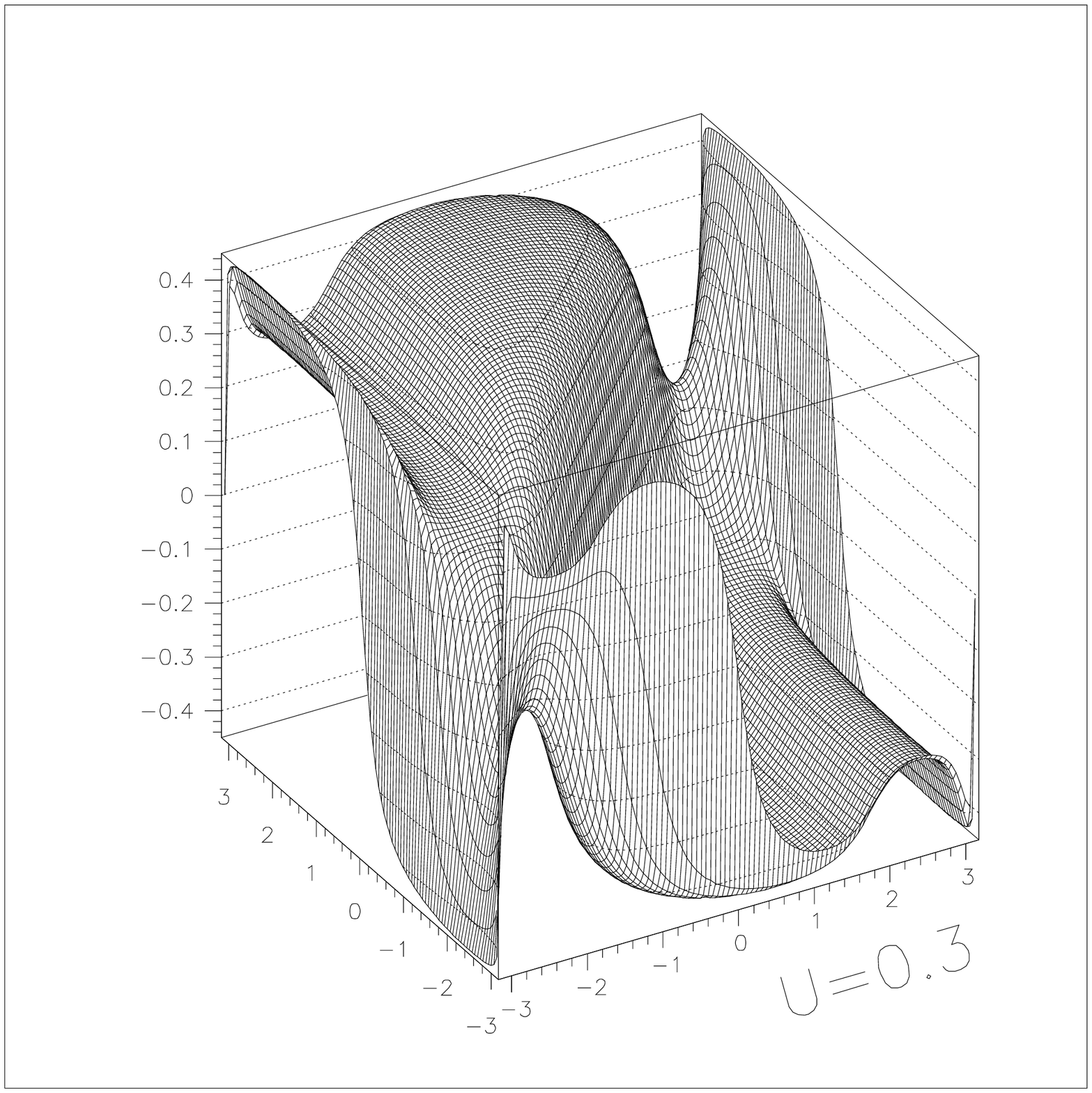}}
\subfigure{\includegraphics[width=7cm,height=7cm]{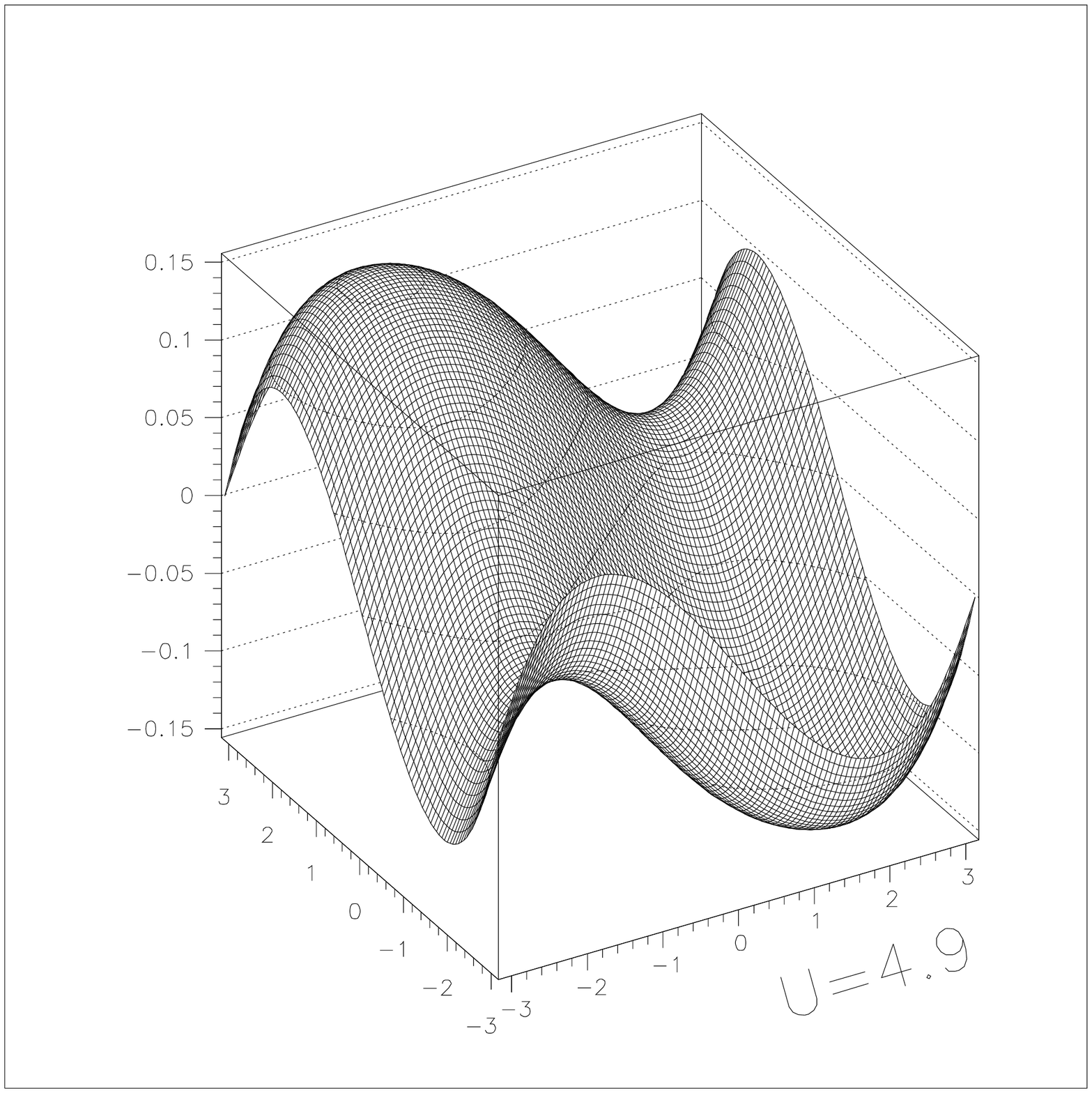}}
\subfigure{\includegraphics[width=7cm,height=7cm]{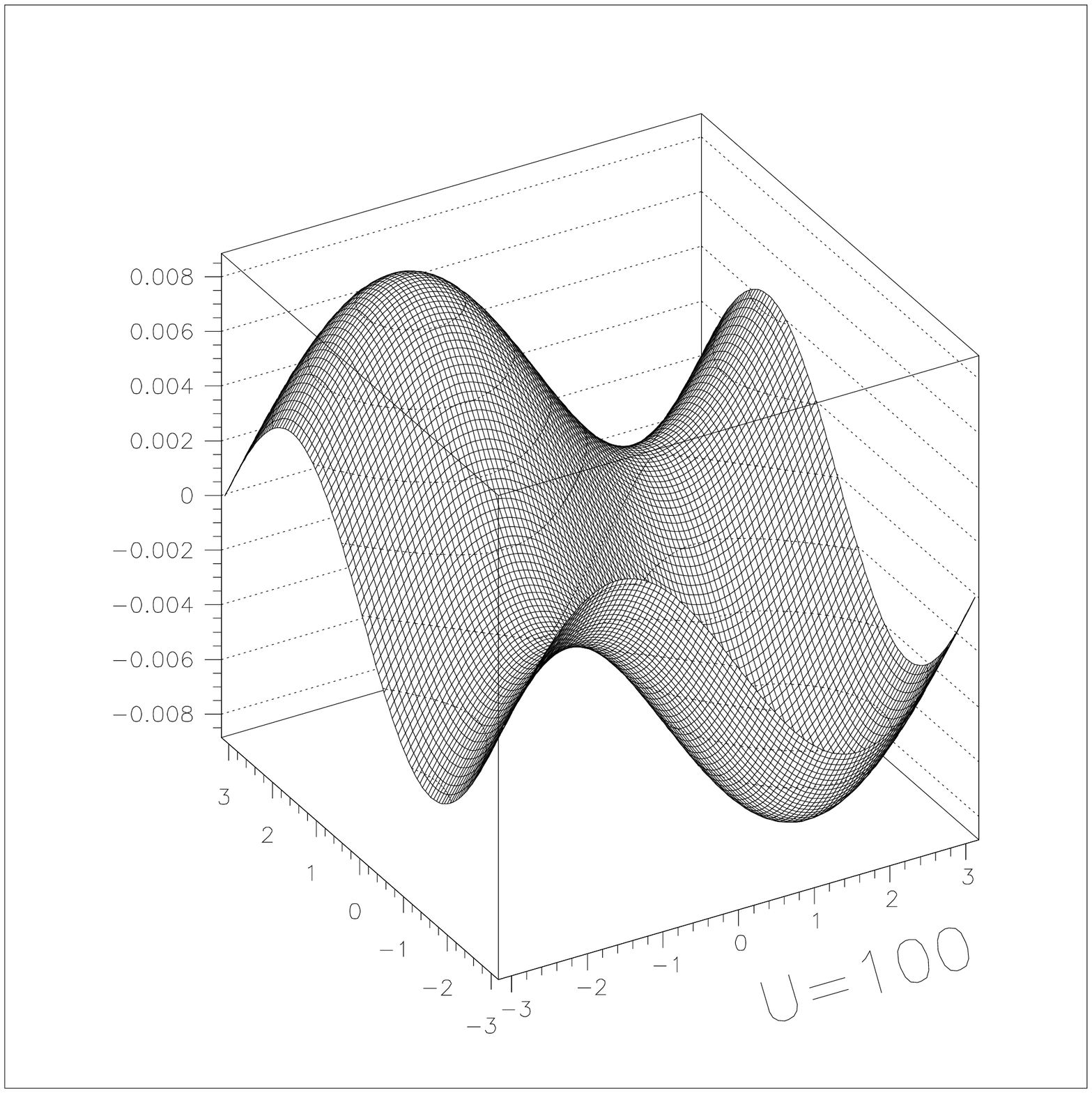}} 
\caption{The elementary two-pseudofermion phase shift $\pi\,\Phi_{c,c}(q,q')$ in units of $\pi$ as a function of
$q$ and $q'$ for $n=0.59$, $m=0$, and (a) $U/t\rightarrow 0$, (b) $U/t=0.3$, (c) $U/t=4.9$, and (d)
$U/t= 100$. The scatterer and scattering-center bare-momentum values $q$ and $q'$, respectively, correspond to the right and left 
axis of the figures, respectively.}
\label{F1}
\end{figure}

The dressed $S$ matrix commutative factorization is that consistent with the form of the 
$\beta$ pseudofermion canonical-momentum occupancy configurations that
describe the PS excited energy eigenstates. Such canonical-momentum occupancy configurations are superpositions of 
local occupancy configurations with the same number of $\beta$ pseudofermions. Those elementary
objects have different positions in each configuration. Hence the number of $\beta$ pseudofermions belonging to $\beta$ branches with
finite occupancy in the virtual state is the same for all occupancy configurations. The relative position of these quantum objects is different in
each such a configuration. Consider that under a specific ground-state - excited-energy-eigenstate
transition a $\beta$ pseudofermion or $\beta$ pseudofermion hole moves once around the lattice ring. It then 
scatters the same $\beta'=c,\alpha\nu$ pseudofermion or $\beta'=c,s1$ pseudofermion-hole scattering centers, but in
different order for different occupancy configurations. However, it is required that the
phase factor $e^{iQ_{\beta}(q_j)}$ acquired by the $\beta$ pseudofermion 
or $\beta$ pseudofermion hole should be the same, independently of that order. This is consistent with the commutativity 
of the dressed $S$-matrix factors $S_{\beta ,\beta'} (q_j, q_{j'})$ in the overall
dressed $S$ matrix $S_{\beta} (q_j)$, Eq. (\ref{San}). 
Such a commutativity follows from the elementary $S$ matrices $S_{\beta,\beta'} (q_j, q_{j'})$, Eq. (\ref{Sanan}), 
being simple phase factors, rather than matrices of dimension larger than one. 

\section{The pseudofermion phase shifts}
\label{Phaseshifts}

In this section we study the $\beta$ pseudofermion phase shifts associated with the
dressed $S$ matrix introduced above in Section \ref{Pseudofermion}. The effects 
of the pseudofermion transformation laws under the electron - rotated-electron
unitary transformation on such phase shifts and corresponding scattering properties and the relation of those
to the PDT are issues also addressed in this section. 

\subsection{Phase-shift definition}
\label{PhSh-def}

As above, our analysis refers to periodic boundary conditions and $N_a\gg1$.
The $c$ effective lattice considered in Ref. \cite{paper-I} equals the original lattice. On the other hand, the spacing,
\begin{equation}
a_{\alpha\nu} = {L\over N_{a_{\alpha\nu}}} = 
{N_a\over N_{a_{\alpha\nu}}}\, a \, , \hspace{0.25cm} 
N_{a_{\alpha\nu}} \geq 1 \, ,
\label{a-a-nu}
\end{equation}
of the remaining $\beta =\alpha\nu$ effective lattices also considered in that reference 
is for $n\neq 1$ and $m\neq 0$ larger than that of the original lattice. According to its definition, Eq. (\ref{a-a-nu}),
the corresponding length, $L =N_a\,a =N_{a_{\beta}}\,a_{\beta}$, equals
that the latter lattice. (In general we use in this paper units of lattice constant $a$, for which $N_a=L$.) 

Depending on the asymptote coordinate choice, there are two possible definitions of the $\beta$ pseudofermion or
$\beta$ pseudofermion hole phase shifts.
Both are associated with the dressed $S$ matrix $S_{\beta} (q_j)$, Eq. (\ref{San}). Indeed,
the uniquely defined quantity is that dressed $S$ matrix. The two choices of asymptote coordinates for the $\beta$ pseudofermion or
$\beta$ pseudofermion hole correspond to $x\in [-L/2,+L/2]$ and $x\in [0,+L]$, respectively.

If when moving around the lattice ring, the $\beta$ pseudofermion or $\beta$ pseudofermion hole departures
from the point $x=-L/2$ and arrives to $x=L/2$, one finds that,
\begin{equation}
\lim_{x\rightarrow L/2}\,\bar{q}\,x = q\,L/2 +Q_{\beta}^0/2 + Q^{\Phi}_{\beta}
(q)/2 = q\,L/2 + \delta_{\beta} (q) \, , \label{qr2}
\end{equation}
where
\begin{equation}
\delta_{\beta} (q) = Q_{\beta} (q)/2 = Q_{\beta}^0/2 + Q^{\Phi}_{\beta}(q)/2\, . 
\label{danq}
\end{equation}
For this asymptote coordinate choice, $\delta_{\beta} (q)$ is the overall $\beta$
pseudofermion or $\beta$ pseudofermion-hole phase shift. 
Moreover, from analysis of Eqs. (\ref{qcan1j}) and (\ref{Qcan1j}) it follows that
$\pi\,\Phi_{\beta,\beta'}(q_j,q_{j'})$ is an elementary two-pseudofermion phase shift. 
This phase-shift definition corresponds to that used in standard quantum non-relativistic scattering 
theory \cite{Taylor}. It is such that the dressed $S$ matrix $S_{\beta} (q_j)$, Eq. (\ref{San}), 
can be expressed as,
\begin{equation}
S_{\beta} (q_j) = e^{i2\delta_{\beta}(q_j)} \, , \hspace{0.5cm} j=1,...,
N_{a_{\beta}} \, . \label{San3}
\end{equation}
The factor $2$ appearing here in the exponential argument corresponds to
the usual form of the $S$ matrix for that theory. This phase-shift definition is consistent 
with an exact Theorem due to Fumi \cite{Mahan}. Within that definition, Eq. (\ref{danq}),
$Q_{\beta}^0/2=0,\mp\pi/2$ corresponds to the scatter-less term $-l\pi/2$ of the
three-dimensional partial-wave problem of orbital angular momentum $l$ \cite{Taylor,Mahan}.
Although the orbital angular momentum vanishes in 1D, the scatter-less phase shift,
Eq. (\ref{pican}), plays a similar role.
\begin{figure}
\subfigure{\includegraphics[width=7cm,height=7cm]{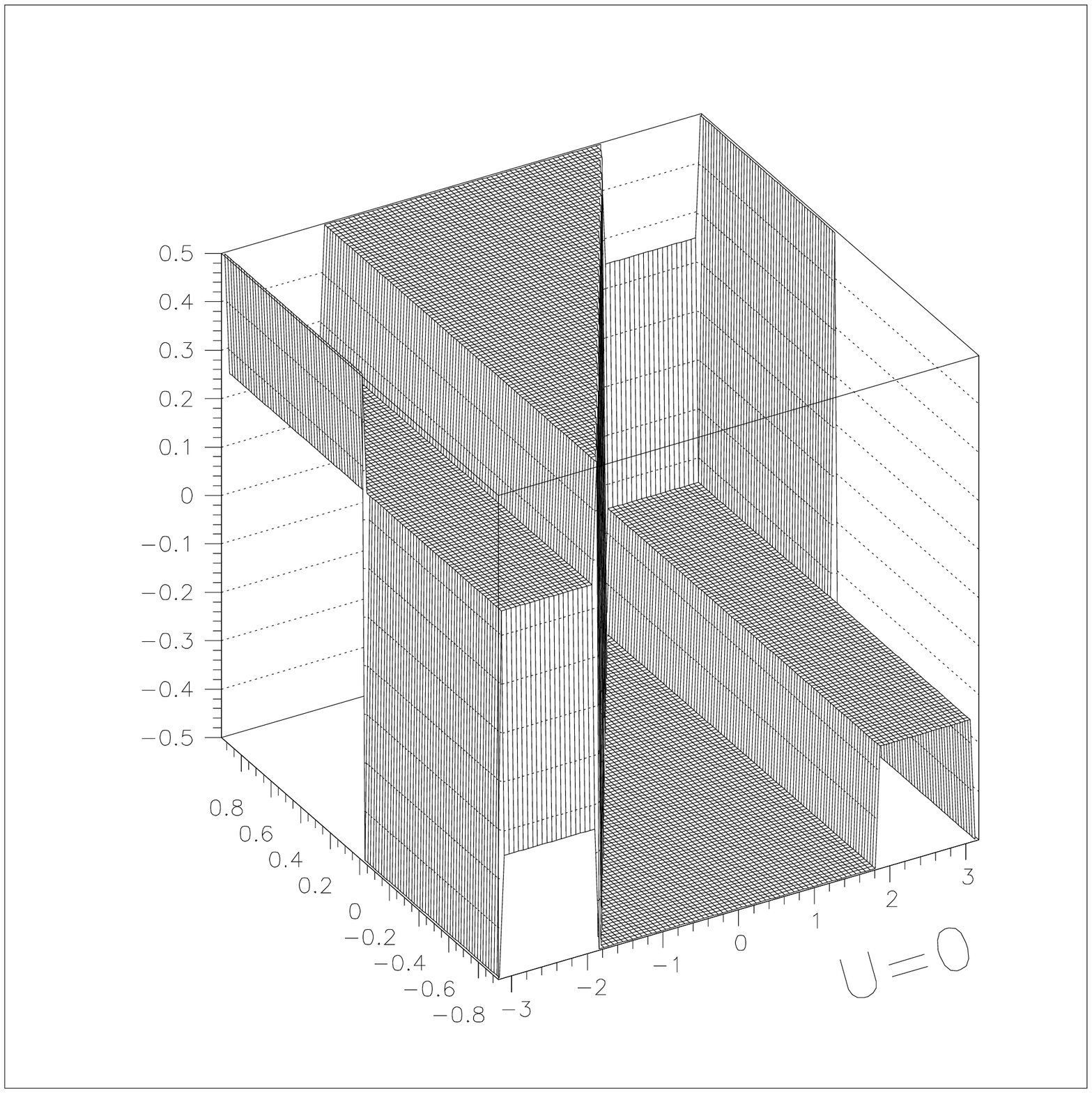}}
\subfigure{\includegraphics[width=7cm,height=7cm]{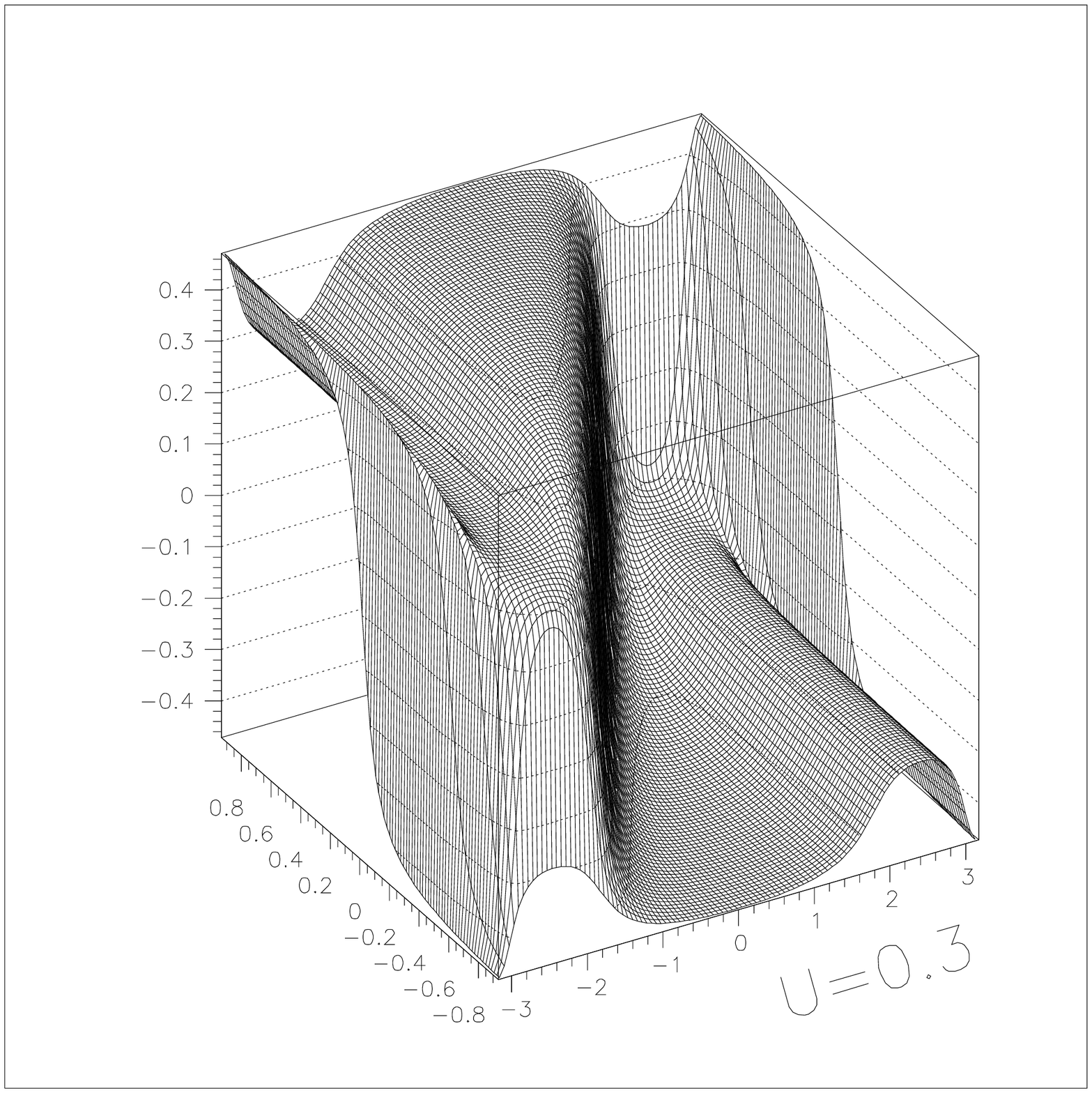}}
\subfigure{\includegraphics[width=7cm,height=7cm]{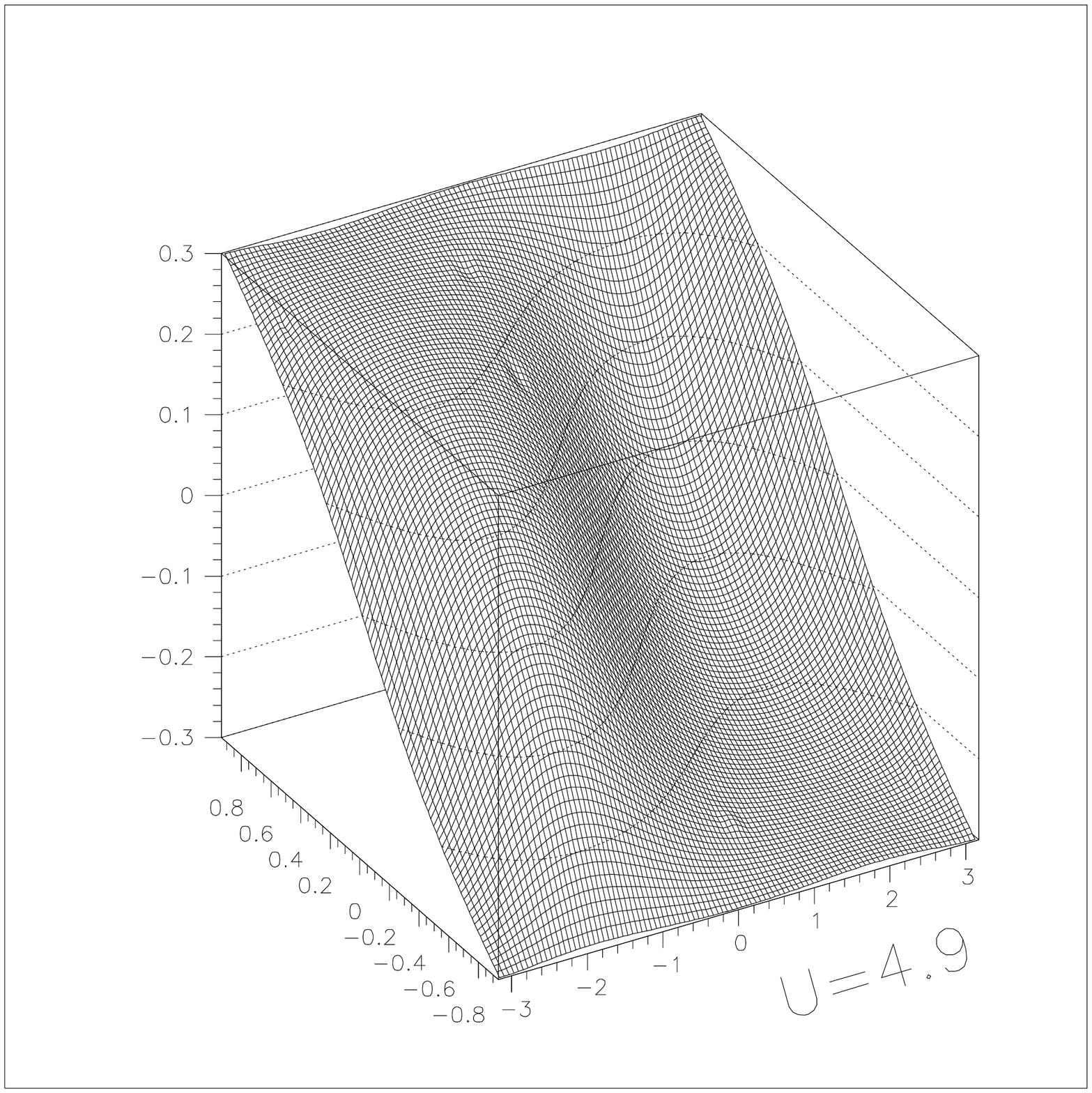}}
\subfigure{\includegraphics[width=7cm,height=7cm]{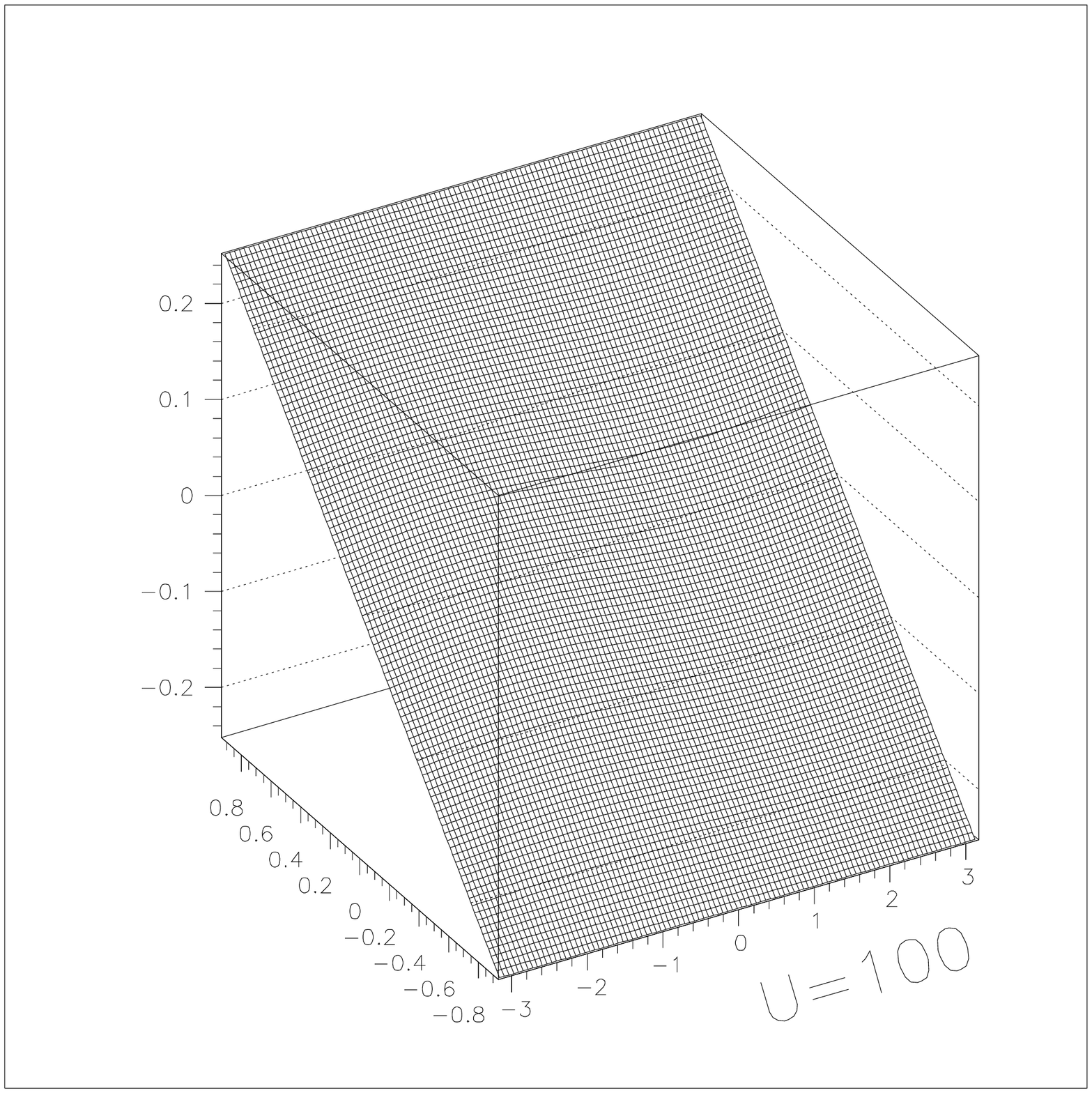}} 
\caption{The elementary
two-pseudofermion phase shift $\pi\,\Phi_{c,s1}(q,q')$ in units of $\pi$ as a function of
$q$ and $q'$ for the same densities and $U/t$ values as in Fig. \ref{F1}. Alike in that figure,  
the scatterer and scattering-center bare-momentum values $q$ and $q'$, respectively, correspond to the right and left 
axis, respectively.}
\label{F2}
\end{figure}

In this paper we follow the definition of the standard
quantum non-relativistic scattering theory and choose the overall $\beta$
pseudofermion phase shift definition, $Q_{\beta} (q)/2$, associated with Eq. (\ref{qr2}).
On the other hand, if when moving around the lattice ring the $\beta$ pseudofermion (or $\beta$ pseudofermion hole)
departures from the point $x=0$ and arrives to $x=L$, one finds that,
\begin{equation}
\lim_{x\rightarrow L}\,\bar{q}\,x = q\,L +Q_{\beta}^0 + Q^{\Phi}_{\beta} (q) =
q\,L + Q_{\beta} (q) \, , \label{qr1}
\end{equation}
where $q$ refers to the ground state. For this asymptote coordinate choice,
$Q_{\beta} (q)$ is the overall $\beta$ pseudofermion or $\beta$ pseudofermion hole phase shift 
and $2\pi\,\Phi_{\beta,\beta'}(q_j,q_{j'})$ is an elementary two-pseudofermion phase shift.

The overall pseudofermion phase-shift choice, 
$Q_{\beta} (q)=Q_{\beta}^0 + Q^{\Phi}_{\beta} (q)$, is associated
with the asymptote condition, Eq. (\ref{qr1}). It corresponds to a generalization of the conventional 
phase-shift definitions previously used in some of the BA literature. Examples are the holon-spinon
scattering theories of Refs. \cite{S-0,S} and \cite{Natan}, respectively. (All
the discussions and analysis presented below in this paper also apply to the phase-shift definition,
$Q_{\beta} (q) =Q_{\beta}^0 + Q^{\Phi}_{\beta} (q)$, provided that the
$\beta$ phase shifts, $\delta_{\beta} (q) = Q_{\beta} (q)/2$, are multiplied by two.)
\begin{figure}
\subfigure{\includegraphics[width=7cm,height=7cm]{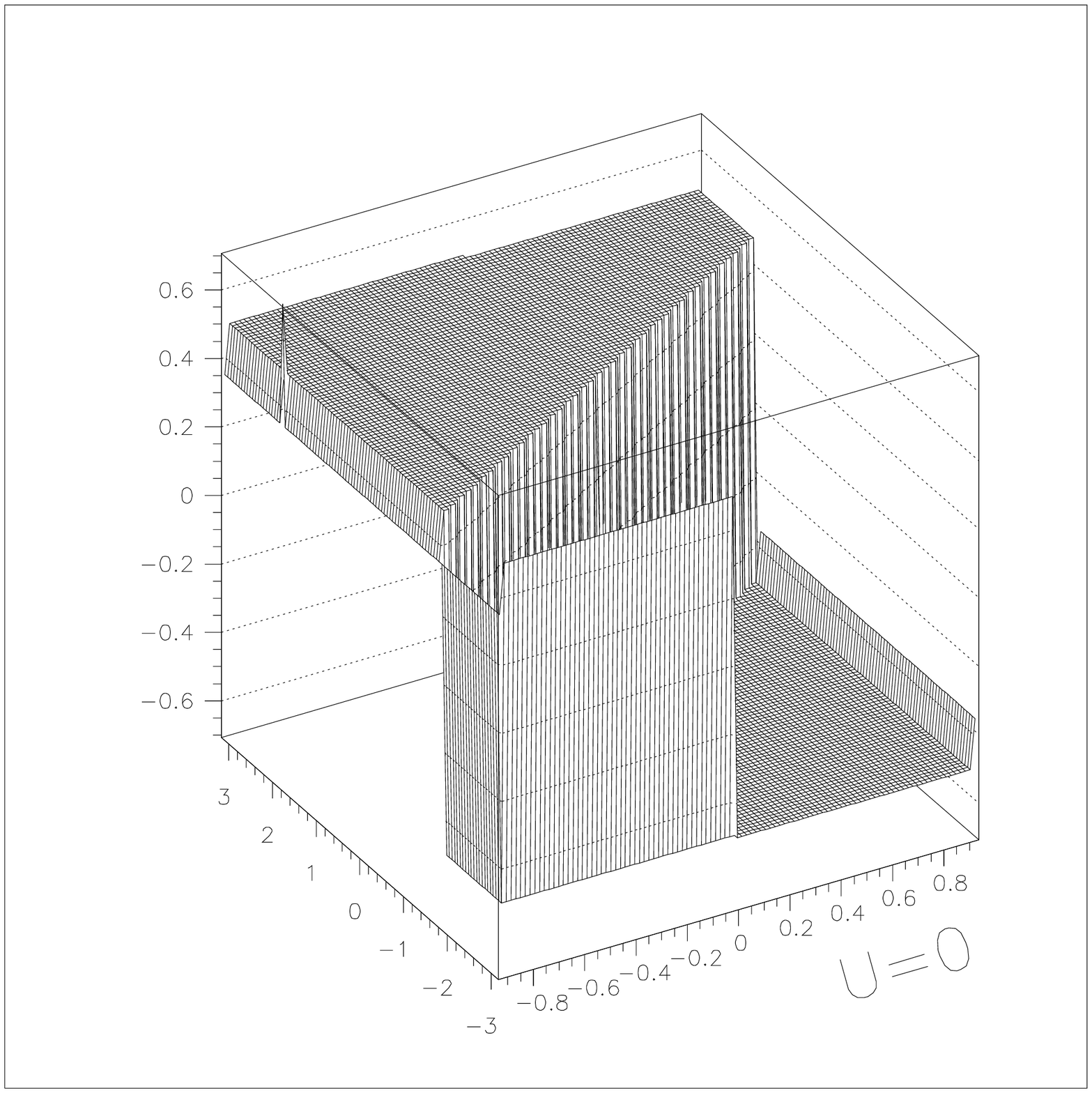}}
\subfigure{\includegraphics[width=7cm,height=7cm]{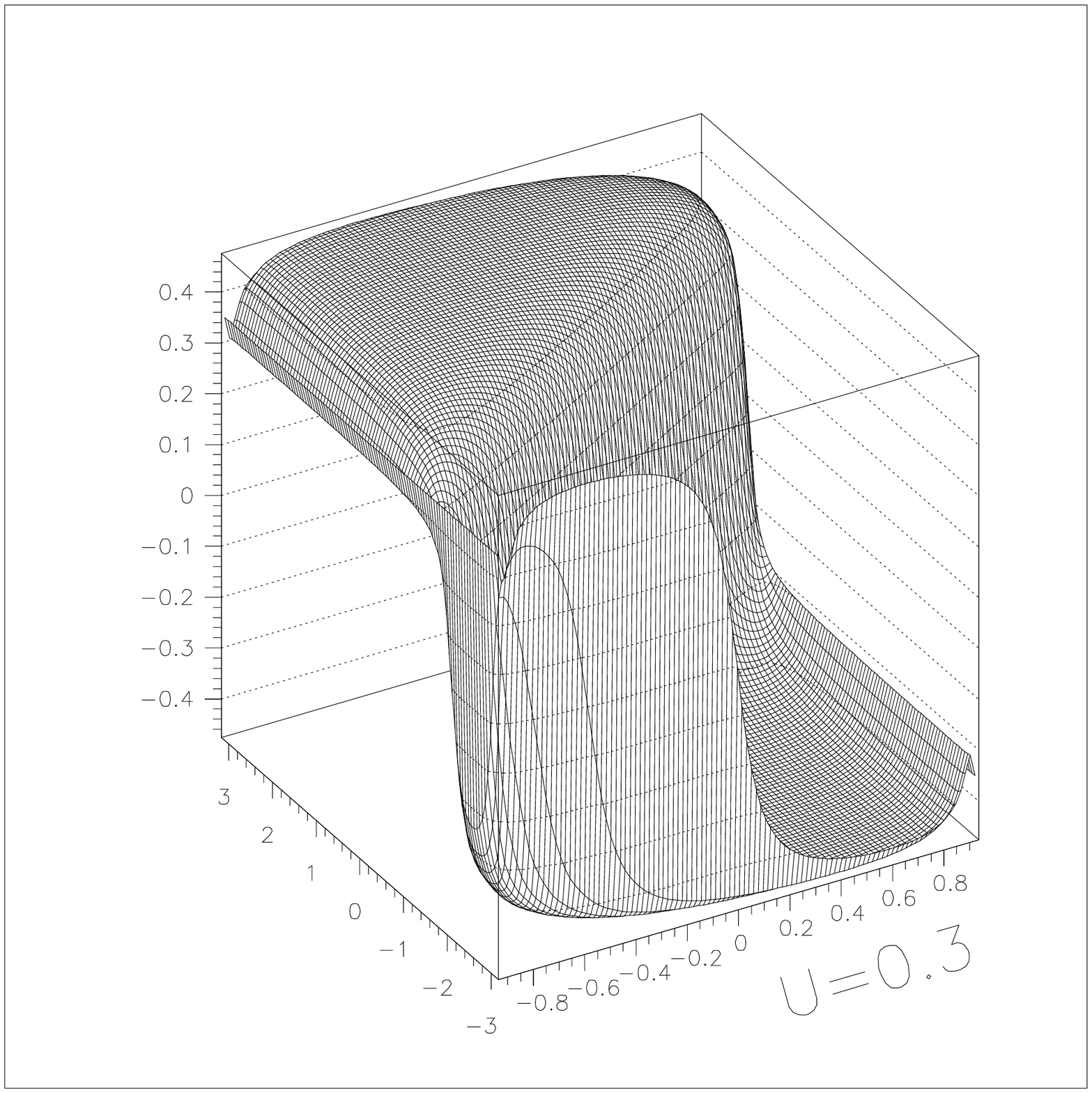}}
\subfigure{\includegraphics[width=7cm,height=7cm]{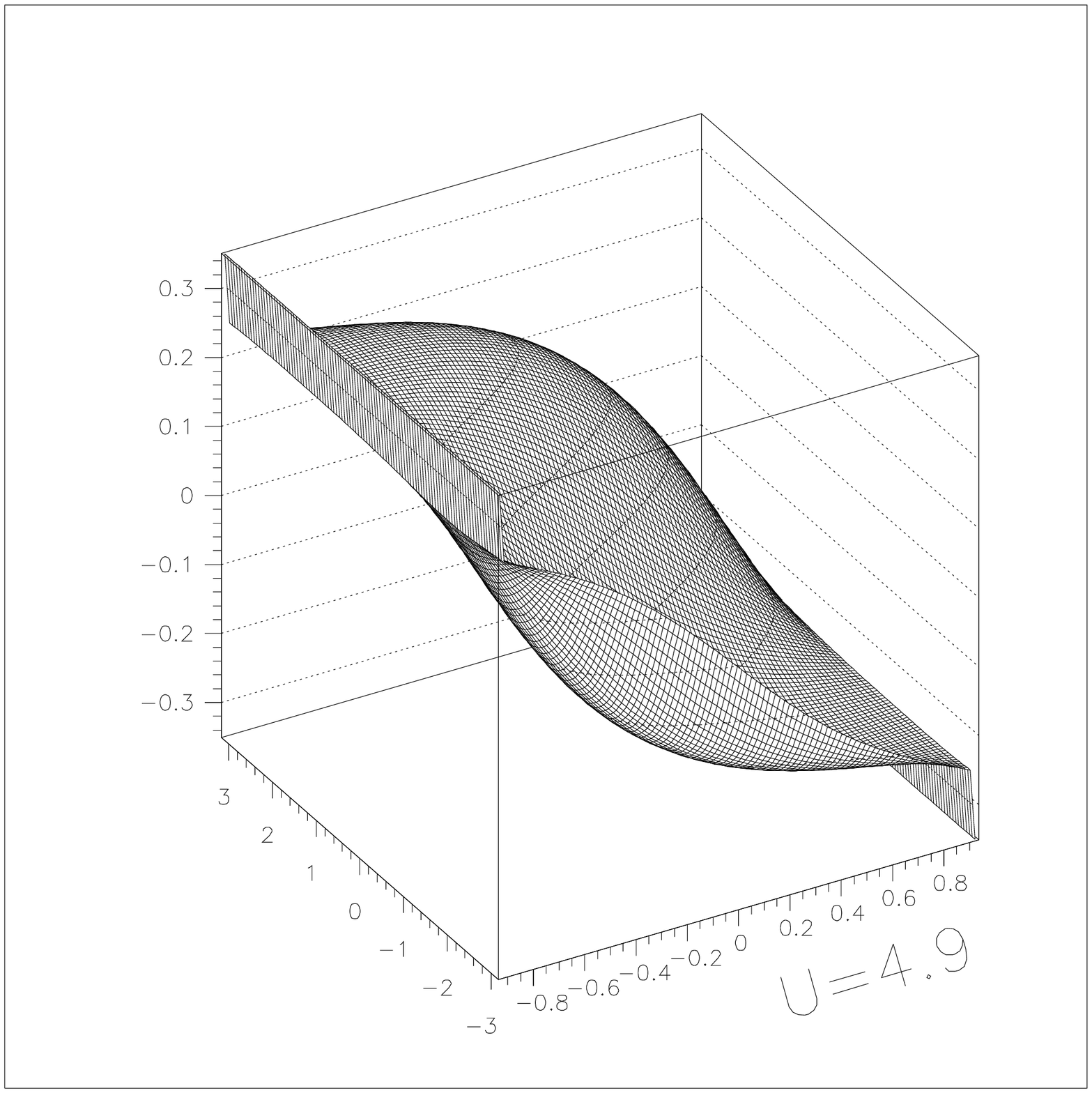}}
\subfigure{\includegraphics[width=7cm,height=7cm]{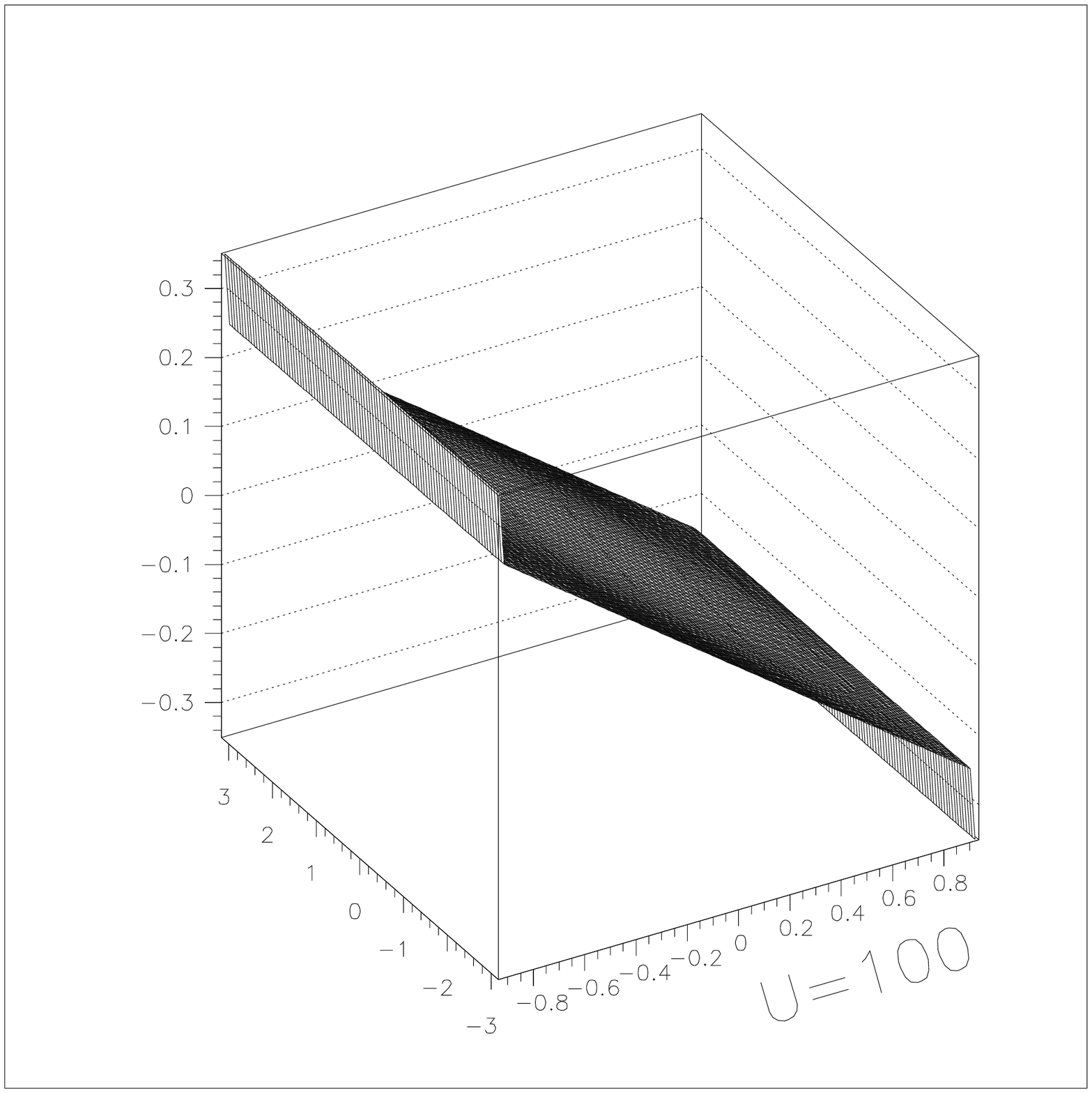}} 
\caption{The elementary
two-pseudofermion phase shift $\pi\,\Phi_{s1,c}(q,q')$ in units of $\pi$ as a function of
$q$ and $q'$ for the same densities and $U/t$ values as in Fig. \ref{F1}. Alike in that figure,  
the scatterer and scattering-center bare-momentum values $q$ and $q'$, respectively, correspond to the right and left 
axis, respectively.}
\label{F3}
\end{figure}

\subsection{The two-pseudofermion phase shifts: Bare-momentum dependence and
Levinson's Theorem}
\label{Levi}

The bare-momentum two-pseudofermion phase shifts, $\pi\,\Phi_{\beta,\beta'}(q,q')$, 
are related to the rapidity two-pseudofermion phase shifts, 
$\pi\,\bar{\Phi }_{\beta,\beta'} (r ,r')$, by Eq. (\ref{Phi-barPhi}). The latter phase 
shifts are defined by the integral equations, Eqs. (\ref{Phis1c-m})-(\ref{Phisnsn-m}) of Appendix \ref{Ele2PsPhaShi}.
Corresponding simplified $m=0$ two-pseudofermion phase-shift equations, closed-form expressions valid 
at densities $n=1$ and $m=0$, and $m=0$ analytical expressions valid for $U/t\rightarrow 0$ 
are provided in Appendix \ref{Consequences}.
\begin{figure}
\subfigure{\includegraphics[width=7cm,height=7cm]{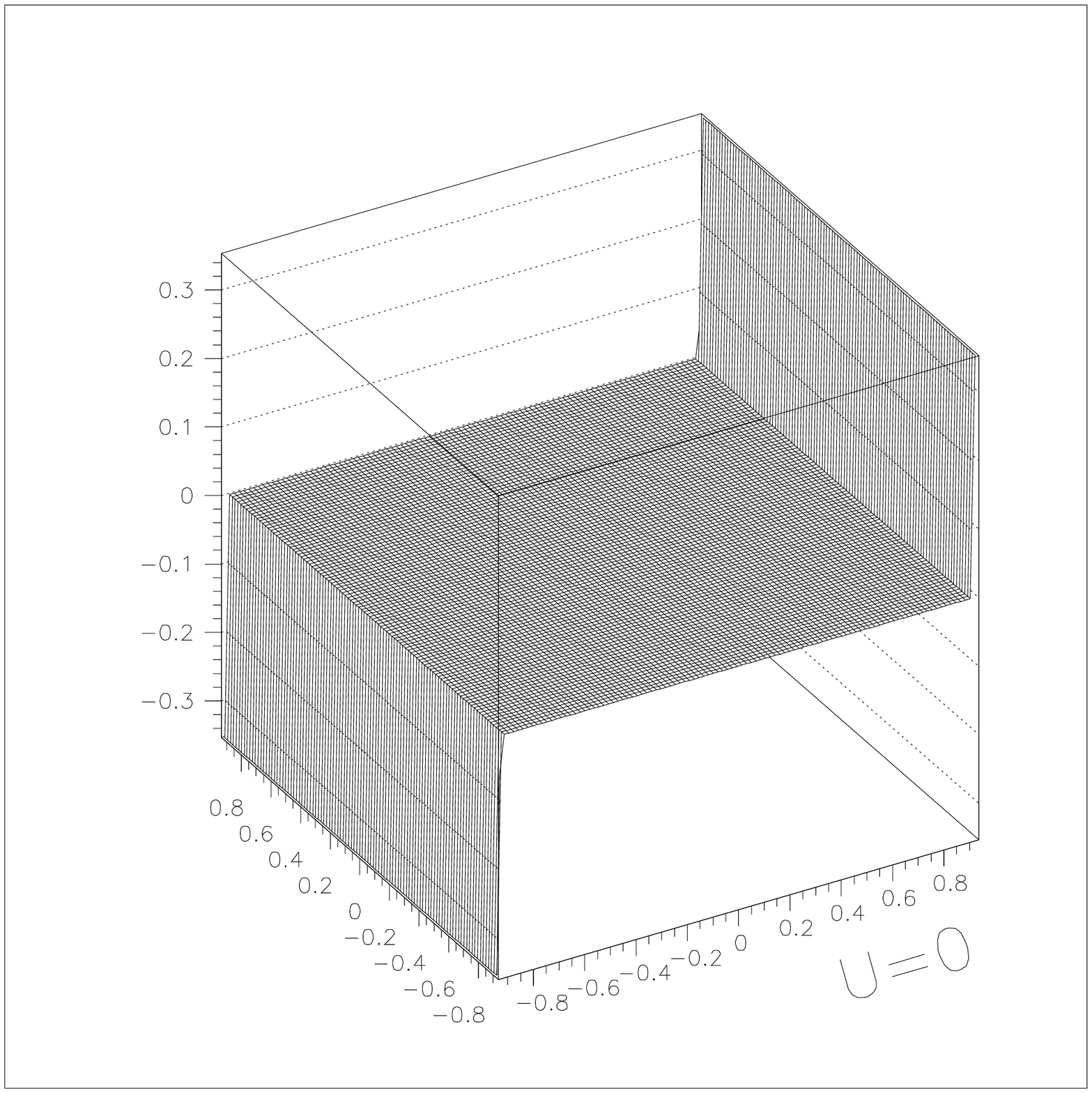}}
\subfigure{\includegraphics[width=7cm,height=7cm]{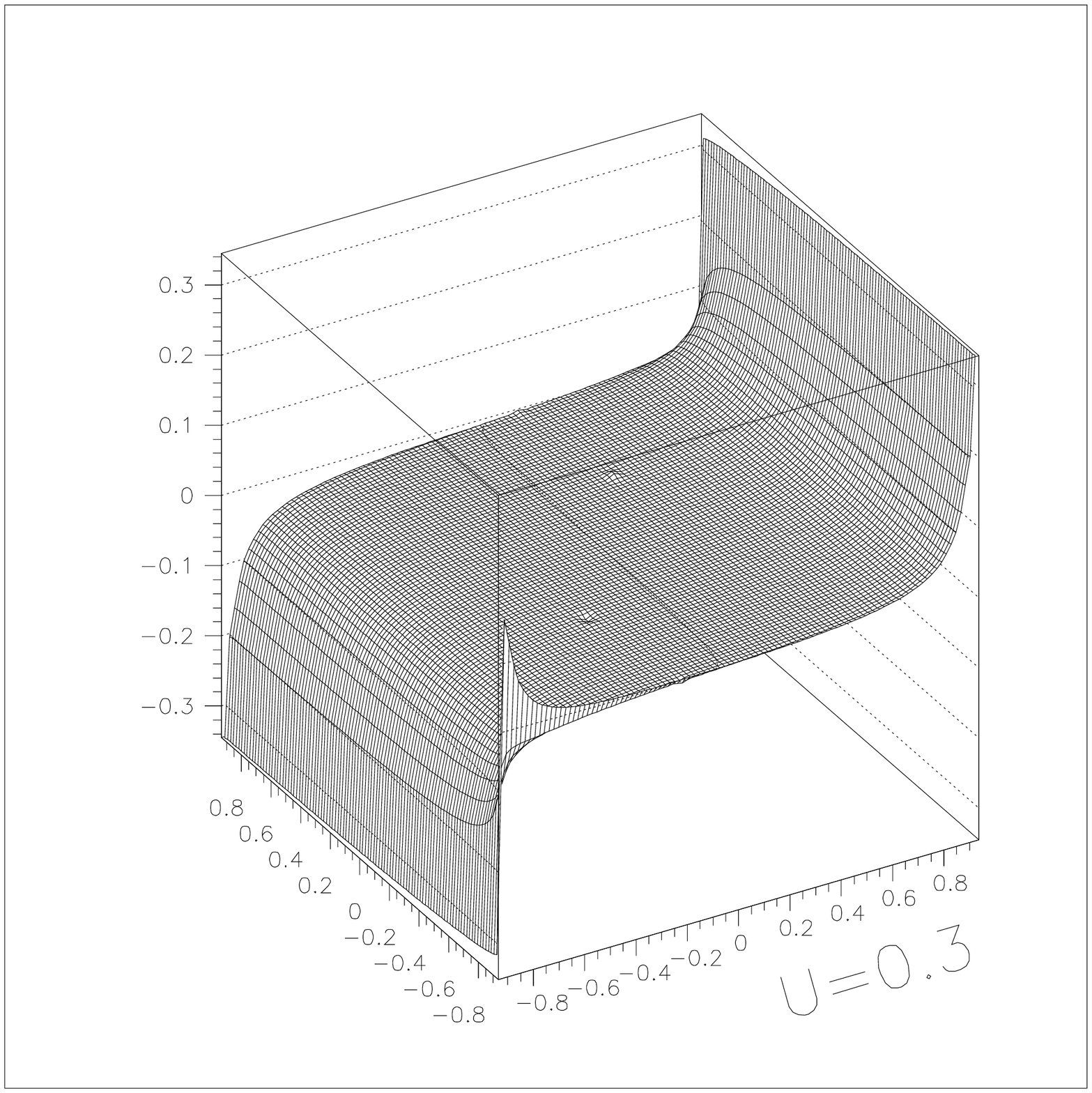}}
\subfigure{\includegraphics[width=7cm,height=7cm]{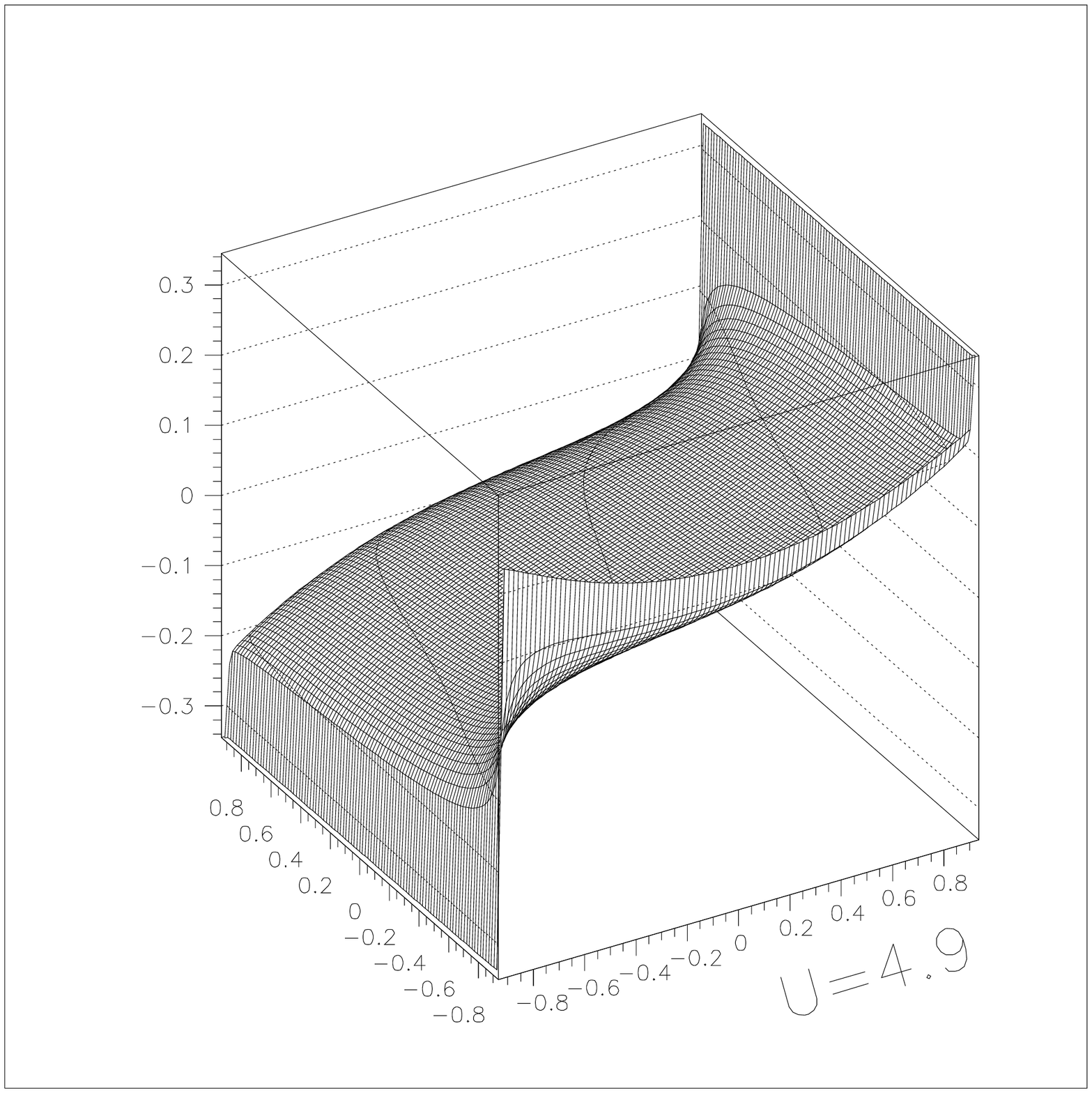}}
\subfigure{\includegraphics[width=7cm,height=7cm]{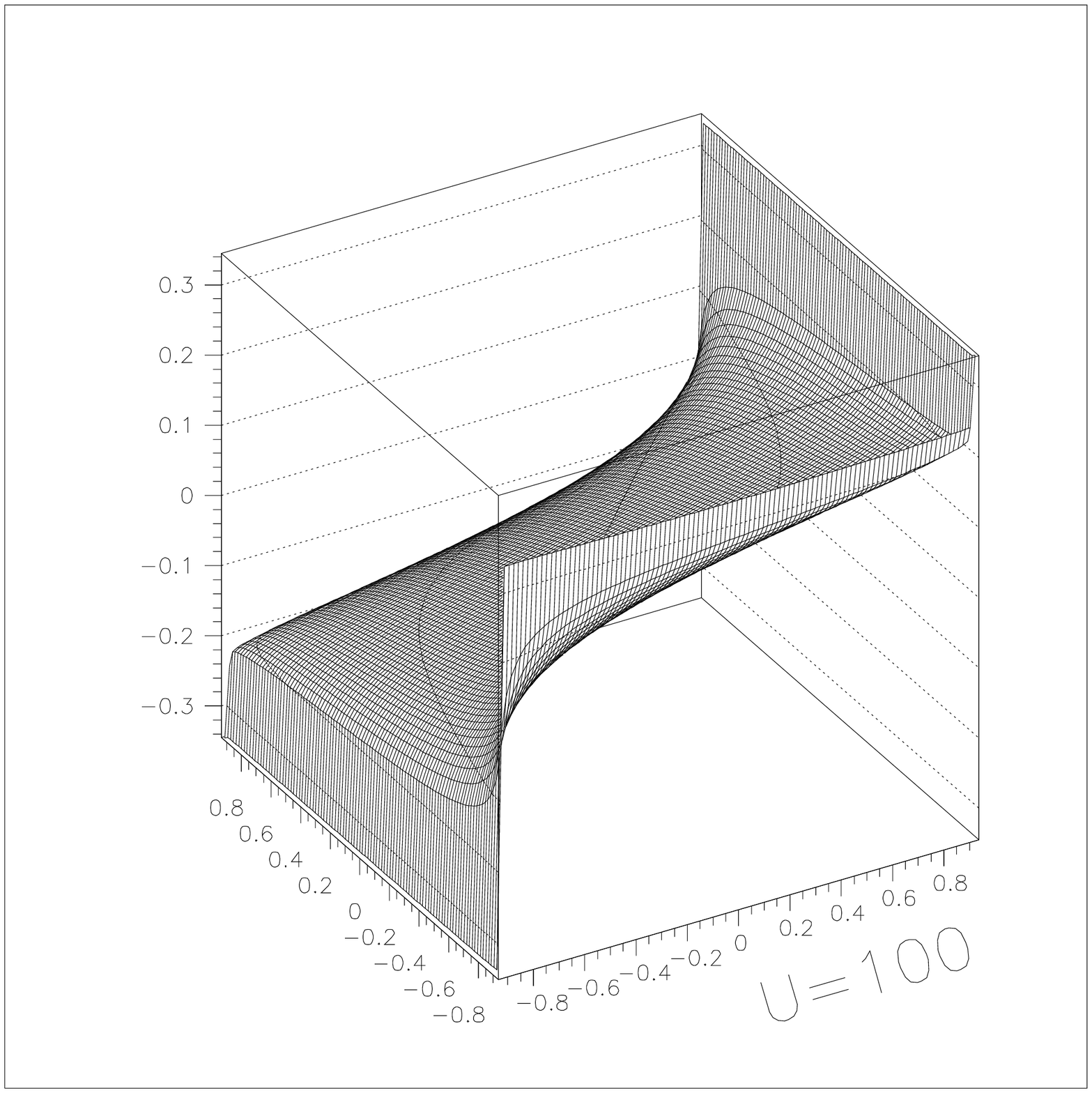}} 
\caption{The elementary
two-pseudofermion phase shift $\pi\,\Phi_{s1,s1}(q,q')$ in units of $\pi$ as a function of
$q$ and $q'$ for the same densities and $U/t$ values as in Fig. \ref{F1}. Alike in that figure,  
the scatterer and scattering-center bare-momentum values $q$ and $q'$, respectively, correspond to the right and left 
axis, respectively.}
\label{F4}
\end{figure}

The two-pseudofermion phase shifts, $\pi\,\Phi_{\beta,\beta'}(q,q')$, control the expression of
the $f$ functions, Eq. (\ref{ff}) of Appendix \ref{pseudo-repre}.
Moreover, analysis of the expressions, Eqs. (\ref{e-0-bands}) and (\ref{epsilon-q}) of that Appendix, reveals that
also the $\beta$ energy dispersions involve such phase shifts.
Specifically, according toEq. (\ref{e-0-bands}) of Appendix \ref{pseudo-repre}, the energy 
associated with the creation of one $\beta$ pseudofermion has a contribution from phase shifts felt
by all ground-state $c$ Fermi sea $c$ pseudofermions, as a result of its creation. Consistent, 
the $\int_{-Q}^{Q}dk$ integration in such expressions may be rewritten in terms of 
a corresponding bare-momentum integration,
$\int_{-q_{Fc}}^{q_{Fc}}dq'=\int_{-2k_F}^{2k_F}dq'$. Alternatively, it may be written as a sum,
$\sum_{j'=1}^{N_a} N^0_c (q_{j'})$. Here $N^0_c (q_{j'})=0$ and $N^0_c (q_{j'})=1$
for $\vert q_{j'}\vert>2k_F$ and $\vert q_{j'}\vert<2k_F$, respectively.
Under such a variable transformation, $\bar{\Phi }_{c,\beta}$ is replaced by a bare-momentum 
two-pseudofermion phase shift in units of $\pi$, ${\Phi }_{c,\beta} (q',q)={\Phi }_{c,\beta} (q_{j'},q_j)$,
Eq. (\ref{Phi-barPhi}). In it the $c$  pseudofermion and the $\beta$ pseudofermion are
the scatterer and the scattering center, respectively. (Here we have used the notation
${\Phi }_{c,\beta} (q_{j'},q_j)$ rather than our usual notation ${\Phi }_{c,\beta} (q_{j},q_{j'})$. That choice
is that consistent with the argument of the corresponding energy dispersion $\varepsilon_{\beta}^0 (q_j)$ 
in Eq. (\ref{e-0-bands}) of Appendix \ref{pseudo-repre} being denoted by $q_j$.)

\begin{figure}
\subfigure{\includegraphics[width=7cm,height=7cm]{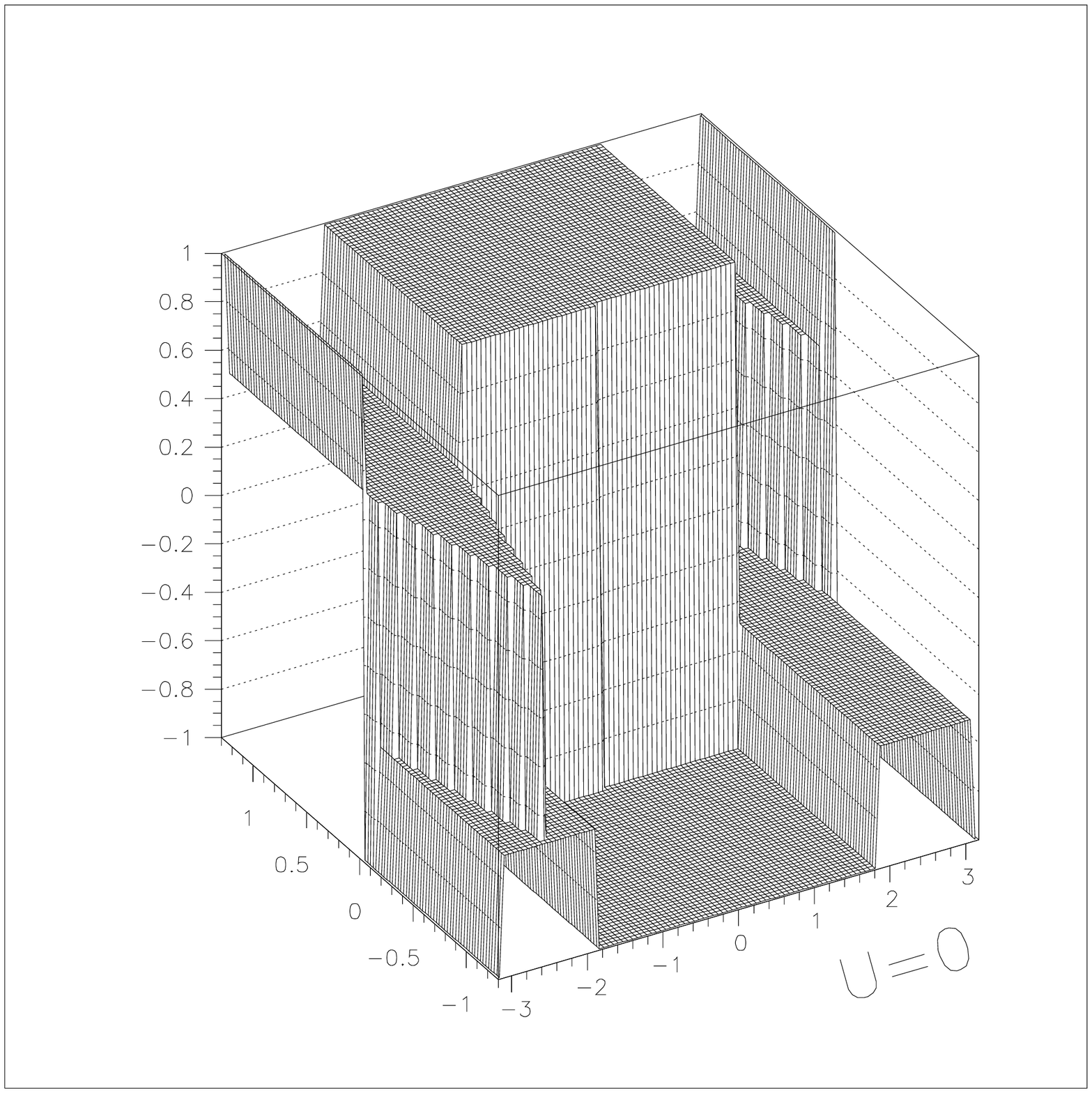}}
\subfigure{\includegraphics[width=7cm,height=7cm]{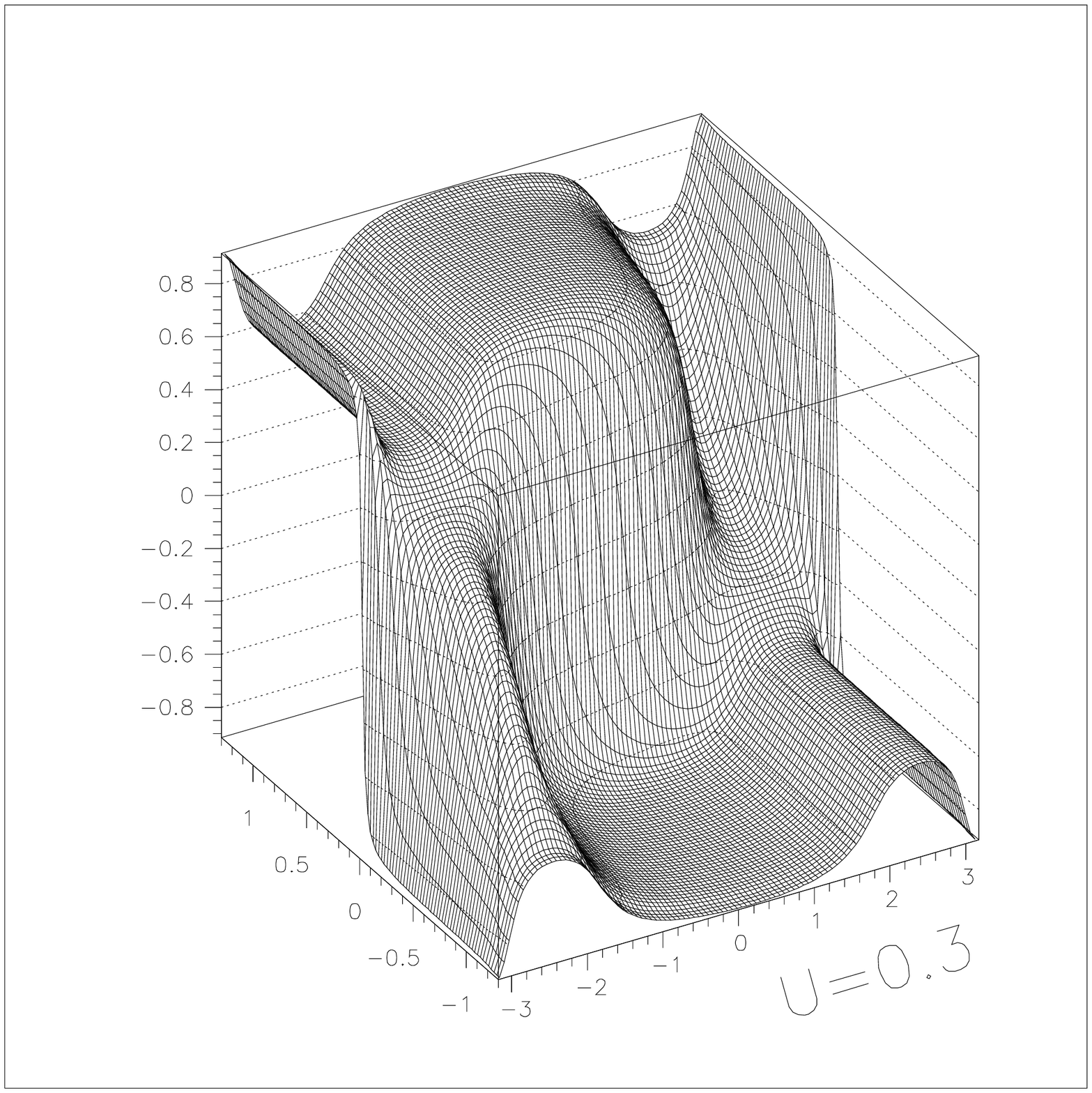}}
\subfigure{\includegraphics[width=7cm,height=7cm]{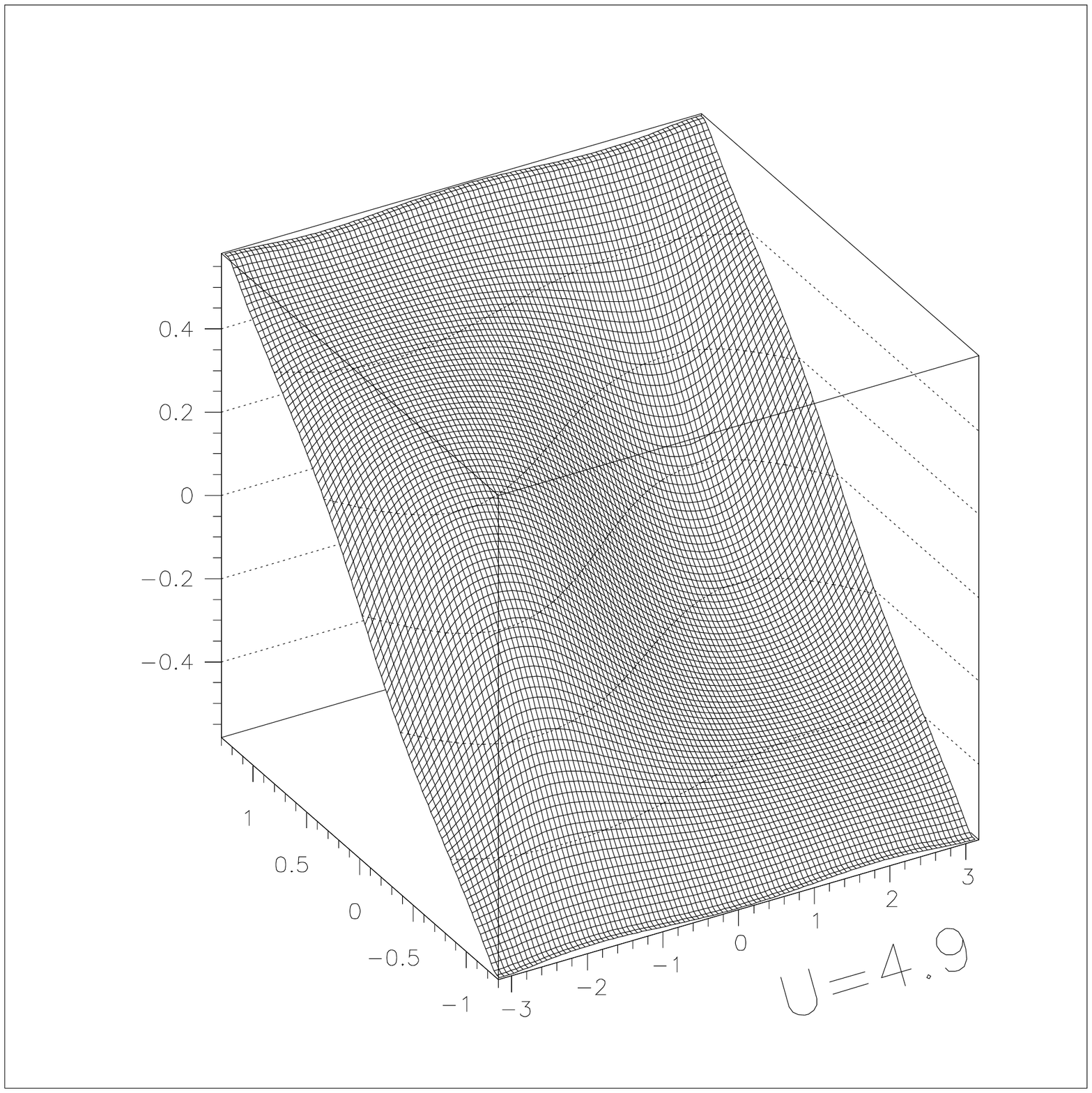}}
\subfigure{\includegraphics[width=7cm,height=7cm]{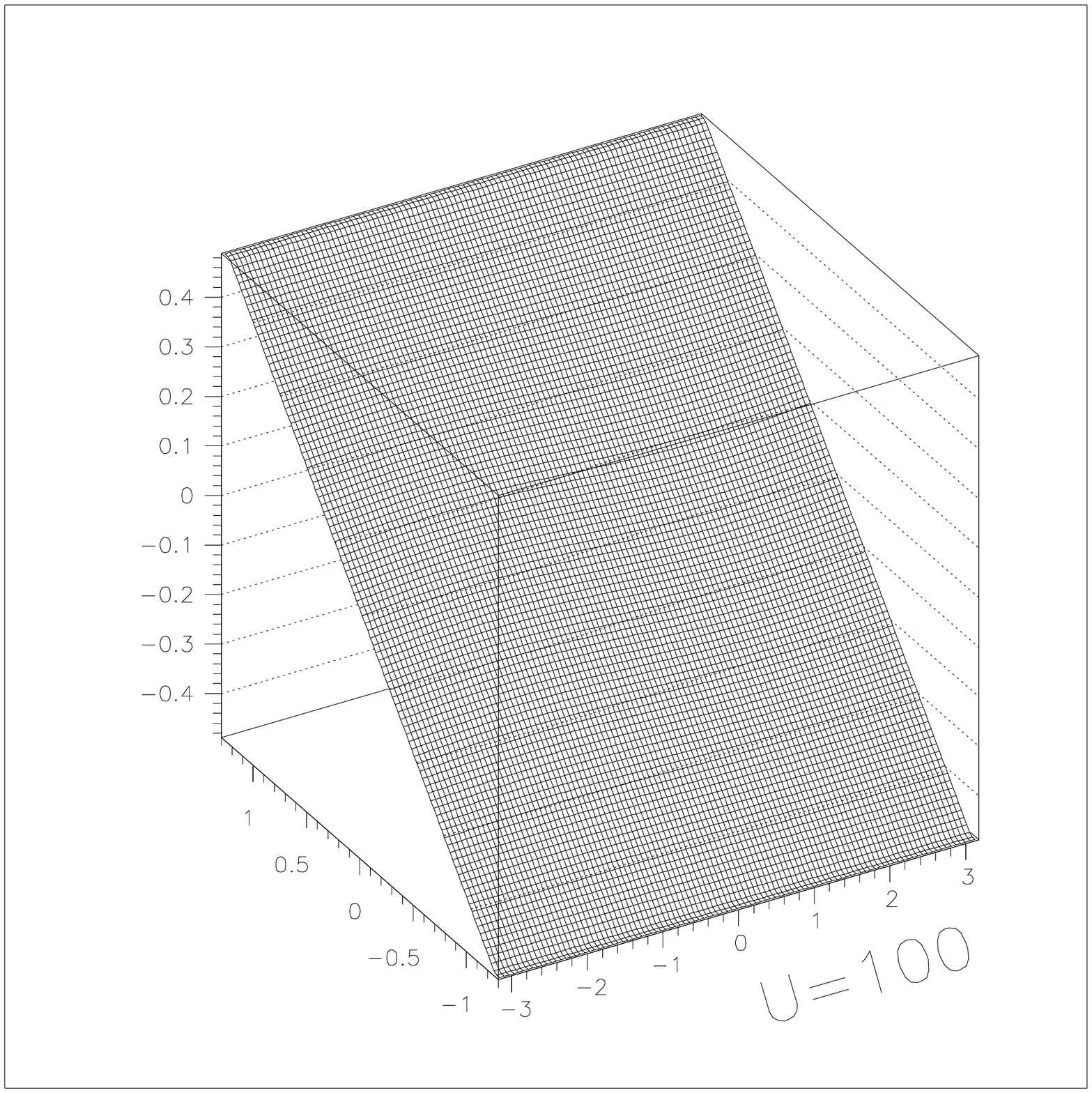}} 
\caption{The elementary
two-pseudofermion phase shift $\pi\,\Phi_{c,\eta 1}(q,q')$ in units of $\pi$ as a function of
$q$ and $q'$ for the same densities and $U/t$ values as in Fig. \ref{F1}. Alike in that figure,  
the scatterer and scattering-center bare-momentum values $q$ and $q'$, respectively, correspond to the right and left 
axis, respectively.}
\label{F5}
\end{figure}

In Figs. \ref{F1}-\ref{F6} some of the two-pseudofermion phase shifts, Eq. (\ref{Phi-barPhi}), in the
expressions of the overall phase shifts $Q_{\beta}(q_j)/2$, Eqs. (\ref{qcan1j}) and (\ref{Qcan1j}), 
associated with the excited states that mostly contribute to the PDT one- and two-electron spectral weights are plotted. They
are plotted are as a function of the scatterer and scattering-center bare momenta $q$ and $q'$, respectively,
for electronic density $n=0.59$, spin density $m\rightarrow 0$, 
and the $U/t$ values $U/t\rightarrow 0$ and $U/t= 0.3,4.9$,$100$. 
The analytical expressions of the $m=0$ two-pseudofermion phase shifts plotted 
in Figs. \ref{F1}-\ref{F6} for $U/t\rightarrow 0$ are given in Eq. (\ref{PhiccU0})-(\ref{PhistU0}) 
of Appendix \ref{Consequences}. The electronic density 
$n=0.59$ and the $U/t=4.9$ value are those used in Refs. \cite{TTF,spectral,spectral-06} for the
description of the TCNQ photoemission dispersions observed in the quasi-1D organic
compound TTF-TCNQ (tetrathiafulvalene tetracyanoquinodimethane). 

The two-pseudofermion phase shift $\pi\,\Phi_{c,\eta 1}(q,q')$, plotted in Fig. \ref{F5} in units of $\pi$,
has values in the domain $\Phi_{c,\eta 1}(q,q')\in [-1,1]$. Within the standard quantum non-relativistic scattering theory phase 
shift definition given in Eq. (\ref{qr2}), this corresponds to 
the phase-shift range $\pi\,\Phi_{c,\eta 1}(q,q')\in [-\pi,\pi]$. Note that the width of this domain is $2\pi$,
whereas the definition of Eq. (\ref{qr1}) would lead to a domain of width $4\pi$.

The phase shifts of Refs. \cite{Carmelo91,Carmelo92} correspond to 
a particular case of the general elementary two-pseudofermion phase shifts considered here. 
If one considers densities in the ranges $n\in [0,1[$ and $m \in ]0,n]$ and
the $c$ and $s1$ excitation branches two-component PSs spanned by energy eigenstates such that $N_{\beta}=0$ 
for the $\beta\neq c,s1$ branches and $M^{un}_{\alpha,-1/2}=0$ for $\alpha =\eta,s$, the general integral equations,
Eqs. (\ref{Phis1c-m})-(\ref{Phisnsn-m}) of Appendix \ref{Ele2PsPhaShi}, reduce to the integral equations,
Eqs. (20)-(23) of Ref. \cite{Carmelo91} and Eqs. (32)-(38) of Ref. \cite{Carmelo92}. 

As shown in the following, there are no
$\beta$ pseudofermion bound states. This ensures that the corresponding phase
shifts $\pi\,\Phi_{\beta,\beta'}(q,q')$, which are associated with the elementary
two-pseudofermion scattering events,
obey Levinson's Theorem  \cite{Ohanian}. Such a
theorem states that when in the reference frame of the scattering center the 
momentum of the scatterer tends to zero, the phase shift is given by $\pi N_b$. Here 
$N_b$ is the number of bound states. In that frame, the phase shift
$\pi\,\Phi_{\beta,\beta'}(q,q')$ reads $\pi\,\Phi_{\beta,\beta'}(q-q',0)$. 

The two requirements, of absence of $\beta$ pseudofermion bound states and validity of Levinson's Theorem, are 
fulfilled provided that the two-pseudofermion phase shifts of the present theory
obey the following equation,
\begin{equation}
\lim_{q-q'\rightarrow 0}\pi\,\Phi_{\beta,\beta'}(q-q',0) = 0 \, . \label{LT}
\end{equation}
To check whether this equation is obeyed, after a straightforward algebra involving the integral equations,
Eqs. (\ref{Phis1c-m})-(\ref{Phisnsn-m}) of Appendix \ref{Ele2PsPhaShi}, we have found that
$\pi\,\bar{\Phi}_{\beta,\beta'}\left(r,r'\right)=-\pi\,\bar{\Phi
}_{\beta,\beta'}\left(-r,-r'\right)$. This result combined with the use of Eq.
(\ref{Phi-barPhi}) and the odd character of the ground-state rapidity functions, 
$\Lambda_{\beta}^0 (q)=-\Lambda_{\beta}^0 (-q)$, then leads to,
\begin{equation}
\pi\,\Phi_{\beta,\beta'}(q,q ') = - \pi\,\Phi_{\beta,\beta'}(-q,-q') \, . \label{antiPP}
\end{equation}
This latter symmetry implies that $\pi\,\Phi_{\beta,\beta'}(q-q',0)$ is an odd
function of $q-q'$. This confirms both the validity of Levinson's Theorem, Eq. (\ref{LT}),
and the absence of $\beta$ pseudofermion bound states.

\subsection{Effects of the pseudofermion transformation laws on the scattering properties}
\label{Pseudo-invariance}

The $\beta$ pseudofermions have the same transformation laws under the electron - rotated-electron
unitary transformation as the corresponding $\beta$ pseudoparticles. For the latter objects, such transformation laws 
and corresponding invariances are studied in Ref. \cite{paper-I}.
Here the issue of the effects of the pseudofermion transformation laws under the electron - rotated-electron
unitary transformation on their scattering properties is addressed. Our analysis focuses on the  pseudofermions that
are invariant under that transformation in the cases of PSs whose ground states have
densities in the ranges $n \in [0,1[$ and $m\in ]0,n]$. The same problem is addressed
in Appendix \ref{Consequences} for PSs of ground states with densities $n=1$ and/or $m=0$. 
As discussed below, the effects of the pseudofermion invariance under the electron - rotated-electron unitary transformation are 
found to be different for these two PS types.

The $\beta$ pseudofermions are not in general invariant under the electron - rotated-electron unitary transformation. 
The exception is as the canonical momentum $\bar{q}$ of the composite $\beta =\alpha\nu$ pseudofermions
approaches its limiting values, $\bar{q}\rightarrow \bar{q} (\pm q_{\alpha\nu})$. Alike the corresponding $\alpha\nu$ pseudoparticles \cite{paper-I}, 
they are invariant under the electron - rotated-electron unitary transformation provided that the corresponding
group velocity vanishes, $v_{\alpha\nu} (\pm q_{\alpha\nu}) =0$.

In contrast to the band limiting momentum values of usual band particles and
Fermi-liquid quasi-particles, the $\alpha\nu$ band bare-momentum limiting values 
$\pm q_{\alpha\nu}$, Eq. (\ref{qcan-range}) of Appendix \ref{pseudo-repre}, of the 
composite $\alpha\nu$ pseudoparticles can change due to shake-up effects. Those are caused 
by the ground-state - excited-state transitions. Specifically $\pm \delta q_{\alpha\nu}=\pm\pi\,\delta N_{a_{\alpha\nu}}/N_a$, 
so that such an exotic behavior occurs when the deviation $\delta N_{a_{\alpha\nu}}$, Eq. (\ref{DN*s1an}), generated by 
the ground-state - excited-state transition is finite. 

Interestingly, it is shown in Appendix \ref{Ele2PsPhaShi} that the functional ${\bar{q}}_j = {\bar{q}} (q_j)$, Eq. (\ref{barqan}), is such
that the ground-state limiting $\alpha\nu$ band canonical-momentum values 
are not shifted by the ground-state - excited-state transitions provided that $v_{\alpha\nu} (\pm q_{\alpha\nu})=0$.
Whenever $v_{\alpha\nu} (\pm q_{\alpha\nu})=0$, the scattering phase shift leads to a 
canonical-momentum shift $Q^{\Phi}_{\alpha\nu} (\pm q_{\alpha\nu})/N_a=\mp \delta q_{\alpha\nu}$ that 
exactly cancels the bare-momentum shift $\pm \delta q_{\alpha\nu}$.
Hence the  limiting $\alpha\nu$ band canonical-momentum values remain those of the ground state, 
$\bar{q} (\pm q_{\alpha\nu}) = \pm q_{\alpha\nu}$. Here
$q_{\alpha\nu}$ is given in Eq. (\ref{qcanGS}) of Appendix \ref{pseudo-repre}.
The condition $v_{\alpha\nu} (\pm q_{\alpha\nu})=0$ is met for all $\alpha\nu\neq s1$ branches in all PSs. For the $s1$ branch it is met only for
PSs whose ground state has a small or vanishing number, $N_{s1}$, of $s1$ pseudofermions. Hence, provided that
$N_{s1}/N_a\rightarrow 0$ as $N_a\rightarrow\infty$. 

At the limiting momentum values, $\bar{q} (\pm q_{\alpha\nu}) =\pm q_{\alpha\nu}$, the energies of the
$\alpha\nu$ pseudofermions equal those of the corresponding
$\alpha\nu$ pseudoparticles and read \cite{paper-I},
\begin{eqnarray} 
\varepsilon_{\eta\nu} (\pm q_{\eta\nu}) & = & \nu\,[\varepsilon_{\eta,-1/2} 
+\varepsilon_{\eta,+1/2}] = (1-\delta_{n,1})\,2\nu\vert\mu\vert + \delta_{n,1}\,2\nu\mu^0 \, ; \hspace{0.50cm}
\varepsilon_{\eta\nu}^0 (\pm q_{\eta\nu}) = 0 \, ,
\nonumber \\ 
\varepsilon_{s\nu} (\pm q_{s\nu}) & = & \nu\,[\varepsilon_{s,-1/2} 
+\varepsilon_{s,+1/2}] = 2\nu\mu_B\,\vert H\vert \, ; \hspace{0.50cm}
\varepsilon_{s\nu}^0 (\pm q_{s\nu}) = 0 \, .
\label{invariant-V}
\end{eqnarray}
The energy scale $2\mu^0$ is the $n=1$ Mott-Hubbard gap, whose limiting behaviors 
are given in Eq. (\ref{DMH}) of Appendix \ref{energies}. For the densities and $u$ dependences of the chemical potential $\mu$ 
and magnetic energy scale $2\mu_B\,H$ appearing here, see Eqs. (\ref{mu-muBH})-(\ref{mBHc-n0-n1}) of that Appendix.
Furthermore, $\varepsilon_{\eta,\pm 1/2}$ and $\varepsilon_{s,\pm 1/2}$ denote in Eq. (\ref{invariant-V}) the following
energies of the unbound $\eta$-spinons and spinons \cite{paper-I},
\begin{eqnarray}
\varepsilon_{\eta,\pm1/2} & = & 2\vert\mu\vert \, ; \hspace{0.50cm} \varepsilon_{\eta,\mp 1/2} = 0
\, , \hspace{0.25cm} {\rm sgn}\{(1-n)\}1 = \mp 1 \, ,
\nonumber \\
\varepsilon_{s,\pm1/2} & = & 2\mu_B\,\vert H\vert \, ; \hspace{0.50cm} \varepsilon_{s,\mp 1/2} = 0
\, , \hspace{0.25cm} {\rm sgn}\{m\}1 = \mp 1 \, ,
\label{energy-eta-s}
\end{eqnarray}
respectively.

As found in Ref. \cite{paper-I},
$\varepsilon_{\eta\nu}^0 (\pm q_{\eta\nu}) = 0$ (and $\varepsilon_{s\nu}^0 (\pm q_{s\nu}) = 0$) are in Eq. (\ref{invariant-V}) the
$\eta\nu$ pseudofermion anti-binding energy (and $s\nu$ pseudofermion binding energy.)
Indeed, the $2\nu=2,4,...,\infty$ $\eta$-spinons (and spinons) that are part of one composite $\eta\nu$ pseudofermion (and 
one composite $s\nu'$ pseudofermion) lose their anti-bound (and bound) character as $\bar{q}\rightarrow \pm q_{\eta\nu}$
(and as $\bar{q}\rightarrow \pm q_{s\nu}$). That property plus the vanishing of such composite elementary objects 
group velocity are directly related to the invariance of such elementary objects under the electron - rotated-electron unitary transformation.

{\it Active scatterers} are those whose overall phase shifts generated by the ground-state - 
excited-state transitions lead to a shift of the corresponding canonical-momentum 
values. Thus, the $\alpha\nu$ pseudofermions of limiting canonical
momentum $\bar{q}=\pm {\bar{q}}_{\alpha\nu}$ and vanishing velocity $v_{\alpha\nu} (\pm {\bar{q}}_{\alpha\nu})=0$ 
are not active scatterers. Next we investigate the properties of such $\alpha\nu$ pseudofermions 
as scattering centers. By use of the integral equations, 
Eqs. (\ref{Phis1c-m})-(\ref{Phisnsn-m}) of Appendix \ref{Ele2PsPhaShi}, 
and Eqs. (A.11)-(A.14) of Ref. \cite{LE}, we find after some algebra that 
for $\alpha'\nu'\neq s1$ branches the two-pseudofermion phase shift 
$\pi\,\Phi_{\beta,\alpha'\nu'}(q,\iota\,q_{\alpha'\nu'})$
may be expressed as follows,
\begin{eqnarray}
\pi\,\Phi_{\beta,\alpha'\nu'}(q,\iota\,q_{\alpha'\nu'}) & = &
{\iota\,\pi\over 2}\Bigl[\delta_{\beta ,c} -\delta_{\beta ,\alpha\nu}\delta_{\alpha ,\alpha'}
(-\delta_{\nu,\nu'} + \nu +\nu' -\vert\nu -\nu'\vert )\Bigr] \nonumber \\
& + & {\iota\over 2}\sum_{\iota'=\pm 1}\iota'\Bigl[\pi\,\Phi_{\beta,c} (q,\iota'\,2k_F) 
-\delta_{\alpha',s}\,2\pi\,\Phi_{\beta,s1} (q,\iota'\,k_{F\downarrow})
\Bigr] \, , \hspace{0.25cm} \iota =\pm 1 \, . 
\label{inde}
\end{eqnarray}
For $\alpha'\nu'=s1$, this expression is valid only in the limit of
vanishing velocity, $v_{s1} (\pm q_{s1})=0$, within which $2\pi\,\Phi_{\beta,s1} (q,\iota'\,k_{F\downarrow})=
2\pi\,\Phi_{\beta,s1} (q,\iota'\,0)=0$. In that expression the label $\beta$ refers to
a $\beta$ pseudofermion or $\beta$ pseudofermion hole scatterer whose $\beta$ band
has finite pseudofermion occupancy for the excited state under consideration. 
If $\beta =\alpha'\nu$, then the values of $q$ are such that $\vert q\vert <q_{\alpha'\nu'}$ . 
Otherwise they can have any value and thus correspond to any excited-state active $\beta$ 
scatterer. 

The form of the two-pseudofermion phase-shift expression, Eq. (\ref{inde}), 
reveals that, except for the constant phase-shift terms, creation of one $\eta\nu'$ 
pseudofermion at canonical momentum $q_{\eta\nu'}=\iota\,[\pi -2k_F]$
is felt by the $\beta$ pseudofermion or $\beta$ pseudofermion-hole active scatterer as a 
$c$ pseudofermion Fermi-points current excitation. Such an excitation is associated with a 
shift, $\iota\pi/N_a$, of both $c$ bare-momentum Fermi points. 
Thus, such a scatterer effectively feels it is scattered by
$c$ Fermi-point current shifts, rather than by the $\eta\nu'$ 
pseudofermion created at canonical momentum $q_{\eta\nu'}=\iota\,[\pi -2k_F]$
under the transition to the excited state.

Similarly, Eq. (\ref{inde}) demonstrates
that, again, except for the constant phase-shift terms, creation of one $s\nu'\neq s1$ pseudofermion
at canonical momentum $q_{s\nu'}=\iota\,[k_{F\uparrow}-k_{F\downarrow}]$
is felt by a $\beta$ pseudofermion or $\beta$ pseudofermion-hole active scatterer as a 
$c$ and $s1$ pseudofermion Fermi-points current excitation. It corresponds
to a shift, $\iota\pi/N_a$, of both $c$ bare-momentum Fermi points 
and a shift, $-\iota 2\pi/N_a$, of both $s1$ bare-momentum Fermi points. 
Such a scatterer effectively feels it is scattered by
$c$ and $s1$ Fermi-point current shifts, rather than by the $s\nu'\neq s1$
pseudofermion created at canonical momentum $q_{s\nu'}=\iota\,[k_{F\uparrow}-k_{F\downarrow}]$.
The same applies to creation of one $s1$ pseudofermion at canonical momentum
$q_{s1}=\iota\,2k_F$ as $k_{F\uparrow}\rightarrow 2k_F$, $k_{F\downarrow}\rightarrow 0$,
and thus $v_{s1} (\iota\,q_{s1})=0$.

{\it Active scattering centers} are those that contribute to the scattering phase 
shift, Eq. (\ref{qcan1j}). In spite of the exotic properties reported above,
a vanishing-velocity $\alpha\nu$ pseudofermion created under a transition
to an excited state at a limiting canonical momentum is an active scattering center.
On the other hand, small-momentum and low-energy
$c$ and $s1$ pseudofermion particle-hole processes in the vicinity of
the corresponding $\beta =c,s1$ Fermi points, called elementary processes (C) below in Section \ref{PDT},
do not generate active scattering centers. Within the latter processes,
the phase shifts generated by the $\beta =c,s1$ pseudofermion particle-like 
excitations exactly cancel those produced by the  $\beta =c,s1$ pseudofermion hole-like excitations. 

Above the effects of the pseudofermion transformation laws under the electron - rotated-electron unitary transformation
were studied for PSs of $S_{\alpha}>0$ ground states. Such effects are of a different type
in the case of a $\alpha\nu\neq s1$ pseudofermion created onto a $S_{\alpha}=0$ ground state.
It a simple exercise to show that such an elementary object energy obeys the equality, 
Eq. (\ref{invariant-V}). Consistent with its invariance under the 
electron - rotated-electron unitary transformation, for $u>0$ creation of one $\eta\nu$ pseudofermion (and one $s\nu'\neq s1$
pseudofermion) onto a $S_{\eta}=0$ (and $S_{s}=0$) ground state leads to a change $\nu =1,2,...$ in the number of lattice sites 
doubly occupied both by electrons and rotated electrons (and singly 
occupied both by spin-down electrons and spin-down rotated electrons). 
The $2\nu=2,4,6,...$ $\eta$-spinons (and $2\nu'=2,4,6,...$ spinons) of that vanishing-velocity $\eta\nu$ pseudofermion 
(and vanishing-velocity $s\nu'$ pseudofermion) are not energetically anti-bound (and bound).
In spite of their energetic non-anti-binding (and unbinding) character, such
$2\nu$ $\eta$-spinons (and $2\nu'$ spinons) remain being in a collective $\eta$-spin-singlet
(and spin-singlet) configuration. 

The corresponding $\alpha\nu\neq s1$ momentum band does not exist for a $S_{\alpha}>0$ ground state.
Hence one has that $Q_{\alpha\nu} (0)/N_a=Q^{\Phi}_{\alpha\nu}(0)/N_a$ for 
an excited state whose generation from that state involves
creation of a single $\alpha\nu\neq s1$ pseudofermion.
Moreover, the invariance under the electron - rotated-electron transformation of that
$\alpha\nu\neq s1$ pseudofermion implies that it is not an active scatterer. For the above excited state, 
a necessary condition for such an object not being an active scatterer 
is that it is not a scatterer. The $\alpha\nu\neq s1$ bare-momentum band
of the corresponding ``in" state has a single value at $q=\pm q_{\alpha\nu}=0$. Thus it is required that the  
canonical-momentum band of the ``out" state has also a single value at $\bar{q}=\pm q_{\alpha\nu}=0$. 
This implies that $Q_{\alpha\nu} (0)/N_a=Q^{\Phi}_{\alpha\nu}(0)/N_a=0$, and thus that 
the $\alpha\nu\neq s1$ pseudoparticle remains invariant under the pseudoparticle - pseudofermion 
unitary transformation. This is why the corresponding $\alpha\nu\neq s1$ 
pseudofermion is not a scatterer.

On the one hand, the following two-pseudofermion phase shifts vanish for the
PS excited states of a $S_{\eta}=0$ (and $S_{s}=0$) ground state: $\pi\,\Phi_{s1,\eta\nu}(q,0) =\pi\,\Phi_{\eta\nu,s1}(0,q') = 0$ 
(and $\pi\,\Phi_{c,s\nu'}(q,0) =\pi\,\Phi_{s\nu',c}(0,q') = 0$ for $\nu'>1$.) On the other hand,
both the effective phase shift $\pi\,\Phi_{\eta\nu,c}(0,q')$ (and $\pi\,\Phi_{s\nu',s1}(0,q')$ for
$\nu'>1$), Eq. (\ref{Phis}) of Appendix \ref{Consequences}, and the two-pseudofermion 
phase shift $\pi\,\Phi_{c,\eta\nu}(q,0)$ (and $\pi\,\Phi_{s1,s\nu'}(q,0)$ for $\nu'>1$) of a $c$ (and $s1$) scatterer
with bare momentum $q$ originated from its collision with the $\eta\nu$ (and $s\nu'\neq s1$) pseudofermion scattering center, respectively,
Eq. (\ref{Phis-2}) of that Appendix, have an interesting property: In addition to the $c$ (and $s1$) scattering-center bare
momentum $q'$ or scatterer bare-momentum $q$, the
two-pseudofermion phase shifts $\pi\,\Phi_{\eta\nu,c}(0,q')$ and $\pi\,\Phi_{c,\eta\nu}(q,0)$
(and $\pi\,\Phi_{s\nu',s1}(0,q')$ and $\pi\,\Phi_{s1,s\nu'}(q,0)$ for $\nu'>1$)
are a function of the set of $2\nu=2,4,6,...$ (and $2(\nu'-1)=2,4,6,...$) bare-momentum values
$\{q_h\}$ of the corresponding $2\nu$ (and $2(\nu'-1)=2,4,6,...$) neutral-shadow
$c$ (and $s1$) pseudofermion holes considered in Section \ref{PSPS}.

The requirement that $Q^{\Phi}_{\alpha\nu} (0)/N_a=0$ is shown in Appendix \ref{Consequences}
to impose a specific form to the corresponding two-peudofermion phase shifts
$\pi\,\Phi_{\alpha\nu,\beta'}(0,q')$. Since the $\alpha\nu\neq s1$ pseudofermion is not a scatterer,
$\pi\,\Phi_{\alpha\nu,\beta'}(0,q')$ is an effective virtual phase shift with no requirement 
to obey Levinson's Theorem. The consequences of the above symmetries physically appear 
within the interplay of the $\eta\nu$ (and $s\nu'\neq s1$) pseudofermion with its $2\nu= 2,4,6,...$
(and $2(\nu'-1)=2,4,6,...$) neutral-shadow $c$ (and $s1$) pseudofermion holes. As discussed in Section \ref{PSPS},
$2\nu=2,4,6,...$ (and $2(\nu'-1)=2,4,6,...$) neutral-shadow $c$ (and $s1$) pseudofermion holes emerge under a transition to 
an excited state for each $\eta\nu$ (and $s\nu'\neq s1$) pseudofermion created under it. The set
of $2\nu=2,4,6,...$ (and $2(\nu'-1)=2,4,6,...$) virtual charge (and spin) elementary currents 
carried by such $2\nu=2,4,6,...$ (and $2(\nu'-1)=2,4,6,...$) neutral-shadow $c$ (and $s1$) pseudofermion holes
of the $\eta\nu$ (and $s\nu'\neq s1$) pseudofermion
exactly cancel that carried by itself. On the other hand, for the present excited state of a $S_{\eta}=0$ (and $S_{s}=0$) ground state,
the $2\nu= 2,4,6,...$ (and $2(\nu'-1)= 2,4,6,...$) virtual two-pseudofermion phase shifts of the  $\eta\nu$ (and $s\nu'\neq s1$) pseudofermion
due to its collisions with the corresponding  $2\nu= 2,4,6,...$ (and $2(\nu'-1)= 2,4,6,...$) neutral-shadow $c$ (and $s1$) pseudofermion-hole
scattering centers also exactly cancel each other, so that $Q^{\Phi}_{\eta\nu} (0)=0$ (and $Q^{\Phi}_{s\nu'} (0)=0$.)

\subsection{Role of the pseudofermion scattering phase shifts in the model dynamical and spectral properties}
\label{PDT}

In this section we revisit some of the spectral weights of the PDT of Refs. \cite{V,VI,LE,TTF}. Our aim
is to specify which pseudofermion microscopic scattering processes considered in the studies
of this paper contribute to and control such spectral weights. Note though that the weights given in 
the following have been previously derived in Refs. \cite{V,VI}. 

An important property specific to the pseudofermion scattering theory is that the phase shift 
$\delta_{\beta} (q_j) = Q_{\beta}(q_j)/2 = Q_{\beta}^0/2 + Q^{\Phi}_{\beta} (q_j)/2$, Eqs. 
(\ref{Qcan1j}) and (\ref{danq}), refers to all $j=1,...,N_{a_{\beta}}$
$\beta$-band scatterers of the excited state. This includes both those created under the transition from the
ground state to the excited state and the scatterers that pre-existed in the ground state. That
phase-shift scattering term, $Q^{\Phi}_{\beta} (q_j)/2$, is expressed in Eq. (\ref{qcan1j})
in terms of a suitable superposition of two-pseudofermion phase shifts, $\pi\,\Phi_{\beta,\beta'}(q_j,q_{j'})$.
Such a superposition is uniquely defined for each excited state. The non-scattering term, $Q_{\beta}^0/2$, 
in the phase shift $\delta_{\beta} (q_j) = Q_{\beta}(q_j)/2$, Eqs. (\ref{Qcan1j}) and (\ref{danq}), results from 
the BA quantum numbers shifts, Eq. (\ref{pican}). 

That the phase shift $\delta_{\beta} (q_j) = Q_{\beta}(q_j)/2$ refers to all $j=1,...,N_{a_{\beta}}$
$\beta$-band scatterers of the excited state is
a property that does not hold for the holon and spinon scattering theories of Refs. \cite{Natan,S-0,S}. Their
phase shifts refer only to the scatterers created under the transition from the ground state to the excited states. Those 
scatterers play as well the role of scattering centers. As discussed in the following, this 
prevents such theories of describing most of the elementary processes that control the model dynamical 
and spectral properties. 

The $N_{a_{\beta}}$ $\beta$-band scatterers are the $N_{\beta}$ $\beta$ pseudofermions and $N_{\beta}^h$
$\beta$ pseudofermion holes that populate the excited state. Here $N_{a_{\beta}}=[N_{\beta}+N_{\beta}^h]$, Eq. (\ref{N*}) of Appendix 
\ref{pseudo-repre}. For ground states with densities in the ranges $n\in [0,1]$ and $m\in [0,n]$
such numbers are for instance for the $\beta =c,s1$ bands of the PS excited states given by $N_{c} = [N + \delta N_c]$, $N_{c}^h =[N_a-N -  \delta N_c]$,
$N_{s1} = [N_{\uparrow} + \delta N_{s1}]$, and $N_{s1}^h =[N_{\uparrow}-N_{\downarrow}+ \delta N_{s1}^h]$ where
$\delta N_c/N_a\rightarrow 0$, $\delta N_{s1}/N_a\rightarrow 0$, and $\delta N_{s1}^h/N_a\rightarrow 0$ as
$N_a\rightarrow\infty$. 

Thus the pseudofermion scattering theory provides the phase shifts of a macroscopic number $N_a$ of $c$ band scatterers
and $N + \delta N_{a_{s1}}$ of $s1$ band scatterers. Such phase shifts result from the scattering of such
scatterers with the centers that emerge in one or several $\beta'$ bands, under the transition to the excited state. The
number of $\beta'$ scattering centers created under such a transition is
much smaller than that of scatterers, Eq. (\ref{DimSm}). It is given by,
\begin{equation}
N_{\rm centers} =\sum_{\beta'}\,\sum_{j'=1}^{N_{a_{\beta'}}}\,\vert\delta N_{\beta'}(q_{j'})\vert \, ,
\label{N-centers}
\end{equation}
where the deviations $\delta N_{\beta'}(q_{j'})$ are those on the right-hand side of the scattering phase-shift
expression, Eq. (\ref{qcan1j}). They refer to the creation of one $\beta' =c,\alpha\nu$
pseudofermion scattering center of momentum $q_{j'}$ for $\delta N_{\beta'}(q_{j'})=1$ and one $\beta'=c,s1$ pseudofermion-hole 
scattering center of momentum $q_{j'}$ for $\delta N_{\beta'}(q_{j'})=-1$. All $N_{a_{\beta}}=[N_{\beta}+N_{\beta}^h]$ $\beta $-band scatterers acquire a phase
shift $\pi\,\Phi_{\beta,\beta'}(q_j,q_{j'})$ due to their interaction with each created $\beta'$ scattering center. This
is confirmed by the form of the scattering phase shift $Q^{\Phi}_{\beta} (q_j)/2$, Eq. (\ref{qcan1j}).

Both the pseudofermion scattering theory studied in this paper and the related
PDT \cite{V,VI,LE,TTF} account for the effects of
the changes in the $\beta$ pseudofermion occupancy configurations that occur
under the ground-state - excited-state transitions. Within the PDT, the elementary processes that generate the PS excited
energy eigenstates from the ground state have three types:
\vspace{0.25cm}

(A) Finite-energy and finite-momentum elementary $\beta =c,s1$ pseudofermion (and $\beta =\alpha\nu\neq s1$
pseudofermion, unbound $\eta$-spinon, and unbound spinon) processes involving creation or annihilation (and creation) of one or a finite number of
pseudofermions (and pseudofermions, unbound $-1/2$ $\eta$-spinons, and unbound $-1/2$ spinons)
with canonical momentum values ${\bar{q}}_j\neq \pm {\bar{q}}_{F\beta}$ (and canonical momentum values ${\bar{q}}_j\neq  \pm q_{\beta}$
and momentum values $q_{\eta,-1/2}=\pi$ and $q_{s,-1/2}=0$, respectively);
\vspace{0.25cm}

(B) Zero-energy and finite-momentum processes that change the
number of $\beta =c,s1$ pseudofermions at the $\iota=+1$ right and 
$\iota=-1$ left $\beta =c,s1$ Fermi points. 
\vspace{0.25cm}

(C) Low-energy and small-momentum elementary $\beta =c,s1$
pseudofermion particle-hole processes in the vicinity of the
$\iota=+1$ right and $\iota=-1$
left $\beta =c,s1$ Fermi points, relative to the
excited-state $\beta = c,s1$ pseudofermion 
momentum occupancy configurations generated by the above
elementary processes (A) and (B).
\vspace{0.25cm}

According to the PDT, the elementary processes (A), (B), and (C) lead to 
qualitatively different contributions to the spectral-weight distributions. 
The PDT studies of Refs. \cite{V,VI} considered that creation of
$\eta\nu$ pseudofermions and $s\nu\neq s1$ pseudofermions at the limiting 
bare-momentum values is felt by both the $c$ and $s1$
scatterers as effective $c$ scattering centers and effective $c$ 
and $s1$ scattering centers, respectively. However, such studies
applied only to $\beta=c$ scatterers and $\beta =s1$ scatterers of 
bare-momentum value $q=\pm q_{F\beta}$. Hence
the general two-pseudofermion expression, Eq. (\ref{inde}), generalizes 
that result to {\it all} active $\beta =c,\alpha\nu$ scatterers of arbitrary bare momentum $q$. 
In Appendix \ref{Ele2PsPhaShi}, the separate contributions of the PDT processes 
(A) and (B) to the overall scattering phase shift $Q^{\Phi}_{\beta} (q)/2$, Eq. (\ref{qcan1j}), 
are specified. 

The pseudoparticles have zero-momentum forward-scattering energy interaction terms. Those are the $f$-function terms in 
Eqs. (\ref{DE-fermions}) of Appendix \ref{pseudo-repre}. The PDT relies on the corresponding pseudofermions having no 
energy interactions for $u>0$, as given in Eq. (\ref{DE}). 
This allows the expressions of the $u>0$ one- and two-electron spectral functions in terms of a sum of terms, each of which 
is a convolution of a $c$ pseudofermion and a $s1$ pseudofermion spectral function. The latter spectral
functions account for the effects of the Anderson's orthogonality catastrophes \cite{Anderson-67}. Those result from 
the discrete canonical-momentum overall shifts, $Q_{\beta}(q_j)/N_a$. Such effects play a major role in
the one- and two-electron matrix elements quantum overlaps that control the corresponding
spectral-weight distributions \cite{V,VI,LE,TTF}. They emerge in the
$c$ and $s1$ pseudofermion spectral functions through the exotic anticommutation relations, Eq. (\ref{pfacrGS}),
which involve the overall phase-shift functional $Q_{\beta}(q_j)$, Eq. (\ref{Qcan1j}). Such spectral functions have the
general form \cite{V,TTF},
\begin{eqnarray}
B_{Q_{\beta}} (k,\omega) & = & {N_a\over 2\pi}\sum_{m_{\beta,\,+1};m_{\beta,\,-1}}\,A^{(0,0)}_{\beta}\,a_{\beta} (m_{\beta,\,+1},\,m_{\beta,\,-1})
\nonumber \\
& \times & \delta \Bigl(\omega -{2\pi\over N_a}\,v_{\beta}\sum_{\iota =\pm1} m_{\beta,\iota}\Bigr)\,
\delta \Bigl(k -{2\pi\over N_a}\,\sum_{\iota =\pm1}\iota\,m_{\beta,\iota}\Bigr) 
\, , \hspace{0.25cm} \beta = c,s1 \, .
\label{BQ-gen}
\end{eqnarray}
Here the {\it lowest peak weight} $A^{(0,0)}_{\beta}$ is that associated with the transition from the ground state
to an excited state generated by the processes (A) and (B). The relative weights $a_{\beta}(m_{\beta,\,+1},\,m_{\beta,\,-1})$ 
are generated by the additional processes (C). Such processes occur in the linear part of the $\beta =c,s1$ energy 
dispersions, which corresponds to the vicinity of the $\beta =c,s1$ Fermi points. 
Thus their energy spectra involve the $\beta =c,s1$ Fermi-points velocities, $v_{\beta}$,
Eq. (\ref{velo}) of Appendix \ref{pseudo-repre}. The number of elementary pseudofermion - pseudofermion-hole processes (C)
of momentum $\pm 2\pi/N_a$ in the vicinity of the $\beta,\iota$ Fermi points is denoted
in the summations of Eq. (\ref{BQ-gen}) by $m_{\beta,\iota}=1,2,3,...$.

After a suitable algebra that involves the effective 
pseudofermion anticommutators, Eq. (\ref{pfacrGS}), one finds that the lowest peak weight 
in Eq. (\ref{BQ-gen}) reads \cite{V,VI},
\begin{eqnarray}
A^{(0,0)}_{\beta} & = & \Big({1\over
N_{a_{\beta}}}\Bigr)^{2N^{\odot}_{\beta}}\, \prod_{j=1}^{N_{a_{\beta}}}\,
\sin^2\Bigl({N_{\beta}^{\odot}(q_j)[Q_{\beta}(q_j)
-\pi]+\pi\over 2}\Big) \, \prod_{j=1}^{N_{a_{\beta}}-1}\,
\Big[\sin\Bigl({\pi j\over N_{a_{\beta}}}\Bigr)\Bigr]^{2[N_{a_{\beta}} -j]} \nonumber \\
& \times &
\prod_{i=1}^{N_{a_{\beta}}}\prod_{j=1}^{N_{a_{\beta}}}\,\theta
(j-i)\,\sin^2\Bigl({N_{\beta}^{\odot}({q}_j)
N_{\beta}^{\odot}({q}_i)[Q_{\beta}({q}_j) -
Q_{\beta}({q}_i) + 2\pi (j-i)-\pi N_{a_{\beta}}] + \pi
N_{a_{\beta}}\over
2N_{a_{\beta}}}\Bigr) \nonumber \\
& \times &
\prod_{i=1}^{N_{a_{\beta}}}\prod_{j=1}^{N_{a_{\beta}}}\,{1\over
\sin^2\Bigl(N_{\beta}^{\odot}({q}_i)
N_{\beta}^{\odot}({q}_j)[{\pi (j-i)\over N_{a_{\beta}}} +
{Q_{\beta}({q}_j)\over 2N_{a_{\beta}}} -{\pi\over 2}] +
{\pi\over 2}\Bigr)} \, , \hspace{0.25cm} \beta = c, s1 \, .
\label{A00}
\end{eqnarray}

The number of $\beta$ band discrete momentum values, $N_{a_{\beta}}$, that of
$\beta$ pseudofermions, $N_{\beta}^{\odot}=\sum_{j=1}^{N_{a_{\beta}}}N_{\beta}^{\odot}({q}_j)$,
and the corresponding $\beta$ band momentum distribution function, $N_{\beta}^{\odot}({q}_j)$,
in this expression are those of the excited state generated by the processes (A) and (B). On the other hand,
$Q_{\beta}({q}_j)$ is the present scattering theory phase-shift functional, Eqs. (\ref{qcan1j}), (\ref{Qcan1j}), and (\ref{pican}).
The deviations in that functional expression are those
generated by the corresponding ground-state - excited-state transition.  

Furthermore, the general expression of the relative weights 
$a_{\beta}(m_{\beta,\,+1},\,m_{\beta,\,-1})$ also appearing in Eq. (\ref{BQ-gen}), which result
from transitions to the tower of excited energy eigenstates generated by the processes (C), reads \cite{V,VI},
\begin{equation}
a_{\beta}(m_{\beta,\,+1},
m_{\beta,\,-1})=\Bigl[\prod_{\iota =\pm 1}
a_{\beta,\iota}(m_{\beta,\iota})\Bigr]
\Bigl[1+{\cal{O}}\Bigl({\ln N_a\over N_a}\Bigr)\Bigr] \, ,
\hspace{0.25cm} \beta = c, s1 \, .
\label{aNNDP}
\end{equation}
Here the relative weight $a_{\beta,\iota}(m_{\beta,\iota})$ is given by,
\begin{equation}
a_{\beta,\iota}(m_{\beta,\iota}) = \prod_{j=1}^{m_{\beta,\iota}}
{(2\Delta_{\beta}^{\iota} + j -1)\over j} = \frac{\Gamma (m_{\beta,\iota} +
2\Delta_{\beta}^{\iota})}{\Gamma (m_{\beta,\iota}+1)\,
\Gamma (2\Delta_{\beta}^{\iota})} \, , \hspace{0.25cm}
\beta = c, s1 \, , \hspace{0.25cm} \iota =\pm 1 \, ,
\label{aNDP}
\end{equation}
where $\Gamma (x)$ is the usual gamma function. It follows from Eq. (\ref{aNDP}) that,
\begin{equation}
a_{\beta,\iota}(1) = 2\Delta_{\beta}^{\iota} =[\delta {\bar{q}}_{F\beta}^{\iota}/(2\pi/N_a)]^2 \, ,
\hspace{0.25cm} \beta = c, s1 \, , \hspace{0.25cm} \iota =\pm 1 \, . 
\label{a10DP-iota}
\end{equation}

The four $\beta = c, s1$ and $\iota =\pm 1$ functionals, $2\Delta^{\iota}_{\beta}$, in this equation play a major role in the 
PDT. They are fully controlled by the corresponding four excited-states right ($\iota =+1$) and left ($\iota =-1$) $\beta = c, s1$ 
canonical-momentum Fermi-point deviations $\delta {\bar{q}}_{F\beta}^{\iota}$, Eq. (\ref{dbarqFb}).
It follows from that equation that in units of $2\pi/N_a$ those four $\beta = c, s1$ and $\iota =\pm 1$ deviations 
read $\delta {\bar{q}}_{F\beta}^{\iota}/(2\pi/N_a) =\iota\,\delta N^F_{\beta,\iota}+
Q^{\Phi}_{\beta} (\iota q_{F\beta})/2\pi$. Thus the functionals $2\Delta^{\iota}_{\beta}$
can be written as,
\begin{equation}
2\Delta^{\iota}_{\beta} = \left(\iota\delta N^{0,F}_{\beta,\iota} +
{Q_{\beta}(\iota q_{F\beta})\over 2\pi}\right)^2 = \left(\iota\delta N^{0,F}_{\beta,\iota}
- i {\log S_{\beta} (\iota q_{F\beta})\over 2\pi}\right)^2  \, ,
\hspace{0.25cm} \beta = c, s1 \, ,
\hspace{0.25cm} \iota =\pm 1 \, .
\label{functional}
\end{equation}
The deviations $\delta N^F_{\beta,\iota}$ and $\delta N^{0,F}_{\beta,\iota}$ in the number of $\beta =c,s1$
pseudofermions at the right ($\iota =+1$) and left ($\iota =+1$) $\beta =c,s1$ Fermi points account and
do not account, respectively, for the possible shift of the
BA quantum numbers $I^{\beta}_j$, Eqs. (\ref{q-j}) and (\ref{Ic-an}) of Appendix \ref{pseudo-repre}. 
Here $Q_{\beta}(q_j)/2$ stands again for the pseudofermion scattering theory phase-shift functional, 
Eqs. (\ref{qcan1j}) and (\ref{Qcan1j}), and $S_{\beta} (q_j)=e^{i\,Q_{\beta} (q_j)}$,
Eq. (\ref{San}), is the corresponding $\beta$ pseudofermion dressed $S$ matrix. In the case
of Eq. (\ref{functional}), such a phase shift and $S$ matrix refer to scatterers at the four 
$\beta= c,s1$ and $\iota =\pm 1$ Fermi points, $q_j=\iota q_{F\beta}$.
Consistent with Eq. (\ref{San}), the $\log $ branch 
in Eq. (\ref{functional}) is $\log S_{\beta} (\iota q_{F\beta})=i\,Q_{\beta}(\iota q_{F\beta})$.

According to Eq. (\ref{a10DP-iota}), the functional, Eq. (\ref{functional}), is the relative
weight of the $\alpha,\iota$ pseudofermion spectral function $m_{\beta,\iota}=1$ peak. Moreover,
\begin{equation}
a_{\beta} (1,\,0)= 2\Delta_{\beta}^{+1} \, ; \hspace{0.50cm}
a_{\beta} (0,\,1) = 2\Delta_{\beta}^{-1} \, ,
\hspace{0.25cm} \beta = c, s1 \, . 
\label{a10DP}
\end{equation}

The relative weight, Eq. (\ref{aNDP}), has the following asymptotic behavior,
\begin{equation}
a_{\beta,\iota} (m_{\beta,\iota}) \approx \frac{1}{\Gamma
(2\Delta_{\beta}^{\iota})}
\Bigl(m_{\beta,\iota}\Bigr)^{2\Delta_{\beta}^{\iota}-1}
\, ; \hspace{0.50cm} 2\Delta_{\beta}^{\iota}\neq 0 \, ;
\hspace{0.25cm} \beta = c, s1 \, , \hspace{0.25cm} \iota =
\pm 1 \, . 
\label{f}
\end{equation}

The $\beta= c,s1$ pseudofermion spectral function lowest peak weight $A^{(0,0)}_{\beta}$,
Eq. (\ref{A00}), involves products $\prod_{i=1}^{N_{a_{\beta}}}$ over all $N_{a_{\beta}}=[N_{\beta}+N_{\beta}^h]$
corresponding excited-state $\beta =c,s1$ band scatterers. That refers both to those that pre-exist
in the ground state and are created under the transition to the excited state.
This confirms that the 1D Hubbard model dynamical and spectral properties are within the pseudofermion 
scattering theory controlled by microscopic processes that have contributions from all such $\beta= c,s1$
scatterers. In contrast, the holon and spinon scattering theories of Refs. \cite{Natan,S-0,S}
consider only the few phase shifts associated with scatterers created under the transition to the excited state.
Thus they do not account for most of the microscopic processes that contribute to the important
lowest peak weights $A^{(0,0)}_{\beta}$. 
\begin{figure}
\subfigure{\includegraphics[width=7cm,height=7cm]{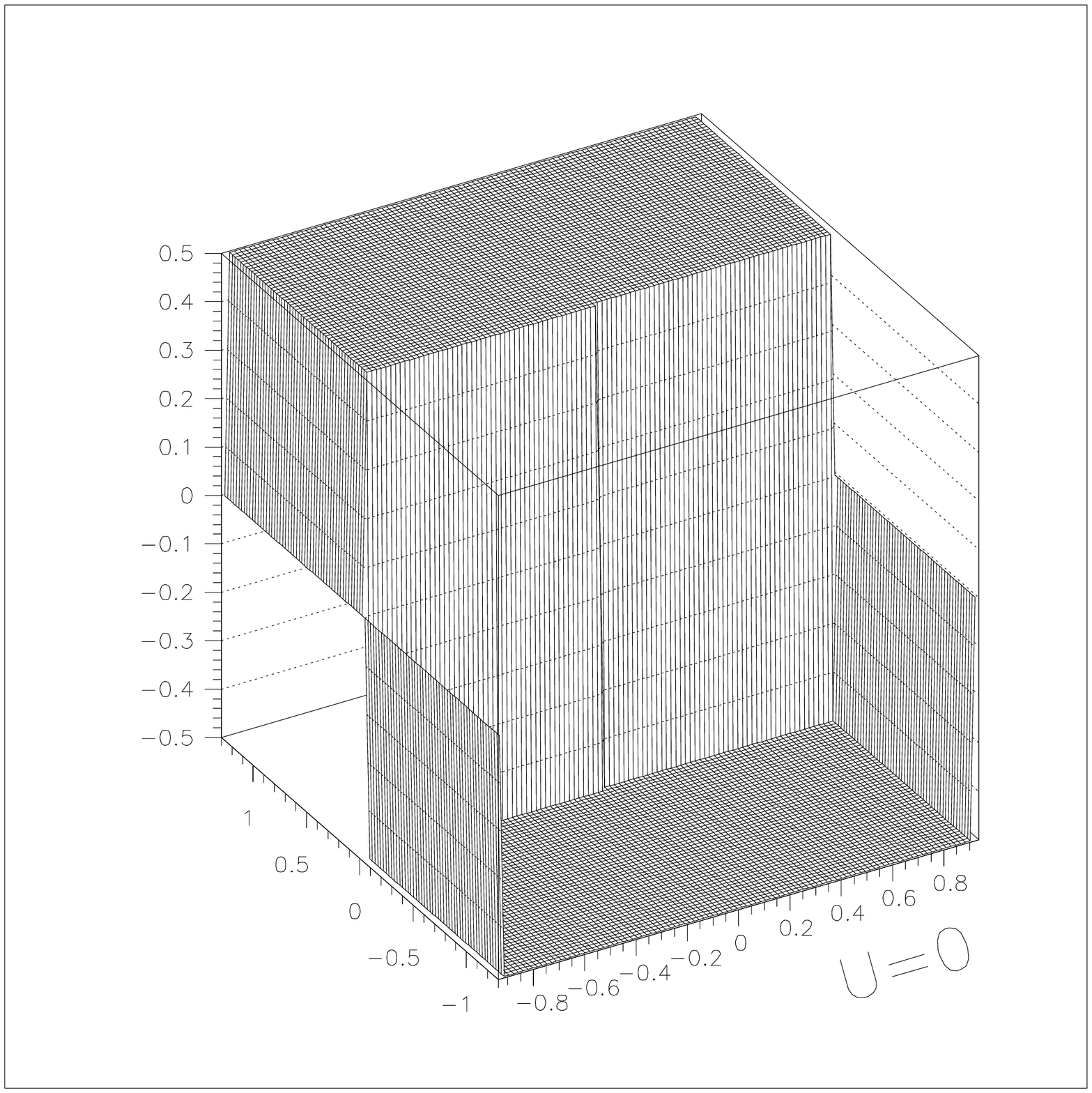}}
\subfigure{\includegraphics[width=7cm,height=7cm]{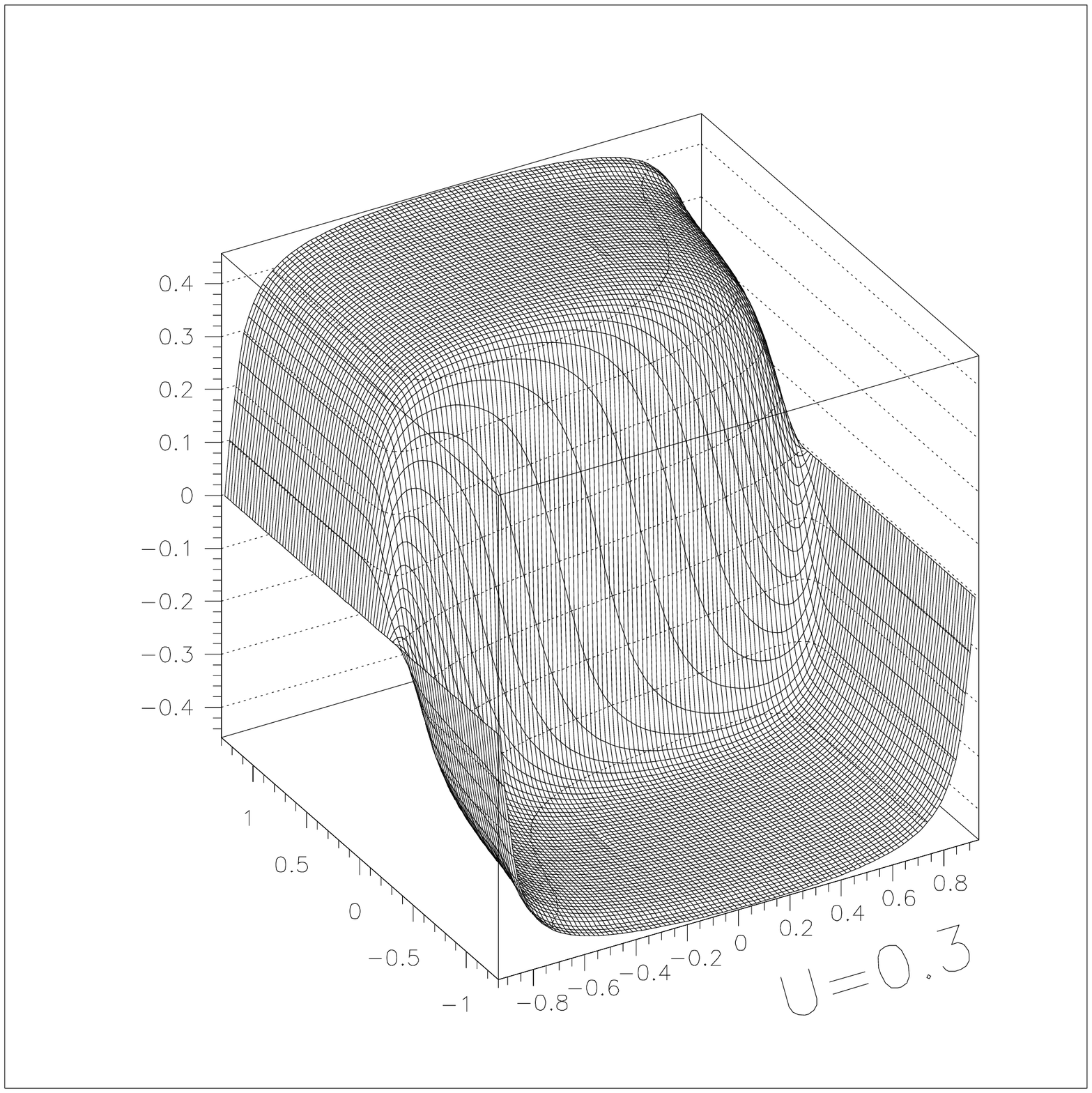}}
\subfigure{\includegraphics[width=7cm,height=7cm]{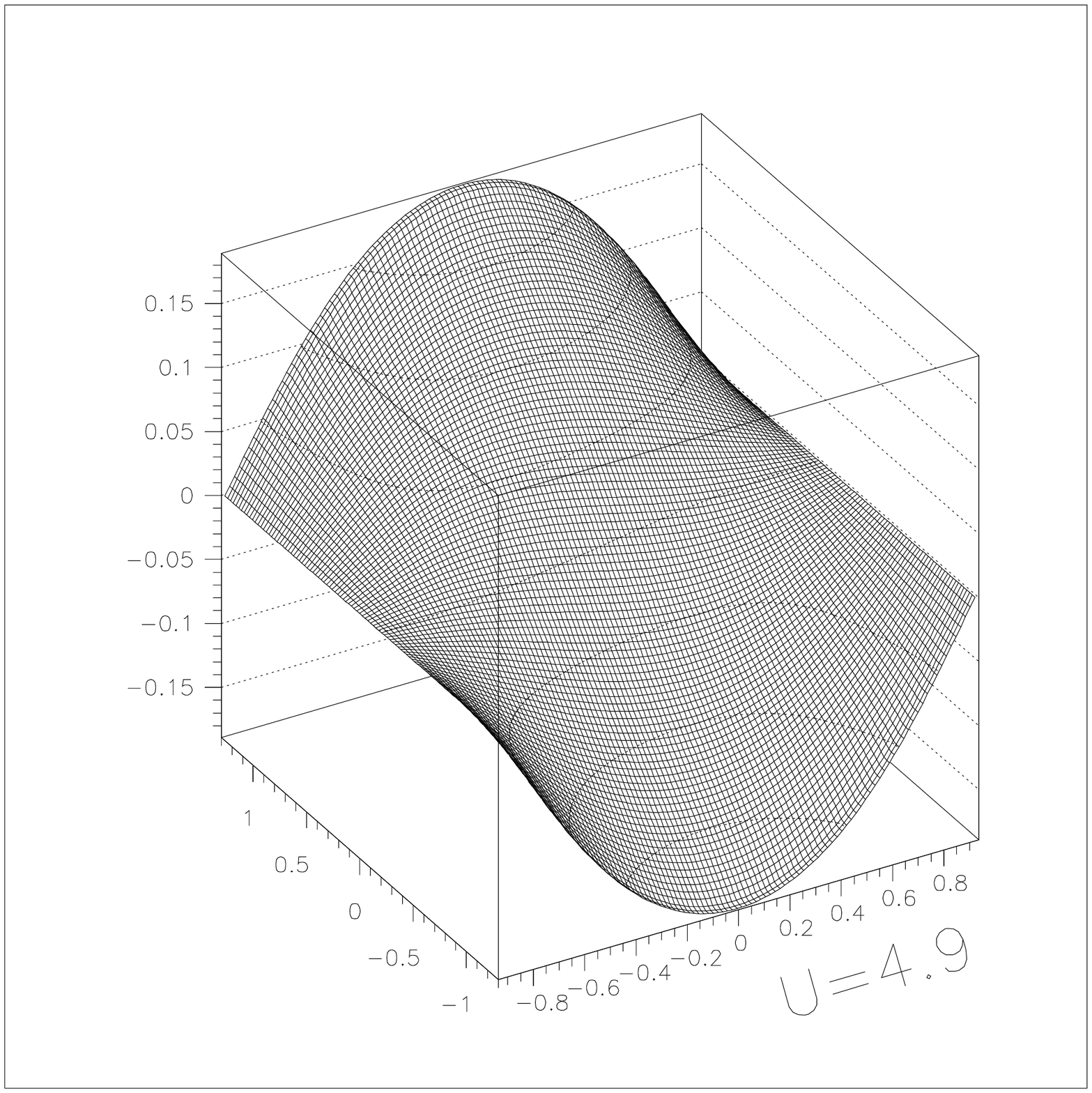}}
\subfigure{\includegraphics[width=7cm,height=7cm]{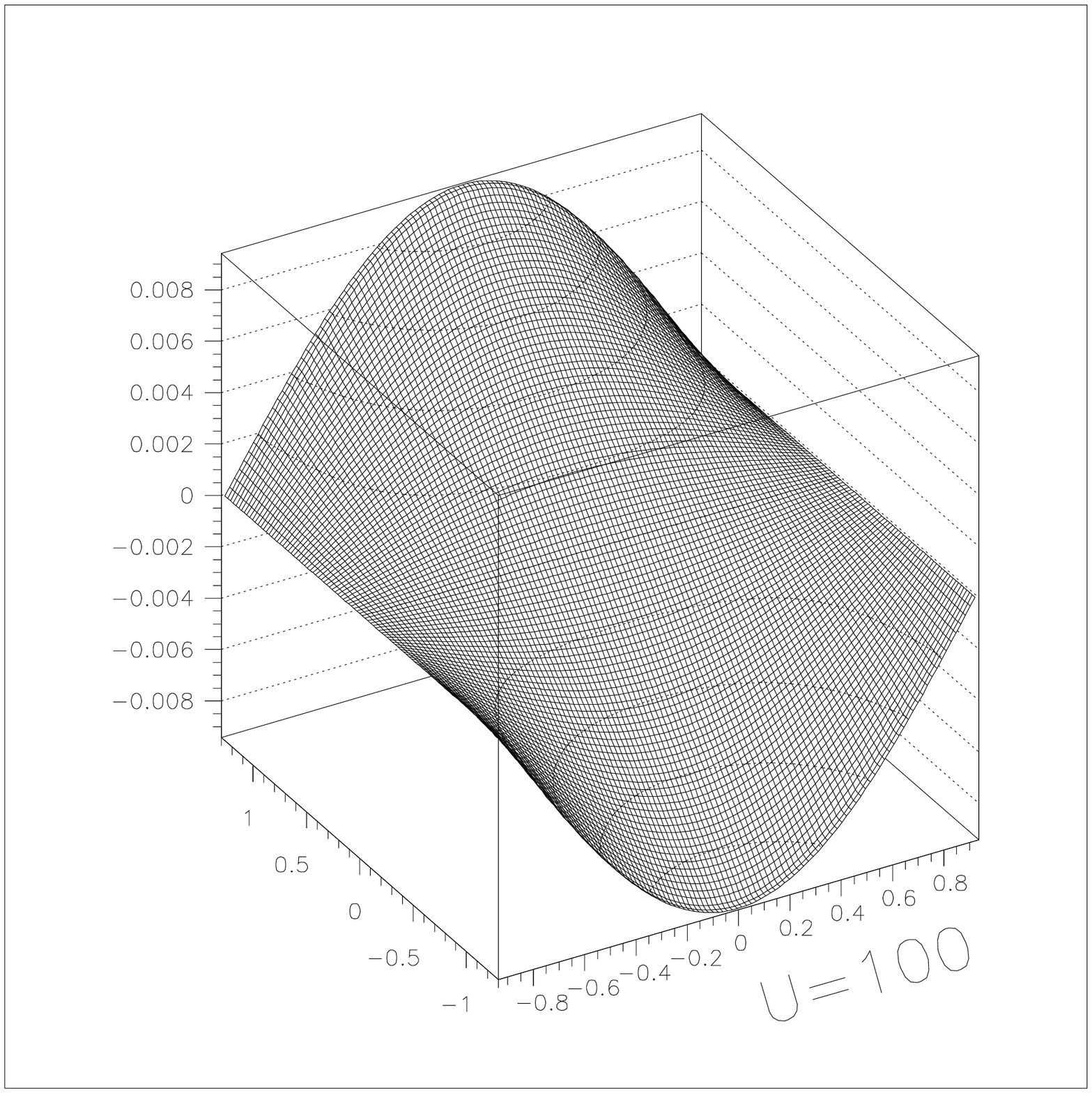}} 
\caption{The elementary two-pseudofermion phase shift $\pi\,\Phi_{s1,\eta 1}(q,q')$ in units of $\pi$ as a function of
$q$ and $q'$ for the same densities and $U/t$ values as in Fig. \ref{F1}. Alike in that figure,  
the scatterer and scattering-center bare-momentum values $q$ and $q'$, respectively, correspond to the right and left 
axis, respectively.}
\label{F6}
\end{figure}

Such holon and spinon scattering theories neither include most of the phase shifts that for $n\neq 1$ 
appear in the expressions of the exponents in the dynamical and spectral function expressions.
Indeed, within the PDT such exponents are linear combinations
of the four $\beta = c, s1$ and $\iota =\pm 1$ phase-shift related functionals, $2\Delta^{\iota}_{\beta}$, Eq. (\ref{functional}).
Specifically, the exponents $\zeta_{\beta} (k)$ and $\zeta$ that control important line-shape singularities in the vicinity of well-defined 
$(k,\omega)$ plane $\beta =c,s1$ branch lines and points have the general form \cite{V,VI,TTF},
\begin{eqnarray}
\zeta_{\beta} (k) & = & -1+\sum_{\beta' = c,s1}\sum_{\iota =\pm 1}2\Delta_{\beta'}^{\iota} (q)\vert_{q=c_0\,[k-P_0]} 
 \, , \hspace{0.25cm} \beta = c,s1 \, .
\nonumber \\
\zeta & = & -2+\sum_{\beta' = c,s1}\sum_{\iota =\pm 1}2\Delta_{\beta'}^{\iota} \, ; \hspace{0.50cm} 
\omega \neq \pm v_{\beta}\,(k-P_0) \, , \hspace{0.25cm} \beta =c,s1 \, ,
\nonumber \\
\zeta & = & -1- 2\Delta_{\beta}^{\mp} +\sum_{\beta' = c,s1}\sum_{\iota =\pm 1}2\Delta_{\beta'}^{\iota} \, ; \hspace{0.50cm} 
\omega \approx \pm v_{\beta}\,(k-P_0) \, , \hspace{0.25cm} \beta =c,s1 \, ,
\label{exponents}
\end{eqnarray}
respectively. Here $c_0=+1$ and $c_0=-1$ refer to $\beta =c,s1$ pseudofermion creation and annihilation, respectively.
The momentum $q$ dependence of $2\Delta_{\beta'}^{\iota} (q)$ in the
$\zeta_{\beta} (k)$ expression stems for a $\beta =c,s1$ branch line
involving a phase-shift contribution, $c_0\,2\pi\,\Phi_{\beta',\beta}(\iota q_{F\beta'},q)$. It emerges within the 
corresponding scattering phase shift
$Q^{\Phi}_{\beta'} (\iota q_{F\beta'})$, Eq. (\ref{qcan1j}). The bare momentum $q$ corresponds to the 
canonical momentum $\bar{q} = \bar{q} (q)$ of the $\beta =c,s1$ pseudofermion branch-line
created or annihilated under the transition to the excited state. The $\beta =c,s1$ energy dispersion
$\varepsilon_{\beta} (q)$, Eq. (\ref{epsilon-q}) of Appendix \ref{pseudo-repre}, of such an
elementary object that determines the $(k,\omega)$ plane shape of the corresponding
$\beta =c,s1$ pseudofermion branch-line \cite{V,VI,LE,TTF}.
Moreover, $P_0$ stands in Eq. (\ref{exponents}) for the following momentum, which may vanish,
\begin{equation}
P_0 = \pi\,M_{\eta,-1/2} + 2k_F\,\delta 2J_c^F + k_{F\downarrow}\,\delta 2J_{s1}^F
+ (\pi -2k_F) 2J_{\eta\nu}^L \, .
\label{P0}
\end{equation}
Here $2J_{\beta}^F=\delta N_{\beta,+1}^F-\delta N_{\beta,-1}^F$ for $\beta =c,s1$ and
$2J_{\eta\nu}^L=N_{\eta\nu,+1}^L-N_{\eta\nu,-1}^L$ where $N_{\eta\nu,\iota}^L$ stands for the
number of $\eta\nu$ pseudofermions created at the $\eta\nu$ band limiting momentum values 
$\iota q_{\eta\nu}=\iota (\pi -2k_F)$, Eq. (\ref{qcanGS}) of Appendix \ref{pseudo-repre}, in the
second expression of Eq. (\ref{DNqF-all}) of Appendix \ref{Ele2PsPhaShi} for $\beta' =\eta\nu$.

Such $c$ and $s1$ branch lines occur, for example, in the one-electron spectral functions \cite{TTF}. 
Under the processes (A) and (B) that generate the one-electron excited states,
one $c$ pseudofermion or one $c$ pseudofermion hole scattering center and one $s1$ pseudifermion hole scattering center are created.
On the other hand, the four scatterers with phase shifts $Q_{c}(\pm q_{Fc})/2$ and $Q_{s1}(\pm q_{Fs1})/2$
in the expression, Eq. (\ref{functional}), of the four functionals 
$2\Delta^{\pm}_{c}$ and $2\Delta^{\pm}_{s1}$, respectively, refer to four scatterers: two $c$ scatterers and two $s1$ scatterers.
Such functionals control the corresponding branch-line
momentum-dependent exponent $\zeta_{\beta} (k)$, Eq. (\ref{exponents}). 
This applies applies as well to the exponents in the low-energy spectral and correlation
functions in the vicinity of a $(k,\omega)$ plane point, $(k,\omega)=(P_0,0)$.
Those also involve the four $\beta =c,s1$ and $\iota =\pm 1$ phase shifts $Q_{\beta}(\iota q_{F\beta})/2$ \cite{LE}.
In this second case, such functionals control the exponent $\zeta$, Eq. (\ref{exponents}). 

In the case of a one-electron branch line, one of the $c$ and $s1$ scattering centers is created at a
momentum away from the corresponding $c$ or $s1$ Fermi points and the other at one of such points.
On the other hand, in the case of the low-energy line shape in the vicinity of a $(k,\omega)$ plane
point, $(k,\omega)=(P_0,0)$, both $c$ and $s1$ scattering centers are created at two of the
corresponding Fermi points. It then follows that for a one-electron branch line (and line shape in the vicinity 
of a $(k,\omega)$ plane point) out of the four $\beta = c, s1$ and $\iota =\pm 1$ scatterers with phase shifts $Q_{\beta}(\iota q_{Fc})/2$,
three (and two) pre-exist in the ground state. 

Concerning the excitation associated with the one-electron branch line (and line shape in the vicinity 
of a $(k,\omega)$ plane point), the scattering theory of the holon-spinon representation of Ref. \cite{Natan}
only provides the phase shift of one scatterer (and two scatterers), out of the four $\beta = c, s1$ and $\iota =\pm 1$ 
phase shifts $Q_{\beta}(\iota q_{Fc})/2$. It is that (and those) of the scatterer (and
two scatterers) that is (and are) created at $\beta =c,s1$ Fermi points under the transition to the
excited state. Hence that theory does not contain the other three (and two) phase shifts that control the branch lines 
(and line shape in the vicinity of $(k,\omega)$ plane points) exponents $\zeta_{\beta} (k)$
(and $\zeta$), Eq. (\ref{exponents}). The same limitations occur in the case of the holon-spinon representation scattering theory of Refs. \cite{S-0,S}, 
which in addition is valid only at half filling. Such results confirm the unsuitability of the two holon-spinon representations 
scattering theories of Refs. \cite{Natan} and \cite{S-0,S} to describe the model dynamical 
and spectral properties. That follows from for them the ground states being mere holon and spinon vacua,
without elementary-object occupancies. 

For further useful information on the relation of the $\beta$ pseudofermion representation to the holon-spinon 
representations of Refs. \cite{Natan} and \cite{S-0,S}, see Ref. \cite{paper-I}. Such representations are shown in
that paper to refer to alternative sets of degenerate energy eigenstates that span well-defined model reduced subspaces. 

The whole one-electron $(k,\omega)$ plane spectral-weight distribution is controlled only by the $\beta$
pseudofermion scattering theory general phase shifts $Q_{\beta} (q_j)/2$, Eqs. (\ref{qcan1j}) and (\ref{Qcan1j}).
The corresponding dynamical theory state summations are in general a very complex numerical problem.
In the $u\rightarrow\infty$ limit such phase shifts become independent of $q_j$. That much simplifies
the computation of the one-electron spectral functions over the whole $(k,\omega)$ plane. They are plotted
in Fig. 1 of Ref. \cite{Penc-96} for electronic density $n=0.5$ and spin density $m=0$.

Numerical studies on spectral functions as those of Ref. \cite{Essler-10}, which do not rely directly on a specific elementary-object 
representation, implicitly account for the microscopic processes and mechanisms of the
pseudofermion scattering theory and PDT, respectively. 

\section{Concluding remarks}
\label{Concluding}

In this paper it has been shown that a set of 1D Hubbard model natural scatterers 
emerges from the rotated-electron occupancy configurations degrees of
freedom separation, as defined in Refs. \cite{I,II,paper-I}. Such scatterers are the elementary 
objects previously used within the PDT studies of the one- and two-electron spectral-weight
distributions \cite{V,VI,LE,TTF}: The $c$ pseudofermions,
the spin-neutral composite $2\nu$-spinon $s\nu$ pseudofermions, and the $\eta$-spin-neutral composite 
$2\nu$-$\eta$-spinon $\eta\nu$ pseudofermions. Those of such elementary objects created under a transition
to an excited energy eigenstate play as well the role of the theory scattering centers.
On the other hand, the $\eta$-spin-$1/2$ unbound $\eta$-spinons and spin-$1/2$ unbound spinons, respectively,
are scattering-less elementary objects as far as their $\eta$-spin and spin, respectively, $SU(2)$ symmetry internal 
degrees of freedom is concerned. Hence all the $\beta =c,\alpha\nu$ pseudofermion scattering theory 
dressed $S$ matrices have dimension one. 

The theory scatterers and scattering centers that naturally emerge from rotated-electron
configurations refer to elementary objects whose occupancy configurations generate
exact energy eigenstates. That is why such occupancy configurations have been constructed
inherently to account for an infinite number of conservation laws \cite{CM-86,CM,Prosen}, which are associated
with the model integrability \cite{Lieb,Takahashi,Woy,Martins}. As a result, the theory
scatterers and scattering centers undergo only zero-momentum forward scattering events. This is
in contrast to the underlying very involved non-perturbative many-electron processes.

There is a $\beta$ pseudofermion scattering theory for each ground state
and corresponding PS. The theory is valid for PSs of ground states with 
arbitrary values of the electronic density $n$ and spin density $m$.
Within each of such large subspaces, the $c$ pseudofermions,
$s\nu$ pseudofermions, and $\eta\nu$ pseudofermions emerge from the
$c$ pseudoparticles, $s\nu$ pseudoparticles, and $\eta\nu$ pseudoparticles, respectively.
This occurs through a well-defined unitary transformation. Under it, the $\beta$ pseudoparticles 
discrete momentum values are slightly shifted. This
leads to the corresponding $\beta$ pseudofermions discrete canonical momentum values,
which are phase-shift dependent. Importantly, that renders the $\beta$ pseudofermion
spectrum without energy interaction terms. This property drastically simplifies the expression 
of electron spectral functions in terms of pseudofermion spectral functions. 
Otherwise, the $\beta$ pseudoparticles and $\beta$ pseudofermions
have the same properties.

Concerning the elementary objects that participate
in the theory scattering events, for densities $n\neq 1$ and $m\neq 0$ the ground states are populated only 
by $c$ and $s1$ pseudofermions and $c$ and $s1$ pseudofermion holes. (The $c$ [and $s1$] momentum
bands of $S_{\eta}=0;n=1$ [and $S_{s}=0;m=0$] ground states are full, so that such states have no 
$c$ [and $s1$] pseudofermion holes.) For all densities, ground states are not populated by $\alpha\nu\neq s1$
pseudofermions. Therefore, only in the case of the $c$ and $s1$ bands do the
discrete canonical-momentum overall shifts, $Q_{\beta}(q_j)/N_a$, lead to
AndersonÕs orthogonality catastrophes \cite{Anderson-67}. Those play a major role in
the one- and two-electron matrix elements quantum overlaps that control the corresponding
spectral-weight distributions \cite{V,VI,LE,TTF}.
    
The ``in" and ``out" states of the pseudofermion scattering theory are exact excited energy eigenstates. 
They can be written as a simple direct product of ``in" asymptote and ``out" asymptote
one-pseudofermion scattering states, respectively. Such a property combined with the also simple 
form of the $\beta$ pseudofermion and $\beta$ pseudofermion hole dressed $S$ matrices is behind the suitability of the
present representation to the describe the finite-energy spectral and correlation properties
of the model metallic phase. As discussed in Section \ref{PDT}, the pseudofermion microscopic scattering 
processes control the PDT one- and two-electron spectral-weight distributions \cite{V,VI,LE,TTF}.
This applies to finite values of the on-site repulsion, at both low and finite excitation energy.

The relation of the elementary objects of the representation used in the studies of this
paper to the holons and spinons of the scattering theories of Ref. \cite{Natan} and Refs. \cite{S-0,S}, respectively, is
an issue that has been clarified in Ref. \cite{paper-I}. As discussed in Section \ref{PDT},
the major advantage of the theory considered in this paper 
relative to such holon and spinon scattering theories refers to the explicit description
of the microscopic processes that control the model dynamical and spectral properties. Such an advantage
follows from the present theory accounting for both the phase shifts of the scatterers that pre-exist in the ground
state and those that are created under the transitions to the excited states. 
In contrast, the holon and spinon scattering theories account only for the few phase
shifts of their scatters, which are those created under the transitions to the excited
states. That is consistent with the holon and spinon vacuum os such theories being the
ground state, whose structure and occupancies are not accounted for.

Several properties predicted by the 1D Hubbard model have been observed in low-dimensional complex
materials \cite{polyace}. The investigations presented in Refs. \cite{TTF,spectral0,spectral,spectral-06} confirm that 
the PDT describes successfully the unusual finite-energy spectral features observed by angle-resolved 
photoelectron spectroscopy in quasi-1D organic metals. Combination of such investigations results with those of this paper 
confirms that the scattering centers of the pseudofermion scattering theory are observed in such materials.

\begin{acknowledgments}
We thank N. Andrei, S.-J. Gu, H. Q. Lin, A. Moreno, A. Muramatsu, and K. Penc
for illuminating discussions and the hospitality and support of the
Beijing Computational Science Research Center. J. M. P. C. thanks the hospitality  of the Institut f\"ur Theoretische Physik III, Universit\"at Stuttgart,
and the financial support by the FEDER through the COMPETE Program, Portuguese FCT both 
in the framework of the Strategic Project PEST-C/FIS/UI607/2011 and under SFRH/BSAB/1177/2011, 
German transregional collaborative research center SFB/TRR21, and Max Planck Institute for Solid State Research.
\end{acknowledgments}
\appendix

\section{Useful information on the pseudoparticle representation}
\label{pseudo-repre}

Here some general results useful for the studies of the paper are presented. 
Our BA representation refers to LWSs of both the spin
and $\eta$-spin $SU(2)$ algebras. For such Bethe states the numbers,
$M_{\eta,-1/2}^{un}$ and $M_{s,-1/2}^{un}$, Eq. (\ref{L-L}),
of $\eta$-spin projection $-1/2$ unbound $\eta$-spinons and spin projection $-1/2$ unbound spinons,
respectively, vanish. Moreover, we use the notation, $\vert l_{\rm r},l_{\eta s},u\rangle$, for the energy eigenstates. 
Within it, the Bethe states are written as
$\vert l_{\rm r},l_{\eta s}^0,u\rangle $. A non-LWS energy eigenstate $\vert l_{\rm r},l_{\eta s},u\rangle$
is generated from the corresponding Bethe state $\vert l_{\rm r},l_{\eta s}^0,u\rangle $ as follows,
\begin{equation}
\vert l_{\rm r},l_{\eta s},u\rangle = \prod_{\alpha=\eta,s}\left[\frac{1}{
\sqrt{{\cal{C}}_{\alpha}}}({\hat{S}}^{+}_{\alpha})^{M_{s,-1/2}^{un}}\right]\vert l_{\rm r},l_{\eta s}^0,u\rangle \, .
\label{Gstate-BAstate}
\end{equation}
Here,
\begin{eqnarray}
{\cal{C}}_{\alpha} & = & \langle l_{\rm r},l_{\eta s}^0,u\vert ({\hat{S}}^{-}_{\alpha})^{M_{s,-1/2}^{un}}({\hat{S}}^{+}_{\alpha})^{M_{s,-1/2}^{un}}\vert l_{\rm r},l_{\eta s}^0,u\rangle
\nonumber \\
& = & [M_{\alpha,-1/2}^{un}!]\prod_{j'=1}^{M_{\alpha,-1/2}^{un}}[\,M_{\alpha}^{un}+1-j'\,] \, , 
\hspace{0.25cm} M_{\alpha,-1/2}^{un}=1,...,M_{\alpha}^{un}
\hspace{0.15cm}  \, ,  \alpha = \eta,s  \, ,
\label{Calpha}
\end{eqnarray}
are normalization constants, ${\hat{S}}_{\alpha}^{+}$ and ${\hat{S}}^{-}_{\alpha}$ are
the $\eta$-spin ($\alpha =\eta$) and spin ($\alpha =s$) off-diagonal generators,
Eq. (\ref{Scs-rot}) for $i=\pm $ and $\alpha =\eta,s$, 
$M_{\alpha}^{un}=2S_{\alpha}$ denotes the total number of unbound $\eta$-spinons ($\alpha =\eta$) and 
unbound spinons ($\alpha =s$), and $l_{\eta s}$ 
and $l_{\eta s}^0$ stand for the set of numbers $[S_{\eta},S_{s},M_{\eta,-1/2}^{un},M_{s,-1/2}^{un}]$
and $[S_{\eta},S_{s},0,0]$, respectively. Within our notation, $l_{\eta s}^0$
refers to limiting values of the general label $l_{\eta s}$ such that $M_{\eta,-1/2}^{un}=M_{s,-1/2}^{un}=0$,
which are those suitable to the Bethe states. The label $\eta s$
in $l_{\eta s}$ and $l_{\eta s}^0$ symbolizes the two $SU (2)$ symmetries associated with the
numbers $S_{\eta},n_{\eta}$ and $S_{s},n_s$, respectively, in $[S_{\eta},S_{s},M_{\eta,-1/2}^{un},M_{s,-1/2}^{un}]$. Moreover, 
$l_{\rm r}$ stands for all remaining quantum numbers
beyond $l_{\eta s}$ needed to uniquely define an energy eigenstate $\vert l_{\rm r},l_{\eta s},u\rangle$.
This includes the eigenvalue $2S_c$ of the $c$ hidden $U(1)$ symmetry generator.

The energy eigenvalues of the Bethe states 
$\vert l_{\rm r},l_{\eta s}^0,u\rangle $ are given by, 
\begin{eqnarray}
E_{l_{\rm r},l_{\eta s}^0} & = & E_{symm}  + \mu\,M^{un}_{\eta,+1/2} 
+\mu_B\,H\,M^{un}_{s,+1/2} \, ,
\nonumber \\
E_{symm} & = & -2t\sum_{j=1}^{N_a}\left[N_{c} (q_j)\,\left(\cos k (q_j) + {u}\right) - u/2\right]
\nonumber \\
& + & 4t\sum_{\nu=1}^{\infty}\sum_{j=1}^{N_{a_{\eta\nu}}}\,N_{\eta\nu} (q_j)
\left[{\rm Re}\,\left\{\sqrt{1 - (\Lambda_{\eta\nu} (q_j) + i \nu\,u)^2}\right\} - {\nu\,u}\right] \, .
\label{E-BA}
\end{eqnarray}
For the non-LWSs, the unbound $\eta$-spinon number $M^{un}_{\eta,+1/2}$ and
unbound spinon number $M^{un}_{s,+1/2}$ appearing in the first expression
given here are replaced by $[M^{un}_{\eta,+1/2}-M^{un}_{\eta,-1/2}]$ and
$[M^{un}_{s,+1/2}-M^{un}_{s,-1/2}]$, respectively. 

The energy spectrum, Eq. (\ref{E-BA}), involves the rapidity functions $k (q_j)$ and
$\Lambda_{\eta\nu} (q_j)$. Those are related to the rapidity function $\Lambda_{s\nu} (q_j)$
through a set of coupled thermodynamic BA equations.
For the Bethe states, such rapidity functions are for each energy eigenstate obtained
by solution of those equations. In functional form, they read,
\begin{eqnarray}
k_c (q_j) & = & q_j - {2\over N_a}\sum_{\nu =1}^{\infty}
\sum_{j'=1}^{N_{a_{s\nu}}}\,N_{s\nu}(q_{j'})\arctan\left({\sin
k_c (q_j)-\Lambda_{s\nu}(q_{j'}) \over \nu u}\right)
\nonumber \\
& - & {2\over N_a}\sum_{\nu =1}^{\infty}
\sum_{j'=1}^{N_{a_{\eta\nu}}}\, N_{\eta\nu}(q_{j'}) \arctan\left({\sin
k_c (q_j)-\Lambda_{\eta\nu}(q_{j'}) \over \nu u}\right) 
\, , \hspace{0.25cm} j = 1,...,N_a \, , 
\label{Tapco1}
\end{eqnarray}
and
\begin{eqnarray}
k_{\alpha\nu} (q_j) & = & q_j 
- {2e^{i\pi\delta_{\alpha,\eta}}\over N_a} \sum_{j'=1}^{N_{a}}\,
N_{c}(q_{j'})\arctan\left({\Lambda_{\alpha\nu}(q_j)-\sin k_c (q_{j'})
\over \nu u}\right)\nonumber \\
& + & {1\over N_a}\sum_{\nu' =1}^{\infty}\sum_{j'=1}^{N_{a_{\alpha\nu'}}}\, N_{\alpha\nu'}(q_{j'})\Theta_{\nu, \,\nu'}
\left({\Lambda_{\alpha\nu}(q_j)-\Lambda_{\alpha\nu'}(q_{j'})\over u}\right) \, ,
\nonumber \\
\nu & = &1,...,\infty \, , \hspace{0.25cm} j = 1,...,N_{a_{\alpha\nu}} 
\, , \hspace{0.25cm} \alpha = \eta, s \, .
\label{Tapco2}
\end{eqnarray}
Here 
\begin{equation}
k_{\alpha\nu}(q_j) = \delta_{\alpha ,\eta}\,2\,{\rm Re}\,\{\arcsin (\Lambda_{\eta\nu} (q_j) + i \nu
u)\} \, , \hspace{0.25cm} \nu = 1,...,\infty \, , \hspace{0.25cm} j = 1,...,N_{a_{\alpha\nu}} 
\, , \hspace{0.25cm} \alpha = \eta, s \, , 
\label{kcn}
\end{equation}
is the $\alpha\nu$ rapidity-momentum functional, which vanishes for $\alpha =s$. The
function $\Theta_{\nu,\nu'}(x)$ is given in Eq. (\ref{Theta}) of Appendix \ref{Ele2PsPhaShi}.

The discrete momentum values $q_j$ appearing in Eqs. (\ref{E-BA})-(\ref{kcn}) are given by,
\begin{equation}
q_j = {2\pi\over N_a}\,I^c_j \, , \hspace{0.25cm} j=1,...,N_{a_c}
\, ; \hspace{0.5cm}
q_j = {2\pi\over N_a}\,I^{\alpha\nu}_j \, , \hspace{0.25cm} j=1,...,N_{a_{\alpha\nu}} \, .
\label{q-j}
\end{equation} 
Here the number $N_{a_c}$ of $c$ band discrete momentum values and that $N_{a_{\alpha\nu}}$ of $\alpha\nu$ band discrete 
momentum values read,
\begin{equation}
N_{a_c} = [N_{c} + N^h_{c}] = N_a \, ,
\, ; \hspace{0.5cm}
N_{a_{\alpha\nu}} = [N_{\alpha\nu} + N^h_{\alpha\nu}] \, ,
\label{N*}
\end{equation}
where $N_c = 2S_c$ and $N_{\alpha\nu}$ denote the number of occupied $c$ and $\alpha\nu$ band discrete
momentum values, respectively. Those of unoccupied values are given by,
\begin{eqnarray}
N_c^h & = & 2S_c^h = [N_a-2S_c] \, ,
\nonumber \\
N^h_{\alpha\nu} & = & 
[2S_{\alpha}+2\sum_{\nu'=\nu+1}^{\infty}(\nu'-\nu)N_{\alpha\nu'}] =
[M^{un}_{\alpha}+2\sum_{\nu'=\nu+1}^{\infty}(\nu'-\nu)N_{\alpha\nu'}] \, , 
\hspace{0.15cm} \alpha=\eta,s \, .
\label{N-h-an}
\end{eqnarray}
Furthermore, the BA quantum numbers $I^{c}_j$ and $\{I^{\alpha\nu}_j\}$ in Eq. (\ref{q-j}) 
are either integers or half-odd integers. For simplicity, we often denote such numbers by
$I_j^{\beta}$ where $\beta = c,\alpha\nu$, $\alpha =\eta,s$, and $\nu=1,...,\infty$. Their
values are uniquely determined by the following boundary conditions,
\begin{eqnarray}
I_j^{\beta} & = & 0,\pm 1,\pm 2,... \hspace{0.25cm}{\rm for}\hspace{0.15cm}G_{\beta}\hspace{0.15cm}{\rm even} \, ,
\nonumber \\
& = & \pm 1/2,\pm 3/2,\pm 5/2,... \hspace{0.25cm}{\rm for}\hspace{0.15cm}G_{\beta}\hspace{0.15cm}{\rm odd} \, ,
\label{Ic-an}
\end{eqnarray}
where
\begin{equation}
G_{\beta} =\delta_{\beta,c}\sum_{\alpha =\eta,s}\sum_{\nu=1}^{\infty}N_{\alpha\nu}
+ \delta_{\beta,\alpha\nu} [2S_c + N_{\alpha\nu} -1] \, , \hspace{0.25cm} \alpha = \eta, s \, .
\label{F-beta}
\end{equation}

Within the $N_a\rightarrow\infty$ limit that the thermodynamic BA equations, Eqs. (\ref{Tapco1})-(\ref{kcn}), refer to
it is often convenient to replace the $\beta$ pseudoparticle
discrete momentum values $q_j$, such that $q_{j+1}-q_j=2\pi/N_a$, by corresponding 
continuous momentum variables $q$. Those belong to domains $q\in [-q_{\beta},+q_{\beta}]$, whose 
limiting absolute values $q_{\beta}$ read,
\begin{eqnarray}
q_{c} & =  & \pi \, ,
\nonumber \\
q_{\alpha\nu} & = &{\pi\over N_a}(N_{a_{\alpha\nu}}-1) \hspace{0.25cm}{\rm for}\hspace{0.15cm}N_{a_{\alpha\nu}}\hspace{0.15cm}{\rm odd} \, ,
\nonumber \\
& = & {\pi\over N_a}N_{a_{\alpha\nu}} \hspace{0.25cm}{\rm for}\hspace{0.15cm}N_{a_{\alpha\nu}}\hspace{0.15cm}{\rm even} \, .
\label{qcan-range}
\end{eqnarray}
For the $\beta =\alpha\nu$ branches the $\beta$ discrete-momentum
values distribution is symmetrical and bound by the momentum values $\pm q_{\alpha\nu}$,
with $q_{\alpha\nu}$ given in Eq. (\ref{qcan-range}). On the other hand, if one accounts
for corrections of order $1/N_a$, the $c$ band $q$ range becomes $q\in [q^{-}_{c},q^{+}_{c}]$ where
$q^{\pm}_{c} = \pm \pi$ plus $1/N_a$ corrections, as given in Eqs. (B.15)-(B.17) of Ref. \cite{I}.

For the present ground states with densities $n\in[0,1]$ and $m\in [0,n]$, the limiting momentum values,
Eq. (\ref{qcan-range}), have simple expressions,
\begin{equation}
q_{c} = \pi \, ; \hspace{0.5cm} q_{s1} = k_{F\uparrow} \, ; \hspace{0.5cm} q_{s\nu} =
[k_{F\uparrow}-k_{F\downarrow}] \, , \hspace{0.3cm} \nu >1  \, ; \hspace{0.5cm}
q_{\eta\nu} = [\pi -2k_F] \, . 
\label{qcanGS}
\end{equation}
Here we have again ignored corrections of $1/N_a$ order. 

The pseudofermion scattering theory studied in this paper refers to transitions
from ground states to PS excited energy eiegenstates.
Ground states are described by compact $c$ and $s1$ pseudofermion finite occupancies. Those
correspond to bare-momentum ranges $q\in [-q_{F\beta},+q_{F\beta}]$ where $\beta = c,s1$. 
For densities in the ranges $n\in [0,1]$ and $m\in [0,n]$,
the ground-state unbound $+1/2$ $\eta$-spinon and unbound $+1/2$ spinon numbers are
$M_{\eta,+1/2}^{un}=M_{\eta}^{un}=N_c^h= N_a-N$ and
$M_{s,+1/2}^{un}=M_{s}^{un}=N_{s1}^h= N_{\uparrow}-N_{\downarrow}$, respectively.
On the other hand, such a ground state is not populated by unbound $-1/2$ $\eta$-spinon and 
unbound $-1/2$ spinon, $M_{\eta,-1/2}^{un}=M_{s,-1/2}^{un}=0$. Moreover,
the $\alpha\nu\neq s1$ branches have
vanishing ground-state occupancy. For such a ground state the $c$ and $s1$ Fermi momenta are given by,
\begin{equation}
q_{Fc} = 2k_F \, ; \hspace{0.5cm} q_{Fs1} = k_{F\downarrow} \, . 
\label{q0Fcs}
\end{equation}
Here we have again ignored $1/N_a$ order corrections, which are provided in Eqs. (C.4)-(C.11) of Ref. \cite{I}. 

We denote by $k_{c}^{0} (q_j)$ and $\Lambda^{0}_{\beta} (q_j)$ the specific rapidity-function solutions
of the thermodynamic integral equations, Eqs. (\ref{Tapco1})-(\ref{kcn}), that refer to a ground state. Those play an important role
both in the PDT of Refs. \cite{V,VI,LE} and pseudofermion scattering theory studied
in this paper. Upon suitable manipulations of such thermodynamic BA equations, the ground-state functions 
$k_{c}^{0} (q_j)$ and $\Lambda^{0}_{\beta} (q_j)$ 
may be defined in terms of their inverse functions, $q_j = q_j (\Lambda^{0}_{\beta,j})$, as follows,
\begin{eqnarray}
q_j & = & F_{\beta}^0 (\Lambda^{0}_{\beta,} (q_j)) + (-1)^{\delta_{\beta,\eta\nu}}
\int_{-Q}^{Q}dk\,\bar{\Phi }_{c,\beta}
\left({\sin k\over u}, {\Lambda^{0}_{\beta} (q_j)\over u}\right) \, , \hspace{0.25cm}
j = 1,...,N_{a_{\beta}} \, ,
\nonumber \\
F_c^0 (\Lambda^{0}_{c} (q_j)) & = & \arcsin(\Lambda^{0}_{c} (q_j)) = k_{c}^{0} (q_j) \, ,
\nonumber \\
F_{\alpha\nu}^0 (\Lambda^{0}_{\alpha\nu} (q_j)) & = & 
\delta_{\alpha,\eta}\,2{\rm Re}\left[\arcsin(\Lambda^0_{\eta\nu} (q_j) + i \nu\,u)\right] \, .
\label{GS-R-functions}
\end{eqnarray}
Here $N_{a_{\beta}}=N_a$ is given in Eq. (\ref{N*}) and (\ref{N-h-an}) and the two-pseudofermion phase
shifts $\bar{\Phi }_{c,\beta} (r,r')$ are defined in Eqs. (\ref{Phicc-m})- (\ref{Phicsn-m}) of 
Appendix \ref{Ele2PsPhaShi}. Moreover, the parameter $Q$ and the related 
parameters $B$, $r_c^0$, and $r_{s}^0$ in other quantities of this paper 
may be expressed in terms of the ground-state rapidity functions 
$k^{0}_c (q)$ and $\Lambda^{0}_{s1}(q)$ at the corresponding $c$ and $s1$ Fermi points, Eq. (\ref{q0Fcs}), respectively, as follows,
\begin{equation}
Q \equiv k^{0}_c (2k_F) \, ; \hspace{0.5cm} B \equiv
\Lambda^{0}_{s1}(k_{F\downarrow}) \, ; \hspace{0.50cm}
r_c^0 = {\sin Q \over u}  \, ; \hspace{0.5cm} r_s^0 = {B\over u} \, .
\label{QB-r0rs}
\end{equation}

For the $S_{\eta}=0;n=1$ and $S_{s}=0;m=0$ absolute ground state the $\beta \neq c,s1$ bands do not exist. On
the other hand, for it the equations given in Eq. (\ref{GS-R-functions}) have an analytical solution for the $\beta =c,s1$ branches
in terms of the inverse of the rapidity functions $k_c^0 (q_j)$ 
(such that $\Lambda^{0}_{c} (q_j) =\sin k_{c}^0 (q_j) $) and $\Lambda^{0}_{s1} (q_j)$,
\begin{eqnarray}
q_j & = & k_{c}^0 (q_j)+ 2\int_0^{\infty}d\omega {\sin(\omega\,\sin k_{c}^0 (q_j))\over\omega\,(1+e^{2\omega\,u})}\,J_0 (\omega) 
\, , \hspace{0.25cm} j = 1,...,N_{a} \, ,
\nonumber \\
q_j & = & \int_0^{\infty}d\omega {\sin(\omega\,\Lambda^{0}_{s1} (q_j))\over\omega\,\cosh
(\omega\,u)}\,J_0 (\omega) \, , \hspace{0.25cm} j = 1,...,N_{a_{s1}} \, . 
\label{Bessel}
\end{eqnarray}
Here $J_0 (\omega)$ is a Bessel function.

The $\beta$ pseudofermion energy functional, Eq. (\ref{DE}), second-order in the $\beta$ band canonical-momentum-distribution function 
deviations terms exactly vanish. That is not so for the corresponding $\beta$ pseudoparticle energy functional,
which to second order in such deviations reads,
\begin{eqnarray}
\delta E_{l_{\rm r},l_{\eta s}} & = & \sum_{\beta}\sum_{j=1}^{N_{a_{\beta}}}\varepsilon_{\beta} (q_j)\delta N_{\beta} (q_j) 
+ 2\vert\mu\vert\,M_{\eta,-1/2}^{un} + 2\mu_B\,\vert H\vert\,M_{s,-1/2}^{un} 
\nonumber \\
& + & {1\over N_a}\sum_{\beta}\sum_{\beta'}\sum_{j=1}^{N_{a_{\beta}}}\sum_{j'=1}^{N_{a_{\beta'}}}
{1\over 2}\,f_{\beta,\beta'} (q_j,q_{j'})\,\delta N_{\beta} (q_j)\delta N_{\beta'} (q_{j'}) \, ,
\nonumber \\ 
& = & \sum_{\beta =c,s1}\sum_{j=1}^{N_{a_{\beta}}}\varepsilon_{\beta} (q_j)\delta N_{\beta} (q_j) 
+  \sum_{\alpha\nu\neq s1}\sum_{j=1}^{N_{a_{\alpha\nu}}}\varepsilon^0_{\alpha\nu} (q_j)\delta N_{\alpha\nu} (q_j) 
\nonumber \\
& + & 2\vert\mu\vert\,M_{\eta,-1/2} + 2\mu_B\,\vert H\vert\,[\delta M_{s,-1/2} -\delta N_{s1}]
\nonumber \\
& + & {1\over N_a}\sum_{\beta}\sum_{\beta'}\sum_{j=1}^{N_{a_{\beta}}}\sum_{j'=1}^{N_{a_{\beta'}}}
{1\over 2}\,f_{\beta,\beta'} (q_j,q_{j'})\,\delta N_{\beta} (q_j)\delta N_{\beta'} (q_{j'}) \, . 
\label{DE-fermions}
\end{eqnarray}
The $\beta$ band bare-momentum-distribution function deviations appearing here are given by, 
\begin{equation}
\delta N_{\beta} (q_j)  = N_{\beta} (q_j) - N^0_{\beta} (q_j) \, , \hspace{0.25cm}
j = 1,...,N_{a_{\beta}}  \, , \hspace{0.25cm} \beta = c, \alpha\nu \, ,
\label{DNq}
\end{equation}
where $N_{\beta} (q_j)$ and $N^0_{\beta} (q_j)$ are the corresponding 
excited-state and ground-state $\beta$ pseudoparticle momentum-distribution functions, respectively.
Furthermore, $M_{\eta,\pm 1/2}^{un}$ and $M_{s,\pm 1/2}^{un}$ are in Eq. (\ref{DE-fermions}) the numbers of
$\eta$-spin projection $\pm 1/2$ $\eta$-spinons and spin projection $\pm 1/2$ spinons, respectively, Eq. (\ref{L-L}). 

Related important conserving numbers for the finite-energy physics are those of rotated-electron doubly occupied sites, $D_r$, and 
spin-down rotated-electron singly occupied sites whose rotated electrons are not associated with
$s1$ pseudoparticles, $S_{r}$. They read,
\begin{eqnarray}
D_r & \equiv & M_{\eta,-1/2} = [M^{un}_{\eta,-1/2} +M^{bo}_{\eta}/2] 
= [M^{un}_{\eta,-1/2} +\sum_{\nu =1}^{\infty}\nu N_{\eta\nu}] \, ,
\nonumber \\
S_r & \equiv & [M_{s,-1/2}-N_{s1}] = [M^{un}_{s,-1/2} +M^{bo}_s/2-N_{s1}]
= [M^{un}_{s,-1/2} +\sum_{\nu =2}^{\infty}\nu N_{s\nu}] \, ,
\label{DrS-r}
\end{eqnarray}
respectively.

The energy dispersions $\varepsilon_{\beta}^0 (q_j)$ in Eq. (\ref{DE-fermions}) can be expressed in terms of 
the two-pseudofermion phase shifts $\bar{\Phi }_{c,\beta} (r,r')$, Eqs. (\ref{Phicc-m})- (\ref{Phicsn-m}) of 
Appendix \ref{Ele2PsPhaShi}, and of the ground-state rapidity functions $\Lambda^{0}_{\beta} (q_j)$ (with 
$\Lambda^{0}_{c} (q_j)\equiv \sin k_c^{0} (q_j)$ for $\beta =c$), which are  
defined by their inverse functions in Eq. (\ref{GS-R-functions}), as follows, 
\begin{eqnarray}
\varepsilon_{\beta}^0 (q_j) & = & E_{\beta}^0 (q_j) + 2t\int_{-Q}^{Q}dk\,\bar{\Phi }_{c,\beta}
\left({\sin k\over u}, {4t\Lambda^{0}_{\beta} (q_j)\over U}\right)\sin k  \, , \hspace{0.25cm}
j = 1,...,N_{a_{\beta}} \, ,
\nonumber \\
E_c^0 (q_j) & = & -{U\over 2} -2t\cos k_c^{0} (q_j) \, ,
\nonumber \\
E_{\alpha\nu}^0 (q_j) & = & \delta_{\alpha,\eta}\left\{-\nu\,U + 4t{\rm Re}\left[\sqrt{1 - (\Lambda^0_{\eta\nu} (q_j) + i \nu\,u)^2}\right]\right\} \, .
\label{e-0-bands}
\end{eqnarray}
The related energy dispersions $\varepsilon_{\beta} (q_j)$ also appearing in Eq. (\ref{DE-fermions}) are given by,
\begin{eqnarray}
\varepsilon_{c} (q_j) & = & \varepsilon_{c}^0 (q_j) - \varepsilon_{c}^0 (q_{Fc}) = \varepsilon_{c}^0 (q_j) +
(1-\delta_{n,1})\,\vert\mu\vert + \delta_{n,1}\,\mu^0 - \mu_B\,\vert H\vert \, , 
\nonumber \\
\varepsilon_{s1} (q_j) & = & \varepsilon_{s1}^0 (q_j) - \varepsilon_{s1}^0 (q_{F{s1}}) = 
\varepsilon_{s1}^0 (q_j) + 2\mu_B\,\vert H\vert  \, , 
\nonumber \\
\varepsilon_{s\nu} (q_j) & = & \varepsilon_{s\nu}^0 (q_j) + 2\nu\mu_B\,\vert H\vert  \, , 
\nonumber \\
\varepsilon_{\eta\nu} (q_j) & = & \varepsilon_{\eta\nu}^0 (q_j) + (1-\delta_{n,1})\,2\nu\vert\mu\vert + \delta_{n,1}\,2\nu\mu^0   \, ,
\label{epsilon-q}
\end{eqnarray} 
where $2\mu^0$ is the $n=1$ Mott-Hubbard gap. Its limiting behaviors
are provided in Eq. (\ref{DMH}) of Appendix \ref{energies}.

The zero-energy level of the energy bands in Eq. (\ref{DE-fermions}) is such that,
\begin{eqnarray}
\varepsilon_{c} (\pm 2k_F) & = & \varepsilon_{s1} (\pm k_{F\downarrow})= 0 \, ,
\nonumber \\
\varepsilon_{\eta\nu}^0 (\pm [\pi -2k_F]) & = & \varepsilon_{s\nu}^0 (\pm [k_{F\uparrow}-k_{F\downarrow}])=0 \, .
\label{eplev0}
\end{eqnarray}
For the $\beta =\alpha\nu\neq s1$ and $\beta =c,s1$ bands it refers to the limiting momenta,
Eq. (\ref{qcanGS}), and $\beta =c,s1$ Fermi momenta, Eq. (\ref{q0Fcs}), respectively.

The $f$ function in the second-order terms of the $\beta$ pseudoparticle energy functional, Eqs. (\ref{DE-fermions}), reads
\cite{paper-I,Carmelo92},
\begin{eqnarray}
f_{\beta,\beta'}(q_j,q_{j'}) & = & v_{\beta}(q_{j})\,2\pi \,\Phi_{\beta,\beta'}(q_{j},q_{j'})+
v_{\beta'}(q_{j'})\,2\pi \,\Phi_{\beta',\beta}(q_{j'},q_{j}) \nonumber \\
& + & {1\over 2\pi}\sum_{\beta''=c,s1} \sum_{\iota =\pm 1} v_{\beta''}\,
2\pi\,\Phi_{\beta'',\beta}(\iota q_{F\beta''},q_{j})\,2\pi\,\Phi_{\beta'',\beta'} (\iota q_{F\beta''},q_{j'}) \, .
\label{ff}
\end{eqnarray}
Within the continuum momentum representation, the group velocities $v_{\beta} (q_j)$ appearing here are given by,
\begin{eqnarray}
v_{\beta} (q) & = & {\partial\varepsilon_{\beta} (q)\over \partial q} = {\partial\varepsilon^0_{\beta} (q)\over \partial q} 
\, , \hspace{0.25cm} \beta = c, \eta\nu, s\nu \, , \hspace{0.25cm} \nu = 1,...,\infty \, ,
\nonumber \\
v_{\beta} & \equiv & v_{\beta} (q_{F\beta}) \, , \hspace{0.25cm} \beta = c, s1 \, .
\label{velo}
\end{eqnarray}

In Appendix \ref{energies} limiting behaviors of the energy dispersions, Eqs. (\ref{e-0-bands}) and (\ref{epsilon-q}), 
along with other useful energy scales are given.

\section{Two-pseudofermion phase shifts}
\label{Ele2PsPhaShi}

In this Appendix we first derive the integral equations that define the two-pseudofermion phase shifts $\pi\,\bar{\Phi}_{\beta,\beta'}\left(r,r'\right)$
on the right-hand side of Eq. (\ref{Phi-barPhi}) for densities in the ranges $n\in [0,1[$ and $m\in ]0,n]$. To achieve that goal, we solve 
the thermodynamic BA equations, Eqs. (\ref{Tapco1})-(\ref{kcn}) of Appendix \ref{pseudo-repre}, 
up to first order in the deviations $\delta N_{\beta} (q_j)$, Eq. (\ref{DNq}) of that Appendix. Although the
two-pseudofermion phase-shift integral equations given below refer to densities $n\in [0,1[$ and $m\in ]0,n]$, within the limit $n\rightarrow 1$ (and
$m\rightarrow 0$) they provide correct $n=1$ (and $m=0$) results. Those are given in Appendix
\ref{Consequences}.

Moreover, in this Appendix the two-pseudofermion phase-shift integral equations are used 
to show that the functional ${\bar{q}}_j = {\bar{q}} (q_j)$, Eq. (\ref{barqan}), is such
that the ground-state limiting $\alpha\nu$ band canonical-momentum values 
$\bar{q} (\pm q_{\alpha\nu})$ are not shifted by the 
ground-state - excited-state transitions provided that $v_{\alpha\nu} (\pm q_{\alpha\nu})=0$.
Finally, we consider the separate contributions of the PDT processes (A) and (B) defined
in Section \ref{PDT} to the overall scattering phase shift $Q^{\Phi}_{\beta} (q)/2$, Eq. (\ref{qcan1j}).

A first group of two-pseudofermion phase shifts obey integral equations by their own,
\begin{equation}
\pi\,\bar{\Phi }_{s1,c}\left(r,r'\right) = -{\rm
arc}{\rm tan}(r-r') + \int_{-r^0_s}^{r^0_s}
dr''\,G(r,r'')\,\pi\,{\bar{\Phi }}_{s1,c}\left(r'',r'\right) \, ,
\label{Phis1c-m}
\end{equation}
\begin{equation}
\pi\,\bar{\Phi }_{s1,\eta\nu}\left(r,r'\right) =  -
{1\over{\pi}}\int_{-r^0_c}^{r^0_c} dr''{{\rm arc}{\rm tan}
\Bigl({r''-r'\over\nu}\Bigr)\over{1+(r-r'')^2}} +
\int_{-r^0_s}^{r^0_s} dr''\,G(r,r'')\,\pi\,{\bar{\Phi
}}_{s1,\eta\nu}\left(r'',r'\right) \, , 
\label{Phis1cn-m}
\end{equation}
and
\begin{eqnarray}
\pi\,\bar{\Phi }_{s1,s\nu}\left(r,r'\right) & = & \delta_{1
,\nu}\,\arctan\Bigl({r-r'\over 2}\Bigl) + (1-\delta_{1
,\nu})\Bigl\{ \arctan\Bigl({r-r'\over \nu-1}\Bigl) +
\arctan\Bigl({r-r'\over
\nu+1}\Bigl)\Bigr\} \nonumber \\
& - &  {1\over{\pi}}\int_{-r^0_c}^{r^0_c} dr''{{\rm arc}{\rm
tan} \Bigl({r''-r'\over\nu}\Bigr)\over{1+(r-r'')^2}} +
\int_{-r^0_s}^{r^0_s} dr''\,G(r,r'')\,\pi\,{\bar{\Phi
}}_{s1,s1}\left(r'',r'\right) \, . 
\label{Phis1sn-m}
\end{eqnarray}
The parameters $r_c^0$ and $r_s^0$ appearing in such equations are
given in Eq. (\ref{QB-r0rs}) of Appendix \ref{pseudo-repre}. The kernel $G(r,r')$ reads,
\begin{equation}
G(r,r') = - {1\over{2\pi}}\left[{1\over{1+((r-r')/2)^2}}\right]
\left[1 - {1\over 2}
\left(t(r)+t(r')+{{l(r)-l(r')}\over{r-r'}}\right)\right] \, .
\label{G}
\end{equation}
Here
\begin{equation}
t(r) = {1\over{\pi}}\left[{\rm arc}{\rm tan}(r + r^0_c) - {\rm
arc}{\rm tan}(r -r^0_c)\right] \, , \label{t}
\end{equation}
and
\begin{equation}
l(r) = {1\over{\pi}}\left[ \ln (1+(r + r^0_c)^2) - \ln (1+(r
-r^0_c)^2)\right] \, . \label{l}
\end{equation}
The kernel, Eqs. (\ref{G})-(\ref{l}), was first
introduced in Ref. \cite{Carmelo92} within the $c$ and $s\equiv s1$ pseudoparticle 
two-component PS considered in that reference.

A second group of two-pseudofermion phase shifts are expressed in
terms of the basic functions, Eqs. (\ref{Phis1c-m})-(\ref{Phis1sn-m}), as follows,
\begin{equation}
\pi\,\bar{\Phi }_{c,c}\left(r,r'\right) =
{1\over{\pi}}\int_{-r^0_s}^{r^0_s} dr''{\pi\,\bar{\Phi
}_{s1,c}\left(r'',r'\right) \over {1+(r-r'')^2}} \, ,
\label{Phicc-m}
\end{equation}
\begin{equation}
\pi\,\bar{\Phi }_{c,\eta\nu}\left(r,r'\right) = -{\rm
arc}{\rm tan}\Bigl({r-r'\over \nu}\Bigr) +
{1\over{\pi}}\int_{-r^0_s}^{r^0_s} dr''{\pi\,\bar{\Phi
}_{s1,\eta\nu}\left(r'',r'\right) \over {1+(r-r'')^2}} \, ,
\label{Phiccn-m}
\end{equation}
and
\begin{equation}
\pi\,\bar{\Phi }_{c,s\nu}\left(r,r'\right) = -{\rm
arc}{\rm tan}\Bigl({r-r'\over \nu}\Bigr) + {1\over{\pi}}
\int_{-r^0_s}^{r^0_s} dr''{\pi\,\bar{\Phi
}_{s1,s\nu}\left(r'',r'\right) \over {1+(r-r'')^2}} \, .
\label{Phicsn-m}
\end{equation}

Finally, the remaining two-pseudofermion phase shifts can be
expressed either only in terms of the functions, Eqs.
(\ref{Phicc-m})-(\ref{Phicsn-m}), 
\begin{equation}
\pi\,{\bar{\Phi }}_{\eta\nu,c}\left(r,r'\right) = {\rm arc}{\rm
tan}\Bigl({r-r'\over {\nu}}\Bigr) - {1\over{\pi}}\int_{-r_c^0}^{+r_c^0} dr''{\pi\,{\bar{\Phi
}}_{c,c}\left(r'',r'\right) \over {\nu[1+({r-r''\over {\nu}})^2]}} \, , 
\label{Phicnc-m}
\end{equation}
\begin{equation}
\pi\,\bar{\Phi }_{\eta\nu,\eta\nu'}\left(r,r'\right) = {1\over{2}}\,\Theta_{\nu,\nu'}(r-r') -
{1\over{\pi}}\int_{-r_c^0}^{+r_c^0} dr''{\pi\,\bar{\Phi }_{c,\eta\nu'}\left(r'',r'\right) \over
{\nu[1+({r-r''\over\nu})^2]}} \, , 
\label{Phicncn-m}
\end{equation}
\begin{equation}
\pi\,\bar{\Phi }_{\eta\nu,s\nu'}\left(r,r'\right) = - {1\over{\pi}}\,\int_{-r_c^0}^{+r_c^0}
dr''{\pi\,\bar{\Phi }_{c,s\nu'}\left(r'',r'\right) \over {\nu[1+({r-r''\over\nu})^2]}} \, , 
\label{Phicnsn-m}
\end{equation}
or both in terms of the basic functions, Eqs.
(\ref{Phis1c-m})-(\ref{Phis1sn-m}), and of the phase shifts,
Eqs. (\ref{Phicc-m})-(\ref{Phicsn-m}),
\begin{equation}
\pi\,{\bar{\Phi }}_{s\nu ,c}\left(r,r'\right) = - {\rm arc}{\rm
tan}\Bigl({r-r'\over {\nu}}\Bigr) +
{1\over{\pi}}\int_{-r^0_c}^{r^0_c} dr''{\pi\,{\bar{\Phi
}}_{c,c}\left(r'',r'\right) \over {\nu[1+({r-r''\over
\nu})^2]}} - \int_{-r^0_s}^{r^0_s} dr''\pi\,{\bar{\Phi
}}_{s1,c}\left(r'',r'\right)
{\Theta^{[1]}_{\nu,1}(r-r'')\over{2\pi}} \, ; \hspace{0.5cm} \nu
> 1 \, , \label{Phisnc-m}
\end{equation}
\begin{equation}
\pi\,{\bar{\Phi }}_{s\nu ,\eta\nu'}\left(r,r'\right) =
{1\over{\pi}}\int_{-r^0_c}^{r^0_c} dr''{\pi\,{\bar{\Phi
}}_{c,\eta\nu'}\left(r'',r'\right) \over {\nu[1+({r-r''\over
\nu})^2]}} - \int_{-r^0_s}^{r^0_s} dr''\pi\,{\bar{\Phi
}}_{s1,\eta\nu'}\left(r'',r'\right)
{\Theta^{[1]}_{\nu,1}(r-r'')\over {2\pi}} \, ; \hspace{0.5cm}
\nu > 1 \, , \label{Phisncn-m}
\end{equation}
\begin{equation}
\pi\,{\bar{\Phi }}_{s\nu ,s\nu'}\left(r,r'\right) =
{1\over{2}}\,\Theta_{\nu,\nu'}(r-r') +
{1\over{\pi}}\int_{-r^0_c}^{r^0_c} dr''{\pi\,{\bar{\Phi
}}_{c,s\nu'}\left(r'',r'\right) \over {\nu[1+({r-r''\over
\nu})^2]}} - \int_{-r^0_s}^{r^0_s} dr''\pi\,{\bar{\Phi
}}_{s1,s\nu'}\left(r'',r'\right)
{\Theta^{[1]}_{\nu,1}(r-r'')\over{2\pi}} \, . 
\label{Phisnsn-m}
\end{equation}

In the above two-pseudofermion phase shift expressions,
$\Theta_{\nu,\nu'}(x)$ is the function,
\begin{eqnarray}
\Theta_{\nu,\nu'}(x) & = & \delta_{\nu
,\nu'}\Bigl\{2\arctan\Bigl({x\over 2\nu}\Bigl) + \sum_{l=1}^{\nu
-1}4\arctan\Bigl({x\over 2l}\Bigl)\Bigr\} + (1-\delta_{\nu
,\nu'})\Bigl\{ 2\arctan\Bigl({x\over \vert\,\nu-\nu'\vert}\Bigl)
\nonumber \\
& + &  2\arctan\Bigl({x\over \nu+\nu'}\Bigl) +
\sum_{l=1}^{{\nu+\nu'-\vert\,\nu-\nu'\vert\over 2}
-1}4\arctan\Bigl({x\over \vert\, \nu-\nu'\vert +2l}\Bigl)\Bigr\} \, , 
\label{Theta}
\end{eqnarray}
and $\Theta^{[1]}_{\nu,\nu'}(x)$ is its derivative,
\begin{eqnarray}
\Theta^{[1]}_{\nu,\nu'}(x) & = & {d\Theta_{\nu,\nu'}(x)\over
dx} = \delta_{\nu ,\nu'}\Bigl\{{1\over \nu[1+({x\over 2\nu})^2]}+
\sum_{l=1}^{\nu -1}{2\over l[1+({x\over 2l})^2]}\Bigr\} +
(1-\delta_{\nu ,\nu'})\Bigl\{ {2\over |\nu-\nu'|[1+({x\over
|\nu-\nu'|})^2]} \nonumber \\
& + & {2\over (\nu+\nu')[1+({x\over \nu+\nu'})^2]} +
\sum_{l=1}^{{\nu+\nu'-|\nu-\nu'|\over 2} -1}{4\over
(|\nu-\nu'|+2l)[1+({x\over |\nu-\nu'|+2l})^2]}\Bigr\} \, .
\label{The1}
\end{eqnarray}

Next it is shown that the ground-state limiting $\alpha\nu$ band canonical-momentum values 
$\bar{q} (\pm q_{\alpha\nu})$ are not shifted by the ground-state - excited-state transitions provided that $v_{\alpha\nu} (\pm q_{\alpha\nu})=0$.
From the use of the integral equations, Eqs. (\ref{Phicnc-m})-(\ref{Phisnsn-m}), for the two-pseudofermion phase shifts
$\pi\,\Phi_{\alpha\nu,\beta'}(q,q')$ involving $\alpha\nu\neq s1$ scatterers, it is found that,
\begin{equation}
\pi\,\Phi_{\alpha\nu,\beta'}(\iota\,q_{\alpha\nu},q) =
{\iota\,\pi\over 2}\Bigl[\delta_{\beta' ,c}(\delta_{\alpha ,\eta}-\delta_{\alpha ,s}) 
+\delta_{\beta' ,\alpha'\nu'}\delta_{\alpha ,\alpha'} (-\delta_{\nu ,\nu'} + \nu +\nu' -\vert\nu -\nu'\vert )\Bigr]  
\, , \hspace{0.25cm} \iota =\pm 1 \, . 
\label{Phi-nas}
\end{equation}
If $\beta'=\alpha\nu'$, this two-pseudofermion phase-shift expression is valid provided that $q\neq\iota\,q_{\alpha\nu'}$. 
From the use of the integral equations, Eqs. (\ref{Phis1c-m})-(\ref{Phis1sn-m}), for the two-pseudofermion phase shifts $\pi\,\Phi_{s1,\beta'}(q,q')$
we find that in the $r^0_s\rightarrow 0$ limit the two-pseudofermion phase shifts $\pi\,\Phi_{s1,\beta'}(\iota\,q_{s1},q)$ 
are as well given by Eq. (\ref{Phi-nas}). Such a $r^0_s\rightarrow 0$ limit is equivalent to the limit, $N_{s1}/N_a\rightarrow 0$ as 
$N_a\rightarrow\infty$, within which the condition $v_{s1} (\iota\,q_{s1})=0$ is fulfilled.

The use of Eq. (\ref{Phi-nas}) in the scattering phase-shift expression, Eq. (\ref{qcan1j}), leads to,
\begin{equation}
{Q^{\Phi}_{\alpha\nu} (\iota q_{\alpha\nu})\over 2} = 
-{\iota\,\pi\over 2}\Bigl[\delta N_{\alpha\nu} - [\delta_{\alpha ,\eta}
- \delta_{\alpha ,s}]\delta N_{c} -\sum_{\nu'=1}^{\infty}
(\nu +\nu' -\vert\nu -\nu'\vert )\delta N_{\alpha\nu'}\Bigr] \, , \hspace{0.25cm} \iota =\pm 1 \, . 
\label{QPhi-limiting}
\end{equation}
Comparison of this expression with that of 
$\iota\delta q_{\alpha\nu}=\iota\pi\,\delta N_{a_{\alpha\nu}}/N_a$ where
$N_{a_{\alpha\nu}}=[N_{\alpha\nu}+N^h_{\alpha\nu}]$ and $N^h_{\alpha\nu}$ 
is provided in Eq. (\ref{N-h-an}) of Appendix \ref{pseudo-repre}, confirms that $Q^{\Phi}_{\alpha\nu} (\iota q_{\alpha\nu})/N_a
=-\iota\delta q_{\alpha\nu}$. (For the $s1$ pseudofermions such results are valid again in the
above limit within which $v_{s1} (\iota\,q_{s1})=0$.)

As mentioned above, for a $\beta'=\alpha\nu'$ pseudofermion scattering center
the two-pseudofermion phase shift expression, Eq. (\ref{Phi-nas}), does not apply at $q=\iota\,q_{\alpha\nu'}$.
However, from the form $\Lambda_{\alpha\nu}(q) = \Lambda_{\alpha\nu}^0({\bar{q}}
(q))$ of the PS excited energy-egenstates rapidity function, Eq. (\ref{FL}), 
one confirms that the relation $Q^{\Phi}_{\alpha\nu} (\iota q_{\alpha\nu})/N_a=\iota\delta q_{\alpha\nu}$ 
is valid for all PS excited states provided that $v_{\alpha\nu} (\iota\,q_{\alpha\nu})=0$. 
Indeed, Eq. (\ref{Phi-nas}) reveals that the rapidity 
functions of the excited energy eigenstates of a given ground state
equal those of the latter state with in the argument of such 
functions the ground-state bare momentum replaced by the excited-state 
canonical momentum. This property implies that the corresponding bare-momentum 
and canonical-momentum bands have precisely the same momentum width.
Hence one has that $Q^{\Phi}_{\alpha\nu} (\iota q_{\alpha\nu})/N_a
=-\iota\delta q_{\alpha\nu}$ for all PS excited states provided that $v_{\alpha\nu} (\iota\,q_{\alpha\nu})=0$.

The PDT processes (A) and (B) considered in Section \ref{PDT} lead to separate
contributions to the overall scattering phase shift $Q^{\Phi}_{\beta} (q)/2$, Eq. (\ref{qcan1j}).
In the following we consider such two separate contributions. 
The part of the $\beta'$ bare-momentum distribution-function 
deviation generated by $\beta'$ scattering centers can be written 
as $\delta N^{NF}_{\beta'} (q') + \delta N^{F}_{\beta'} (q')
+ \delta N^{L}_{s1} (q')\times\delta_{v_{s1} (q_{s1}),0}$ for the
$\beta'=c,s1$ branches and $\delta N^{NF}_{\beta'} (q') + \delta N^{L}_{\beta'} (q')$
for the $\beta'=\alpha'\nu'\neq s1$ branches.
Here $\delta N^{NF}_{\beta'} (q')$ is generated by the processes
called elementary processes (A) in Section \ref{PDT}. Those create and annihilate 
(and create) $\beta' = c,s1$ pseudofermions (and $\beta'=\alpha'\nu'\neq 
s1$ pseudofermions) away from the $\beta' = c,s1$ Fermi points and
$s1$ limiting values $\pm q_{s1}=\pm 2k_F$ when $k_{F\uparrow}\rightarrow 2k_F$ and $k_{F\downarrow}\rightarrow 0$
(and from the limiting values $\pm q_{\alpha'\nu'}$). On the other hand, 
$\delta N^{F}_{\beta'} (q')$ and $\delta N^{L}_{\beta'} (q')$ are generated by the processes 
called elementary processes (B) in that section. Those involve the creation and 
annihilation (and creation) $\beta' = c,s1$ pseudofermions (and $\beta'=\alpha'\nu'$ 
pseudofermions of vanishing velocity) at the $\beta' = c,s1$ Fermi points (and at the limiting 
momentum values $\pm q_{\alpha'\nu'}$). 

The general expressions of the deviations $\delta N^{F}_{\beta'} (q')$ and $\delta N^{L}_{\beta'} (q')$
associated with scattering centers generated by processes (B) read,
\begin{eqnarray}
\delta N^{F}_{\beta'} (q') & = & \sum_{\iota =\pm 1}
\Bigl[{\delta N^{F}_{\beta'}\over 2}+
\iota\,\delta J^{F}_{\beta'}\Bigr]\,
\delta_{q',\iota\,q_{F\beta'}}
\, , \hspace{0.25cm}
\beta' = c,s1 \, ; \nonumber \\
\delta N^{L}_{\beta'} (q') & = & \sum_{\iota =\pm 1}
\Bigl[{\delta N^{L}_{\beta'}\over 2}+
\iota\,\delta J^{L}_{\beta'}\Bigr]\,
\delta_{q',\iota\,q_{\beta'}} \, , \hspace{0.25cm}
\beta'=\alpha'\nu' \, ,
\label{DNqF-all}
\end{eqnarray}
respectively. (For the $\beta'=s1$ branch, the deviation $\delta N^{L}_{\beta'} (q')$
applies only to the vanishing-velocity $s1$ pseudofermion scattering centers
with $s1$ limiting values $\pm q_{s1}=\pm 2k_F$; Those correspond to
the $k_{F\uparrow}\rightarrow 2k_F$ and $k_{F\downarrow}\rightarrow 0$ limit.)

The deviation numbers $\delta N^F_{\beta'}$ (and
$\delta N_{\beta'}^{L}$) are in Eq. (\ref{DNqF-all}) such that $\delta N_{\beta'} = 
\delta N^F_{\beta'}+\delta N^{NF}_{\beta'}$ (and $\delta N_{\beta'} 
= \delta N_{\beta'}^{L} + \delta N_{\beta'}^{NF}$). They can be expressed
as $\delta N^F_{\beta'}=\delta N^F_{\beta' ,+1}+\delta N^F_{\beta' ,-1}$ 
(and $\delta N^{L}_{\beta'} = \delta N^{L}_{\beta' ,+1}+ \delta N^{L}_{\beta' ,-1}$). 
Here $\delta N^F_{\beta' ,\pm 1}$ is the deviation in the number of 
$\beta'=c,s1$ pseudofermions at the right $(+1)$ and left $(-1)$ $\beta'=c,s1$ Fermi 
point (and $\delta N_{\beta',\iota}^{L}$ is the deviation in the number of $\beta'=\alpha'\nu'$ 
pseudofermions created at $\iota\,q_{\beta'}$ with $\iota =\pm 1$).
The deviation current numbers read 
$2\delta J^F_{\beta'}=[\delta N^F_{\beta' ,+1}-\delta 
N^F_{\beta' ,-1}]$ for $\beta'=c,s1$ pseudofermions (and $2\delta J^L_{\beta'}=[\delta N^L_{\beta' ,+1}-
\delta N^L_{\beta',-1}]$ for vanishing-velocity $\beta'=\alpha'\nu'$ pseudofermions). 
For the $\beta'=\alpha'\nu'\neq s1$ pseudofermions the deviations
$\delta N_{\beta'} = \delta N_{\beta'}^{L} + \delta N_{\beta'}^{NF}$ equal the corresponding numbers, $N_{\beta'} 
= N_{\beta'}^{L} + N_{\beta'}^{NF}$. Indeed the $\beta'=\alpha'\nu'\neq s1$ pseudofermion
occupancy vanishes for the ground states.

From the linearity in the deviations of the overall scattering phase shift,
Eq. (\ref{qcan1j}), one can write $Q^{\Phi}_{\beta} (q)/2= 
[Q^{\Phi (NF)}_{\beta} (q)/2+Q^{\Phi (F)}_{\beta} (q)/2]$. Also the part of the total 
momentum deviation, Eq. (\ref{DP}), associated with the elementary processes 
(A) and (B) can be written as $\delta P^{NF}+\delta P^{F}$. After some 
algebra involving the use of Eqs. (\ref{qcan1j}), (\ref{DP}), (\ref{inde}), 
and (\ref{DNqF-all}), we reach the following expressions for such
quantities,
\begin{equation}
{Q^{\Phi (NF)}_{\beta} (q)\over 2} = \sum_{\beta'}\,
\sum_{q'}\,\pi\,\Phi_{\beta,\beta'}(q,q')\, \delta
N^{NF}_{\beta'}(q') \, , \label{qcan1j-NF}
\end{equation}
\begin{eqnarray}
{Q^{\Phi (F)}_{\beta} (q)\over 2} & = & \sum_{\beta'=c,s1}\,
\sum_{\iota'=\pm 1}\pi\,\Phi_{\beta,\beta'}(q,\iota'\,q_{F\beta'})\, 
{\delta N^{F}_{\beta'}\over 2} \nonumber \\
& + & \sum_{\iota'=\pm 1}\iota'\,\pi\,\Phi_{\beta,c} (q,\iota'\,2k_F)
\Bigl[\delta J^F_{c} + \sum_{\nu' =1}^{\infty}\,J^L_{\eta\nu'} +
\sum_{\nu' =2}^{\infty}\,J^L_{s\nu'} + \delta J^L_{s1}\times\delta_{v_{s1} (q_{s1}),0}\Bigr] 
\nonumber \\
& + & \sum_{\iota'=\pm 1}\iota'\,\pi\,\Phi_{\beta,s1} (q,\iota'\,k_{F\downarrow})
\Bigl[\delta J^F_{s1} -
2\sum_{\nu' =2}^{\infty}\,J^L_{s\nu'}\Bigr] \nonumber \\
& + & \sum_{\beta'=\alpha'\nu'\neq s1} \pi\Bigl[\delta_{\beta,c}
-\delta_{\beta,\alpha\nu}\delta_{\alpha,\alpha'}\,(-\delta_{\nu,\nu'} + \nu +\nu' -\vert\nu -\nu'\vert )\Bigr] J^L_{\alpha'\nu'}
\nonumber \\
& + & \pi\Bigl[\delta_{\beta,c} -\delta_{\beta,s\nu}\,(2-\delta_{\nu,1})\Bigr] \delta J^L_{s1}\times\delta_{v_{s1} (q_{s1}),0}
\, , \label{qcan1j-J-F-q}
\end{eqnarray}
and
\begin{eqnarray}
\delta P^{NF} & = & \sum_{\beta'=c,s\nu'}\,
\sum_{q'}\, q'\, \delta N^{NF}_{\beta'}(q') + 
\sum_{\eta\nu'}\,\sum_{q'}\, [\pi -q']\, 
\delta N^{NF}_{\eta\nu'}(q') \, ; 
\nonumber \\
\delta P^{F} & = & \pi\,[M^{un}_{\eta,-1/2}+\sum_{\nu' =1}^{\infty}\nu' N^L_{\eta\nu'}]
\nonumber \\
& + & 4k_F\Bigl[\,\delta J^F_{c} + \sum_{\nu' =1}^{\infty}\,J^L_{\eta\nu'} + \sum_{\nu'
=2}^{\infty}\,J^L_{s\nu'} + \delta J^L_{s1}\times\delta_{v_{s1} (q_{s1}),0}\Bigr] + 2k_{F\downarrow}\Bigl[\,\delta J^F_{s1} - 2\sum_{\nu'
=2}^{\infty}\,J^L_{s\nu'}\Bigr] \, . \label{PFI}
\end{eqnarray}
In the above expressions we used that $\delta J_{\beta'} = J_{\beta'}$
for the $\beta'=\alpha'\nu'\neq s1$ scattering centers.

The general expression of the phase shift $Q^{\Phi (F)}_{\beta} (q)/2$,  
Eq. (\ref{qcan1j-J-F-q}), is valid for all active $\beta$ scatterers, as defined in Section \ref{Pseudo-invariance}.
In the $\delta P^{F}$ expression, Eq. (\ref{PFI}), we have included the contribution from the 
unbound $-1/2$ $\eta$-spinons. (The momentum contributions from the unbound $+1/2$ $\eta$-spinons 
and unbound $\pm 1/2$ spinons vanish.) Note that the current contributions 
to the momentum spectrum $\delta P^{F}$, Eq. (\ref{PFI}), which multiply
$4k_F$ and $2k_{F\downarrow}$, are identical to the current contributions to
the scattering phase shift, Eq. (\ref{qcan1j-J-F-q}), which multiply the 
phase shifts $\pi\,\Phi_{\beta,c} (q,\iota'\,2k_F)$ and
$\pi\,\Phi_{\beta,s1} (q,\iota'\,k_{F\downarrow})$, respectively.

\section{Pseudofermion scattering theory for densities $n=1$ and/or $m=0$}
\label{Consequences}

The general pseudofermion scattering theory studied in this paper
also applies to PSs of $S_{\eta}=0;n=1$ and/or $S_s=0;m=0$ ground states, 
provided that the specific issues addressed here 
are accounted for. 

The general two-pseudofermion phase-shift
$\pi\,\Phi_{\beta,\beta'}(q_j,q_{j'})$ expression, Eq. (\ref{Phi-barPhi}), 
applies to ground states with densities in the ranges $n\in [0,1[$ and $m\in ]0,n]$.
On the other hand, for excited states belonging to PSs of $S_{\eta}=0;n=1$ (and $S_{\eta}=0;m=0$) 
ground states and $\beta =\eta\nu$ scatterers and/or
$\beta' =\eta\nu'$ scattering centers (and $\beta =s\nu\neq s1$ scatterers and/or $\beta' =s\nu'\neq s1$ scattering centers)
the rapidity function $\Lambda^{0}_{\beta}(q_j)$ and/or $\Lambda^{0}_{\beta'}(q_{j'})$
in the argument of the phase-shift $\pi\,\bar{\Phi }_{\beta,\beta'} (r ,r')$,
Eq. (\ref{Phi-barPhi}), must be replaced by 
a corresponding excited-state functional $\Lambda_{\beta}(q_j)$ and/or $\Lambda_{\beta'}(q_{j'})$,
respectively. Fortunately, however, the $n\rightarrow 1$ (and $m\rightarrow 0$) limit of the above
equations defining the two-pseudofermion phase-shifts $\pi\,\bar{\Phi }_{\beta,\beta'} (r ,r')$
defines as well such phase shifts for PSs of an $S_{\eta}=0;n=1$ (and $S_{\eta}=0;m=0$) ground state.

We start by using such a property in the case of PSs of an $m=0$ ground state with
electronic density in the range $n\in [0,1]$. (The correct phase-shift results for electronic density $n=1$ 
are reached by taking the $n\rightarrow 1$ limit in the following equations within 
which $r_c^0 \rightarrow 0$.) The rapidity two-pseudofermion phase
shifts $\pi\,{\bar{\Phi }}_{\beta,\beta'}(r,r')$ are defined by the integral
equations, Eqs. (\ref{Phis1c-m})-(\ref{Phisnsn-m}) of Appendix \ref{Ele2PsPhaShi}. 
By Fourier transforming such equations after accounting for that $B\rightarrow\infty$ and 
thus $r^0_s = 4t\,B/U\rightarrow\infty$ for $u$ finite within the $m\rightarrow 0$ limit,
we arrive to the following equations valid for $m =0$ and $u$ finite,
\begin{equation}
\pi\,{\bar{\Phi }}_{c,c}(r,r') = - B (r-r') + \int_{-r_c^0}^{+r_c^0}dr''\,A
(r-r'')\,\pi\,{\bar{\Phi }}_{c,c}(r'',r') \, , 
\label{Phicc}
\end{equation}
\begin{equation}
\pi\,{\bar{\Phi }}_{c,s1}(r,r') = -{1\over 2}\,{\rm arc}\tan\Bigl(\sinh
\Bigl({\pi\over 2}(r-r')\Bigr)\Bigr) + \int_{-r_c^0}^{+r_c^0}dr''\,A (r-r'')\,\pi\,{\bar{\Phi}}_{c,s1}(r'',r') \, , 
\label{Phics}
\end{equation}
\begin{eqnarray}
\pi\,{\bar{\Phi }}_{s1,c}(r,r') & = & -{1\over 2}\,{\rm arc}\tan\Bigl(\sinh
\Bigl({\pi\over 2}(r-r')\Bigr)\Bigr) + {1\over 4}\int_{-r_c^0}^{+r_c^0}dr''\,{\pi\,{\bar{\Phi
}}_{c,c}(r'',r')\over \cosh \Bigl({\pi\over 2}(r-r'')\Bigr)} \, ;
\hspace{0.5cm} r \neq \pm \infty \nonumber \\
& = & -{{\rm sgn} (r)\pi\over 2\sqrt{2}} \, ; \hspace{0.5cm} r = \pm \infty \, ,
\label{Phisc}
\end{eqnarray}
\begin{eqnarray}
\pi\,{\bar{\Phi }}_{s1,s1}(r,r') & = & B (r-r') + {1\over
4}\int_{-r_c^0}^{+r_c^0}dr''\,{\pi\,{\bar{\Phi }}_{c,s1}(r'',r')\over \cosh \Bigl({\pi\over
2}(r-r'')\Bigr)} \, ; \hspace{0.3cm} r\neq\pm\infty \nonumber \\
& = & {{\rm sgn} (r)\pi\over 2\sqrt{2}} \, ; \hspace{0.5cm} r = \pm\infty \, ,
\hspace{0.5cm} r' \neq r \nonumber
\\
& = & [{\rm sgn} (r)]\Bigl({3\over 2\sqrt{2}}-1\Bigr)\pi \, ; \hspace{0.5cm} r = r' =
\pm\infty \, , \label{Phiss}
\end{eqnarray}
\begin{equation}
\pi\,{\bar{\Phi }}_{c,\eta\nu}(r,r') = -{\rm arc}\tan\Bigl({r-r'\over
\nu}\Bigr) + \int_{-r_c^0}^{+r_c^0}dr''\,A (r-r'')\,\pi\,{\bar{\Phi }}_{c,\eta\nu}(r'',r') \, , 
\label{Phiccn}
\end{equation}
\begin{equation}
\pi\,{\bar{\Phi }}_{c,s\nu}(r,r') = 0 \, ; \hspace{0.5cm} \nu
>1 \, , \label{Phicsn}
\end{equation}
\begin{equation}
\pi\,{\bar{\Phi }}_{s1,\eta\nu}(r,r') =  {1\over 4}\int_{-r_c^0}^{+r_c^0}dr''\,{\pi\,{\bar{\Phi
}}_{c,\eta\nu}(r'',r')\over \cosh \Bigl({\pi\over 2}(r-r'')\Bigr)} \, , 
\label{Phiscn}
\end{equation}
\begin{eqnarray}
\pi\,{\bar{\Phi }}_{s1,s\nu}(r,r') & = & {\rm arc}\tan\Bigl({r-r'\over \nu
-1}\Bigr) \, ; \hspace{0.5cm} r\neq \pm\infty \, ; \hspace{0.5cm} \nu
>1 \, , \nonumber \\
& = & \pm {\pi\over\sqrt{2}} \, ; \hspace{0.5cm} r= \pm\infty \, ; \hspace{0.5cm} \nu
>1 \, , \label{Phissn}
\end{eqnarray}
\begin{equation}
\pi\,{\bar{\Phi }}_{s\nu,c}\left(r,r'\right) = \pi\,\bar{\Phi }_{s\nu,\eta\nu'}\left(r,r'\right)
= 0 \, ; \hspace{0.5cm} \nu
> 1 \, , \label{Phisnc}
\end{equation}
\begin{equation}
\pi\,{\bar{\Phi }}_{s\nu,s1}\left(r,r'\right) = {\rm arc}{\rm
tan}\Bigl({r-r'\over {\nu -1}}\Bigr) \, ; \hspace{0.5cm} \nu > 1 \, , \label{Phisns1}
\end{equation}
\begin{equation}
\pi\,\bar{\Phi }_{s\nu,s\nu'}\left(r,r'\right) = {1\over{2}}\,\Theta_{\nu,\nu'}(r-r') -
{\rm arc}{\rm tan}\Bigl({r-r'\over {\nu +\nu' -2}}\Bigr) -
{\rm arc}{\rm tan}\Bigl({r-r'\over {\nu +\nu'}}\Bigr) \, , \hspace{0.5cm}
\nu ,\nu' > 1  \, . 
\label{Phisnsn}
\end{equation}
The two-pseudofermion phase shifts $\pi\,{\bar{\Phi }}_{\eta\nu,c}\left(r,r'\right)$, $\pi\,\bar{\Phi }_{\eta\nu,\eta\nu'}\left(r,r'\right)$,
and $\pi\,\bar{\Phi }_{\eta\nu,s\nu'}\left(r,r'\right)$ remain being given by 
Eqs. (\ref{Phicnc-m})-(\ref{Phicnsn-m}) of Appendix \ref{Ele2PsPhaShi}. 

In the above expressions, the function $\Theta_{\nu,\nu'}(x)$ is defined in Eq. (\ref{Theta})
of that Appendix,
\begin{equation}
B (r) = \int_{0}^{\infty} d\omega{\sin (\omega\,r)\over \omega
(1+e^{2\omega})} = {i\over 2} \ln {\Gamma \Bigl({1\over 2}+i{r\over 4}\Bigr)\,\Gamma
\Bigl(1-i{r\over 4}\Bigr)\over \Gamma \Bigl({1\over 2}-i{r\over 4}\Bigr)\,\Gamma
\Bigl(1+i{r\over 4}\Bigr)} \, , \label{Br}
\end{equation}
and
\begin{equation}
A (r) = {1\over\pi}{d B (r)\over dr} = {1\over\pi}\int_{0}^{\infty} d\omega{\cos (\omega\,r)\over
1+e^{2\omega}} \, , 
\label{Ar}
\end{equation}
where $\Gamma (x)$ is the usual $\Gamma$ function. The four equations, Eq. (\ref{Phicc})-(\ref{Phiss}) 
with $s1 =s$, $r =x$, and $r_c^0 =x_0$, are equivalent
to Eqs. (A9)-(A12) of Ref. \cite{Carmelo92}. (For the phase shifts, Eqs. (\ref{Phisc})
and (\ref{Phiss}), this equality refers to values of $r$ such that $r\neq\infty$.)

The two-pseudofermion phase-shift expressions defined here for $m = 0$ are those
of some of the two-pseudofermion phase shifts plotted in Figs. \ref{F1}-\ref{F6}. Specifically,
the two-pseudofermion phase shifts $\pi\,\Phi_{c,c}(q,q')$, $\pi\,\Phi_{c,s1}(q\,q')$, $\pi\,\Phi_{s1,c}(q,q')$, 
$\pi\,\Phi_{s1,s1}(q,q')$, $\pi\,\Phi_{c,\eta 1}(q,q')$, and $\pi\,\Phi_{s1,\eta 1}(q,q')$, Eq. (\ref{Phi-barPhi}), are plotted in 
such figures in units of $\pi$ for $n=0.59$, $m=0$, and several $U/t$ values.  

As mentioned above, by taking the $n\rightarrow 1$ and $r_c^0 \rightarrow 0$ limits
the phase-shift equations, Eqs. (\ref{Phicsn}) and (\ref{Phissn})-(\ref{Phisnsn}), apply to 
PS excited states generated by transitions from a $S_{\eta}=0;n=1$ and $S_s=0;m=0$ ground state.
Such a procedure leads to,
\begin{equation}
\pi\,{\bar{\Phi }}_{c,c}(r,r') = - B (r-r') = - {i\over 2} \ln {\Gamma \Bigl({1\over 2}+i{(r-r')\over 4}\Bigr)\,\Gamma
\Bigl(1-i{(r-r')\over 4}\Bigr)\over \Gamma \Bigl({1\over 2}-i{(r-r')\over 4}\Bigr)\,\Gamma
\Bigl(1+i{(r-r')\over 4}\Bigr)} \, , \label{Phicc-n1}
\end{equation}
\begin{equation}
\pi\,{\bar{\Phi}}_{c,s1}(r,r') = -{1\over 2}\,{\rm arc}\tan\Bigl(\sinh \Bigl({\pi\over
2}(r-r')\Bigr)\Bigr) = {i\over 2} \ln \left(-i\,{1+ie^{{\pi\over2}(r-r')}\over 1-ie^{{\pi\over2}(r-r')}}\right) 
\, , \label{Phics1}
\end{equation}
\begin{eqnarray}
\pi\,{\bar{\Phi }}_{s1,c}(r,r') & = & -{1\over 2}\,{\rm arc}\tan\Bigl(\sinh
\Bigl({\pi\over 2}(r-r')\Bigr)\Bigr) = {i\over 2} \ln \left(-i\,{1+ie^{{\pi\over2}(r-r')}\over 1-ie^{{\pi\over2}(r-r')}}\right) \, ;
\hspace{0.5cm} r \neq \pm \infty  \nonumber \\
& = & -{{\rm sgn} (r)\pi\over 2\sqrt{2}} \, ; \hspace{0.5cm} r = \pm \infty \, ,
\label{Phiseta1}
\end{eqnarray}
\begin{eqnarray}
\pi\,{\bar{\Phi }}_{s1,s1}(r,r') & = & B (r-r') = {i\over 2} \ln {\Gamma \Bigl({1\over 2}+i{(r-r')\over 4}\Bigr)\,\Gamma
\Bigl(1-i{(r-r')\over 4}\Bigr)\over \Gamma \Bigl({1\over 2}-i{(r-r')\over 4}\Bigr)\,\Gamma
\Bigl(1+i{(r-r')\over 4}\Bigr)} \, ; \hspace{0.3cm} r\neq\pm\infty  
\nonumber
\\
& = & {{\rm sgn} (r)\pi\over 2\sqrt{2}} \, ; \hspace{0.5cm} r = \pm\infty \, ,
\hspace{0.5cm} r' \neq r \nonumber
\\
& = & [{\rm sgn} (r)]\Bigl({3\over 2\sqrt{2}}-1\Bigr)\pi \, ; \hspace{0.5cm} r = r' =
\pm\infty \, , \label{Phiss1}
\end{eqnarray}
\begin{equation}
\pi\,{\bar{\Phi }}_{c,\eta\nu}(r,r') = -{\rm arc}\tan\Bigl({r-r'\over
\nu}\Bigr) \, , \hspace{0.5cm} 
\pi\,{\bar{\Phi }}_{s1,\eta\nu}(r,r') = \pi\,{\bar{\Phi }}_{c,s\nu'}(r,r') = 0 
\, , \hspace{0.5cm} \nu' >1 \, , 
\label{Phiccn-csn1}
\end{equation}
\begin{eqnarray}
\pi\,{\bar{\Phi }}_{s1,s\nu}(r,r') & = & {\rm arc}\tan\Bigl({r-r'\over \nu
-1}\Bigr) \, ; \hspace{0.5cm} r\neq \pm\infty \, ; \hspace{0.5cm} \nu
>1 \, , \nonumber \\
& = & \pm {\pi\over\sqrt{2}} \, ; \hspace{0.5cm} r= \pm\infty \, ; \hspace{0.5cm} \nu
>1 \, , \label{Phicn-ssn1}
\end{eqnarray}
\begin{equation}
\pi\,{\bar{\Phi }}_{\eta\nu,c}\left(r,r'\right) = {\rm arc}{\rm
tan}\Bigl({r-r'\over {\nu}}\Bigr) \, ; \hspace{0.5cm} \pi\,\bar{\Phi
}_{\eta\nu,\eta\nu'}\left(r,r'\right) = {1\over{2}}\,\Theta_{\nu,\nu'}(r-r') \, ;
\hspace{0.5cm} \pi\,\bar{\Phi }_{\eta\nu,s\nu'}\left(r,r'\right) = 0 \, , \hspace{0.5cm} \nu > 0 \, . 
\label{Phicnc-cncn-cnsn1}
\end{equation}

The $u\rightarrow 0$ two-pseudofermion phase shifts plotted in Figs. \ref{F1}-\ref{F6} have 
analytical expressions that we provide in the following. Such expressions refer to densities $n\in [0,1[$ and $m=0$.
The form of the general two-pseudofermion phase-shift expression, Eq. (\ref{Phi-barPhi}), 
reveals that the evaluation of such $u\rightarrow 0$ analytical expressions requires that of the 
ground-state rapidity functions $\Lambda^{0}_{\beta}(q)$ for $\beta =c,s1,\eta\nu$ where $\nu =1,...,\infty$.
(For a $m=0$ ground state there are no $s\nu\neq s1$ bands so that the corresponding
rapidity functions $\Lambda^{0}_{s\nu}(q)$ are undifined for $\nu >1$.) 

The $\beta =c,s1,\eta\nu$ ground-state rapidity functions $\Lambda^{0}_{\beta}(q)$ are
in Eq. (\ref{GS-R-functions}) defined in terms of their inverse functions. 
For a $m=0$ ground state with electronic density in the range $n\in [0,1[$ we find
in the $u\rightarrow 0$ limit from use of the latter equation the following closed-form expressions for the ground-state 
rapidity functions $k_c^0 (q)$, $\Lambda_{c}^0 (q)$, $\Lambda_{\eta\nu}^0 (q)$, and $\Lambda_{s1}^0 (q)$,
\begin{eqnarray}
k_c^0 (q) & = & {q\over 2} \, , \hspace{0.25cm} \vert q\vert \leq 2k_F \, , 
\nonumber \\ 
& = & {\rm sgn} (q)\,[\vert q\vert - k_F] \, ,
\hspace{0.25cm} 2k_F \leq \vert q\vert < \pi 
\nonumber \\ 
& = & {\rm sgn} (q)\,\pi \, , \hspace{0.25cm} \vert q\vert = \pi
\, , \hspace{0.25cm} u\rightarrow 0 \, , 
\label{k0lim}
\end{eqnarray}
\begin{eqnarray}
\Lambda^0_{c} (q) & = & \sin\Bigl({q\over 2}\Bigr) \, , \hspace{0.25cm} \vert q\vert \leq 2k_F \, , 
\nonumber \\ 
& = & {\rm sgn}(q)\,\sin\Bigl((\vert q\vert - k_F)\Bigr) \, , \hspace{0.25cm} 2k_F \leq \vert q\vert < \pi \, , 
\nonumber \\ 
& = & 0 \, , \hspace{0.25cm} \vert q\vert = \pi 
\, , \hspace{0.25cm} u\rightarrow 0 \, , 
\label{Gclim}
\end{eqnarray}
\begin{eqnarray}
\Lambda^0_{\eta\nu} (q) & = & {\rm sgn} (q)\,\sin\Bigl({(\vert q\vert + \pi n)\over 2}\Bigr)
\, , \hspace{0.25cm}  0<\vert q\vert< (\pi -2k_F) \, ,
\nonumber \\ 
& = & 0 \, , \hspace{0.25cm} q = 0 \, , 
\nonumber \\ 
& = & \pm\infty \, , \hspace{0.25cm} q = \pm (\pi -2k_F) 
\, , \hspace{0.25cm} u\rightarrow 0 \, , 
\label{Gcnlim}
\end{eqnarray}
and
\begin{eqnarray}
\Lambda^0_{s1} (q) & = & \sin (q) \, , \hspace{0.25cm} \vert q\vert < k_F \, ,
\nonumber \\ 
& = & \pm\infty \, , \hspace{0.25cm} q =\pm k_F
\, , \hspace{0.25cm} u\rightarrow 0 \, , 
\label{Gs1lim}
\end{eqnarray}
respectively.

Next, we use Eqs. (\ref{k0lim})-(\ref{Gs1lim}) in the integral equations, Eqs. (\ref{Phis1c-m})-(\ref{Phisnsn-m}), 
which define the $m=0$ two-pseudofermion phase shifts. By manipulation of these equations,
we find the following two-pseudofermion phase-shift expressions for the $u\rightarrow 0$ limit 
and at $u= 0$, if the corresponding expression is different,
\begin{equation}
\pi\,\Phi_{c,c}(q,q') = - {\rm sgn} \Bigl(\sin k^0_{c} (q)-\sin k^0_{c}
(q')\Bigr){\pi\over C_c (q)} + \delta_{\vert q\vert ,2k_F}\delta_{q,q'}[{\rm sgn}
(q)]\Bigl({3\over 2\sqrt{2}}-1\Bigr)\pi \, , 
\label{PhiccU0}
\end{equation}
\begin{equation}
\pi\,\Phi_{c,s1}(q,q') = - {\rm sgn} \Bigl(\sin k^0_{c} (q)-c_{s1}(q')\sin
(q')\Bigr){\pi\over C_c (q)} \, , 
\label{PhicsU0}
\end{equation}
\begin{eqnarray}
\pi\,\Phi_{s1,c}(q,q') & = & - {\rm sgn} \Bigl(\sin (q)-\sin k^0_{c} (q')\Bigr){\pi\over 2}
\, , \hspace{0.25cm} q \neq \pm k_F \nonumber \\
& = & -{{\rm sgn} (q)\pi\over 2\sqrt{2}} \, , \hspace{0.25cm} q = \pm k_F
\, , \hspace{0.25cm} u\rightarrow 0  \nonumber \\
& = & 0 \, ; \hspace{0.5cm} q = \pm k_F \, , \hspace{0.25cm} u= 0 \, , 
\label{Phisc-n1}
\end{eqnarray}
\begin{eqnarray}
\pi\,\Phi_{s1,s1}(q,q') & = & 0 \, ,
\hspace{0.25cm} q \neq \pm k_F \nonumber \\
& = & {{\rm sgn} (q)\pi\over 2\sqrt{2}}\Bigl[1 + \delta_{q,q'}2(1-\sqrt{2})\Bigr] \, ,
\hspace{0.25cm} q = \pm k_F
\, , \hspace{0.5cm} u\rightarrow 0  \nonumber \\
& = & 0 \, , \hspace{0.25cm} q = \pm k_F \, , \hspace{0.25cm} u= 0 \, , 
\label{PhissU0}
\end{eqnarray}
\begin{equation}
\pi\,\Phi_{c,\eta 1}(q,q') = - {\rm sgn} \Bigl(\sin k^0_{c} (q)-c_{\eta 1}(q')\sin k^0_{\eta 1}
(q')\Bigr){2\pi\over C_c (q)} \, , 
\label{PhictU0}
\end{equation}
and
\begin{equation}
\pi\,\Phi_{s1,\eta 1}(q,q') = - \theta (k_F -\vert q\vert )\,{\rm sgn} \Bigl(\sin
q-\sin k^0_{\eta 1} (q')\Bigr){\pi\over 2} \, . 
\label{PhistU0}
\end{equation}
Here the sign function is such that ${\rm sgn} (0)=0$ and the $\theta (x)$ function reads
$\theta (x) = 1$ for $x> 0$ and $\theta (x) = 0$ for $x\leq 0$, alike that of Eq. (\ref{DimSm}).
Hence, $\pi\,\Phi_{s1,\eta 1}(\pm k_F,q')=0$.
Moreover, in the above equations, $k^0_{c} (q)=\lim_{u\rightarrow 0}k_c^0 (q)$ where the
$u\rightarrow 0$ value of $k_c^0 (q)$ is given in Eq. (\ref{k0lim}),
\begin{eqnarray}
k^0_{\eta 1} (q) & = & {q\over 2} +{\rm sgn} (q)\,k_F \, , \hspace{0.25cm} 0<\vert q\vert
\leq [\pi -2k_F] \nonumber \\
& = & 0 \, , \hspace{0.25cm} q=0 \, , \label{k0eta1}
\end{eqnarray}
\begin{equation}
C_c (q) = 2\Bigl[\theta (2k_F -\vert q\vert ) + \sqrt{2}\,\delta_{\vert q\vert ,2k_F}
+2\,\theta (\pi -\vert q\vert )\theta (\vert q\vert -2k_F) + \delta_{\vert q\vert
,\pi}\Bigr] \, , \label{Ccsq}
\end{equation}
and
\begin{eqnarray}
c_{s1} (q) & = & 1 \, , \hspace{0.25cm} \vert q\vert <k_F \, , \hspace{0.25cm} c_{s1} (q) =
\infty \, , \hspace{0.25cm} q=\pm k_F \nonumber \\ c_{\eta 1} (q) & = & 1 \, , \hspace{0.25cm}
\vert q\vert <[\pi -2k_F] \, , \hspace{0.25cm} c_{\eta 1} (q) = \infty \, , \hspace{0.25cm}
q=\pm [\pi -2k_F] \, . \label{ccsq}
\end{eqnarray}

Finally, we address the issue of why in spite of the lack of ground-state $\eta\nu$ (and $s\nu\neq s1$) pseudofermion bands, 
the pseudofermion scattering theory can be generalized to the PSs of $S_{\eta}=0;n=1$ (and $S_s=0;m=0$) ground states. 
The point is that the ``in" asymptote one-pseudofermion 
scattering states do not contribute to the direct-product expression of the 
ground state but rather to that of the ``in" state, as defined in Section \ref{PSPS}. 
We start by considering excited energy 
eigenstates of an $S_{\eta}=0;n=1$ (and $S_s=0;m=0$) ground state with a single $\eta\nu$ 
pseudofermion, $N_{\eta\nu}=1$, and vanishing number values, $N_{\eta\nu'}=0$, for $\nu'>\nu$ branches 
(and a  single $s\nu\neq s1$ pseudofermion, $N_{s\nu}=1$, and vanishing number values, $N_{s\nu'}=0$, 
for $\nu'>\nu$ branches). For these excited states the use of Eqs. (\ref{N*}) and (\ref{N-h-an}) of Appendix \ref{pseudo-repre} leads to
$N_{a_{\eta\nu}}=1$ (and $N_{a_{s\nu}}=1$). The corresponding $\eta\nu$ (and $s\nu$) 
band reduces to the bare momentum $q=\pm q_{\eta\nu}=0$ (and $q=\pm q_{s\nu}=0$.) 
Consistent, one finds from Eq. (\ref{qcanGS}) of Appendix \ref{pseudo-repre} that at $n=1$ (and $m=0$),
$\pm q_{\eta\nu} = \pm [\pi -2k_F] =0$ (and $\pm q_{s\nu} =
\pm [k_{F\uparrow}-k_{F\downarrow}]=0$.)

For ground states with densities $n\in [0,1[$ and $m\in ]0,n]$,
except for non-physical higher-order $1/N_a$ contributions,
the scattering canonical-momentum shift $Q^{\Phi}_{\beta} (q)/N_a$, 
Eq. (\ref{qcan1j}), has the same value whether one uses the 
ground-state rapidity functions $\Lambda^{0}_{\beta}(q)$
and $\Lambda^{0}_{\beta'}(q')$ or the corresponding ``out"-state
(excited-energy-eigenstate) rapidity functions $\Lambda_{\beta}(q)$ and
$\Lambda_{\beta'}(q')$ in the
two-pseudofermion phase shift expression, Eq. (\ref{Phi-barPhi}). Indeed, these two alternative definitions of the
two-pseudofermion phase shifts lead to the same value of the functional
$Q^{\Phi}_{\beta} (q)/N_a$ up to contributions of $1/N_a$ order. 

Thus the general pseudofermion scattering theory also applies to PSs of $S_{\eta}=0;n=1$ (and $S_s=0;m=0$) 
ground states provided that the two-pseudofermion expression, 
Eq. (\ref{Phi-barPhi}), is replaced by
$\pi\,\Phi_{\beta,\beta'}(q,q') = \pi\,\bar{\Phi }_{\beta,\beta'}
(\Lambda_{\beta}(q)/u,\Lambda_{\beta'}(q')/u)$. Here the 
rapidity function $\Lambda_{\beta}(q)$ (and $\Lambda_{\beta'}(q')$)
is that of the excited state if $\beta =\alpha\nu\neq s1$ (and 
$\beta' =\alpha'\nu'\neq s1$) and of the ground state if $\beta =c,s1$ (and $\beta' =c,s1$).

In Section \ref{Pseudo-invariance} it is shown that for an excited state generated from
a $S_{\alpha}=0$ ground state by creation of a single $\alpha\nu\neq s1$ pseudofermion
that object is not a scatterer. The canonical-momentum band of the corresponding
``out" state has a single value at $\bar{q}=\pm q_{\alpha\nu}=0$. As discussed in that section,
this implies that $Q_{\alpha\nu} (0)/N_a=Q^{\Phi}_{\alpha\nu}(0)/N_a=0$.
The requirement that $Q^{\Phi}_{\alpha\nu} (0)/N_a=0$ 
imposes a specific form to the corresponding two-peudofermion phase shifts
$\pi\,\Phi_{\alpha\nu,\beta'}(0,q')$. Actually, since such a $\alpha\nu\neq s1$ pseudofermion is not a scatterer, the quantities
$\pi\,\Phi_{\alpha\nu,\beta'}(0,q')$ are not two-peudofermion phase shifts. 
They may be seen as mere effective virtual two-peudofermion phase shifts whose values are such that 
the overall scattering phase shift $Q^{\Phi}_{\alpha\nu} (0)/2$ vanishes. 

For simplicity, we consider three types of excited states of the $S_{\eta}=0;S_s=0;2S_c=N_a$ absolute ground state.
Those have no unbound $\eta$-spinons and no unbound spinons, pseudofermion occupancy in the $c$ and $s1$ bands, 
plus (a) one $\eta\nu$ pseudofermion and one $s\nu'\neq s1$ pseudofermion, 
(b) one $\eta\nu$ pseudofermion, and (c) one $s\nu'\neq s1$ pseudofermion. 
Consistent with Eq. (\ref{N-h-c-s1}) and as discussed in Section \ref{PSPS}, the excited energy eigenstates
(a) and (b) have $2\nu$ $c$ pseudofermion holes associated with the 
$\eta\nu$ pseudofermion and the excited states (a) and (c) have $2(\nu'-1)$ $s1$ pseudofermion holes 
associated with the $s\nu'\neq s1$ pseudofermion.

The $S_{\eta}=0;S_s=0;2S_c=N_a$ absolute ground state is described by full $c$ and $s1$ 
pseudofermion bands whose $c$ and $s1$ Fermi momenta read $q_{Fc}=2k_F =q_{c}=\pi$ and 
$q_{Fs1} =k_{F\downarrow}=q_{s1}=k_{F\uparrow} =k_F=\pi/2$, respectively. Thus, from
the use of Eq. (\ref{N-h-an}) of Appendix \ref{pseudo-repre} we find that $\delta N_{c}=-\delta N_{c}^h
=-2\nu$, $\delta N_{s1} = -(\nu + \nu')$, $\delta N^h_{s1} = 2(\nu' - 1)$ for the excited states (a), 
$\delta N_{c}=-\delta N_{c}^h=-2\nu$, $\delta N_{s1} = -\nu$, $\delta N^h_{s1} = 0$ 
for the excited states (b), and $\delta N_{c}=-\delta N_{c}^h =0$, 
$\delta N_{s1} = -\nu'$, $\delta N^h_{s1} = 2(\nu' - 1)$ for the excited states (c). 

It follows from Eqs. (\ref{Phisnc}), (\ref{Phiccn-csn1}), (\ref{Phicn-ssn1}), and 
(\ref{Phicnc-cncn-cnsn1}) that $\pi\,\Phi_{c,s\nu'}(q,0) = 
\pi\,\Phi_{\eta\nu,s\nu'}(q,0) = \pi\,\Phi_{s\nu,\eta\nu'}(q,0) = 0$ for the $s\nu'\neq s1$ and
$\eta\nu'$ scattering centers of the PS excited states of the $S_{\eta}=0;S_s=0;2S_c=N_a$ absolute ground state.
On the other hand, according to Eqs. (\ref{Phisnc}), (\ref{Phisns1}), and
(\ref{Phicnc-cncn-cnsn1}), within the simultaneous 
$m\rightarrow 0$ and $n\rightarrow 1$ limits the
two-pseudofermion phase shifts that contribute to the phase shifts $Q^{\Phi}_{\eta\nu} (0)/2$ and
$Q^{\Phi}_{s\nu'} (0)/2$ of such a PS excited states
simplify to $\pi\,\bar{\Phi }_{\eta\nu,s\nu'}\left(r,r'\right)
=\pi\,{\bar{\Phi }}_{s\nu',c}\left(r',r\right) = \pi\,\bar{\Phi
}_{s\nu',\eta\nu}\left(r',r\right) = 0$, $\pi\,{\bar{\Phi }}_{\eta\nu,c}\left(r,r'\right) =
{\rm arc}{\rm tan}\Bigl({r-r'\over {\nu}}\Bigr)$ and $\pi\,{\bar{\Phi
}}_{s\nu',s1}\left(r',r\right) = {\rm arc}{\rm tan}\Bigl({r'-r\over
{\nu' -1}}\Bigr)$ for $s\nu'\neq s1$. It follows then from the use of Eq. (\ref{qcan1j}) that for the excited energy
eigenstates (a)-(c) the equation $Q^{\Phi}_{\eta\nu} (0)/2=0$ and/or 
$Q^{\Phi}_{s\nu'} (0)/2=0$ leads to,
\begin{eqnarray}
{Q^{\Phi}_{\eta\nu} (0)\over 2} & = & \sum_{h=1}^{2\nu}\pi\,{\Phi }_{\eta\nu,c} (0,q_h)
= \sum_{h=1}^{2\nu}\,{\rm arc}{\rm tan}\left({\Lambda_{\eta\nu}(0)-
\Lambda^{0}_{c}(q_h)\over \nu\,u}\right) =
0 \, , \nonumber \\
{Q^{\Phi}_{s\nu'} (0)\over 2} & = & \sum_{h=1}^{2(\nu'-1)}\pi\,{\Phi }_{s\nu',s1} (0,{q'}_h)
= \sum_{h=1}^{2(\nu'-1)}\,{\rm arc}{\rm tan}\left({\Lambda_{s\nu'}(0)- \Lambda^{0}_{s1}({q'}_h)\over 
(\nu' -1)\,u}\right) = 0 \, ,
\hspace{0.25cm} s\nu'\neq s1 \, . 
\label{QIsnNV}
\end{eqnarray}
Here the first and second equations refer to the $c$ branch of both the states (a) and
(b) and to the $s1$ branch of both the states (a) and (c), respectively. In these
equations, the set of $2\nu= 2,4,6,...$ values $\{q_h\}$ and of $2(\nu'-1)=2,4,6,...$ values $\{{q'}_h\}$
correspond to the above-mentioned excited-energy-eigenstate $c$ pseudofermion holes and $s1$
pseudofermion holes, respectively. As discussed in Section \ref{Pseudo-invariance}, the
physics revealed by these results is that addition of the $2\nu= 2,4,6,...$ effective phase shifts of the $\alpha\nu\neq s1$ pseudofermion
that result from its collisions with its $2\nu= 2,4,6,...$ neutral-shadow scattering centers, as defined in Section \ref{PSPS}, exactly
cancel each other, so that $Q^{\Phi}_{\alpha\nu} (0)=0$. 

The absolute ground-state rapidity functions,
$\Lambda^{0}_{c}(q)=\sin k_0 (q)$ and $\Lambda^{0}_{s1}(q)$, are
defined in terms of their inverse functions in Eq. (\ref{Bessel}) of Appendix \ref{energies}.
The form of the equalities, Eq. (\ref{QIsnNV}), reveals that the corresponding solutions,
$\Lambda_{\eta\nu}(0)=\Lambda_{\eta\nu}(0,\{q_h\})$ and/or
$\Lambda_{s\nu'}(0)=\Lambda_{s\nu'}(0,\{{q'}_h\})$, are functions of the above sets of
bare-momentum values $\{q_h\}$ and $\{{q'}_h\}$, respectively. 

We have obtained here Eq. (\ref{QIsnNV}) from the $Q^{\Phi}_{\alpha\nu} (0)/2=0$ 
condition imposed by the invariance under the electron - rotated-electron 
unitary transformation of the $\alpha\nu\neq s1$ pseudofermion. 
For $\alpha\nu =\eta\nu$ and $\alpha\nu =s\nu'\neq s1$ that equation
is obeyed by the functions $\Lambda_{\eta\nu}(0)=
\Lambda_{\eta\nu}(0,\{q_h\})$ and $\Lambda_{s\nu'}(0)=\Lambda_{s\nu'}(0,\{{q'}_h\})$
for the rapidities $\Lambda_{\eta\nu}(0)$ and $\Lambda_{s\nu'}(0)$, respectively.
For the above excited states, the solution of the thermodynamic BA equations, 
Eqs. (\ref{Tapco1})-(\ref{kcn}) of Appendix \ref{pseudo-repre}, 
leads exactly to the same functions, which thus again obey Eq. (\ref{QIsnNV}). This
confirms that for such excited states the exact solution of the BA equations is 
equivalent to imposing the symmetry requirement $Q^{\Phi}_{\eta\nu} (0)/2=0$ 
and $Q^{\Phi}_{s\nu'} (0)/2=0$. The latter is associated with the invariance under the
electron - rotated-electron unitary transformation and corresponding non-scatterer character
of the $\eta\nu$  pseudofermion and/or $s\nu'$ pseudofermion under consideration, respectively.
Hence the BA solution accounts for that symmetry.

The above functions $\Lambda_{\eta\nu}(0)=\Lambda_{\eta\nu}(0,\{q_h\})$ and
$\Lambda_{s\nu'}(0)=\Lambda_{s\nu'}(0,\{{q'}_h\})$ are to be
used in the following expressions,
\begin{eqnarray}
\pi\,\Phi_{\eta\nu,c}(0,q') & = & \pi\,\bar{\Phi }_{\eta\nu,c}
\Bigl({\Lambda_{\eta\nu}(0,\{q_h\})\over u}, {\Lambda^0_{c}(q')\over u}\Bigr) =
{\rm arc}{\rm tan}\Bigl({\Lambda_{\eta\nu}(0,\{q_h\})-\Lambda^0_{c}(q')\over
{\nu\,u}}\Bigr) \, ; \nonumber \\
\pi\,\Phi_{s\nu',s1}(0,q') & = & \pi\,\bar{\Phi }_{s\nu',s1}
\Bigl({\Lambda_{s\nu'}(0,\{{q'}_h\})\over u}, {\Lambda^0_{s1}(q')\over u}\Bigr) =
{\rm arc}{\rm tan}\Bigl({\Lambda_{s\nu'}(0,\{{q'}_h\})-\Lambda^0_{s1}(q')\over
{(\nu'-1)\,u}}\Bigr)\, . 
\label{Phis}
\end{eqnarray}
They assure that $Q^{\Phi}_{\eta\nu} (0)/2=0$ and/or $Q^{\Phi}_{s\nu'} (0)/2=0$. 
The simplest case corresponds to $\nu =1$ and/or $\nu'=2$. In that case, the solution of Eq. (\ref{QIsnNV}) leads to
$\Lambda_{\eta 1}(0,q_1,q_2)=[\Lambda^{0}_{c}(q_1)+\Lambda^{0}_{c}(q_2)]/2$ and/or
$\Lambda_{s2}(0,{q'}_1,{q'}_2)=[\Lambda^{0}_{s1}({q'}_1)+\Lambda^{0}_{s1}({q'}_2)]/2$,
respectively. Hence, the symmetry requirement that the $\eta\nu$ pseudofermion (and $s\nu\neq s1$ pseudofermion)
considered here is not a scatterer implies that the corresponding rapidity function
$\Lambda_{\eta\nu}(0)=\Lambda_{\eta\nu}(0,\{q_h\})$ (and
$\Lambda_{s\nu'}(0)=\Lambda_{s\nu'}(0,\{{q'}_h\})$) does not in general vanish. Rather, it is
the unique solution of the first (and second) equation in Eq. (\ref{QIsnNV}). 

On the other hand, combination
of this result with the two-pseudofermion phase shift expressions, Eqs. (\ref{Phiccn-csn1}) and
(\ref{Phicn-ssn1}), reveals that the $c$ scatterer two-pseudofermion phase shift,
$\pi\,\Phi_{c,\eta\nu}(q,0)$, and the $s1$ scatterer two-pseudofermion phase shift,
$\pi\,\Phi_{s1,s\nu'}(q,0)$, are such that,
\begin{eqnarray}
\pi\,\Phi_{c,\eta\nu}(q,0) & = & \pi\,\bar{\Phi }_{c,\eta\nu}
\Bigl({\Lambda^0_{c}(q)\over u}, {\Lambda_{\eta\nu}(0,\{q_h\})\over u}\Bigr) =
-{\rm arc}{\rm tan}\Bigl({\Lambda^0_{c}(q)-\Lambda_{\eta\nu}(0,\{q_h\})]\over
{\nu\,u}}\Bigr) \, , 
\nonumber \\
\pi\,\Phi_{s1,s\nu'}(q,0) & = & \pi\,\bar{\Phi }_{s1,s\nu'}
\Bigl({\Lambda^0_{s1}(q)\over u}, {\Lambda_{s\nu'}(0,\{{q'}_h\})\over u}\Bigr) =
{\rm arc}{\rm tan}\Bigl({\Lambda^0_{s1}(q)-\Lambda_{s\nu'}(0,\{{q'}_h\})]\over
{(\nu'-1)\,u}}\Bigr) \, , \hspace{0.25cm} q\neq\pm k_F 
\nonumber \\
& = & \pm {\pi\over\sqrt{2}} \, , \hspace{0.25cm} q= \pm k_F \, . 
\label{Phis-2}
\end{eqnarray}

In addition to the $c$ or $s1$ scatterer bare-momentum $q$, the
two-pseudofermion phase shifts provided in Eq. (\ref{Phis-2}) 
are functions of the set of $2\nu=2,4,6,...$ bare-momentum values
$\{q_h\}$ or $2(\nu'-1)=2,4,6,...$ bare-momentum values $\{{q'}_h\}$ of the $2\nu=2,4,6,...$ 
neutral-shadow $c$ pseudofermion-hole scattering centers or $2(\nu'-1)=2,4,6,...$ neutral-shadow $s1$ pseudofermion-hole 
scattering centers, respectively. As discussed in Section \ref{PSPS}, the latter emerge under the ground-state - 
excited-state transition so that their virtual $\alpha$ elementary currents plus those of the 
$\alpha\nu\neq s1$ pseudofermion exactly cancel.

This confirms that as a result of the creation of one $\eta\nu$ pseudofermion (and 
one $s\nu'\neq s1$ pseudofermion), the $c$ (and $s1$) scatterers 
acquire the phase shift $\pi\,\Phi_{c,\eta\nu}(q,0)$ (and $\pi\,\Phi_{s1,s\nu'}(q,0)$).
Its value is fully controlled by the $2\nu=2,4,6,...$ (and $2(\nu'-1)=2,4,6,...$) bare-momentum values
of the corresponding $2\nu=2,4,6,...$ (and $2(\nu'-1)=2,4,6,...$) neutral-shadow $c$ (and $s1$) pseudofermion-hole scattering centers. 
Thus, through the $\{q_h\}$ (and $\{{q'}_h\}$) momentum dependence of the 
two-pseudofermion phase shift $\pi\,\Phi_{c,\eta\nu}(q,0)$  (and $\pi\,\Phi_{s1,s\nu'}(q,0)$), the $c$ 
(and $s1$) scatterers feel the $\eta\nu$ pseudofermion (and the $s\nu'\neq s1$ pseudofermion) created under the transition to the excited state
as $2\nu=2,4,6,...$ corresponding neutral-shadow $c$ effective scattering centers (and $2(\nu'-1)=2,4,6,...$ corresponding neutral-shadow $s1$ effective 
scattering centers). 

Similar results are obtained for PS excited states of $S_{\eta}=0;n=1$ and/or $S_s=0;m=0$ ground states 
with occupancy of a larger finite number of $\beta$ 
pseudofermions belonging to several $\beta=\alpha\nu\neq s1$ branches.
In that general case, the number of equations defining the set of rapidities $\{\Lambda_{\alpha\nu}\}$
is in general larger than above. Moreover, each of these equations is more involved than the 
two equations, Eq. (\ref{QIsnNV}). And the equations that result from the $Q^{\Phi}_{s\nu'} (0)/2=0$ and
$Q^{\Phi}_{\eta\nu} (0)/2=0$ symmetry requirements involve additional phase shifts
$\pi\,{\bar{\Phi }}_{s\nu' ,s\nu''}\left(r',r''\right)$ and $\pi\,\bar{\Phi}_{\eta\nu,\eta\nu'}\left(r,r'\right)$,
Eq. (\ref{Phisnsn-m}) of Appendix \ref{Ele2PsPhaShi} and Eq. (\ref{Phicnc-cncn-cnsn1}),
respectively. The corresponding $\beta = \alpha\nu\neq s1$ pseudofermions are also invariant under both 
the electron - rotated-electron unitary transformation and pseudoparticle - pseudofermion unitary transformation. 

Finally, the expression for $Q^{\Phi}_{\eta\nu} (0)/2$ (and $Q^{\Phi}_{s\nu'} (0)/2$),
Eq. (\ref{QIsnNV}), is also valid for the type of excited states of $S_{\eta}=0;n=1$ ground states
with spin density $m\neq 0$ (and of $S_{s}=0;m=0$ ground states
with electronic density $n\neq 1$) with the following occupancies:
no unbound $\eta$-spinons and no unbound spinons, 
finite pseudofermion occupancy in the $c$ and $s1$ bands, plus  
one $\eta\nu$ pseudofermion (and one $s\nu'\neq s1$ pseudofermion). 
Indeed, within the $n\rightarrow 1$ limit for $m\neq 0$
(and the $m\rightarrow 0$ limit for $n\neq 1$) the two-pseudofermion phase shifts that contribute to 
$Q^{\Phi}_{\eta\nu} (0)/2$ (and $Q^{\Phi}_{s\nu'} (0)/2$) also simplify
to $\pi\,\bar{\Phi }_{\eta\nu,s\nu'}\left(r,r'\right) = 0$ and $\pi\,{\bar{\Phi }}_{\eta\nu,c}\left(r,r'\right) =
{\rm arc}{\rm tan}\Bigl({r-r'\over {\nu}}\Bigr)$
(and to $\pi\,{\bar{\Phi }}_{s\nu',c}\left(r',r\right) = \pi\,\bar{\Phi
}_{s\nu',\eta\nu}\left(r',r\right) = 0$ and $\pi\,{\bar{\Phi
}}_{s\nu',s1}\left(r',r\right) = {\rm arc}{\rm tan}\Bigl({r'-r\over
{\nu' -1}}\Bigr)$ for $s\nu'\neq s1$.) Similarly, the expressions for
$\pi\,\Phi_{\eta\nu,c}(0,q')$ and $\pi\,\Phi_{c,\eta\nu}(q,0)$ 
(and $\pi\,\Phi_{s\nu',s1}(0,q')$ and $\pi\,\Phi_{s1,s\nu'}(q,0)$),
Eqs. (\ref{Phis}) and (\ref{Phis-2}), respectively, are valid
as well for the above type of excited states of $S_{\eta}=0;n=1$ ground states
with spin density $m\neq 0$ (and excited states of $S_{s}=0;m=0$ ground states
with electronic density $n\neq 1$).

\section{Useful energy scales}
\label{energies}

In this Appendix we introduce several energy scales that are extracted from the BA solution
and play an important role in the studies of this paper. 

For electronic densities $n\neq 1$, the values of the chemical potential $\mu = \mu (n)$ and magnetic-field energy 
$2\mu_B\,H = 2\mu_B\,H (m)$ are fully controlled by the energy dispersions of the 
$\beta =c,s1$ bands, Eq. (\ref{e-0-bands}) of Appendix \ref{pseudo-repre}. They read \cite{Carmelo91A},
\begin{equation}
\mu = {\rm sgn}\{(1-n)\}\,\left[\varepsilon_{c}^0 (q_{Fc}) + {1\over 2}\varepsilon_{s1}^0 (q_{F{s1}})\right] \, ;
\hspace{0.50cm}  
2\mu_B\,H = {\rm sgn}\{m\}\varepsilon_{s1}^0 (q_{F{s1}}) \, , \hspace{0.25cm} n \neq 1 \, .
\label{mu-muBH}
\end{equation}
The expressions given here are valid for the whole range of densities $n\neq 1$ and $m$.

The corresponding chemical-potential 
dependence on the hole concentration $x=(1-n)$ is such that,
\begin{eqnarray}
\mu (x) & = & -\mu (-x) \, ; \hspace{0.50cm} 
\mu (x) \in [\mu^0,\mu^1] \, , \hspace{0.25cm} x\in [-1,0]
\, ; \hspace{0.50cm}
\mu \in [-\mu^0,\mu^0] \, , \hspace{0.25cm} x = 0 \, ,
\nonumber \\
\mp \mu^1 & = & \lim_{x\rightarrow \pm 1} \mu (x)
\, ; \hspace{0.50cm}
\mp \mu^0 = \lim_{x\rightarrow 0^{\pm}} \mu (x)
\, , \hspace{0.25cm} \mu^1 > \mu^0 \, .
\label{chemi-range}
\end{eqnarray}
The energy scales $\mu^0$ and $\mu^1$ are defined in the following.

The finiteness for $u>0$ of the half-filling Mott-Hubbard gap, $2\mu^0$, implies that the chemical potential 
curve $\mu = \mu (x)$ has a well-defined discontinuity at $n=1$.
For $u\gg 1$ and $m=0$ the corresponding
energy scale $2\mu = 2\mu (x)$ is of the form,
\begin{eqnarray}
2\mu (x) & = & -{\rm sgn}\{x\}[U-4t\cos (\pi x)] \, , \hspace{0.25cm} x\in [-1,0]
\hspace{0.10cm}{\rm and} \hspace{0.10cm}  x\in [0,1] \, ,
\nonumber \\
& \in & [-(U-4t),(U-4t)] \, , \hspace{0.25cm} x = 0 \, .
\label{chem-ugg}
\end{eqnarray}

For $u>0$ and $n=1$ the Mott-Hubbard gap remains finite for all spin densities $m\in [-1,1]$. It is
an even function of $m$. For instance, at $m=0$ and $m=-1,1$ it is given by \cite{Lieb},
\begin{eqnarray}
2\mu^0 & = & U -4t + 8t\int_0^{\infty}d\omega {J_1 (\omega)\over\omega\,(1+e^{\omega\,2u})} 
= {16\,t^2\over U}\int_1^{\infty}d\omega {\sqrt{\omega^2-1}\over\sinh\left({2\pi t\omega\over U}\right)} 
\, , \hspace{0.25cm} m = 0 \, ,
\nonumber \\
& = & \sqrt{(4t)^2+U^2} - 4t \, , \hspace{0.25cm} m = -1,1 \, ,
\label{2mu0}
\end{eqnarray}
respectively, where $J_1 (\omega)$ is a Bessel function. For $u\ll 1$ and $u\gg 1$, this energy scale behaves as,
\begin{eqnarray}
2\mu^0 & \approx & {8\over \pi}\sqrt{t\,U}\,e^{-2\pi \left({t\over U}\right)} 
\, , \hspace{0.25cm} m = 0 
\, ; \hspace{0.50cm}
2\mu^0 \approx {U^2\over 8t} \, , \hspace{0.25cm} m = -1,1
\, , \hspace{0.25cm} u\ll 1 \, ,
\nonumber \\
2\mu^0 & \approx & [U - 4t] \, , \hspace{0.25cm} m \in [-1,1]
\, , \hspace{0.25cm} u\gg 1  \, .
\label{DMH}
\end{eqnarray}

The energy scale $2\mu^1$ associated with the minimum $-\mu^1$
and maximum $\mu^1$ chemical-potencial values is for all $m$ magnitudes given by,
\begin{equation}
2\mu^1 =  [U + 4t] \, .
\label{mu-1}
\end{equation}

On the other hand, the magnetic energy scale $2\mu_B\,H$, Eq. (\ref{mu-muBH}),
dependence on the spin density $m$ is such that,
\begin{eqnarray}
2\mu_B\,H (m) & = & -2\mu_B\,H (-m) \, ; \hspace{0.50cm} 
2\mu_B\,H (m) \in [0,2\mu_B\,H_c] \, , \hspace{0.25cm} m \in [-(1-\vert x\vert),0] \, ,
\nonumber \\
2\mu_B\,H (0 ) & = & 0 \, ; \hspace{0.50cm} 
\mp 2\mu_B\,H_c = \lim_{m\rightarrow \pm [1-\vert x\vert]} 2\mu_B\,H (m) 
 \, , \hspace{0.25cm} x = (1-n) \, .
\label{H-range}
\end{eqnarray}

A closed-form expression for the dependence on $U$, $t$, and density $n$ of the energy 
scale $2\mu_B\,H_c$ associated with the critical field $H_c$ can be derived from the 
general $2\mu_B\,H$ expression, Eq. (\ref{mu-muBH}) \cite{Carmelo91A}. 
Since $2\mu_B\,H_c$ is a even function of the hole concentration $x=(1-n)$, we expressed it in terms of it ,
\begin{eqnarray}
2\mu_B\,H_c & = & {1\over 2}\sqrt{(4t)^2+U^2}\left[1+{2\over\pi}\arccot\left({\sqrt{(4t)^2+U^2}\over U}
\tan (\pi \vert x\vert)\right)\right]
\nonumber \\
& - & U\,(1-\vert x\vert) + {4t\over\pi}\cos (\pi x) \arctan\left({4t\sin (\pi \vert x\vert)\over U}\right) 
\, , \hspace{0.25cm} x = (1-n) \, .
\label{mBHc}
\end{eqnarray}
At $u=0$ and for $u\gg 1$ its limiting behaviors are,
\begin{eqnarray}
2\mu_B\,H_c & = & 4t\sin^2 \left({\pi x\over 2}\right) \, , \hspace{0.25cm} u = 0 \, , \hspace{0.25cm} x = (1-n) \, , 
\nonumber \\
& = & {8\,(1-\vert x\vert)t^2\over U}\left[1+{\sin (2\pi \vert x\vert)\over 2\pi (1-\vert x\vert)}\right] \, , \hspace{0.25cm} u \gg 1 
\, , \hspace{0.25cm} x = (1-n) \, .
\label{mBHc-U0-Ul}
\end{eqnarray}
As a function of $n$, it has for instance the following values,  
\begin{eqnarray}
2\mu_B\,H_c & = & 0 \, , \hspace{0.25cm} n = 0,2 \, ,
\nonumber \\
& = & {1\over 2}\left[\sqrt{(4t)^2+U^2} - U\right] \, , \hspace{0.25cm} n = {1\over 2},{3\over 2}  \, ,
\nonumber \\
& = & \sqrt{(4t)^2+U^2} - U \, , \hspace{0.25cm} n = 1 \, .
\label{mBHc-n0-n1}
\end{eqnarray}

Other energy scales involved in the studies of this paper are the $c$ and $\alpha\nu$ pseudoparticle energy dispersions,
Eqs. (\ref{e-0-bands}) and (\ref{epsilon-q}) of Appendix \ref{pseudo-repre}. Both the momentum 
widths $2\pi (1-n)$ and $2\pi m$ of the $\eta\nu$ and $s\nu\neq s1$ momentum bands, 
respectively, and their energy-dispersion bandwidths, $[\varepsilon^0_{\alpha\nu} (q_{\alpha\nu})-\varepsilon^0_{\alpha\nu} (0)]$, 
where $\alpha\nu =\eta\nu$ and $\alpha\nu =s\nu\neq s1$, vanish in the $n\rightarrow 1\, ; m\rightarrow 0$ limit.
Consistent, for the $S_{\eta}=0;n=1;S_s=0;m=0;2S_c=N_a$ absolute ground state the 
corresponding $\alpha\nu\neq s1$ pseudoparticle energy dispersions do not exist. On the
other hand, the $c$ pseudoparticle energy dispersions $\varepsilon_{c}^0 (q)$ and $\varepsilon_{c} (q)$ and 
$s1$ pseudoparticle energy dispersions $\varepsilon_{s1}^0 (q)$ and $\varepsilon_{s1} (q)$ have closed-form 
expressions, which read \cite{Carmelo92},
\begin{eqnarray}
\varepsilon_{c}^0 (q) & = & -{U\over 2} -2t\cos k_c^0 (q) - 4t\int_0^{\infty}d\omega {\cos(\omega\,\sin k_c^0
(q))\over\omega\,(1+e^{\omega\,U/2t})}\,J_1 (\omega) \, , \hspace{0.25cm} q\in [-\pi,\pi] \, ,
\nonumber \\
\varepsilon_{c} (q) & = & \varepsilon_{c}^0 (q) + \mu^0 \, , \hspace{0.25cm} q\in [-\pi,\pi] \, , 
\nonumber \\
\varepsilon_{s1} (q) & = & \varepsilon^0_{s1} (q) = -2t\int_0^{\infty}d\omega {\cos(\omega\,\Lambda^{0}_{s1}(q))\over\omega\,\cosh
(\omega\,u)}\,J_1 (\omega) \, , \hspace{0.25cm} q\in [-\pi/2,\pi/2] \, .
\label{dispersions-n1}
\end{eqnarray}
The absolute ground state rapidity functions $k_c^0 (q)$ 
(such that $\Lambda^{0}_{c}(q) = \sin k_c^0 (q)$) and $\Lambda^{0}_{s1}(q)$ are defined in terms of 
their inverse functions as follows \cite{Carmelo92},
\begin{eqnarray}
q & = & k_c^0 (q) + 2\int_0^{\infty}d\omega {\sin(\omega\,\sin k_c^0
(q))\over\omega\,(1+e^{2\omega\,u})}\,J_0 (\omega) \, , \hspace{0.25cm} q\in [-\pi,\pi] \, ,
\nonumber \\
q & = & \int_0^{\infty}d\omega {\sin(\omega\,\Lambda^{0}_{s1}(q))\over\omega\,\cosh
(\omega\,u)}\,J_0 (\omega) \, , \hspace{0.25cm} q\in [-\pi/2,\pi/2] \, . 
\label{Bessel}
\end{eqnarray}

On the other hand, for $m\rightarrow 0$ and $n\in [0,1]$ all $s\nu\neq s1$ bands momentum and energy bandwidths 
vanish. Consistent, for a $S_s;m=0$ ground state the corresponding $s\nu\neq s1$ pseudoparticle energy dispersions,
Eqs. (\ref{e-0-bands}) and (\ref{epsilon-q}) of Appendix \ref{pseudo-repre}, do not exist. 
For densities $n\in [0,1]$ and $m=0$ the $c$ pseudoparticle, $s1$ pseudoparticle, and 
$\eta\nu$ pseudoparticle energy dispersions $\varepsilon_c (q)$, $\varepsilon_{s1} (q)$, and
$\varepsilon^0_{\eta\nu} (q)$, respectively, defined in such equations have the following
limiting behaviors for $u\rightarrow 0$ and $u\gg 1$,
\begin{eqnarray}
\varepsilon_c (q) & = & -4t\,\left[\cos\left({q\over 2}\right)-\cos\left({\pi n\over 2}\right)\right] 
\, , \hspace{0.25cm} \vert q\vert\leq 2k_F = \pi\,n \, , \hspace{0.25cm} u\rightarrow 0 \, ,  
\nonumber \\
& = & -2t\,\left[\cos\left(\vert q\vert - {\pi n\over 2}\right)-\cos \left({\pi n\over 2}\right)\right] 
\, , \hspace{0.25cm} 2k_F = \pi\,n \leq \vert q\vert \leq \pi \, , \hspace{0.5cm} u\rightarrow 0 \, , 
\nonumber \\ 
& = & -2t\,\left[\cos (q)-\cos (\pi n)\right] \nonumber \\
& - & {8\,n\,t^2\over U}\,\ln (2)\,\left[\,\sin^2 (q)-\sin^2 (\pi n) \right] 
\, , \hspace{0.25cm} \vert q\vert \leq \pi \, , \hspace{0.25cm} u\gg 1 \, ,
\label{limec}
\end{eqnarray}
\begin{eqnarray}
\varepsilon_{s1} (q) & = & \varepsilon_{s1}^0 (q) = -2t\,\left[\cos (q) -\cos \left({\pi n\over 2}\right)\right] 
\, , \hspace{0.25cm} \vert q\vert \leq k_F = \pi\,n/2 
\, , \hspace{0.25cm} u\rightarrow 0 \, ,
\nonumber \\ 
& = & -{2\pi\,n\,t^2\over U}\,\left[1 - {\sin (2\pi\,n)\over 2\pi\,n}\right]\,\cos \left({q\over n}\right) 
\, , \hspace{0.25cm} \vert q\vert \leq k_F = \pi\,n/2 \, , \hspace{0.2cm} u\gg 1 \, , 
\label{limes1}
\end{eqnarray}
\begin{eqnarray}
\varepsilon^0_{\eta\nu} (q) & = & 4t\,\cos\left({\vert q\vert + \pi n \over 2}\right) 
\, , \hspace{0.25cm}  \vert q\vert \leq (\pi -2k_F) = \pi\,(1-n) \, , \hspace{0.25cm} u\rightarrow 0 \, ,
\nonumber \\ 
& = & {8t^2 (1-n)\over \nu\,U}\,\left[1 - {\sin (2\pi (1-n))\over 2\pi\,(1-n)}\right]\,\cos^2 \left({q\over 2(1-n)}\right)
\, , \hspace{0.25cm} \vert q\vert \leq  (\pi -2k_F) = \pi\,(1-n) \, , \hspace{0.25cm} u\gg 1 \, .
\label{limeetan}
\end{eqnarray}

For the excited states of a fully polarized ground state, which for electronic densities $n\in [0,1]$ corresponds to $m\rightarrow n$, 
the energy dispersions $\varepsilon_{c}^0 (q)$ and $\varepsilon_{c} (q)$,
$\varepsilon_{s\nu}^0 (q)$ and $\varepsilon_{s\nu} (q)$, and $\varepsilon_{\eta\nu}^0 (q)$ and $\varepsilon_{\eta\nu} (q)$,
Eqs. (\ref{e-0-bands}) and (\ref{epsilon-q}) of Appendix \ref{pseudo-repre}, have closed-form expressions, 
\begin{eqnarray}
\varepsilon_{c}^0 (q) & = & -{U\over 2} -2t\cos q \, , \hspace{0.25cm} q\in [-\pi,\pi] \, ,
\nonumber  \\
\varepsilon_{c} (q) & = & -2t[\cos q - \cos (\pi n)]  \, , \hspace{0.25cm} q\in [-\pi,\pi] \, ,
\nonumber  \\
\varepsilon_{s\nu}^0 (q) & = &
- {2t\over \pi}\int_{-\pi n}^{\pi n}d k \sin k 
\arctan \left({[\sin k - \Lambda_{s\nu}^0 (q)]\over \nu u}\right)  \, , \hspace{0.25cm} q\in [-2k_F,2k_F] =  [-\pi n,\pi n]\, ,
\nonumber \\
\varepsilon_{s\nu} (q) & = & \varepsilon_{s\nu}^0 (q) + W_{s\nu} \, , \hspace{0.25cm} q\in [-2k_F,2k_F] = [-\pi n,\pi n] \, ,
\nonumber \\
\varepsilon_{\eta\nu}^0 (q) & = &
-\nu\,U + 4t{\rm Re}\left[\sqrt{1 - (\Lambda^0_{\eta\nu} (q_j) + i \nu\,u)^2}\right]
\nonumber \\
& - & {2t\over \pi}\int_{-\pi n}^{\pi n}d k \sin k 
\arctan \left({[\sin k - \Lambda_{\eta\nu}^0 (q)]\over \nu u}\right) \, , \hspace{0.25cm} q\in [-(\pi -2k_F),(\pi -2k_F)] = [-\pi (1-n),\pi (1-n)] \, ,
\nonumber \\
\varepsilon_{\eta\nu} (q) & = & \varepsilon_{\eta\nu}^0 (q) + W_{\eta\nu} \, , \hspace{0.25cm} q\in [-(\pi -2k_F),(\pi -2k_F)] = [-\pi (1-n),\pi (1-n)]  \, .
\label{e-0-bands-m-n}
\end{eqnarray}
In this limit, $k^0_c (q) = q$, whereas the rapidity functions
$\Lambda_{\alpha\nu}^0 (q)$ appearing in the above expressions are defined by their inversion functions,
\begin{eqnarray}
q & = & 2{\rm Re}[\arcsin (\Lambda^0_{\eta\nu} (q_j) + i \nu\,u)]
\nonumber \\
& + & {1\over \pi}\int_{-\pi n}^{\pi n}d k 
\arctan\left({[\sin k - \Lambda_{\eta\nu}^0 (q)]\over \nu u}\right) 
\, , \hspace{0.25cm} q\in [-(\pi -2k_F),(\pi -2k_F)] = [-\pi (1-n),\pi (1-n)]  \, ,
\nonumber \\
q & = & - {1\over \pi}\int_{-\pi n}^{\pi n}d k  
\arctan \left({[\sin k - \Lambda_{s\nu}^0 (q)]\over \nu u}\right) 
\, , \hspace{0.25cm} q\in [-2k_F,2k_F] = [-\pi n,\pi n] \, .
\label{rapidi-m-n}
\end{eqnarray}

Moreover, for the electronic density range $n\in [0,1]$ and spin density $m\rightarrow n$ 
the $s\nu$ and $\eta\nu$ energy dispersion bandwidths appearing in some of the expressions 
provided in Eq. (\ref{e-0-bands-m-n}) are given by,
\begin{eqnarray}
W_{s\nu} & = & {1\over 2}\sqrt{(4t)^2+(\nu U)^2}\left[1-{2\over\pi}\arccot\left({\sqrt{(4t)^2+(\nu U)^2}\over (\nu U)}
\tan (\pi n)\right)\right]
\nonumber \\
& - & \nu U\,n - {4t\over\pi}\cos (\pi n) \arctan\left({4t\sin (\pi n)\over \nu U}\right) 
\, , \hspace{0.25cm} n \in [0,1] \, , \hspace{0.25cm} m\rightarrow n \, ,
\label{Ws-nu}
\end{eqnarray}
and
\begin{eqnarray}
W_{\eta\nu} & = & {1\over 2}\sqrt{(4t)^2+(\nu U)^2}\left[1+{2\over\pi}\arccot\left({\sqrt{(4t)^2+(\nu U)^2}\over (\nu U)}
\tan (\pi n)\right)\right]
\nonumber \\
& - & \nu U\,(1-n) + {4t\over\pi}\cos (\pi n) \arctan\left({4t\sin (\pi n)\over \nu U}\right) 
\, , \hspace{0.25cm} n \in [0,1] \, , \hspace{0.25cm} m\rightarrow n \, ,
\label{Weta-nu}
\end{eqnarray}
respectively. The energy bandwidths $W_{s\nu}$ and $W_{\eta\nu}$ given here are an increasing and decreasing function of $n$,
respectively. For instance, for $n\rightarrow 0$, $n=1/2$, and $n=1$ they read,
\begin{eqnarray}
W_{s\nu} & = & 0 \, , \hspace{0.25cm} n\rightarrow 0 \, , \hspace{0.25cm} m\rightarrow n \, ,
\nonumber \\
& = & {1\over 2}\left[\sqrt{(4t)^2+(\nu U)^2} - U\right] \, , \hspace{0.25cm} n = {1\over 2} \, , \hspace{0.25cm} m\rightarrow n \, ,
\nonumber \\
& = & \sqrt{(4t)^2+(\nu U)^2} - \nu U \, , \hspace{0.25cm} n = 1 \, , \hspace{0.25cm} m\rightarrow n \, ,
\label{Ws-nu-0-1}
\end{eqnarray}
and
\begin{eqnarray}
W_{\eta\nu} & = &  \sqrt{(4t)^2+(\nu U)^2} - \nu U \, , \hspace{0.25cm} n\rightarrow 0 \, , \hspace{0.25cm} m\rightarrow n \, ,
\nonumber \\
& = & {1\over 2}\left[\sqrt{(4t)^2+(\nu U)^2} - U\right] \, , \hspace{0.25cm} n = {1\over 2} \, , \hspace{0.25cm} m\rightarrow n \, ,
\nonumber \\
& = & 0 \, , \hspace{0.25cm} n = 1 \, , \hspace{0.25cm} m\rightarrow n \, ,
\label{Weta-nu-0-1}
\end{eqnarray}
respectively. 

For spin density $m=0$ and electronic density $n=1$ all
energy bandwidths $W_{\alpha\nu}$ vanish except that of the
$s1$ pseudoparticle energy dispersion, which reads,
\begin{equation}
W_{s1} = 2t\int_0^{\infty}d\omega {J_1 (\omega)\over\omega\,\cosh
(\omega\,u)} \, , \hspace{0.25cm} n=1 \, , \hspace{0.25cm} m= 0 \, .
\label{Ws1-m0-n0-1}
\end{equation}

Finally, for $u\rightarrow 0$ and $u\gg 1$, electronic densities $n\in [0,1]$, and 
spin densities $m=0$ and $m\rightarrow n$ the $s\nu$ and $\eta\nu$ dispersion energy bandwidths 
have the following limiting behaviors,
\begin{eqnarray}
W_{s\nu} & = & \delta_{\nu,1}\left\{2t\left[1-\cos\left({\pi\over 2}n\right)\right]\right\} \, , \hspace{0.25cm} m = 0\, , \hspace{0.25cm} u \rightarrow 0 \, ,
\nonumber \\
& = & 4t\sin^2 \left({\pi n\over 2}\right) = 2\mu_B\,H_c \, , \hspace{0.25cm} m \rightarrow n \, , \hspace{0.25cm} u \rightarrow 0 \, ,
\nonumber \\
& = & \delta_{\nu,1}\,\left\{{2\pi n\,t^2\over U}\left[1-{\sin (2\pi n)\over 2\pi n}\right]\right\} =  \delta_{\nu,1}\,{\pi\over 4}\left(2\mu_B\,H_c\right)
\, , \hspace{0.25cm} m = 0 \, , \hspace{0.25cm} u \gg 1 \, , 
\nonumber \\
& = & {8n\,t^2\over \nu U}\left[1-{\sin (2\pi n)\over 2\pi n}\right] =  {1\over \nu}\,2\mu_B\,H_c
\, , \hspace{0.25cm} m \rightarrow n \, , \hspace{0.25cm} u \gg 1 \, , 
\label{Ws1-U0-Ul}
\end{eqnarray}
and
\begin{eqnarray}
W_{\eta\nu} & = & 4t\cos\left({\pi\over 2}n\right) = 2\vert\mu\vert \, , \hspace{0.25cm} m = 0\, , \hspace{0.25cm} u \rightarrow 0 \, ,
\nonumber \\
& = & 4t\left[1-\sin^2 \left({\pi n\over 2}\right)\right] = 2\vert\mu\vert 
\, , \hspace{0.25cm} m \rightarrow n \, , \hspace{0.25cm} u \rightarrow 0 \, ,
\nonumber \\
& = & {8(1-n)\,t^2\over \nu U}\left[1-{\sin (2\pi (1-n))\over 2\pi (1-n)}\right]
\, , \hspace{0.25cm} m = 0 \, , \hspace{0.25cm} u \gg 1 \, , 
\nonumber \\
& = & {8n\,t^2\over \nu U}{\sin (2\pi\,n)\over 2\pi\,n} 
\, , \hspace{0.25cm} m \rightarrow n \, , \hspace{0.25cm} u \gg 1 \, ,
\label{Wetanu-UU}
\end{eqnarray}
respectively.



\begin{references}
\bibitem{Gutzwiller}
	M. C. Gutzwiller, Phys. Rev. Lett. 10 (1963) 159.
\bibitem{Hubbard}	
	J. Hubbard, Proc. Roy. Soc. (London) A 276 (1963) 238.
\bibitem{Lieb}
        Elliott H. Lieb, F. Y. Wu, Phys. Rev. Lett. 20 (1968) 1445;\\
        Elliott H. Lieb, F. Y. Wu, Physica A 321 (2003) 1.      
\bibitem{Takahashi}
        Minoru Takahashi, Progr. Theor. Phys 47 (1972) 69.    
\bibitem{Woy}
	F. Woynarovich, J. Phys. C 15 (1982) 85;\\
	F. Woynarovich, J. Phys. C 15 (1982) 97.
\bibitem{Martins}
	P. B. Ramos, M. J. Martins, J. Phys. A 30 (1997) L195;\\
        M. J. Martins, P. B. Ramos, Nucl. Phys. B 522 (1998) 413.     
\bibitem{Schulz}
        H. J. Schulz, Phys. Rev. Lett. 64 (1990) 2831.	 
\bibitem{Voit}
	J. Voit, Rep. Prog. Phys. 58 (1995) 977.    
\bibitem{Lederer}
        K.-V. Pham, M. Gabay, P. Lederer, 
        Phys. Rev. B 61 (2000) 16 397.		
\bibitem{Woy-89}
	F. Woynarovich, J. Phys. A 22 (1989) 4243.
\bibitem{Frahm}        
        H. Frahm, V. E. Korepin, Phys. Rev. B 42 (1990) 10553. 
\bibitem{V}
        J. M. P. Carmelo, K. Penc, D. Bozi, Nucl. Phys. B 725 (2005) 421; \\
        J. M. P. Carmelo, K. Penc, D. Bozi, Nucl. Phys. B 737 (2006) 351, Erratum.
\bibitem{VI}
	J. M. P. Carmelo, K. Penc, Eur. Phys. J. B 51 (2006) 477.
\bibitem{LE}
        J. M. P. Carmelo, L. M. Martelo, K. Penc, Nucl. Phys. B 737 (2006) 237.
\bibitem{TTF}    
        J. M. P. Carmelo, D. Bozi, K. Penc,
        J. Phys.: Cond. Mat. 20 (2008) 415103;\\
        D. Bozi, J. M. P. Carmelo, K. Penc, P. D. Sacramento,
        J. Phys.: Cond. Mat. 20 (2008) 022205.                 
\bibitem{Penc-96}
        Karlo Penc, Karen Hallberg, Fr\'ed\'eric Mila, Hiroyuki Shiba,
        Phys. Rev. Lett. 77 (1996) 1390.
\bibitem{Penc-97}        
        Karlo Penc, Karen Hallberg, Fr\'ed\'eric Mila, Hiroyuki Shiba, Phys. Rev. B 55 (1997) 15 475.
\bibitem{Glazman}
	A. Imambekov, L. I. Glazman, Science 323 (2009) 228;\\
         A. Imambekov, L. I. Glazman, Phys. Rev. Lett. 100 (2008) 206805.
\bibitem{Glazman-09}         
         A. Imambekov, L. I. Glazman, Phys. Rev. Lett. 102 (2009) 126405. 
\bibitem{Glazman-10}  
	Thomas L. Schmidt, Adilet Imambekov, Leonid I. Glazman,
	Phys. Rev. Lett. 116 (2010) 116403;\\
	Thomas L. Schmidt, Adilet Imambekov, Leonid I. Glazman,
	Phys. Rev. B, 82 (2010) 245104;\\
	A. Shashi, L. I. Glazman, J.-S. Caux, A. Imambekov,
	Phys. Rev. B, 84 (2011) 045408.
\bibitem{Glazman-12}  
	Adilet Imambekov, Thomas L. Schmidt, Leonid I. Glazman,
	Rev. Mod. Phys. 84 (2012) 1253.
\bibitem{Affleck}
         R. G. Pereira, S. R. White, I. Affleck, Phys. Rev. Lett. 100 (2008) 027206;         
         S. R. White, I. Affleck, Phys. Rev. B 77 (2008) 134437.
\bibitem{Affleck-09}         
         R. G. Pereira, S. R. White, I. Affleck, Phys. Rev. B 79 (2009) 165113.	
\bibitem{Essler-10}        
        Fabian H. L. Essler, Phys. Rev. B 81 (2010) 205120.
\bibitem{DSF-n1}            
        R. G. Pereira, K. Penc, S. R. White, P. D. Sacramento J. M. P. Carmelo,
	Phys. Rev. B 85 (2012) 165132.  
\bibitem{I}
        J. M. P. Carmelo, J. M. Rom\'an, K. Penc, Nucl. Phys. B
        683 (2004) 387.
\bibitem{II}
        J. M. P. Carmelo, P. D. Sacramento, Phys. Rev. B 68
        (2003) 085104.               
\bibitem{paper-I}     
	J. M. P. Carmelo, to appear in Physics Reports.
\bibitem{PST-05}      	
	J. M. P. Carmelo, J. Phys.: Condens. Matter 17 (2005) 5517.	
\bibitem{bipartite}
	J. M. P. Carmelo, S. \"Ostlund, M. J. Sampaio,
	Ann. Phys. 325 (2010) 1550.	
\bibitem{spectral0}
        M. Sing, U. Schwingenschl\"ogl, R. Claessen, P.
        Blaha, J. M. P. Carmelo, L. M. Martelo, P. D. Sacramento, M.
        Dressel, C. S. Jacobsen, Phys. Rev. B 68 (2003) 125111.
\bibitem{spectral}
        J. M. P. Carmelo, K. Penc, L. M. Martelo, P. D. Sacramento,
        J. M. B. Lopes dos Santos, R. Claessen, M. Sing,
        U. Schwingenschl\"ogl, Europhys. Lett. 67 (2004) 233.
\bibitem{spectral-06}         
        J. M. P. Carmelo, K. Penc, P. D. Sacramento, M. Sing, R. Claessen, M. Sing,
        J. Phys.: Cond. Mat. 18 (2006) 5191.
\bibitem{polyace}
        Dionys Baeriswyl, Jos\'e Carmelo, Kazumi Maki, Synth. Met. 21 (1987) 271.
\bibitem{Zoller}
        D. Jaksch, P. Zoller, Ann. Phys. 315 (2005) 52.   
\bibitem{Natan}
        N. Andrei, {\it Integrable Models in Condensed Matter Physics,
        in Series on Modern Condensed Matter Physics -
        Vol. 6}, 458, World Scientific, Lecture Notes of ICTP Summer
        Course, Editors: S. Lundquist, G. Morandi, Yu Lu [cond-mat/9408101].	
\bibitem{S-0}
        F. H. L. Essler, V. E. Korepin, Phys. Rev. Lett. 72 (1994) 908.
\bibitem{S}
        F. H. L. Essler, V. E. Korepin,
        Nucl. Phys. B 426 (1994) 505.
\bibitem{CM-86}
        B. Sriram Shastry, Phys. Rev. Lett. 56 (1986) 1529;\\
        B. Sriram Shastry, Phys. Rev. Lett. 56 (1986) 2453. 
\bibitem{CM}
        B. Sriram Shastry, J. Stat. Phys. 50 (1988) 57. 
\bibitem{Prosen}
        J. M. P. Carmelo, T. Prosen, D. K. Campbell, Phys. Rev. B 63 (2001) 205114.
\bibitem{Completeness} 
 	F. H. L. Essler, V. E. Korepin, K. Schoutens, Phys. Rev. Lett. 67 (1991) 3848.
\bibitem{HL}
        O. J. Heilmann, E. H. Lieb, Ann. N. Y. Acad. Sci. 172
        (1971) 583.
\bibitem{Lieb89}
	E. H. Lieb, Phys. Rev. Lett. 62 (1989) 1201.
\bibitem{Yang}
        C. N. Yang, Phys. Rev. Lett. 63 (1989) 2144;\\
	C. N. Yang, S. C. Zhang, Mod. Phys. Lett. B 4 (1990) 759;\\    
	S. C. Zhang, Phys. Rev. Lett. 65 (1990) 120.
\bibitem{U(1)-NL} 
	Stellan \" Ostlund, Eugene Mele, 
	Phys. Rev. B 44 (1991) 12413. 
\bibitem{92-00}
	J. M. P. Carmelo, P. D. Sacramento, N. M. R. Peres, 
	Phys. Rev. Lett. 84 (2000) 4673;\\
	J. M. P. Carmelo, J. M. E. Guerra, J. M. B. Lopes
	dos Santos, A. H. Castro Neto, Phys. Rev. Lett. 83 (1999) 3892;\\ 
	J. M. P. Carmelo, P. Horsch, A. A. Ovchinnikov, D. K. Campbell, A. H. Castro Neto, 
	N. M. R. Peres, Phys. Rev. Lett.  81 (1998) 489;\\ 
	J. M. P. Carmelo, A. H. Castro Neto, Phys. Rev. Lett. 70 (1993) 1904;\\  
	J. M. P. Carmelo, A. H. Castro Neto, D. K. Campbell,
	Phys. Rev. Lett. 73 (1994) 926 and Erratum 74 (1995) 3089;\\
	J. M. P. Carmelo, P. Horsch, A. A. Ovchinnikov, Phys. Rev. Lett. 
	68 (1992) 871.	
\bibitem{currents}
	J. M. P. Carmelo, A. H. Castro Neto, D. K. Campbell, Phys. Rev. B 50 (1994) 3667;\\ 
        J. M. P. Carmelo, A. H. Castro Neto, D. K. Campbell, Phys. Rev. B 50 (1994) 3683;\\           
        J. M. P. Carmelo, P. Horsch, D. K. Campbell, A. H. Castro Neto,
        Phys. Rev. B (RC) 48 (1993) 4200;\\   
        N. M. R. Peres, J. M. P. Carmelo, 
        D. K. Campbell, A. W. Sandvik, Z. Phys. B 103 (1997) 217. 
\bibitem{Natan-79}
	N. Andrei, J. Lowenstein, Phys. Rev. Lett. 43 (1979) 1698.
\bibitem{S-Natan}
	N. Andrei, J. Lowenstein, Phys. Lett. 91B (1980) 401.
\bibitem{S-Hein}
        L. D. Faddeev, L. A. Takhtajan, Phys. Lett. 85A (1981) 375;\\
        L. D. Faddeev, L. A. Takhtadzhyan, J. Math. Sciences 24 (1984) 241.
\bibitem{Taylor} 
	John R. Taylor, {\em Scattering theory: the quantum theory of
        nonrelativistic collisions} (Robert E. Krieger Publishing Company, Malabar,
        Florida, 1987).
\bibitem{Carmelo91}
        J. Carmelo and A. A. Ovchinnikov, J. Phys.: Condens.
        Matter 3 (1991) 757.
\bibitem{Carmelo92}
        J. M. P. Carmelo, P. Horsch, A. A. Ovchinnikov,
        Phys. Rev. B 45 (1992) 7899.
\bibitem{Mahan} 
	Gerald D. Mahan, {\em Many-particle physics} (Kluwer Academic/Plenum
        Publishers, New York, 2000), Chapter 4; F. G. Fumi, Philos. Mag. 46 (1955) 1007.
\bibitem{Ohanian} 
	H. C. Ohanian, {\em Principles of quantum mechanics}
        (Prentice Hall, Englewood Cliffs, New Jersey, 1993). 
\bibitem{Anderson-67}
	P. W. Anderson,
	Phys. Rev. Lett. 18 (1967) 1049.	
\bibitem{Carmelo91A}
        J. M. P. Carmelo, P. Horsch, P. A. Bares, A. A. Ovchinnikov,
        Phys. Rev. B 44 (1991) 9967.      
\end{references}
\end{document}